\newif\ifrmp \rmptrue

\newif\ifhires \hiresfalse

\ifrmp
\documentclass[aps,rmp,twocolumn]{revtex4}
\usepackage{graphicx,amsmath,amssymb,float,bm,times}
\else
\documentclass[]{tADP2e}
\fi

\newcommand{\w}{\omega}
\newcommand{\qaf}{\vec{Q}_{\rm AF}}
\newcommand{\Tch}{T_{\rm ch}}
\newcommand{\Tsp}{T_{\rm sp}}
\newcommand{\Tc}{T_{\rm c}}

\newcommand{\Qch}{{\vec Q}_{\rm c}}
\newcommand{\Qchx}{{\vec Q}_{{\rm c}x}}
\newcommand{\Qchy}{{\vec Q}_{{\rm c}y}}
\newcommand{\Qsp}{{\vec Q}_{\rm s}}
\newcommand{\Qspx}{{\vec Q}_{{\rm s}x}}
\newcommand{\Qspy}{{\vec Q}_{{\rm s}y}}
\newcommand{\xich}{{\xi}_{\rm ch}}

\newcommand{\Qp}{{\vec Q}_{\rm p}}

\newcommand{\ybco}{YBa$_2$Cu$_3$O$_{6+\delta}$}

\newcommand{\lbco}{La$_{2-x}$Ba$_x$CuO$_4$}
\newcommand{\lbcoo}{La$_{15/8}$Ba$_{1/8}$CuO$_4$}
\newcommand{\lsco}{La$_{2-x}$Sr$_x$CuO$_4$}
\newcommand{\lesco}{La$_{2-x-y}$Eu$_y$Sr$_x$CuO$_4$}
\newcommand{\lnsco}{La$_{2-x-y}$Nd$_y$Sr$_x$CuO$_4$}
\newcommand{\bscco}{Bi$_2$Sr$_2$CaCu$_2$O$_{8+\delta}$}
\newcommand{\ccoc}{Ca$_{2-x}$Na$_x$CuO$_2$Cl$_2$}

\ifrmp
\begin{document}
\else
\begin{document}
\doi{10.1080/0001873YYxxxxxxxx}
\issn{1460-6976}
\issnp{0001-8732}  \jvol{00} \jnum{00} \jyear{2008} \jmonth{June}
\markboth{M. Vojta}{Lattice symmetry breaking}
\articletype{REVIEW}
\fi

\title{
Lattice symmetry breaking in cuprate superconductors:\\
Stripes, nematics, and superconductivity
}

\ifrmp
\author{Matthias Vojta}
\affiliation{
Institut f\"ur Theoretische Physik, Universit\"at zu K\"oln,
Z\"ulpicher Stra\ss e 77, 50937 K\"oln, Germany
}
\else
\author{Matthias Vojta$^{\ast}$\thanks{$^\ast$Email: vojta@thp.uni-koeln.de\vspace{6pt}}
\\\vspace{6pt}
{\em{Institut f\"ur Theoretische Physik, Universit\"at zu K\"oln,
Z\"ulpicher Stra\ss e 77, 50937 K\"oln, Germany}}
\\\vspace{6pt}
\received{May 15, 2009}
}
\maketitle
\fi


\begin{abstract}
This article will give an overview on both theoretical and experimental developments
concerning states with lattice symmetry breaking in the cuprate high-temperature
superconductors. Recent experiments have provided evidence for states with broken
rotation as well as translation symmetry, and will be discussed in terms of nematic and
stripe physics.
Of particular importance here are results obtained using the techniques of neutron and
x-ray scattering and scanning tunneling spectroscopy.
Ideas on the origin of lattice-symmetry-broken states will be reviewed,
and effective models accounting for various experimentally observed phenomena
will be summarized.
These include both weak-coupling and strong-coupling approaches,
with a discussion on their distinctions and connections.
The collected experimental data indicate that the tendency toward uni-directional
stripe-like ordering is common to underdoped cuprates, but becomes weaker
with increasing number of adjacent CuO$_2$ layers.
\ifrmp
\else
\bigskip
\bigskip
\fi
\end{abstract}

\ifrmp
\date{May 15, 2009}
\maketitle
\tableofcontents
\fi


\section{Introduction}

High-temperature superconductivity in the copper oxides constitutes one of the
most fascinating and challenging problems in modern condensed matter physics.
Since its discovery in 1986 by Bednorz and M\"uller \cite{bednorz}
it has influenced and inspired a vast variety of both experimental and theoretical developments,
ranging from the tremendous improvements in techniques like photoemission and scanning tunneling
microscopy over the development of theoretical tools for strongly correlated and low-dimensional
systems to the discovery of fundamentally new states of matter.

\subsection{Doped Mott insulators}

Superconductivity in the copper oxides arises from doping of half-filled Mott
insulators.\footnote{More precisely, in terms of a three-band model of
Cu $3d_{x^2-y^2}$ and O $2p_{x,y}$ orbitals, the CuO$_2$ planes at half-filling
are charge-transfer, rather than Mott, insulators.}
While a number of low-temperature properties of the superconducting state
are compatible with BCS-like pairing \cite{bcs} of $d$-wave symmetry,
the normal state is highly unconventional and appears to violate Fermi-liquid
properties (except, perhaps, at strong overdoping),
in contrast to most other superconductors where pairing derives from
a Fermi-liquid metallic state.

Understanding doped Mott insulators is at the heart of the high-$\Tc$ problem \cite{ssrmp,leermp},
however, here both our conceptual and methodological toolboxes are still rather limited.
Phenomenologically, the pseudogap regime in underdoped compounds is not understood,
as is the doping evolution of the normal-state Fermi surface and the fate of
local magnetic moments. Further puzzles are connected to
the linear-in-$T$ resistivity around optimal doping above $\Tc$ and
the asymmetry between electron- and hole-doped materials \cite{carlson_rev,norman_ap,timusk}.

It has become clear that doped Mott insulators are characterized by a plethora
of competing ordering tendencies, with commensurate antiferromagnetism and
superconductivity being the most prominent ones.
Others are incommensurate spin and charge density waves, orbital magnetism
with circulating currents, and exotic fractionalized states with topological
order \cite{ssrmp,leermp,lee_rev}.

Incommensurate uni-directional spin and charge density waves,
often dubbed ``stripes'' \cite{pnas,jan,antonio_rev,brom_rev,kiv_rmp,oles_rev},
play an interesting role: Those states were first predicted in
mean-field studies of Hubbard models \cite{za89,poil89,schulz89,machida89}
and later observed in \lbco\ and \lnsco\ \cite{jt95,jt96}, belonging to the
so-called 214 family of cuprate compounds.
Those stripe states break the discrete translation and rotation symmetries of the
square lattice underlying the CuO$_2$ planes.
A conceptually related state is an (Ising) ``nematic'' which only breaks the
lattice rotation symmetry and which may occur as intermediate state upon
melting of stripe order \cite{kiv_rmp,KFE98}.

Stripes and nematics, together with the associated phase transitions and fluctuation
phenomena, constitute the focus of this review article.
While stripe states were originally thought to be a very special
feature of the 214 compounds, this view has changed.
The last few years have seen exciting experimental progress:
Signatures of states with lattice symmetry breaking have been identified
in a number of other cuprates and in a number of different probes,
and the dependence of the ordering on external parameters has been mapped out in more
detail.
In parallel, stripe-like charge ordering has been identified and investigated in
other correlated oxides, most importantly nickelates \cite{cheong97,jt98} and
manganites \cite{mori98}.
While there are some important differences between those systems and cuprates,
it is likely that the phenomena are related.

Taken together, the observations suggest that stripe and nematic states
are an integral part of the physics of doped Mott insulators, qualifying them
an interesting subject of fundamental research on their own right.
However, their role for superconductivity in cuprates is unclear at present.
On the one hand, interesting theoretical ideas of stripe-driven or stripe-enhanced superconductivity
have been put forward. On the other hand, in the cuprates there seems to be an anticorrelation
between $\Tc$ and the strength of stripe ordering phenomena, and no (static) stripe
signatures have been identified in the cuprates with the highest $\Tc$.
I will return to this discussion towards the end of the article.

\begin{figure*}
\begin{center}
\includegraphics[width=2.9in]{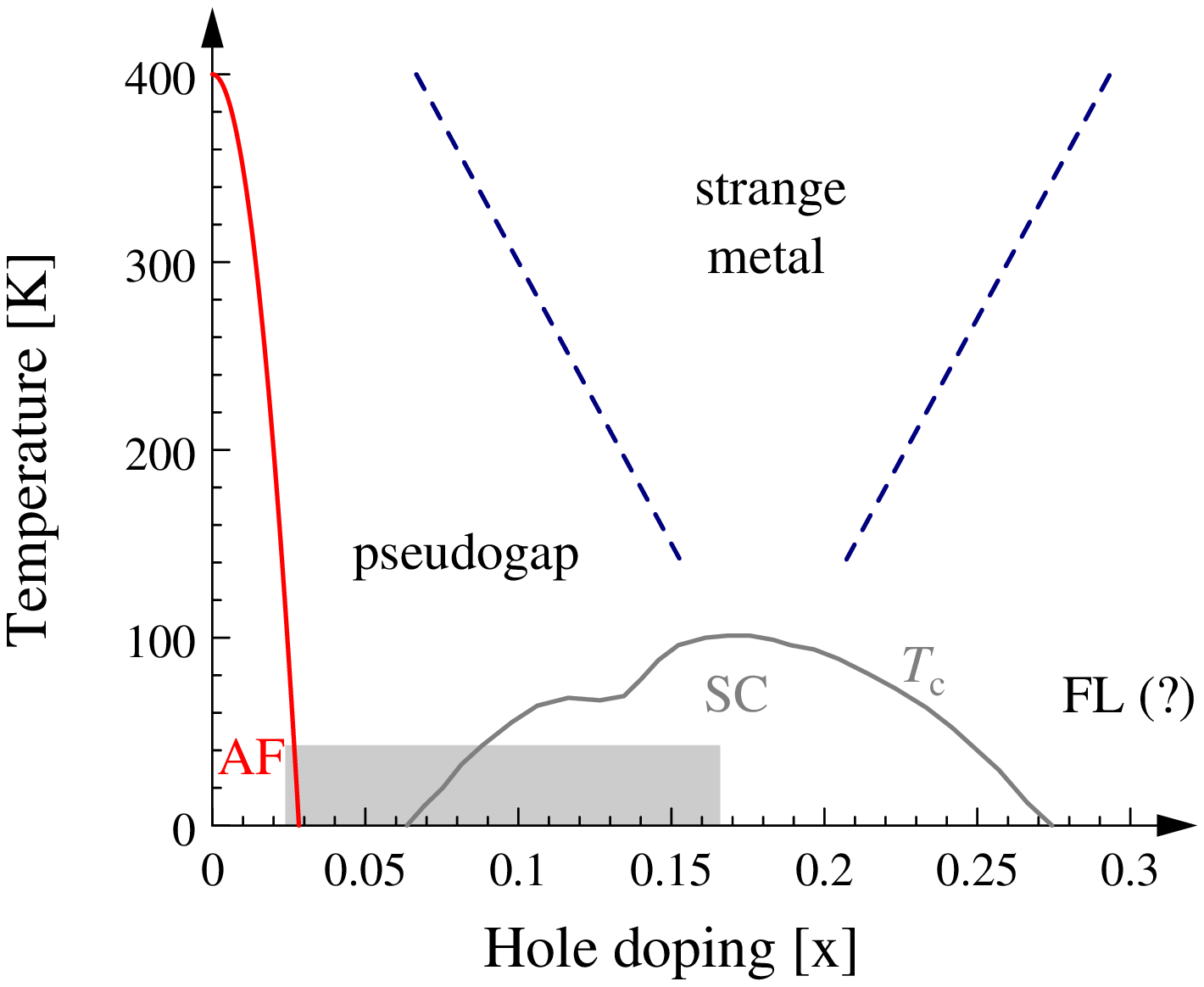}
\hspace*{30pt}
\includegraphics[width=1.6in]{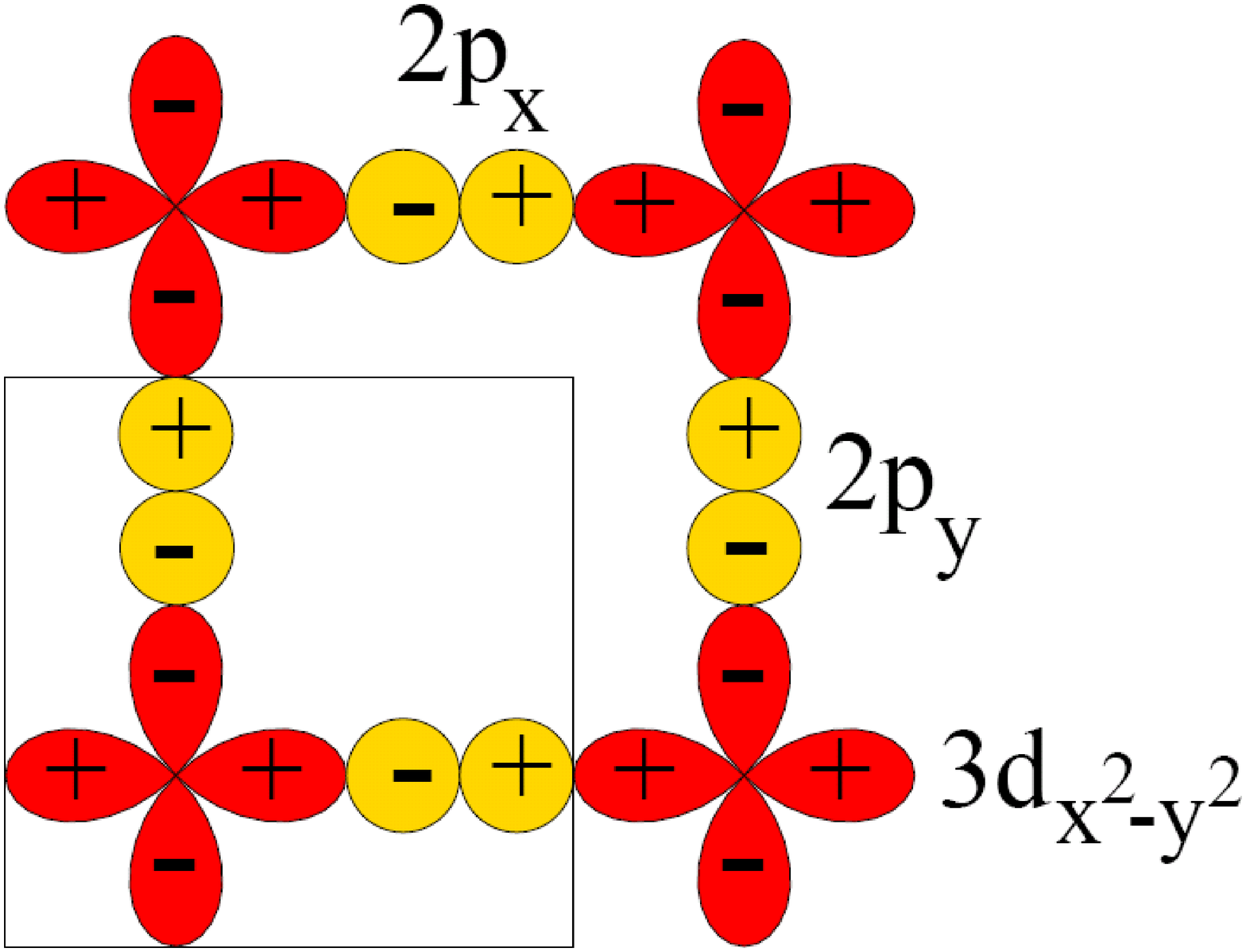}
\caption{
Left:
Schematic phase diagram of hole-doped cuprates in the temperature--doping plane.
Shown are the phases with N\'{e}el-like antiferromagnetic order (AF) and superconductivity
(SC) as well as the pseudogap, strange-metal, and overdoped Fermi-liquid (FL) regimes.
The phases with lattice symmetry breaking, discussed in this article, occur primarily in the
underdoped regime at low temperatures (shaded).
Various features of the phase diagram are not settled, such as shape of the crossover lines
near the superconducting dome, the character of the pseudogap line, and the character of
the overdoped FL-like regime.
Not shown are additional crossovers associated with the onset
pairing fluctuations and with spin glass behavior.
Right:
Electronic structure of the CuO$_2$ planes, with a unit cell
containing a Cu $3d$ and two O $2p$ orbitals.
}
\label{fig:pd1}
\end{center}
\end{figure*}


\subsection{Why another article on stripes?}

Both experimental and theoretical aspects of stripes have been
subject of other reviews in the past \cite{pnas,jan,antonio_rev,brom_rev,kiv_rmp,oles_rev}.
In addition, stripes have been discussed in more general
introductory or review articles on cuprate superconductors,
both from experimental \cite{timusk,zxshen_rmp,yoshida_rev,stm_rmp} and
theoretical \cite{carlson_rev,ssrmp,so5rmp,norman_ap,leermp,lee_rev,ogata_rop}
perspectives.
The most recent and comprehensive review closely related to the present subject
is the one by Kivelson {\em et al.} \cite{kiv_rmp}
on ``How to detect fluctuating stripes in the high-temperature superconductors'',
published in 2003.

The main motivation for a new article is that a number of key experiments were performed in the
last few years (i.e. after 2003), such that those and subsequent theoretical developments
are not covered in the above articles. I feel that the recent developments have
changed the perception of the community with regard to states with
broken lattice symmetries, making this old subject a very timely one.


\subsection{Focus}

The goal of this article is to review concepts, experimental results, and
theoretical ideas relevant to lattice-symmetry-broken states in doped cuprate
superconductors.
The emphasis is on uni-directional spin and charge density waves (stripes)
and on nematic states with broken rotational symmetry. These types of order
can co-exist with bulk superconductivity, and modulated superconducting states will naturally
appear in the discussion.

The insulating regime of zero and very small doping will be discussed only briefly,
as the ordered states occurring there can likely be described by quasi-classical concepts
and are of less relevance to cuprate superconductivity (with caveats to be noted later).

An exciting upcoming topic is that of circulating-current
states, of the $d$-density-wave type \cite{ddw} and of the type proposed
by Varma \cite{varma99,varma02}.
Here, experiments and interpretations are partially controversial,
such that I believe, at present, it is too early for an extensive review.
Therefore, the coverage of circulating-current states will be brief and restricted
to a few important experimental and theoretical results.

Even with these restrictions, the amount of published literature is vast. Therefore
I will focus most of the discussion to works which appeared over the last few years,
and the coverage of older results will be restricted to a few key references.


\subsection{Outline}

To set the stage, Sec.~\ref{sec:phases} will introduce the phenomenology of ordered phases
in the language of symmetries, order parameters, and Landau theory. An important aspect
is the coupling of order parameters to structural distortions and to quenched disorder
arising from dopant atoms.
Sec.~\ref{sec:exp} will give on overview on experimental data, which either establish the
existence of symmetry-breaking order or else can be used to extract characteristics of
such order.
Important classes of results, providing {\em direct} evidence for order, are those from neutron
and x-ray scattering and from scanning tunneling microscopy and spectroscopy.

The remainder of the article is devoted to theory: In Sec.~\ref{sec:mic}, microscopic
considerations in the language of Hubbard and $t$--$J$ models (or variants thereof) will be
reviewed, with the focus on identifying mechanisms and conditions for symmetry breaking.
Sec.~\ref{sec:eff} then describes concrete theoretical efforts in modeling experimental
characteristics of such symmetry-broken phases -- due to the difficulties with microscopic
or ab-initio approaches, most published work in this area is based on
effective phenomenological models which take some input from experiment.
Finally, Sec.~\ref{sec:impl} contains attempts to discuss the ``global picture'', i.e.,
the implications of both experiments and effective theories for our understanding of
cuprate high-temperature superconductivity.
Conclusions and an outlook will wrap up the article.

To shorten the notations, the following abbreviations for cuprate compounds will be
frequently used:
LSCO for \lsco, LBCO for \lbco,
LESCO for \lesco\ with $y\!=\!0.2$,
LNSCO for \lnsco\ with $y\!=\!0.4$,
YBCO for \ybco,
BSCCO or Bi-2212 for \bscco,
Bi-2201 for Bi$_2$Sr$_2$CuO$_{6+\delta}$,
and
CCOC for \ccoc.
The doping level will be specified as LSCO-0.12 or YBCO-6.6.
Moreover YBCO-124 denotes YBa$_2$Cu$_4$O$_{8}$.


\section{Ordered phases: Phenomenology}
\label{sec:phases}

This section will introduce the order parameters (in the Landau sense) of
various ordering phenomena which have been observed or discussed in the cuprates.
The orders break symmetries of the underlying microscopic models,
such as the translation symmetry, the $C_{4h}$ point group symmetry of the two-dimensional (2d)
CuO$_2$ square lattice,\footnote{
For most of the article, only ordering phenomena within the CuO$_2$ planes will be discussed.
Ordering perpendicular to the planes is often short-ranged, with exceptions to be noted
below.
}
and charge conservation.
The coupling between order parameters will be
discussed in terms of Landau theory, which will be used to derive global scenarios for
phase diagrams and sequences of phase transitions.

Importantly, the following discussion will only make reference to symmetries, but not to
{\em mechanisms} of ordering phenomena.
Consequently, terms like ``spin density wave'' or ``charge density wave'' will
be used (throughout the article) for states with spatially modulated spin or charge densities,
independent of whether the underlying physics is of weak-coupling or strong-coupling
nature.\footnote{
An introduction to weak-coupling and strong-coupling approaches of stripe formation is given
in Sec.~\ref{sec:mic}. A discussion of the question which of the two is better suited to
describe actual cuprate experiments appears in Sec.~\ref{sec:wkstr}.
}

An order-parameter field, $\phi(\vec{r},\tau)$, is often introduced such that,
in an ordered phase or close to the ordering transition,
it varies slowly in space and time, i.e., the typical length and time scales of
fluctuations are large compared to microscopic scales.
To this end, modulations of the relevant observable on microscopic scales
are absorbed in the definition of $\phi$, such that $\phi$
is associated with a finite lattice momentum.\footnote{
Occasionally, we will also refer to an order parameter description on the
lattice scale, where the order parameter simply represents a Hubbard-Stratonovich
field.}


Close to an ordering transition, the effective theory for $\phi(\vec{r},\tau)$
can often be written in the form of a $\phi^4$ or Landau-Ginzburg-Wilson (LGW)
model,\footnote{
Throughout this article, Einstein's summation convention will be employed, such that
indices occurring twice are summed over, e.g.,
$(\phi_{\alpha})^2 = \phi_\alpha \phi_\alpha \equiv \sum_\alpha \phi_\alpha \phi_\alpha$.
}
\ifrmp
\begin{eqnarray}
\label{phi4}
{\cal S} \!&=&\! \int\!d^d r d\tau \left[
\frac{c^2}{2} (\vec{\nabla} \phi_\alpha)^2 + \frac{r_0}{2} (\phi_{\alpha})^2 + \frac{u_0}{24}
(\phi_{\alpha}^2)^2 - h\cdot\phi
\right] \nonumber\\
&+&{\cal S}_{\rm dyn},
\end{eqnarray}
\else
\begin{eqnarray}
\label{phi4}
{\cal S} \!&=&\! \int\!d^d r d\tau \left[
\frac{c^2}{2} (\vec{\nabla} \phi_\alpha)^2 + \frac{r_0}{2} (\phi_{\alpha})^2 + \frac{u_0}{24}
(\phi_{\alpha}^2)^2 - h\cdot\phi
\right]
+{\cal S}_{\rm dyn},
\end{eqnarray}
\fi
here for a real $n$-component field $\phi_\alpha(\vec r,\tau)$, and
$h(\vec{r},\tau)$ is the field conjugate to the order parameter.
$r_0$ is the (bare) mass of the order parameter field $\phi$, used to tune the system through the
order--disorder transition, $c$ a velocity, and $u_0$ the quartic self-interaction.
${\cal S}_{\rm dyn}$ contains the $\phi$ dynamics,
which is often of the form $(\partial_\tau \phi)^2$;
exceptions arise from fermionic Landau damping or field-induced magnetic
precession dynamics.


\subsection{Charge and spin density waves}
\label{sec:ph_sdw}
\label{sec:ph_cdw}

A charge density wave (CDW) is a state where the charge density oscillates
around its average value as
\footnote{
Higher harmonics may always be present in the modulation pattern,
but depend on microscopic details.
}
\begin{equation}
\langle \rho ({\vec R}, \tau) \rangle = \rho_{\rm avg} + \mbox{Re} \left[e^{i \Qch
\cdot {\vec R}} \phi_c (\vec R, \tau)  \right] \,.
\label{chargemod}
\end{equation}
The scalar order parameter $\phi_c$ is associated with the finite momentum (or
charge-ordering wavevector) $\Qch$.\footnote{
The lattice spacing $a_0$ (of the Cu square lattice) is set to unity unless
otherwise noted.
}
In general, $\phi_c$ is complex,
with the phase describing a sliding degree of freedom of the density wave;
for the sites of a square lattice,
exceptions are simple modulations with $\Qchx,\Qchy=\pi$,
where $\phi_c$ is real.
For wavevectors $\Qch$ which are incommensurate with the lattice periodicity, all
values of the complex phase of $\phi_c$ are equivalent, whereas the phase will prefer
discrete values in the commensurate case -- the latter fact
reflects lattice pinning of the density wave. In particular, depending on the phase, the
resulting modulation pattern can preserve a reflection symmetry at bonds or sites of the
lattices, then dubbed bond-centered or site-centered density wave.
For stripe order in cuprates we will be mainly concerned with $\Qchx = (2\pi/N,0)$
and $\Qchy = (0,2\pi/N)$ where $N$ is the real-space periodicity, and $N=4$ appears
to be particularly stable experimentally.

The modulation \eqref{chargemod} describes a uni-directional density wave,
with a single wavevector $\Qch$, which also breaks the point group
symmetry.\footnote{
In low-symmetry crystals, a single-$\vec Q$ modulation may not break the point
group symmetry. This applies to stripes in planes with orthorhombic symmetry,
relevant for the LNSCO, LESCO, LBCO, and YBCO, see Sec.~\ref{sec:distort} below.}
Alternatively, multiple modulations with different wavevectors may occur simultaneously.
If a wavevector and its symmetry-equivalent partners occur with equal amplitude, then
lattice point group symmetry is unbroken. This is the case in a checkerboard state
with equal modulations along $\Qchx$ and $\Qchy$.

In this article, we shall use the term ``charge density wave'' in its most general sense,
referring to a periodic spatial modulation in observables which are invariant under spin
rotations and time reversal; apart from the charge density this includes the
local density of states and the magnetic exchange energy, the pairing amplitude, or the
electron kinetic energy, all defined on the links of the square lattice.
Note that even in a strongly ordered state, the modulation in the total charge density
may be unobservably small because longer-range Coulomb interactions tend to suppress
charge imbalance.
This applies e.g. to valence-bond-solid (VBS, or spin-Peierls) order in a Mott insulator.
In the described sense, a columnar VBS state \cite{ssrmp} is a bond-centered
CDW with $\Qch=(\pi,0)$.

A spin density wave (SDW) is specified by a vector order parameter $\phi_{s\alpha}({\vec r},\tau)$,
$\alpha=x,y,z$, and the spin density modulation is given by
\begin{equation}
\langle S_{\alpha} ({\vec R}, \tau) \rangle = \mbox{Re} \left[e^{i {\vec Q}_s
\cdot {\vec R}} \phi_{s\alpha} ({\vec R}, \tau) \right] \,.
\label{spinmod}
\end{equation}
Here $\Qsp\!=\!0$ describes a ferromagnet, while the parent antiferromagnet of the
cuprates has $\Qsp\!=\!(\pi,\pi)$. The field $\phi_{s\alpha}$ transforms as angular
momentum under $O(3)$ spin rotations and is odd under time reversal.

Both collinear and non-collinear SDWs can be described by Eq. \eqref{spinmod}:
\begin{eqnarray}
\mbox{collinear:} &~& \phi_{s\alpha} = e^{i \Theta} n_\alpha ~
\mbox{ with } n_\alpha \mbox{ real}, \\
\mbox{spiral:} &~& \phi_{s\alpha} = n_{1\alpha} + i n_{2\alpha} ~
\mbox{ with } n_{1,2\alpha} \mbox{ real,}\, n_{1\alpha} n_{2\alpha} = 0.
\nonumber
\end{eqnarray}
In cuprates, both spiral and collinear order have been discussed.
In the context of stripe order, our focus will be on collinear order with
$\Qspx = 2\pi(0.5\pm 1/M,0.5)$ and $\Qspy = 2\pi(0.5,0.5\pm 1/M)$,
with $M=2N$ such that $2\Qsp = \Qch$ modulo reciprocal lattice vectors.

Spin anisotropies play an important role, as they can freeze out low-energy fluctuations
in some directions. In the cuprates, a combination of spin-orbit and crystal-field
effects leads to antiferromagnetic moments lying in the a-b plane, and the
Dzyaloshinskii-Moriya (DM) interaction results in a small spin canting out of the
plane in La$_2$CuO$_4$ \cite{thio88}.


\begin{figure*}
\begin{center}
\includegraphics[width=5in]{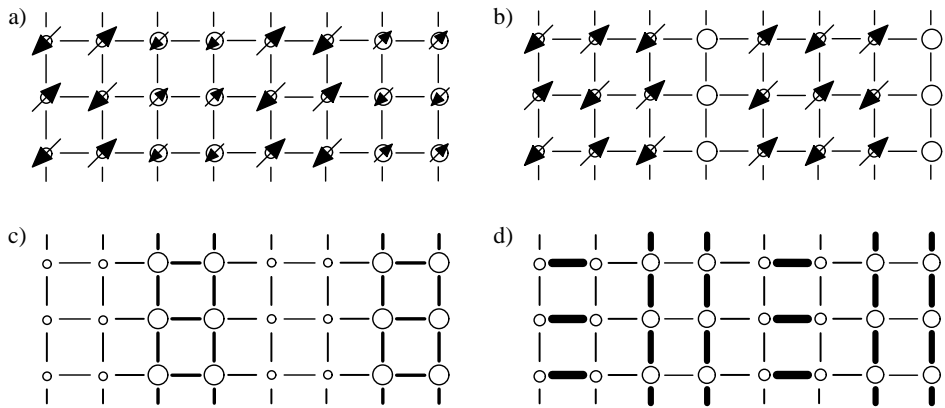}
\caption{
Schematic real-space structure of stripe states with period 4 in the charge sector.
a) Bond-centered and b) site-centered stripes with period-8 magnetic order,
showing the on-site hole densities and magnetizations.
c), d) Bond-centered charge-only stripes, now showing the modulations
in both the on-site hole density and the bond strength (i.e. kinetic energy).
Depending on the form factor $F_c$ in Eq.~\eqref{formfactor},
the modulation in the spin-singlet sector can be
primarily of c) $s$-wave or d) $d$-wave type -- the latter structure
is dominated by {\em bond} modulations.
}
\label{fig:stripe1}
\end{center}
\end{figure*}

Charge and spin ordering can be expressed via expectation values of particle--hole
bilinears. Here, charge and spin order simply correspond to ordering in the spin singlet and
triplet channels of the particle--hole pair.
Assuming a one-band description and neglecting the time dependence, we have for charge order:
\ifrmp
\begin{eqnarray}
\langle c_\sigma^\dagger (\vec r) c_\sigma (\vec r') \rangle &=&
F_{\rm avg}(\vec r-\vec r')
 \nonumber\\&+&
F_c(\vec r-\vec r')
\mbox{Re}\left[e^{i \Qch \cdot {\vec R}}  \phi_c(\vec R)\right]
\label{formfactor}
\end{eqnarray}
\else
\begin{eqnarray}
\langle c_\sigma^\dagger (\vec r) c_\sigma (\vec r') \rangle &=&
F_{\rm avg}(\vec r-\vec r')
 +
F_c(\vec r-\vec r')
\mbox{Re}\left[e^{i \Qch \cdot {\vec R}}  \phi_c(\vec R)\right]
\label{formfactor}
\end{eqnarray}
\fi
where $c_\sigma^\dagger(\vec r)$ is an electron creation operator at coordinate $\vec r$
and spin $\sigma$, and $\vec R = (\vec r+\vec r')/2$.
$F_{\rm avg}$ and $F_c$ are short-ranged functions,
characterizing the unmodulated state and the modulation, respectively.
After Fourier transformation, $\vec k$ in $F_c(\vec k)$ refers to the internal momentum of
the particle--hole pairs (in contrast to their center-of-mass momentum $\Qch$),
it determines the structure of the modulation within a unit cell.
The simplest possible order corresponds to
$F_c(\vec r\!-\!\vec r') \sim \delta_{\vec r\vec r'}$, i.e., $F_c(\vec k)$ is
$\vec k$-independent, and we will loosely refer to this as ``$s$-wave'' order.
However, proposals of order with non-trivial $\vec k$ dependence have also been made.
This applies to the so-called ``$d$-density wave order'' \cite{ddw}
and $d$-wave checkerboard order \cite{dwave_checker1,dwave_checker2},
where the modulations are on lattice bonds, while the modulation on
sites vanishes identically.
(Note that the $d$-density wave state breaks time reversal whereas the $d$-wave checkerboard does not.)
$d$-wave order parameters have the interesting feature that
they couple primarily to antinodal quasiparticles, leaving nodal quasiparticles along the
Brillouin zone diagonal unaffected.
In general, the symmetry classification of $F_c$ has to be done according to the
lattice point group of the ordered state: For a square lattice and
uni-directional CDW order with $\vec Q$ along the (1,0) or (0,1) direction,
$s$-wave and $d_{x^2-y^2}$-wave representations always mix,
because the ordering wavevector $\Qch$ breaks the symmetry from $C_4$ to $C_2$.
Nevertheless, stripes may be dominated by either the $s$-wave or the $d$-wave component
of $F_c$, the latter leading to valence-bond stripes \cite{mvor08}, with little
modulation on the square-lattice sites and  little coupling to nodal quasiparticles.

Fig.~\ref{fig:stripe1} schematically shows the real-space structure of
stripes with period 4 in the charge sector, illustrating bond-centered and site-centered
stripes. The STM results of Kohsaka {\em et al.} \cite{kohsaka07} are most consistent with
bond-dominated stripe order as in Fig.~\ref{fig:stripe1}d.

As noted above, the terms ``charge density wave'' and ``spin density wave''
will be used without reference to the underlying {\em cause} of
the modulation, hence ``stripe'' is equivalent to ``uni-directional charge density wave''.
CDW and SDW states can be insulating (like Wigner crystals), or metallic, or superconducting.


\subsection{Superconducting pairing}

For singlet superconducting pairing, the order parameter is a
charge-2 scalar field $\psi(\vec{R},\tau)$, where $\vec{R}$ and $\tau$ are the
center-of-mass coordinates of the Cooper pair.
Most superconducting states have a zero-momentum condensate, i.e., an ordering
wavevector $\Qp\!=\!0$.
However, modulated pairing is possible as well, and we may write in analogy to
Eq.~\eqref{formfactor}:
\ifrmp
\begin{eqnarray}
\langle c_\uparrow^\dagger (\vec r) c_\downarrow^\dagger (\vec r') \rangle &=&
F_p(\vec r-\vec r')
\left[e^{i \Qp \cdot {\vec R}} \psi(\vec R) + e^{-i \Qp \cdot {\vec R}} \bar{\psi}(\vec R)\right] .
\nonumber\\
\label{pdw}
\end{eqnarray}
\else
\begin{eqnarray}
\langle c_\uparrow^\dagger (\vec r) c_\downarrow^\dagger (\vec r') \rangle &=&
F_p(\vec r-\vec r')
\left[e^{i \Qp \cdot {\vec R}} \psi(\vec R) + e^{-i \Qp \cdot {\vec R}} \bar{\psi}(\vec R)\right] .
\label{pdw}
\end{eqnarray}
\fi
Because the superconducting order parameter is complex, two fields $\psi$ and $\bar{\psi}$
are required to implement the sliding degree of freedom of the ``pair density wave''.
The form factor $F_p$ determines the internal structure of a Cooper pair:
For homogeneous pairing, $\Qp\!=\!0$, the angular dependence of $F_p(\vec r)$
leads to the usual classification into $s$-wave, $d_{x^2-y^2}$-wave etc.
In the cuprates, there is good evidence for the pairing symmetry being $d_{x^2-y^2}$
\cite{uni_dwave}.

Finite-momentum pairing has first been proposed by Fulde and Ferrell \cite{FF} and Larkin
and Ovchinnikov \cite{LO}, as a state realized for large Zeeman splitting of the Fermi
surfaces in an external field.
More than 40 years later, evidence for such a FFLO state has indeed been found in the
organic superconductor $\kappa$-(BEDT-TTF)$_2$Cu(NCS)$_2$ \cite{wosnitza07}.
In cuprates, finite-momentum pairing in zero field has been proposed in the context of stripe
phases \cite{himeda02,berg07}, see Sec.~\ref{sec:antiphase}.


\subsection{Nematics}
\label{sec:ph_nematic}

States which spontaneously break the (discrete or continuous) real-space rotation
symmetry of the underlying Hamiltonian are often called ``nematic'' states.
The discussion here will be restricted to homogeneous ($\vec Q\!=\!0$), time-reversal invariant,
and spin-symmetric nematic order.\footnote{
Modulated nematic order is equivalent to density wave order.
Spin-antisymmetric nematic order and ``spin nematic'' order,
corresponding to spontaneously broken spin rotation symmetry,
will not be of interest in this article.}
Moreover, the focus will be on electronically driven symmetry breaking: While lattice
distortions invariably follow the broken symmetry, transitions involving simple structural
distortions only shall not be considered ``nematic''.

The term ``nematic'' originates in the physics of liquid crystals where it refers to
a phase with orientational, but without translational, order of molecules.
An electronic nematic phase can arise as a Pomeranchuk instability of a Fermi
liquid \cite{pomer}, in which case the ordered phase possesses a full Fermi surface,
but can also be driven by strong correlations and even occur in an insulator, i.e., in
the absence of a Fermi surface.
In particular, a nematic phase may occur upon melting of stripe order,
namely when the rotation and the translation symmetry are broken at separate transitions.
Then, a nematic is the intermediate phase between the stripe and disordered
phases \cite{KFE98}.\footnote
{
In the physics of liquid crystals, phases are termed ``liquid'', ``nematic'', ``smectic'', and
``crystalline'', according to their broken spatial symmetries. Here, a smectic breaks
translation symmetry in one direction, whereas a crystal breaks it in all directions.
Some workers have applied this terminology to cuprates, identifying a conducting stripe
with a ``smectic'' and an insulating Wigner crystal with a ``crystal''.
To avoid terminology problems with e.g. multiple-$\vec Q$ modulated states, this article will employ
the terms ``CDW'' and ``SDW'' instead, as described above.
}

In the case of a continuous rotation symmetry, the order can be characterized by an
angular momentum $l\!>\!0$;
e.g. for $l\!=\!2$, the order parameter is a director, specifying an axis in real space.
In contrast, on the square lattice with $C_4$ rotation symmetry, $l\!=\!2$ (or $d$-wave) nematic order
corresponds to breaking the rotation symmetry down to $C_2$.
One can distinguish $d_{xy}$ and $d_{x^2-y^2}$ nematic order, here we shall focus on the
latter.
The order parameter is of Ising type; globally, it may be defined via the electronic momentum
distribution according to
\begin{equation}
\phi_n = \sum_{\vec k} d_{\vec k} \langle c_{\vec k\sigma}^\dagger c_{\vec k\sigma}\rangle,~
d_{\vec k} = \cos k_x - \cos k_y.
\end{equation}
A local order parameter can be defined from any spin-singlet {\em bond} observable
which is even under time reversal,
like the bond kinetic energy,
\begin{equation}
\phi_n(\vec r) =
\langle c_\sigma^\dagger(\vec r) c_\sigma(\vec r+x) - c_\sigma^\dagger(\vec r) c_\sigma(\vec
r+y)\rangle,
\end{equation}
or the bond magnetic energy or pairing strength.

The so-defined nematic order parameter $\phi_n$ measures rotation symmetry
breaking. As such, it is finite in a uni-directional density wave state as well:
The density wave will induce a $\phi_n$ proportional to $|\phi_{cx}|^2-|\phi_{cy}|^2$ or
$|\phi_{s\alpha x}|^2-|\phi_{s\alpha y}|^2$, see Sec.~\ref{sec:coupling}.
Strictly speaking, the density wave is not a nematic state, as translation symmetry
is broken as well.
However, here we shall use the term ``nematic order'' synonymously to lattice rotation symmetry
being spontaneously broken.\footnote{
Experimentally establishing the existence of a nematic instead of a smectic
(in the liquid-crystal terminology) would require disproving the existence of translation
symmetry breaking, which is rather difficult.
}

In recent years, nematic phases have been proposed as explanations of some puzzling
experimental observations, such as the enigmatic ordering transition in URu$_2$Si$_2$
\cite{VZ05} and the low-temperature phase near the metamagnetic transition of
Sr$_3$Ru$_2$O$_7$ \cite{meta_pom}.


\subsection{Loop-current order}

While SDW, CDW, and nematic order have been experimentally
identified at least in some cuprates with reasonable confidence,
other symmetry-breaking order parameters (apart from superconductivity)
have been suggested but are more controversial.
Most prominent are various forms of loop-current order, proposed as origin for the
pseudogap behavior of underdoped cuprates. Charge current loops break the time reversal
symmetry and induce orbital moments. In particular, orbital antiferromagnets are
characterized by current loops with directions alternating from plaquette to plaquette on
the lattice, such that the total moment is zero and hence there is no a global edge
current.

Varma \cite{varma99,varma02} has proposed different patterns of current loops within a unit cell of the
CuO$_2$ plane, such that the order exists at wavevector $\vec Q\!=\!0$, but has vanishing
total moment.
The description of these states in general requires a three-band model of the CuO$_2$ plane,
with Cu and O orbitals, however, the symmetry of the state $\Theta_{\rm II}$ of
Ref.~\cite{varma02} can also be represented in a one-band model (case G in Ref.~\cite{vzs00b}).
The magnetic moments are expected to point in a direction perpendicular to the
CuO$_2$ layers, but in principle loop-current order could also involve oxygen orbitals
outside layers, resulting in canted moment directions.

A distinct type of state was proposed by Chakravarty {\em et al.} \cite{ddw} as
resolution of the pseudogap puzzle, namely an alternating
current pattern with $\vec Q=(\pi,\pi)$ with $d$-wave form factor \cite{ddw1,ddw2},
dubbed $d$-density wave.
In fact, symmetrywise this state is a generalization to finite doping of the so-called
staggered flux phase which appeared in early mean-field studies of Hubbard-Heisenberg
models \cite{affmar}.

Recent neutron scattering experiments reported magnetic order in the pseudogap regime at
$\vec Q\!=\!0$, which has been interpreted as evidence for the loop current of $\Theta_{\rm
II}$ type, and we shall come back to these experiments in Sec.~\ref{sec:varma}.
Some other neutron scattering experiments have also been argued to be consistent with
$d$-density wave order, and will be mentioned in Sec.~\ref{sec:exp_elscatt}.
We shall, however, refrain from a detailed discussion of loop-current states in this article,
considering that some aspects of both theory and experiment are either controversial
or not fully settled.


\subsection{Order parameter coupling and global phase diagrams}
\label{sec:coupling}

In the presence of multiple ordering phenomena, the various order parameters are locally coupled.
The general form of this coupling can be deduced from symmetry arguments.
For any two order parameters $\phi_1$ and $\phi_2$, a density--density coupling term
$v|\phi_1(\vec r,\tau)|^2|\phi_2(\vec r,\tau)|^2$ in the LGW theory is generically
present. The sign of $v$ depends on microscopic parameters and decides about repulsion or
attraction between $\phi_1$ and $\phi_2$. For instance, if $\phi_{1,2}$ represent horizontal
and vertical CDW order parameters, then $v>0$ will lead to uni-directional (stripe) order
where $v<0$ results in bi-directional (checkerboard) order.

More interesting are couplings involving one order parameter {\em linearly}.
Those terms are strongly constrained by symmetry and momentum conservation.
A nematic order parameter $\phi_n$ in a tetragonal environment
couples to CDW and SDW according to
\begin{equation}
\lambda_1 \phi_n (|\phi_{cx}|^2-|\phi_{cy}|^2) +
\lambda_2 \phi_n (|\phi_{s\alpha x}|^2-|\phi_{s\alpha y}|^2),
\end{equation}
note that $|\phi_{cx}|^2$ etc. carry vanishing lattice momentum.
A CDW order parameter couples to a spin density wave $\phi_s$, a uniform superconducting
condensate $\psi_0$, and a modulated condensate $\psi$, $\bar{\psi}$, Eq.~\eqref{pdw}, according to
\ifrmp
\begin{eqnarray}
&&\lambda_3 (\phi_c^\ast \phi_s^2 +c.c.) +
\lambda_4 (\phi_c^\ast \psi \bar{\psi}^\ast +c.c.) \nonumber\\
&+&\lambda_5 (\phi_c^\ast \psi_0 \psi^\ast + \phi_c \psi_0 \bar{\psi}^\ast + c.c.).
\label{eqcoupl}
\end{eqnarray}
\else
\begin{equation}
\lambda_3 (\phi_c^\ast \phi_{s\alpha}^2 +c.c.) +
\lambda_4 (\phi_c^\ast \psi \bar{\psi}^\ast +c.c.) +
\lambda_5 (\phi_c^\ast \psi_0 \psi^\ast + \phi_c \psi_0 \bar{\psi}^\ast + c.c.).
\label{eqcoupl}
\end{equation}
\fi
The couplings $\lambda_{3,4}$ are allowed only if the ordering wavevectors obey $\Qch = 2\Qsp$
and $\Qch = 2\Qp$, respectively, whereas $\lambda_5$ is only allowed if $\Qch = \Qp$.

\begin{figure*}
\begin{center}
\includegraphics[width=5in]{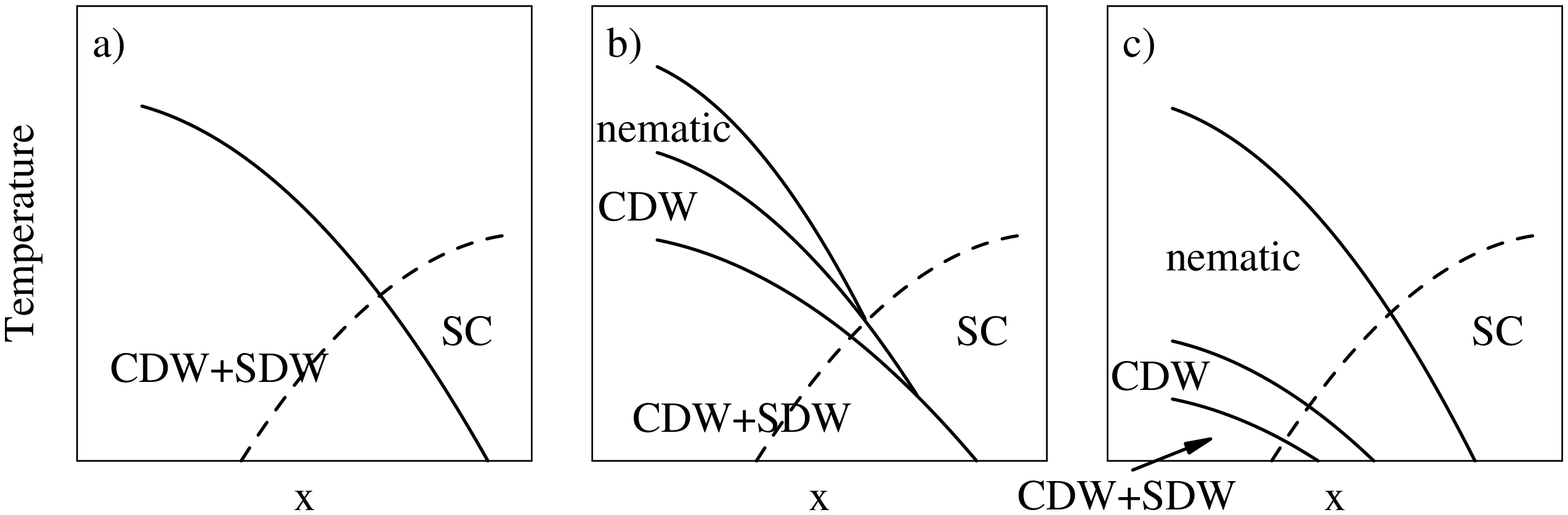}
\caption{
Schematic phase diagrams, illustrating possible transition scenarios into stripe-ordered
states. The vertical axis is temperature, whereas the horizontal axis -- where $x$ may represent doping --
tunes the interplay between superconductivity and spin/charge order, which is assumed to be
competitive.
The solid (dashed) lines are transition into lattice-symmetry-breaking (superconducting)
states.
Case a) corresponds to a weak-coupling scenario, with a single stripe ordering transition,
while c) may be realized at strong coupling, with distinct transitions for nematic,
charge, and spin order. Assuming that increasing $x$ also moves the system from strong
to weak coupling, the intermediate case b) is possible as well.
Effects of lattice pinning and incommensurability are ignored,
as is the interplay between stripe/checkerboard and horizontal/diagonal order.
Quenched disorder will smear out both CDW and nematic transition and likely
turn the SDW into a cluster spin glass \cite{ek93}.
In addition, a structural anisotropy will smear out the nematic transition as well,
see Sec.~\ref{sec:distort}.
}
\label{fig:pd2}
\end{center}
\end{figure*}

The listed terms imply, e.g.,
that a uni-directional density wave induces nematic order via $\lambda_{1,2}$ and
that a collinear SDW with $\Qsp$ induces a CDW with $2\Qsp$ via $\lambda_3$.
As a consequence, a transition into a stripe-ordered state may occur as a direct
transition from a disordered into a CDW+SDW state, or via intermediate nematic and CDW
phases, see Fig.~\ref{fig:pd2}.
The phase diagram of the corresponding Landau theory has been worked out in
detail by Zachar {\em et al.} \cite{zachar98}.
Whether the intermediate phases are realized depends again in microscopic details;
weak-coupling theories typically give a direct transition into a CDW+SDW state.

Similar considerations can be applied to the coupling between stripes and superconducting
condensates arising from $\lambda_{4,5}$. In particular, the co-existence of stripes and
superconductivity induces a modulated component of the condensate.
A detailed analysis can be found in Ref.~\cite{berg08b}.

An external Zeeman magnetic field couples only quadratically to any
of the order parameters, as all have vanishing uniform magnetization.
The orbital part of an external field will have a strong influence primarily on the
superconducting order, by inducing vortices. These considerations will be important in
discussing the field tuning of ordering phenomena in Sec.~\ref{sec:fieldtune}.


\subsection{Concept of ``fluctuating order''}


``Fluctuating order'' describes a situation on the disordered side of, but close to, an
ordering transition, such that precursor effects of the ordering phenomenon are visible
in physical observables. Regarding symmetries or symmetry breaking, ``fluctuating order''
is equivalent to ``no order''. For metallic systems and in the absence of additional
symmetry breaking, a regime of fluctuating order can then be adiabatically connected to
the weakly interacting Fermi liquid.

In weakly interacting and/or high-dimensional systems, the fluctuation regime is usually
tiny and restricted to the immediate vicinity of the critical point. In contrast, large
fluctuation regimes typically occur in strongly coupled and/or low-dimensional systems.

Fluctuating order is characterized by a correlation length $\xi$ and a typical fluctuation
energy (or frequency) $\Delta$. The physical properties at distances larger
than $\xi$ and energies smaller than $\Delta$ will be that of the disordered phase.
On length scales smaller than $\xi$, the system is critical (not ordered, as sometimes
implied) -- this follows directly from standard scaling arguments.
However, if the anomalous dimension $\eta$ of the order parameter is small,
then spectral features of the critical regime are not very different from that of the ordered
phase, e.g., the branch points in the critical spectrum follow a dispersion similar to that of
Goldstone modes of the ordered state.\footnote{
The critical exponent $\eta$ characterizes the propagator of order-parameter correlations
$G(k,\w)$ at criticality. At a quantum critical point with dynamical exponent
$z\!=\!1$, $G$ takes the form $G(k,\w) \propto [k^2-\omega^2]^{(-2 + \eta)/2}$.
}
Close to a quantum critical point, $\Delta$ is typically the gap in the collective-mode
spectrum, and at low temperatures $\xi$ and $\Delta$ scale with the distance to criticality
according to $\xi \propto |r-r_c|^{-\nu}$, $\Delta \propto |r-r_c|^{\nu z}$, where
$r$ is the tuning parameter (e.g. doping), and $\nu$ and $z$ are the correlation length
and dynamical exponents of the transition at $r\!=\!r_c$.

Fluctuating order is invisible to {\em static} probes (unless there is some form of pinning,
see below). Some probes, like elastic neutron scattering or $\mu$SR, are
{\em quasi-static}, i.e., average over a time scale which is large compared to electronic
scales, but can be comparable to the time scale, $1/\Delta$, of collective fluctuations.
As a result, the ``ordering'' temperature as determined by quasi-static probes will not
be unique, but instead depend on the type of probe -- this is a typical signature of a large
fluctuation regime.

A direct probe of fluctuating order is given by the low-frequency dynamic susceptibility,
$\chi(\vec k,\w)$. The imaginary part $\chi''(\vec k,\w)$ will be strongly peaked at the
ordering wavevector, $\vec k\sim \vec Q$, and at $\w \sim \Delta$ (provided sufficient energy resolution
of the experiment), and the real part $\chi'(\vec Q,\w\!=\!0)$ will diverge at the critical
point.
In the spin sector, $\chi''$ can be measured by inelastic neutron scattering; in
the charge sector, electron energy-loss spectroscopy (EELS) is in principle the
appropriate method, however, to date its energy resolution ($>0.1$\,eV) is insufficient to
detect fluctuating stripes in cuprates.
Note that the static structure factor $S(\vec k)$, being an energy-integrated quantity,
is not a suitable probe for fluctuating order near a quantum phase transition, because it
is not directly sensitive to the low-energy part of the fluctuation spectrum.\footnote{
The static structure factor is given by $S(\vec k) \!=\! \int d\w S(\vec k,\w)/(2\pi)$,
and the dynamic structure factor $S(\vec k,\w)$ is related to $\chi''(\vec k,\w)$
via the fluctuation--dissipation theorem, $\chi''(\vec k,\w) \!=\! (1-e^{-\beta\w}) S(\vec k,\w)/2$,
where $\beta=1/T$ is the inverse temperature.
}
(This is different near a classical phase transition, where $S(\vec k)$ and $\chi'(\vec k,\w\!=\!0)$
contain the same information, because the temperature $T$ is much larger than the
relevant fluctuation frequencies.)

While all statements here were for a clean system, the presence of quenched disorder
qualitatively modifies the picture. In particular, disorder can induce pinning of
otherwise slowly fluctuating order, such that it is detectable by static probes.
This will be discussed in more detail in the next subsection.


\subsection{Influence of structural effects and disorder}
\label{sec:distort}
\label{sec:symfield}
\label{sec:dis}

So far, we have discussed the concepts of ``order'' and ``phase transitions''
without taking into account the effects of structurally broken lattice symmetry and of
quenched disorder.
These effects often severely complicate (and also sometimes simplify) the identification
of ordered phases \cite{kiv_rmp}.

\subsubsection{Uni-axial in-plane anisotropy}

For our discussion, the most important structural effect is that of a uni-axial
lattice anisotropy of the CuO$_2$ plane, which breaks the $C_4$ rotation
symmetry.\footnote{
While such an anisotropy can be the {\em result} of an
interaction-driven symmetry breaking of the correlated electron system,
we are here concerned with anisotropies of structural origin.
}
While such anisotropies are absent in the BSCCO\footnote{
BSCCO displays a structural supermodulation along the (1,1) direction of
the Cu lattice, with a wavelength of 4.8 unit cells,
whose origin and properties are not completely understood \cite{supermod}.}
and CCOC compounds,
they are important in YBCO and 214 materials.

In YBCO, in addition to the CuO$_2$ planes there exist chain layers
with CuO chains running parallel to the orthorhombic b axis. 
For doping $\delta\geq0.4$, this results in an orthorhombic
structure with inequivalent a and b axes (with lattice constants $a<b$).
Single crystals commonly are ``twinned'', i.e., contain both orientations
of chains, such that macroscopic measurements average over the inequivalent
a and b directions.
However, it has become possible to produce de-twinned crystals,
and relevant experimental results will be described below.

In the order parameter language, the anisotropic structure results in a
(small) global field coupling linearly to nematic and quadratically to CDW and SDW order.
Hence a transition to a nematic state will be smeared; conceptually, the electronic
nematic state remains well defined only for a large electronic anisotropy in the presence of
a small structural anisotropy.
In contrast, the transition to a stripe state remains sharp, with its critical
temperature being enhanced.

In the 214 compounds, various structural modifications occur as function of temperature.
A transition from a high-temperature tetragonal (HTT) structure to a low-temperature
orthorhombic structure (LTO) occurs at a doping-dependent transition temperature between
200 and 500\,K.
This HTT$\rightarrow$LTO transition occurs as a result of the bond
length mismatch between the CuO$_2$ planes and the La$_2$O$_2$ bi-layers. This mismatch
is relieved by a buckling of the CuO$_2$ plane and a rotation of the CuO$_6$ octahedra.
In the LTO phase, the crystallographic axes are rotated by 45$^\circ$ w.r.t. those of the
HTT phase; in this article, we shall use the coordinate notation of the HTT phase unless
otherwise noted.
In terms of electronic parameters of the Cu lattice, the LTO phase is characterized by inequivalent
diagonals.

While LSCO remains in the LTO phase down to lowest temperatures, the compounds
LBCO, LESCO, and LNSCO display an additional low-temperature tetragonal (LTT) phase.
The LTO$\rightarrow$LTT transition occurs between 50 and 150\,K and is driven by the
smaller radius of e.g. the Nd and Eu ions compared to La.
In the LTT phase, the CuO$_6$ octahedra are rotated such that now the a and b axes of
Cu lattice are inequivalent.
Therefore, in the LTT phase, there is again a field coupling to nematic order in each CuO$_2$ plane,
which is relevant for stabilizing stripe order.
\footnote{
Consequently, electronic stripe ordering in the LTT phase of LBCO, LESCO, and LNSCO
is only accompanied by spontaneous breaking of translation symmetry, i.e.,
those stripes do not possess electronic nematic order.
}
However, the direction of the in-plane anisotropy alternates from plane to plane,
rendering the global crystal symmetry tetragonal and macroscopic in-plane anisotropies
absent.

Concrete numbers for the in-plane anisotropies can be extracted from first-principles calculations.
For the LTT phase of 214 compounds, a simple estimate can be obtained from the octahedral tilt angles.
For a tilt angle of $\alpha\approx 4^\circ\ldots5^\circ$ in LNSCO-1/8, the relation
$t_x/t_y \simeq |\cos (\pi-2\alpha)|$ gives a hopping anisotropy of about
$\Delta t/t \sim 1\%\ldots1.5\%$ \cite{klauss00,kampf01}.
In YBCO the situation is more complicated, as both the structural distortion and the CuO
chains contribute to the (effective) hopping parameters in the planes.
Recent LDA calculations \cite{andersen08} indicate $t_a<t_b$ and
$\Delta t/t \sim 3\ldots4\%$.


In principle, an electronic in-plane anisotropy can also be induced by applying uni-axial
pressure to an otherwise isotropic sample. However, those experiments tend to be
difficult. A few experimental results are available, demonstrating the interplay of
lattice distortions and stripes \cite{takagi_strain}.

\subsubsection{Quenched disorder}

Most superconducting cuprates are ``dirty'' materials, in the sense that chemical doping in
non-stoichiometric composition inevitably introduces disorder
(exceptions are e.g. the oxygen-ordered YBCO-6.5 and YBCO-124
compositions).\footnote{
The relevant dopants are typically located away from the CuO$_2$ planes, therefore the disorder
potential is often assumed to be smooth.}
The physics of disorder in cuprate superconductors is extremely rich and only partially
understood; the reader is referred to a recent review of both experimental and
theoretical aspects \cite{alloul_rev}.

For the discussion of ordering phenomena, it is important that the disorder potential
couples to the charge sector of the CuO$_2$ plane.
From symmetry, this disorder then is of random-mass type for an SDW order parameter,
i.e. an impurity at $\vec x_0$ acts as $\zeta \phi_s^2(\vec x_0)$,
whereas it is of much stronger random-field type\footnote{
The crucial difference between random mass and random field cases is that a
random field breaks the order parameter symmetry whereas the random mass does not.
}
for a CDW order parameter,
$\zeta \phi_c(\vec x_0)$.
For a nematic order parameter, the coupling is also of random-field type,
except for pure site disorder which does not locally select a direction.

Ordering in the presence of randomness is a difficult problem in statistical mechanics,
which is not fully understood even in the classical case.
Some important questions are about
(i) the existence of a true ordered phase, associated with a sharp ordering transition,
(ii) the nature of the phase transition,
(iii) the existence of regimes with anomalous properties near the (putative) phase
transition.
In the following, I shall only touch upon a few aspects relevant to cuprates and the focus of
this article,
and I refer the reader to literature, namely to
Ref.~\cite{harris} for the famous Harris criterion about the stability of phase
transitions in the presence of randomness,
Ref.~\cite{imryma} for the Imry-Ma argument about the stability of ordered phases,
Ref.~\cite{natter} for an overview on the random-field Ising model, and
Ref.~\cite{tvojta} for an overview on rare regions near phase transitions.

In the random-mass case, the ordered phase is characterized by true symmetry breaking and
survives as a distinct phase. The phase transition is expected to be sharp, albeit
perhaps modified compared to the clean case according to the Harris criterion,
with the exception of certain quantum phase transitions with order parameter damping \cite{tvojta}.

In contrast, in the random-field case the effects of quenched disorder are more drastic.
In low dimensions, the ordered phase ceases to exist, because the system breaks up into
domains which are pinned by the local fields \cite{imryma}.
Consequently, there is no sharp phase transition upon cooling, and the correlation length
is finite even in the zero-temperature limit.
This applies in particular to the random-field Ising model in two space dimensions,
and is expected to hold for discrete $Z_N$ symmetries as well.\footnote{
$Z_N$ is the cyclic group of $N$ elements, formed e.g. by
rotations about a single axis with $N$-fold rotation symmetry.
}
As the system is {\em at} its lower critical dimension $d_c^-\!=\!2$, the domains are exponentially
large for weak disorder.
In the case of continuous symmetry the lower critical dimension is shifted to $d_c^-\!=\!4$.
Experimentally, random-field pinning implies that static probes will see a gradual, i.e.
smeared, onset of order.
Rather little is known about dynamical properties in the quantum case.
On the disordered side of a quantum phase transition, randomness will induce in-gap
spectral weight, and it is conceivable that the smeared phase transition is accompanied by
slow glass-like order-parameter dynamics.\footnote{
It has been demonstrated that stripes arising from frustrated interactions may display
glass-like behavior even in the absence of quenched disorder \cite{schmalian00}.}

For stripes, the conclusion is that the charge ordering transition is generically
smeared due to random-field effects (assuming the inter-plane coupling to be weak).
The spin ordering transition could still be sharp, however,
the coupling between spin and charge sectors may invalidate this guess:
Short-range charge order can induce magnetic frustration in the spin sector,
leading the spin-glass behavior. Whether this setting allows for a sharp phase
transition is not known.
Experimentally, both spin and charge ordering transitions appear to be
broad and glassy, see Sec.~\ref{sec:exp_glass}.

Pinning of multiple equivalent collective modes, e.g. horizontal and vertical stripes,
can lead to interesting phenomena \cite{maestro,robertson}.
If the clean system would display uni-directional order, random-field pinning tends to
globally restore the rotation symmetry, because domains of different orientation will
occur. Locally, rotation symmetry will still be broken (unless the repulsion between the
two order parameters is much weaker than the disorder effect).
In contrast, if the clean system has a tendency to bi-directional (checkerboard) order,
pinning may locally induce some anisotropy.
As a consequence, random-field pinning makes it difficult to deduce the nature of
the clean order (stripes vs. checkerboard), in particular for weak order and strong
pinning.\footnote{
A detailed analysis of the impurity-induced patterns in a magnet with a
tendency towards valence-bond order is in Ref.~\cite{metlitski08}.
}


\section{Experimental evidence for lattice symmetry breaking}
\label{sec:exp}

\begin{figure*}
\begin{center}
\includegraphics[width=2.5in]{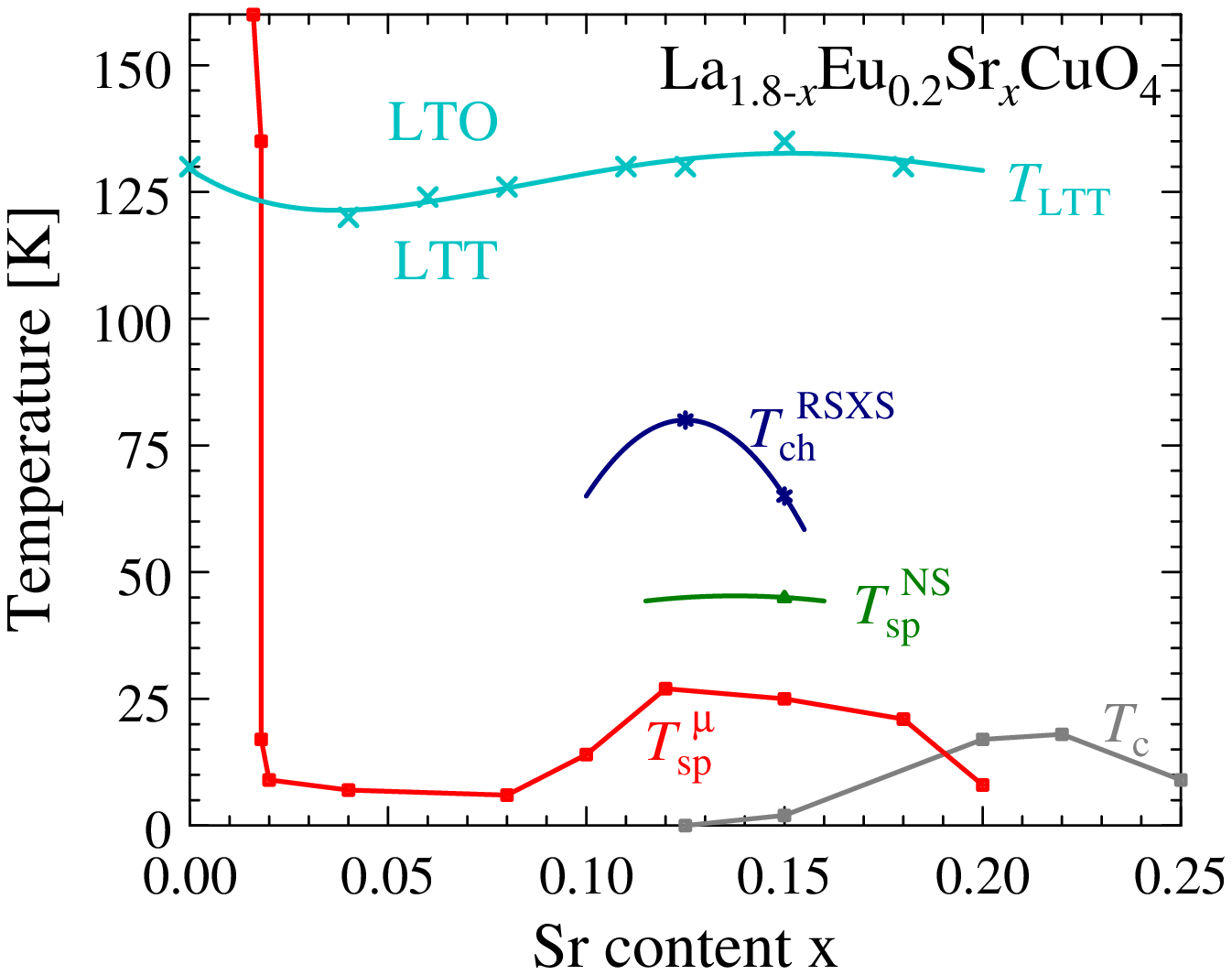}
\hspace*{10pt}
\includegraphics[width=2.5in]{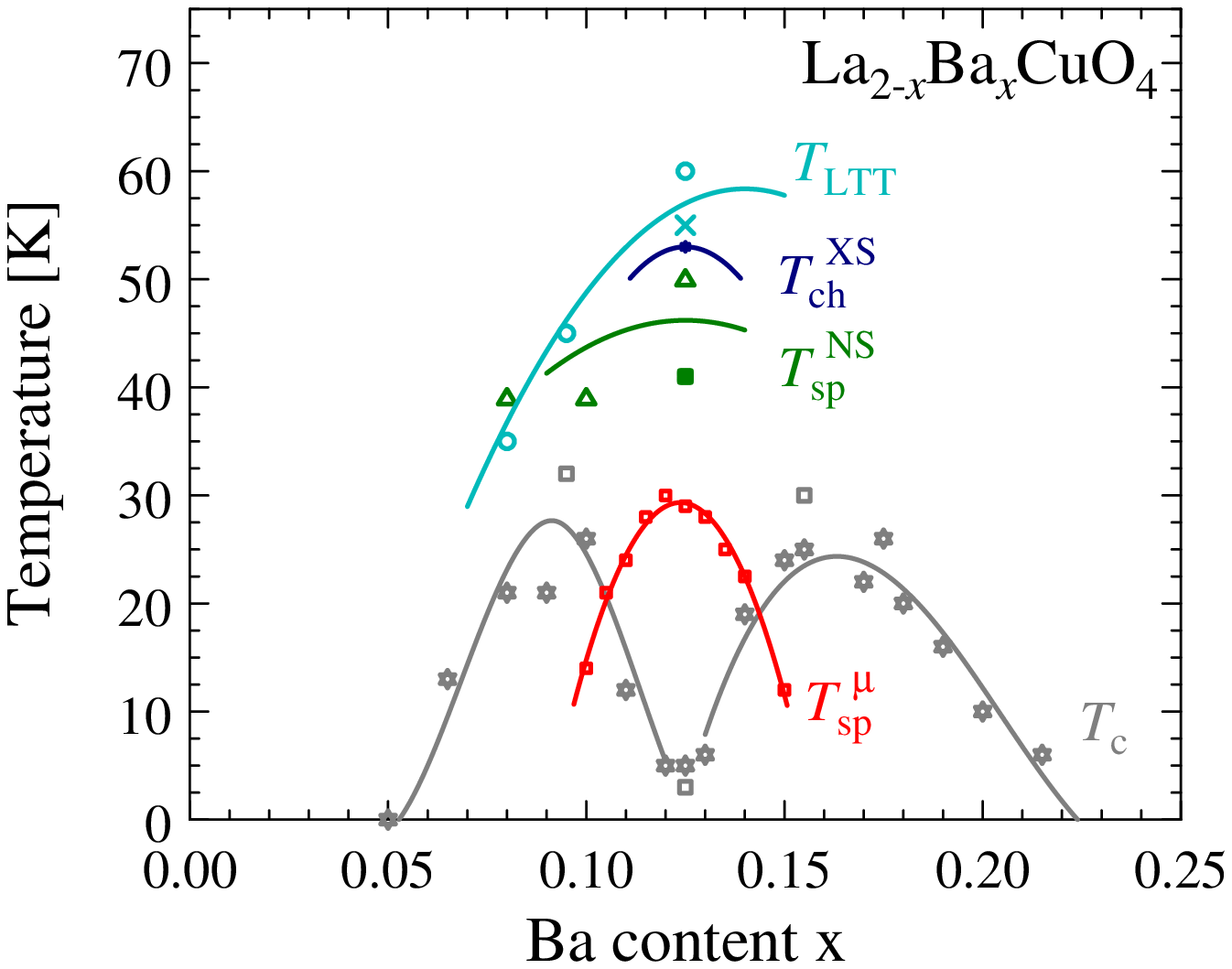}
\caption{
Phase diagrams of LESCO (left) and LBCO (right).
For LESCO, the data points represent the superconducting $\Tc$ \cite{takagi_iso},
the charge-ordering temperature $\Tch$ from resonant soft x-ray scattering \cite{fink08},
the spin-ordering temperature $\Tsp$ from neutron scattering \cite{huecker07}
and from $\mu$SR \cite{klauss00},
and the LTO--LTT transition temperature $T_{\rm LTT}$ from x-ray scattering \cite{klauss00}.
For LBCO, the data are $\Tc$ \cite{mooden88,huecker05} (stars,squares),
$T_{\rm LTT}$ from x-ray scattering \cite{jt08,duns08} (crosses,circles)
and $\Tch$ from x-ray scattering \cite{jt08},
and $\Tsp$ from neutron scattering \cite{jt08,duns08} (filled squares,triangles)
and from $\mu$SR \cite{arai03}.
The lines are guides to the eye only.
The published data display a rather large spread, with transitions often being broad,
which may be due to effects of disorder and/or sample inhomogeneities.
Moreover, data obtained on polycrystals and single crystals may differ substantially:
For LBCO-1/8, $\mu$SR results show $\Tsp \approx 29$\,K for polycrystals \cite{arai03}
and $\Tsp \approx 40$\,K for single crystals \cite{savici05}.
}
\label{fig:pd3}
\end{center}
\end{figure*}

Let me turn to actual results of experiments on cuprate superconductors.
While it seems that detecting a symmetry-breaking order should be a straightforward
undertaking, both fundamental and practical problem complicate matters.

The most important fundamental problem is related to the effect of quenched disorder,
described in Sec.~\ref{sec:dis}. Disorder from chemical doping acts as a random field for
the CDW order parameter, hence it is a relevant perturbation. Thus, the charge ordering
transition is smeared, most likely eliminating thermodynamic singularities. A second
fundamental problem is that macroscopic manifestations of broken rotation symmetry can
only be expected when uni-directional stripes in all CuO$_2$ planes are commonly aligned
in one direction. Such a single-domain situation cannot be expected to be realized,
mainly because of the random-field effects, unless a global symmetry-breaking field
exists which selects one stripe direction, Sec.~\ref{sec:symfield}. The best candidate
here is the YBCO family.

The direct observation of superstructures is nevertheless possible and will be discussed
below. Here, practical issues such as the availability of sufficiently large single
crystals, clean surfaces etc. become important, but those issues have been resolved at
least partially.
Scattering experiments which average over a large spatial area are again faced with the
domain problem: If e.g. domains of both horizontal and vertical stripes are
simultaneously present, distinguishing this from a local superposition of both (i.e.
checkerboard order) requires a careful analysis.

As a result of those efforts, signatures of {\em translation} symmetry breaking have been found
in a variety of hole-doped cuprates, most notably in LNSCO, LESCO, and LBCO.
The phase diagrams of the two latter are shown in Fig.~\ref{fig:pd3},
where the results from different measurement techniques have been collected.
For the three materials,
the order can be consistently interpreted in terms of uni-directional SDW and CDW
order, i.e. stripes, over a wide doping range.
Moreover, static incommensurate SDW order has been established
in La$_2$CuO$_{4+\delta}$, in LSCO for $x<0.13$,
and for YBCO for $\delta\leq 0.45$ (but in the latter case the
order is only short-ranged and of glassy character).
An exciting recent development is the clear observation of stripes in STM experiments
on BSCCO and CCOC compounds, although it has to be kept in mind that STM
probes the surface layer only.
As discussed in more detail in Sec.~\ref{sec:exp_univ}, stripe signatures seem to weaken
with increasing number of adjacent CuO$_2$ layers in hole-doped compounds,
and have not been reported in electron-doped cuprate materials.

Direct observations of stripe order are corroborated by more indirect probes:
NMR and $\mu$SR experiments provide evidence for inhomogeneous magnetism
in 214 compounds, with a temperature dependence similar to that seen in
scattering experiments.
Besides the temperature and doping dependence of the symmetry-breaking order,
its dependence on an applied magnetic field has been studied as well, which allows to
draw conclusions about the relation between superconductivity and stripe order.

In principle, translational symmetry breaking should leave well-defined traces in
the dispersion of all elementary excitations due to band backfolding and the
opening of Bragg gaps.
However, clear-cut experimental signatures are difficult to identify, due to a variety of
complications: Disorder and the simultaneous presence of horizontal and vertical stripes
tend to smear the signal, and matrix element effects do not allow to observe
all bands.
The experimental results for magnetic excitations, phonons, and the single-electron
spectrum as measureed by ARPES will be discussed in subsections below.


Signatures of {\em rotation} symmetry breaking have been most clearly identified in
underdoped YBCO, both in transport and in neutron scattering.
In particular, the magnetic excitation spectrum of YBCO-6.45
was found to develop a spatial anisotropy below about 150\,K,
see Sec.~\ref{sec:exp_nematic}.
Whether these data should be interpreted in terms of a Pomeranchuk instability of
the Fermi surface or in terms of fluctuating stripes is open at present and will
be discussed in Sec.~\ref{sec:th_nematic}.
Locally broken rotation symmetry is clearly visible in STM data obtained
from the surface of underdoped BSCCO and CCOC, where it is accompanied
by stripe formation, see Sec.~\ref{sec:exp_stm}.


\subsection{Static order in neutron and x-ray scattering}
\label{sec:exp_elscatt}

Long-range order accompanied by breaking of lattice translation symmetry
leads to sharp superlattice Bragg peaks in diffraction experiments.
Those can be detected by neutrons or by x-rays.
Experiments require sufficiently large single crystals, which are by now available for
many cuprate families.

\subsubsection{Spin density waves seen by neutron diffraction}

\begin{figure}
\begin{center}
\includegraphics[width=3.3in]{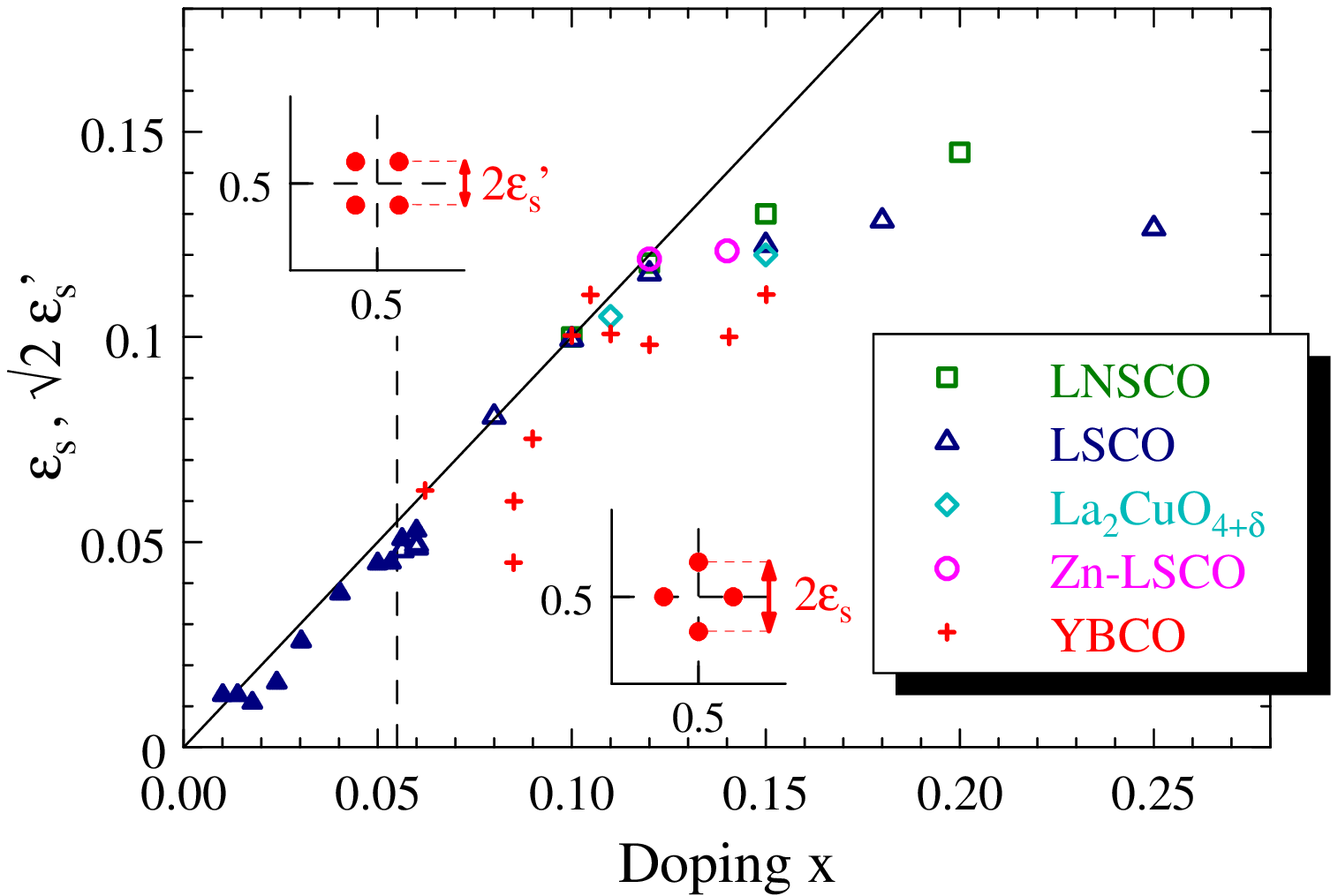}
\caption{
Summary of experimental data illustrating the doping dependence of
incommensurability $\epsilon_s$ in the cuprates (the so-called ``Yamada plot'').
Results have been obtained by different groups:
LNSCO \cite{jt95,jt96,ichikawa00} (squares);
LSCO \cite{yamada98,waki99,waki00,matsuda00,matsuda00a,matsuda02a} (triangles); 
La$_{2}$CuO$_{4+\delta}$ Ref.~\cite{wells,oxy} (rhombi);
Zn-doped LSCO \cite{kim99,jt99a} (circles);
YBCO \cite{dai01,mook02,hinkov08a} (crosses).
For LSCO, the plot also shows the incommensurability $\sqrt{2}\epsilon_s'$ of the
diagonal modulations observed for $x\leq0.06$ (filled triangles).
For both YBCO and La$_{2}$CuO$_{4+\delta}$,
the $x$ values corresponding to the hole doping level are estimates only.
Note that most data points are slightly below the line $\epsilon_s\!=\!x$ (solid).
The vertical dashed line is the boundary between insulating and superconducting
phases in LSCO.
}
\label{fig:yamada}
\end{center}
\end{figure}

Experimental evidence for static stripe-like order was first found in
neutron-scattering experiments on LNSCO with doping level 0.12 \cite{jt95,jt96},
which is a superconductor with an anomalously low $\Tc$ of roughly 5\,K.
Those experiments detected static spin correlations with an onset temperature of
about 55\,K, which were peaked at wavevectors
$\Qspx = 2\pi (0.5\pm\epsilon_s,0.5)$ and $\Qspy = 2\pi (0.5,0.5\pm\epsilon_s)$, i.e.,
at four spots slightly away from the $(\pi,\pi)$ AF order of the
parent antiferromagnet.
At the same time, neutron scattering was used to locate the LTO--LTT structural transition
at 70\,K and the onset of charge order slightly below this temperature,
with ordering wavevectors $\Qchx = 2\pi(\pm\epsilon_c,0)$ and $\Qchy =
2\pi(0,\pm\epsilon_c)$.
Within the experimental accuracy, $\epsilon_s = \epsilon_c/2 = 0.12$,
where $\epsilon_s = \epsilon_c/2$ is expected on symmetry ground for
coexisting collinear SDW and CDW orders, Sec.~\ref{sec:coupling}.
This type of order was later found in LNSCO also for doping levels
$0.08\leq x\leq 0.20$, using both neutron and x-ray scattering \cite{jt97,ichikawa00}.
The incommensurability roughly follows the doping level,
$\epsilon_s \approx x$ for $x\leq 0.12$, whereas it tends to saturate for larger $x$,
with $\epsilon_s \approx 0.14$ at $x\!=\!0.20$ (Fig.~\ref{fig:yamada}).\footnote{
In a picture of charge stripe order, the doping dependence of the incommensurability,
$\epsilon_s(x)$, can be translated into the doping dependence of both stripe distance
and stripe filling. This allows an interesting connection to the doping dependence
of the chemical potential, see Sec.~\ref{sec:chempot}.
}

\begin{figure}
\begin{center}
\includegraphics[width=3in]{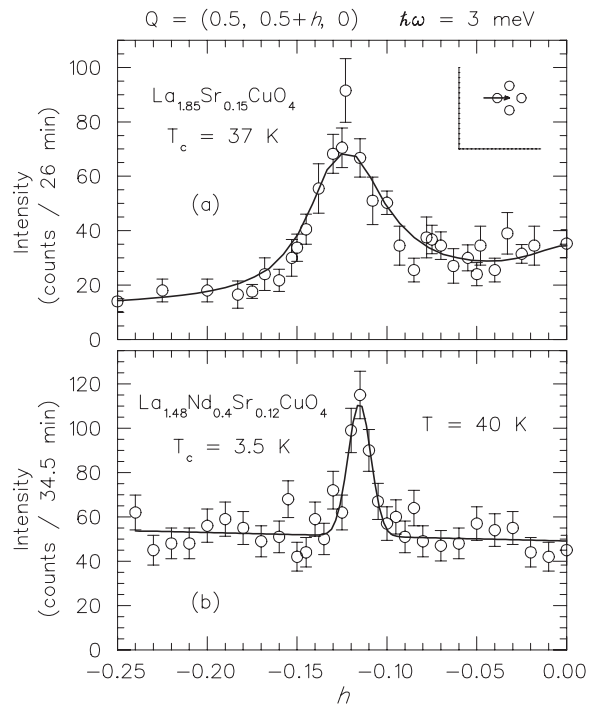}
\caption{
Comparison of constant-energy scans at 3\,meV
through an incommensurate magnetic peak (along
path shown in inset) for
(a) La$_{1.85}$Sr$_{0.15}$CuO$_4$ and
(b) La$_{1.48}$Nd$_{0.4}$Sr$_{0.12}$CuO$_4$.
Both scans are at $T = 40$\,K $> \Tc$
and illustrate the similarity of low-energy fluctuation
in non-stripe-ordered (a) and stripe-ordered (b) 214 compounds
(reprinted with permission from Ref.~\cite{jt99b}, copyright 1999 by the American Physical Society).
}
\label{fig:braggsdw}
\end{center}
\end{figure}

In fact, incommensurate {\em dynamic} correlations were detected much earlier in
LSCO at various doping levels, with a similar wavevector dependence on
doping \cite{cheong91,yamada92}.
Subsequent elastic neutron scattering experiments found quasi-static magnetic
order, with wavevectors obeying $\epsilon_s \approx x$, in superconducting
LSCO for $x<0.13$ \cite{waki99,waki01a}.
This makes clear that LSCO is closely located to a QCP associated with
incommensurate SDW order. Indeed, inelastic neutron scattering on LSCO-0.14 reported
scaling behavior of the magnetic excitation spectrum, consistent with a nearby QCP
\cite{aeppli97}.
Low-energy incommensurate fluctuations then occur as precursors of static order at
these wavevector, see Fig.~\ref{fig:braggsdw} and Sec.~\ref{sec:exp_hour}.

Impurity doping experiments of LSCO, with a few percent of Cu atoms replaced by
non-magnetic Zn, support the idea of nearly ordered stripes in LSCO:
Zn induces elastic intensity for $x\!=\!0.14$ \cite{hirota01},
whereas it somewhat broadens the elastic peaks at $x\!=\!0.12$ as compared to the Zn-free
sample \cite{kim99,jt99a}.
These observations support the notion that Zn impurities tend to
pin stripes, although it should be noted that Zn is also known to
induce magnetic moments in its vicinity \cite{fink90,alloul91,julien00}
which certainly contribute to the slowing down of spin fluctuations.

Static incommensurate SDW order, similar to that of LNSCO, has also been
detected in neutron scattering on La$_2$CuO$_{4+\delta}$ \cite{oxy},
LBCO-1/8 \cite{fujita04}, and LESCO-0.15 \cite{huecker07}.
In the two latter compounds, the full doping range of static magnetic order has been
mapped out by other probes:
In LBCO, magnetization and $\mu$SR measurements have established magnetic
order for $0.095\leq x \leq 0.155$ \cite{huecker07,arai03}, while $\mu$SR measurements give
evidence for magnetic order in LESCO over the entire doping range up to $x\!=\!0.20$
\cite{klauss00}.
The La$_2$CuO$_{4+\delta}$ system,
where interstitial oxygens are positioned in every fourth La$_2$O$_2$ layer
for $\delta\approx0.12$ (the so-called stage-4 structure),
displays an incommensurability of $\epsilon_s \approx 0.12$
at a hole doping level of roughly 0.15.
Remarkable, SDW order and bulk superconductivity appear simultaneously
at $\Tc \approx 42$\,K \cite{oxy}.
However, the behavior of this compound has been attributed to phase separation
into magnetic and superconducting domains \cite{moho06}.\footnote{
With an eye towards La$_2$CuO$_{4+\delta}$, Ref.~\cite{KAE01} discussed a scenario of
competing magnetism and superconductivity, where -- in the case of phase coexistence at low
temperatures -- both orders may set in at the {\em same} temperature.}

To fully characterize the SDW order common to the 214 compounds,
more information is required.
The first question, triggered by the finding of {\em four} magnetic Bragg peaks,
is whether the order consists of two types of uni-directional stripe domains with a single
$\vec Q$ vector each, or whether it is of checkerboard type with two $\vec Q$ vectors.
The experiment did not detect magnetic peaks along the diagonal direction,
i.e., at locations $2\pi(0.5\pm\epsilon',0.5\pm\epsilon')$; this rules out a checkerboard
with modulation directions along (1,0), (0,1), but could be compatible with
a checkerboard of diagonal stripes \cite{jt99a}.
However, such a structure would lead to diagonal CDW peaks, which were not
detected.\footnote{
See Sec.~\ref{sec:coupling} for the Landau theory argument regarding the relative
orientation of spin and charge peaks.}
Hence, the most plausible interpretation is that of two types of large domains
(i.e. horizontal and vertical) in which uni-directional SDW and CDW co-exist.
In fact, if stripes follow the structural distortion pattern of the LTT phase, then the stripe
direction can be expected to rotate by 90$^\circ$ from plane to plane.
\footnote{
In both LSCO-0.12 and La$_2$CuO$_{4+\delta}$, the quasi-elastic magnetic peaks
do not exactly lie along the high-symmetry directions of the crystal, but are rotated by
approximately 3$^\circ$ \cite{kim00,oxy}. This is consistent with the crystal symmetry
being LTO instead of LTT \cite{robertson}.}

The second question concerns the size and orientation of the spin moments in the SDW phase.
The size of the ordered moment per Cu site was determined from neutron scattering
in LSCO to vary between $0.04\ldots0.07\,\mu_B$ for $0.06\leq x\leq 0.1$
and to reach a maximum of $0.1\,\mu_B$ for LSCO-0.12 \cite{waki01a};
in LNSCO-0.12 and La$_2$CuO$_{4+\delta}$ the neutron scattering results
yield moment sizes of 0.1 and $0.15\,\mu_B$ \cite{jt96,oxy}.
In comparison, $\mu$SR experiments suggest a moment size of $0.3\ldots0.35\,\mu_B$,
i.e. roughly 60\% of the value in the undoped parent compound,
for LNSCO, LBCO-1/8, LSCO-0.12, and La$_2$CuO$_{4+\delta}$ \cite{nachumi98,savici02}.
Note that the $\mu$SR moment sizes are usually inferred from a simulation
of the $\mu$SR signal taking into account the spatial stripe structure \cite{nachumi98,klauss04},
which consists of small and large moments, and the quoted numbers denote the maximum (not
average) moment size in the stripe state.
The remaining discrepancy regarding the moment size is not fully understood,
and disorder in the moment directions as well as calibration issues of neutron scattering
may play a role.

The moment orientation was studied on LNSCO-0.12 using polarized neutron scattering by
Christensen {\em et al.} \cite{christensen07}.
The moments are primarily lying in the CuO$_2$ plane, as in the undoped parent compounds,
consistent with conclusions drawn from susceptibility measurements \cite{huecker05}.
However, the neutron result was not entirely conclusive w.r.t. the in-plane orientation:
The polarization signal was found consistent with
either a collinear single-$\vec Q$ (i.e. stripe-like) structure
or a non-collinear two-$\vec Q$ structure,
but was inconsistent with a single-$\vec Q$ spiral or a collinear checkerboard
order.
Together with the simultaneous existence of charge order, the data are again
consistent with collinear ordering, but a direct proof is still missing.

The third question concerns the correlation length of the SDW order, both in-plane
and along the c axis. In 214 compounds, the in-plane order reaches correlation
lengths $\xi_{ab}$ of 200\,\AA\ and beyond, however, is often not resolution limited
(which points toward disorder effects and/or glassy behavior \cite{jt99b}).
In contrast, the correlation length $\xi_c$ in c direction is typically only one
interplane distance, with the exception of the field-induced signal in LSCO-0.10
\cite{lake05} where $\xi_c$ corresponds to 6 interplane distances.

The weak c-axis correlations of stripes can be rationalized considering that, with stripe directions
alternating from plane to plane, interactions between 2nd-neighbor layers are required
to align the stripe pattern along the c axis. Such interactions will be small and
have to compete with pinning forces from defects -- it is no surprise that the latter are dominant.
In contrast, in the case of field-induced order, ``correlated'' pinning by vortex lines can enhance
three-dimensional correlations.

A fourth question is: How important is the LTT distortion for the appearance of static stripes?
A number of studies \cite{fujita02,fujita04,kimura04} have been devoted to the compounds
La$_{15/8}$Ba$_{1/8-x}$Sr$_x$CuO$_4$
with doping level 1/8, where the LTT distortion of LBCO-1/8 weakens with Sr doping and
disappears at $x\approx0.09$.
These studies suggest that static stripe order is tied to the LTT distortion,
i.e., it disappears in favor of stronger superconductivity around $x\approx0.09$.
In contrast, a very recent pressure study on LBCO-1/8 revealed that
stripe domains still occur in the high-pressure high-temperature tetragonal (HTT)
phase above pressures of 2\,GPa \cite{huecker09}.
If verified, this would constitute an exciting example of simultaneous, electronically
driven, spontaneous breaking of rotation and translation symmetry.

In the insulating regime of both LSCO \cite{waki99,matsuda02a} and LNSCO \cite{waki01b}
at very small doping, $x<0.055$, a different type of magnetic
order has been found.
Here, elastic peaks were found at $\Qsp = 2\pi(0.5\pm\epsilon_s',0.5\pm\epsilon_s')$,
i.e., in the diagonal direction of the Cu square lattice.
The incommensurability follows $\epsilon_s' \approx x/\sqrt{2}$ for $0.02<x<0.055$;
thus the peak distance to $(\pi,\pi)$ appears to follow the same linear $x$ dependence
as for superconducting compounds with $0.055<x<0.125$.
This may suggest a common origin of the SDW in both doping ranges,
and consequently, the low-doping SDW order has been interpreted in terms of
diagonal stripes.
However, no evidence for charge order has been reported for $x<0.055$.
An alternative scenario consists of spiral SDW order without accompanying CDW
\cite{shraiman,sushkov06,juricic06}.
To my knowledge, polarized neutron scattering -- which would be able to
distinguish the two alternatives -- has not been performed to date.

In other cuprate families, {\em static} incommensurate spin order has not been detected,
with the exception of strongly underdoped YBCO:
A conclusive set of data stems from de-twinned crystals of YBCO-6.35 and 6.45
where neutron scattering measurements of Hinkov {\em et al.} \cite{hinkov08a}
and Haug {\em et al.} \cite{haug_635}
detected incommensurate quasi-static order at wavevectors $2\pi(0.5\pm\epsilon_s,0.5)$.
However, the data on YBCO-6.45 suggest the order is weak and glassy:
Quasi-static order in neutron scattering sets in below 30\,K,
with a correlation length of about 20\,\AA\ only,
and $\mu$SR measurements detect static order only at 1.5\,K \cite{hinkov08a}.
The findings of Refs. \cite{hinkov08a,haug_635} are broadly consistent with earlier neutron scattering
reports using twinned crystals \cite{mook02,buyers06,buyers08}.
While Mook {\em et al.} \cite{mook02} suggested the presence of stripe order in YBCO-6.35,
Stock {\em et al.} \cite{buyers06,buyers08} reported the existence of a ``central mode''
in quasi-elastic scattering, centered at $(\pi,\pi)$ and corresponding to glassy
short-range order.
Note that the effect of twinning on the in-plane geometry has been nicely demonstrated
in Ref.~\cite{hinkov08a}, i.e., twinning smears an incommensurate signal such that it
appears as a single peak at $(\pi,\pi)$.
In a stripe picture, the incommensurate SDW order in underdoped YBCO would correspond
to stripes along the b axis. In YBCO-6.45, $\epsilon_s \approx 0.045$,
while the nominal hole doping is $x\approx 0.085$.
These values are inconsistent with the relation $\epsilon_s \approx x$,
established for stripe order in LSCO with $x<1/8$.

Finally, we briefly mention the neutron-scattering search for commensurate $(\pi,\pi)$
antiferromagnetic order in superconducting cuprates, which was primarily motivated by the
proposal of staggered loop-current order (dubbed $d$-density-wave) in the pseudogap
regime \cite{ddw}.
There were several observations of AF order in the underdoped YBCO samples
\cite{sidis01,mook01,mook02b}, with ordered moments of about
0.02\ldots0.05 $\mu_B$.
However, the polarization neutron analysis revealed that the order is
dominated by moments aligned in the CuO$_2$ plane, as in the insulating
parent compound. Considering, on the one hand, that other
high-quality single crystals with a similar doping
level do not show a similar order \cite{buyers02} and, on the
other hand, that impurity substitution can induce antiferromagnetic
order at 300\,K even at optimal doping \cite{hodges02},
it is likely that the observed AF order is not a generic property of the
underdoped state.
Subsequently, Mook {\em et al.} \cite{mook04} reported polarized neutron experiments
suggesting the existence of a weak AF quasi-2d order with moments
perpendicular to the CuO$_2$ plane. However, the estimated magnitude of the
ordered moment is 0.0025\,$\mu_B$, which is close to the experimental threshold of
detection \cite{mook04}.
No subsequent experiments have reported conclusive evidence for magnetic order
at or near $(\pi,\pi)$ setting in at the pseudogap scale.

\subsubsection{Charge density waves seen by x-ray scattering}

Although the picture of stripe order, i.e., coexisting SDW and CDW, was already
proposed in the context of the first magnetic neutron scattering results in 1995,
the first unambiguous observation of {\em charge} order was only made by
Abbamonte {\em et al.} in 2005 \cite{abbamonte05}.
The primary reason is that neutron and non-resonant x-ray scattering
can only detect charge order indirectly by the associated lattice distortion,
as these techniques are mainly sensitive to the nuclear scattering and the core electron scattering,
respectively.

Nevertheless, results from neutron and non-resonant x-ray scattering have provided
valuable information on charge ordering, because lattice distortions can be
expected to follow charge order regarding both the amplitude and the temperature
dependence. In particular, high-energy x-rays allow the use of a synchrotron source to
obtain high intensity and high momentum resolution \cite{zimmermann98}.
Using these techniques, charge-order superlattice peaks were found
in LNSCO for doping levels $0.08\leq x\leq 0.20$ \cite{jt96,jt97,zimmermann98,niemoller99,ichikawa00}
and recently in LBCO-1/8 \cite{jt08,kim08a}.

In this context, studies using extended x-ray absorption fine structure (EXAFS)
are worth mentioning, which provide information on local lattice displacememnts.
In LNSCO-0.12, lattice fluctuations have been found to strongly increase below the
charge-ordering temperature \cite{bianconi01}.
A similar effect was found before in LSCO-0.15 \cite{bianconi96}, supporting the idea of
dynamic stripes in this material.


The most direct information on the charge modulation in cuprates can be obtained by
resonant soft x-ray scattering using photon energies at the O $K$ and the Cu $L$ edge
\cite{abbamonte02}.
This technique was applied to stripe-ordered LBCO-1/8 by Abbamonte {\em et al.}
\cite{abbamonte05} and to LESCO at various doping levels by Fink {\em et al.}
\cite{fink08}.

In LBCO-1/8, charge superlattice peaks were detected at in-plane wavevector
$2\pi(0.25\pm0.02,0)$ with an in-plane correlation length $\xi_{ab}$ of about
480\,\AA\ (=\,125 lattice spacings). The dependence on vertical momentum was consistent
with a period-2 order in c-axis direction, with a small correlation length of
$\xi_c \lesssim 2$ interplane distances
(which is consistent with the x-ray study of Ref.~\cite{kim08a}).
The onset temperature of the signal was $\Tch \approx 60$\,K which almost coincides
with the temperature of the LTO--LTT transition in this sample.
The form factor of the scattering signal was used to extract an estimate of the actual
modulation amplitude in the charge sector. Assuming the orbital charge-order pattern
obtained in a slave-boson analysis of stripes in the three-band Hubbard model
\cite{lorenzana02}, the authors obtained a large modulation of oxygen hole densities
varying between 0.03 and 0.12, i.e., by a factor of 4, within a unit cell.
However, it has to be emphasized that this analysis is strongly model-dependent,
and the assumed modulation pattern, although bond-centered, appears inconsistent with that
obtained in STM \cite{kohsaka07}.

\begin{figure*}
\begin{center}
\ifhires
\includegraphics[width=3.8in]{fink1.eps}
\else
\includegraphics[width=3.8in]{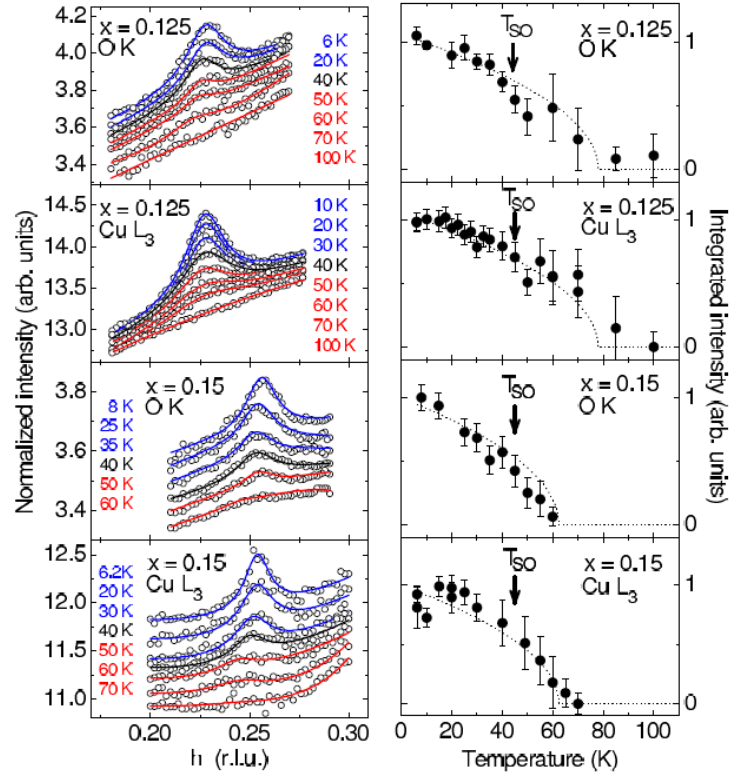}
\fi
\caption{
Resonant soft x-ray scattering results on LESCO
(reprinted with permission from Ref.~\cite{fink08}, copyright 2009 by the American Physical Society).
Left: Temperature dependence of $h$ scans along $(h,0,l)$ showing
superstructure reflections of LESCO-1/8 and LESCO-0.15
using O $K$ ($l\!=\!0.75$) and Cu $L_3$ ($l\!=\!1.6$) photon energies.
The curves are vertically shifted for clarity; the solid lines are fits to the data.
Right: Temperature dependence of the intensities of the superstructure
reflections, normalized to the intensity at $T\!=\!6$\,K.
The dotted line is a $\sqrt{\Tch-T}$ fit.
The estimated spin-ordering temperatures are marked by arrows \cite{huecker07,klauss00}.
}
\label{fig:fink}
\end{center}
\end{figure*}

Recently, resonant soft x-ray scattering was also performed on LESCO \cite{fink08},
with some results shown in Fig.~\ref{fig:fink}.
For doping $x=1/8$, the order appeared at $\Tch = 80\pm 10$\,K
at a wavevector $\Qch = 2\pi(0.228,0)$, whereas
for $x\!=\!0.15$ $\Tch$ was $70\pm10$\,K and $\Qch = 2\pi(0.254,0)$.
The correlation length $\xi_{ab}$ was about 80\ldots100 lattice spacings,
while $\xi_c$ was of order unity.
In contrast to LBCO, the LTO-LTT transition in LESCO occurs at significantly higher
temperature ($T_{\rm LTT} \approx 125$\,K),
while spin ordering has been detected at 25\,K (45\,K) by $\mu$SR (neutrons)
\cite{klauss00,huecker07}.
Thus, LESCO displays a sequence of well-separated phase transitions
with $\Tsp < \Tch < T_{\rm LTT}$.

Taken together, the neutron and x-ray experiments establish a number of
important characteristics of the order in 214 cuprates:
(i) CDW order sets in at higher temperatures than SDW order, and
both coexist at low $T$. Thus, CDW order is unlikely to be only a subleading
consequence of collinear SDW order in the sense of Eq.~\eqref{eqcoupl}.
(Of course, this does {\em not} exclude that stripe formation is driven
by antiferromagnetic exchange.)
(ii) The ordering wavevectors are related, $\epsilon_s=\epsilon_c/2$
within error bars.
(iii) The wavevector dependence on doping $\epsilon_c(x)$ is inconsistent
with a simple nesting scenario of CDW formation, because $\epsilon_c$ increases
with $x$ whereas the distance between the antinodal Fermi surface
decreases (see e.g. Ref.~\cite{valla06}).

It should be noted that no signatures of charge order have been reported for
LSCO, even in the doping regime $x<0.13$ where the SDW order appears static.
Also, resonant x-ray scattering failed to detect charge order in CCOC-1/8
\cite{smadici06}, where STM has established the existence of a period-4 charge
ordering pattern \cite{hanaguri04,kohsaka07}.
Here, two explanations are possible: Either the STM pattern exists at the surface only,
or the correlation length of the charge order is too small to be detectable
by x-rays at present.
Indeed, the intensity of the x-ray signal scales with $\xi_{ab}^2$,
thus assuming $\xi_{ab}$ of 10 lattice spacings \cite{kohsaka07}
renders the signal 100 times smaller than for LBCO-1/8 (for comparable CDW amplitudes)
which is below the sensitivity limit of the experiment.

\subsubsection{Magnetic order at ${\vec Q}=0$}
\label{sec:varma}

Over the last few years, indications for a distinct type of order
have been found in several experiments, which were motivated by the proposal
\cite{varma99,varma02} for the pseudogap phase in terms of a spontaneous loop-current
order within the unit cell of CuO$_2$ planes, with ordering wavevector $\vec Q\!=\!0$.

The most direct indication for circulating-current order comes from recent elastic polarized
neutron scattering experiments \cite{fauque06,fauque08a,fauque08b}.
Such measurements are difficult, as the magnetic signal is located on top of a large
nuclear Bragg peak.
Therefore, the magnetic Bragg peak\footnote{
The actual measurement is done at finite wavevectors $(2\pi,0)$ or $(4\pi,0)$.
}
has to be obtained as a difference between
spin-flip and non-spin-flip neutron scattering signals. As the spin-flip ratio is an
unknown constant, the subtraction is done by suitable re-scaling of one of the signals
such that the difference at high temperatures is zero.
While the initial experiments on YBCO at doping levels $\delta=0.5\ldots0.75$ were
somewhat controversial, the experiment was repeated on a larger sample of YBCO-6.6
\cite{fauque08a}, with consistent results:
The data suggest the magnetic order at wavevector $(0,0)$ at a doping-dependent temperature
which varies from 300\,K for YBCO-6.5 to 170\,K for YBCO-6.75; these values appear to
match the accepted pseudogap temperatures $T^\ast$ for these samples.
(Note that the subtraction procedure renders the determination of a sharp onset temperature
difficult.)
Recently, a similar neutron scattering signature of magnetic order in the pseudogap state
was obtained in the tetragonal single-layer HgBa$_2$CuO$_{4+\delta}$ compound \cite{fauque08b}.

In all cases, the moment amplitude is of order $0.1 \mu_B$, and the moments are
oriented roughly in a 45$^\circ$ angle w.r.t. the planes.
As susceptibility measurements appear to exclude ferromagnetic order of this magnitude,
the ordered moments within a unit cell apparently compensate each other.
One possibility is spin moments on oxygen atoms with opposite directions,
another one is given by the loop-current order of Varma \cite{varma99,varma02}.
In the latter case, however, the moment directions suggest that the current loops
involve out-of-plane oxygen orbitals.

A number of other experiments are worth mentioning:
Early ARPES experiments on BSCCO using circularly polarized photons reported
a dichroic signal indicating time-reversal symmetry breaking in the pseudo-gap
state \cite{kaminski02} (however, at the time, the result was questioned by others).
A search for static fields in the pseudogap regime using $\mu$SR of LSCO gave a null
result \cite{luke08}.
Static screening of the muon charge, leading to a local change in the doping level,
has been invoked to explain the absence of a signal \cite{shekhter08}.
However, a null result was also reported from NMR measurements in the pseudogap
regime of Y$_2$Ba$_4$Cu$_7$O$_{15-\delta}$ \cite{straessle08}.
Very recently, extremely sensitive polar Kerr effect measurements detected signatures of
broken time reversal in the pseudogap regime in a series of underdoped YBCO
crystals \cite{kerr08}.
The straightforward interpretation of the Kerr signal is in terms of {\em ferro}magnetic
order, however, with a tiny magnitude of $10^{-5}\,\mu_B$ per Cu atom.
A puzzling aspect of the Kerr measurements is that memory effects have been
found to survive up to room temperature far above $T^\ast$.
At present, the relation of the Kerr effect to the neutron observations is not settled.
Finally, careful measurements of the magnetic susceptibility in underdoped YBCO
revealed a small kink in samples with $0.4<\delta<0.8$ \cite{monod08}.
The kink may signify a thermodynamic phase transition and occurs at a doping-dependent
temperature which appears to match the onset temperature of the described neutron
signal.

Although these results are exciting, further experiments are needed to check whether
loop-current order is indeed a common feature of high-temperature superconducting cuprates,
and also to establish the relation between different possible experimental signatures.
A brief theoretical discussion of loop-current order is in Sec.~\ref{sec:th_varma}.


\subsection{Inelastic neutron scattering}
\label{sec:exp_inelscatt}

High-resolution inelastic neutron scattering, used to probe the spectrum of magnetic
fluctuations as well as of phonons, has been performed extensively on 214 cuprates, on
YBCO, and, to a lesser extend, on BSCCO.
For other materials, the lack of sufficiently large single crystals limits the available
data.

While the most direct oberservation of translation symmetry breaking is
via Bragg peaks in elastic scattering, the spectrum of finite-energy excitations can provide
information complementary to that of elastic scattering probes,
such as the energy range and character of the fluctuations.
Moreover, the presence of excitations at very low energies and specific wavevectors
is usually a precursor of an ordered state, i.e., incommensurate low-energy spin
excitations will occur close to stripe ordering.

Inelastic scattering is also a suitable probe for rotation symmetry breaking:
An anisotropic fluctuation spectrum must arise from an anisotropic state, as the local
excitation created by the external perturbation cannot change the symmetry.
As explained above, such a situation can only be expected in the presence of a
(structural) anisotropy field, as otherwise both types of domains will be present
with equal weight.
The candidate material is YBCO, where indeed signatures of rotation symmetry breaking
have been identified, see Sec.~\ref{sec:exp_nematic}.

\subsubsection{Magnetic excitations of ordered stripe phases}


\begin{figure*}
\ifhires
\includegraphics[width=2.6in]{cuts_jt04.eps}
\hspace*{10pt}
\includegraphics[width=2.6in]{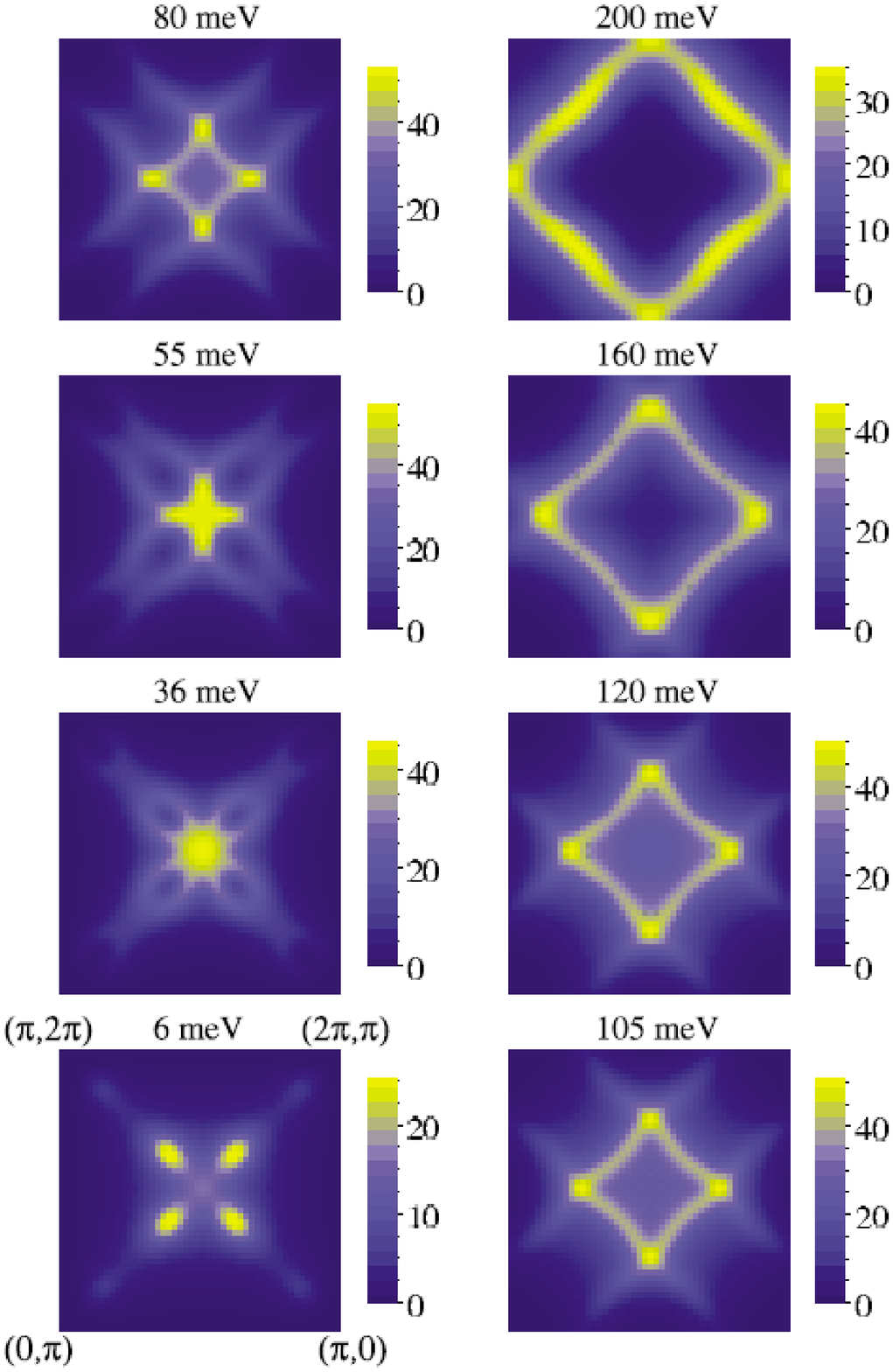}
\else
\includegraphics[width=2.6in]{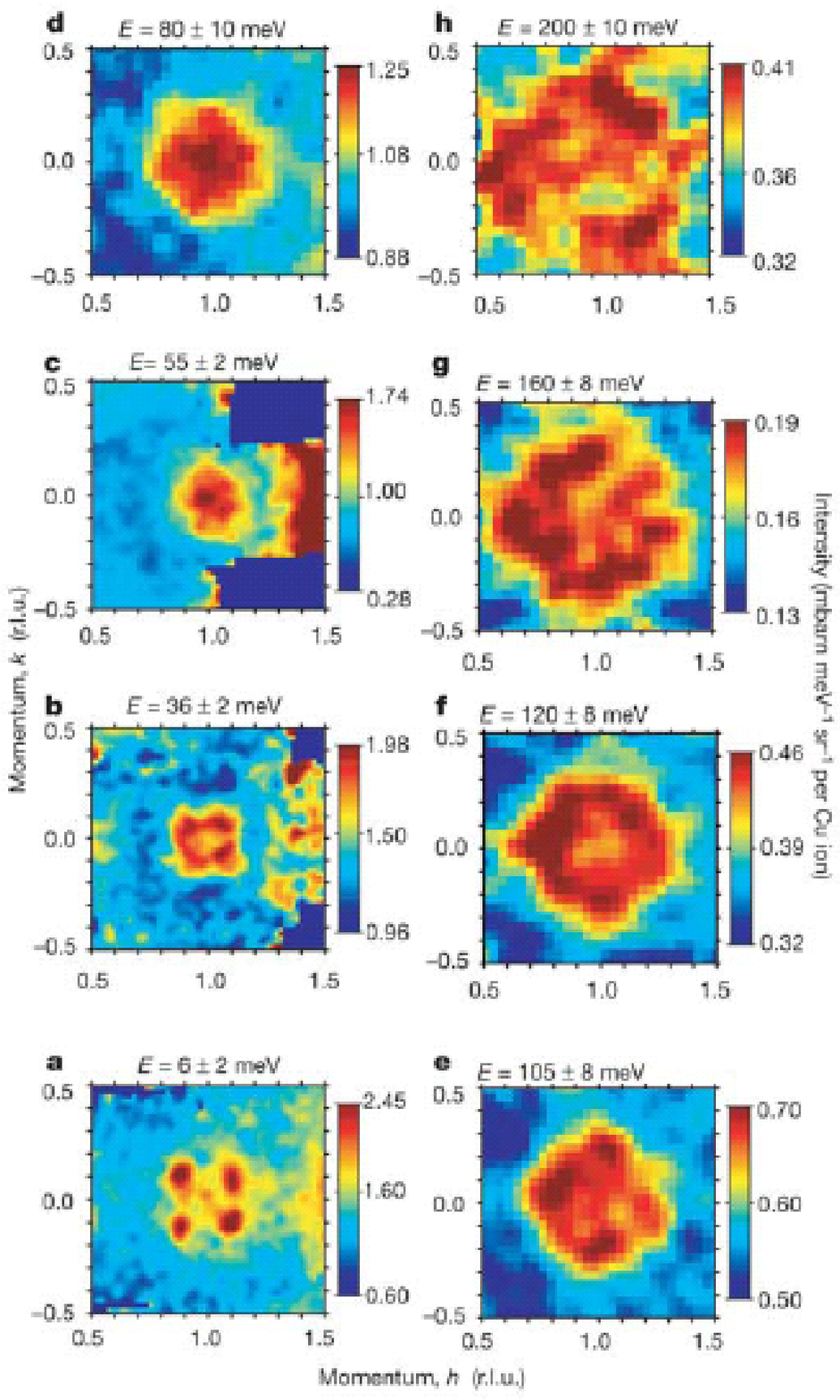}
\hspace*{10pt}
\includegraphics[width=2.6in]{cuts_mv04b.eps}
\fi
\caption{
Left:
Neutron-scattering results from ordered stripes in LBCO-1/8
(from Ref.~\cite{jt04}, reprinted by permission from Macmillan Publishers Ltd:
Nature {\bf 429}, 534, copyright (2004)).
Shown are constant-energy cuts through the magnetic excitation spectrum $\chi_s''({\vec q},\w)$.
Energy has been integrated over the ranges indicated by the error bars.
Panels a--c were measured with an incident neutron energy of 80\,meV,
panels d--g with\,240 meV, and panel h with 500\,meV.
Right: Theoretical result from a model of coupled spin ladders
(reprinted with permission from Ref.~\cite{mvtu04}, copyright 1999 by the American Physical Society),
as described in Sec.~\ref{sec:eff_spin}. The signals from horizontal and vertical stripes
have been added.
(Results obtained in Refs.~\cite{gsu04,seibold05,carlson06} are very similar.)
Note that only the reduced Brillouin zone of $(\pi,\pi)$ magnetic order is shown.
}
\label{fig:ecuts}
\end{figure*}

For static spin stripe order, the existence of low-energy magnetic excitations (spin waves)
follows from the Goldstone theorem.\footnote{
Magnetic anisotropies will induce a small gap for lattice-pinned commensurate order.}
Low-energy incommensurate excitations have been probed in a number of stripe-ordered 214
cuprates, but extensive data over a wide range of energies are available only for LBCO-1/8.

Such data, taken by Tranquada {\em et al.} \cite{jt04}, are shown in
Fig.~\ref{fig:ecuts}. Remarkably, rather well-defined collective excitations are visible
at all energies up to 200\,meV.
At low energies, four spots near $(\pi\pm\pi/4,\pi)$ and $(\pi,\pi\pm\pi/4)$,
i.e. the ordering wavevector, are observed.
The spots are found to disperse towards $(\pi,\pi)$ with increasing energy,
where they meet at around 50\,meV.
This is essentially consistent with spin-wave theory, although the expected
spin-wave cones (i.e. intense ellipses in constant-energy scans) are never observed,
possibly due to a combination of broadening and matrix element effects.
Interestingly, the spectrum at elevated energies does not appear to follow a
simple spin-wave dispersion. Instead, the excitation branch above 50\,meV is very similar
to that of a two-leg spin ladder with an exchange constant $J \approx 100$\,meV \cite{jt04}.
Constant-energy cuts show four intense spots in diagonal direction from $(\pi,\pi)$,
i.e., the scattering pattern has rotated by 45$^\circ$ from low to high energies.
At an energy of 200\,meV, the experimental intensity distribution has reached the
boundary of the magnetic Brillouin zone, but at the same time becomes damped
rather strongly.
The location of the intensity maxima in constant-energy cuts, shown in
Fig.~\ref{fig:ecuts}, trace out an ``hour-glass'' dispersion, Fig.~\ref{fig:hourglass}.


As will be discussed in detail in Sec.~\ref{sec:eff_stripemag}, the low-temperature
excitation spectrum of LBCO-1/8 can be nicely described by simple models of coupled spin
ladders \cite{mvtu04,gsu04,seibold05,carlson06},
provided that the response of horizontal and vertical stripes is summed up, Fig.~\ref{fig:ecuts}.
In the absence of perfect charge order, a model of fluctuating (or disordered) stripes
appears more appropriate \cite{vvk}, which can account for the data as well.

\subsubsection{Incommensurate spin excitations and hour-glass spectrum}
\label{sec:exp_hour}

Investigations of the spin excitation spectrum have played a prominent role in cuprate
research, mainly because spin fluctuations are a candidate for the glue that binds the
cooper pairs.

Historically, the observation \cite{respeak1} of the so-called ``resonance mode'' in optimally doped YBCO
below $\Tc$, located at 41\,meV and $\vec Q=(\pi,\pi)$, triggered enormous activities.
The doping dependence of the resonance energy was mapped out \cite{respeak1b,respeak1c},
and a similar resonance was also found in other cuprates \cite{respeak2,respeak3}.
On the other hand, the 214 family of cuprates displayed low-energy excitations at
incommensurate wavevectors \cite{yamada98,waki99,waki00}.
As a result, it was believed that the underlying magnetism is very different,
and that the behavior of 214 compounds is rather special.

This view was challenged by a series of detailed neutron scattering
experiments \cite{jt04,hinkov04,buyers04,buyers05,hayden04,face,christensen04,vignolle07},
which mapped out the spin excitations of various, mainly underdoped, cuprates over a
wide energy range.
For YBCO, dispersive excitations were found which emanate from the resonance peak and
disperse both upwards and downwards in energy \cite{hinkov04,buyers04,buyers05,hayden04,face}.
For stripe-ordered LBCO, the incommensurate low-energy excitations were found to
merge at $(\pi,\pi)$ at 50\,meV, and an upper excitation branch emerges which is well
described by the spectrum of a spin ladder \cite{jt04}, as discussed above.
For LSCO a similar excitation structure was found, with incommensurate low-energy
excitations dispersing towards $(\pi,\pi)$ and dispersing outwards again above
50\,meV \cite{christensen04,vignolle07,lips08}.
Very recently, signatures of an hour-glass spectrum were reported as well for optimally
doped BSCCO, although the features in constant-energy cuts are relatively broad \cite{xu09}.

\begin{figure}
\begin{center}
\ifhires
\includegraphics[width=2.5in]{hourglass1.eps}
\else
\includegraphics[width=2.5in]{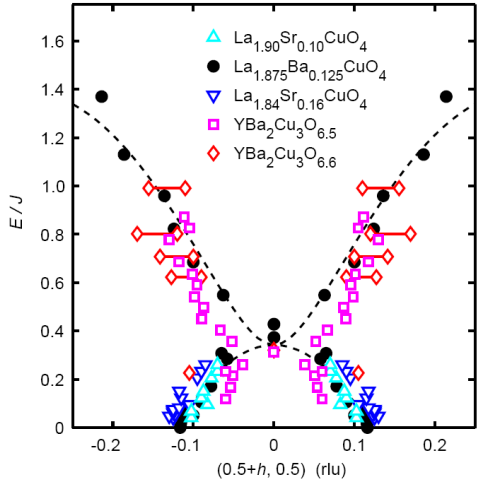}
\fi
\caption{
Universal hour-glass disperison of magnetic excitations
in cuprates
(Fig. 6.3 of Ref.~\cite{jt06}, reproduced with kind permission of
Springer Science and Business Media).
Comparison of measured dispersions along ${\vec Q} = 2\pi
(0.5+h,0.5)$ in LSCO-0.10 (up triangles) and LSCO-0.16 (down
triangles) from Christensen {\it et al.} \cite{christensen04},
in LBCO-1/8 (filled circles) from \cite{jt04},
and in YBCO-6.5 (squares) from Stock {\it et al.} \cite{buyers05}
and YBCO-6.6 (diamonds) from Hayden {\it et al.} \cite{hayden04}.
The energy has been scaled by the superexchange energy $J$ for the appropriate
parent insulator.
For YBCO-6.6, the data at higher energies were fit along the [1,1] direction;
the doubled symbols with bars indicate two different ways of interpolating the
results for the [1,0] direction. The upwardly-dispersing dashed curve
corresponds to the result for a two-leg spin ladder,
with an effective superexchange of $\sim\frac23 J$; the
downward curve is a guide to the eye.
}
\label{fig:hourglass}
\end{center}
\end{figure}

Taken together, these experiments provide evidence for an excitation spectrum of
hour-glass form being common to many cuprates. Indeed, plotting the scattering intensity
maxima as function of momentum and re-scaled energy, i.e., relative to the exchange
constant of parent compound, results in Fig.~\ref{fig:hourglass}, which suggests
{\em universality} of the hour-glass spectrum. In addition, it shows that the relevant
energy scale for magnetic fluctuations is $J$.

This prompts the question for a common microscopic origin of the excitation spectrum.
While a weak-coupling description based on RPA captures some features of the YBCO data,
a picture of stripes is clearly appropriate for 214 compounds where
low-energy incommensurate excitations are clear precursors of ordered stripes,
Fig.~\ref{fig:braggsdw}.
It has been proposed early on that dynamic stripes are responsible for incommensurate
excitations below the resonance energy in YBCO as well \cite{mook00,jt05}, but opposite
views were also put forward \cite{hinkov04}.
I shall return to this discussion in Sec.~\ref{sec:eff_gappedmag}.

Despite the similarities there are, however, a number of important differences between
the cuprate families which should be mentioned:
(i) While YBCO displays a sizeable spin gap at low temperatures except for small dopings
below $\delta<0.5$, the spin gap in 214 materials is small.
(ii) In YBCO, constant-energy cuts at the resonance energy show a very sharp peak at
$(\pi,\pi)$ \cite{hayden04,hinkov04,hinkov07,buyers04,buyers05}. In contrast, in LSCO the
peak is broader and well fit by the sum of two incommensurate peaks
\cite{vignolle07,lips08}.
(iii) In YBCO, again with the exception of small dopings below $\delta<0.5$, the spectra
above and below $\Tc$ are significantly different. At optimal doping, a clear spin gap
opens below $\Tc$ together with the resonance appearing, while most of the structured
response disappears above $\Tc$.
In both YBCO-6.6 and YBCO-6.95, the lower branch of the hour glass is only
present below $\Tc$, but strongly smeared above $\Tc$ \cite{hinkov07,reznik08c}.
In contrast, in
LSCO and LBCO the spectral changes at $\Tc$ are weaker \cite{jt04,jt04b,chang07,lips08}.
(iv) To my knowledge, an hour-glass dispersion has not been measured in other cuprates,
with the exception of optimally doped BSCCO \cite{xu09},
mainly due to the lack of large single crystals required for neutron scattering.

It should be pointed out that the approximate universality of the hour-glass spectrum,
when plotted as in Fig.~\ref{fig:hourglass}, only applies to doping levels between
10\% and optimal doping.
For smaller dopings, both the incommensurability $\epsilon_s$ and the energy $E_{\rm cross}$,
where upward and downward dispersing branches meet, decrease.
In LSCO for $x<1/8$, both vary roughly linearly with doping, and
$E_{\rm cross}/400\,{\rm meV} \approx x$ has been reported.
Interestingly, this also includes the insulating small-doping phase of LSCO,
which is characterized by static order with diagonal incommensurate peaks and
displays an hour-glass-like spectrum as well \cite{matsuda08}.

For LSCO at large doping, incommensurate low-energy spin excitations have been found
to survive throughout the entire superconducting phase, disappearing only at $x\!=\!0.30$
where the sample becomes non-superconducting \cite{waki04}.
In this overdoped regime, the wavevector-integrated magnetic intensity below 60\,meV
drops with doping, and almost vanishes at $x\!=\!0.30$ as well \cite{waki07}.
These findings point to an intriguing relation between magnetism and superconductivity.
Moreover, they strongly argue against simple Fermi surface nesting as source of
magnetic fluctuations even in overdoped cuprates, because the magnetic intensity drops
within a small doping window despite the Fermi surface becoming more well defined with
increasing $x$ \cite{waki07}.

\subsubsection{Phonon anomalies}
\label{sec:phonons}

Charge order naturally couples to the lattice:
Static charge order will lead to periodic atomic displacements and cause backfolding of
the phonon branches. One can expect that slowly fluctuating charge order induces
corresponding precursors, which should be manifest in anomalies in the phonon dispersion
and linewidth near the charge ordering wavevector.
Hence, strong stripe-related effects should be visible in a momentum scan of the phonon dispersion
{\em perpendicular} to the stripe direction, with a strength dictated by the overall
electron-phonon coupling and by matrix elements which depend on the specifics of the charge order
and the phonon mode.

While various reports on phonon anomalies are in the literature, it has not been
conclusively established which of them are related to stripe order. Essentially all
studies focused on optical phonon branches, which exist in the energy range of $60 \ldots
100$ meV and can be easily distinguished from other excitations seen in neutron
scattering.

%
For LSCO at $x\!=\!0.15$, McQueeney {\em et al.} \cite{mcq99} reported a discontinuity in
the longitudinal optical (LO) dispersion branch at 10\,K, which was interpreted as
evidence for unit cell doubling. However, a subsequent study \cite{braden99} of the same
compound arrived at a different conclusion: The data speak in favor of a continuous
dispersion of this bond-stretching mode. In addition, the strongest broadening was
observed near wavevectors $(0.5\pi\ldots0.6\pi,0)$, which can be taken as a precursor to
stripe ordering with a periodicity of $4\ldots3$ lattice spacings. The softening of the
LO phonon at $(\pi/2,0)$ is in fact found in LSCO over a wide doping range, including
strongly overdoped samples at $x\!=\!0.29$ \cite{fukuda05}, suggesting that it is unrelated
to translational symmetry breaking tendencies.
For YBCO, conflicting results on phonon anomalies have been reported
\cite{mook00,pintsch02}, their relation to stripe order being unclear at present.

A comprehensive neutron-scattering analysis of bond-stretching phonons in stripe-ordered
cuprates was performed by Reznik {\em et al.} \cite{reznik06,reznik07}.
Both LBCO and LNSCO at doping $x\!=\!1/8$, known to have static stripe order, displayed strong anomalies
near wavevector $(\pi/2,0)$. The broad lineshapes were interpreted in terms of two (instead
of one) phonon branches. A subsequent x-ray scattering study with higher resolution
\cite{reznik08a} essentially confirmed the broad lineshapes, but showed that the two-peak
interpretation is not justified.
The observed phonon anomaly is strongest at the lowest $T$ of 10\,K, and the phonon lines sharpen with
increasing temperature. The authors extended the measurements to LSCO at doping levels
$x\!=\!0.07$, 0.15, and 0.3, and found broad lines at $(\pi/2,0)$ for the $x\!=\!0.07$ and 0.15
samples, while narrow lines where found in the non-superconducting $x\!=\!0.30$ compound.

It appears plausible to associate the bond-stretching phonon anomalies with the
tendency toward stripe order. A theoretical study of the phonon dynamics in the presence
of static stripes \cite{kaneshita02} seem to support this assertion, although the interplay of
{\em fluctuating} stripes and phonons has not been investigated theoretically.
There are , however, serious caveats with this interpretation:
Static stripes should cause multiple phonon branches due to backfolding,
which are not clearly observed experimentally (this could be related to matrix-element effects).
More importantly, the stripe ordering wavevector in the 214 compounds is known to vary
with doping, both in the spin and charge sector (see Fig.~\ref{fig:yamada}),
but the phonon anomalies do not show a similarly large shift in momentum space.
Also, a somewhat similar LO phonon anomaly was found \cite{reznik08b} in optimally
doped YBCO-6.95, where there is otherwise little evidence for stripe behavior.

An interesting alternative interpretation \cite{mukhin07,reznik08b} of the phonon
anomalies invokes one-dimensional physics: It has been suggested that
the observed anomalies are due to $2k_F$ effects of essentially one-dimensional metallic
stripes, and hence occur in a direction {\em parallel} to the stripes.
This scenario would explain the weak doping dependence of the anomaly wavevector,
as, in such a one-dimensional picture, $k_F$ within a stripe is expected to vary weakly
with $x$ for $x\leq 1/8$. It remains open why no backfolding effects are observed.

In the context of phonons coupling to stripes, thermal conductivity measurements may
provide additional information. The phonon thermal conductivity of LSCO was found to be
strongly suppressed at low temperature, with the suppression being correlated with
superconductivity \cite{heat98}. Remarkably, the suppression was absent in
non-superconducting stripe-ordered LNSCO and LESCO. While these findings were originally
interpreted \cite{heat98} as evidence for phonon scattering off fluctuating stripes, a
careful re-analysis \cite{heat03} showed that soft phonons, caused by the structural
instability of the LTO phase, provide a scattering mechanism which can account for the
observed suppression of heat transport. Conceptually, a separation of soft-phonon-induced
and soft-stripe-induced scattering is difficult.

As there can be no doubt about a strong coupling of stripes to the lattice,
as evidenced e.g. by the large isotope effect \cite{takagi_iso} in stripe-ordered LESCO,
more experimental and theoretical work is required to elucidate the interplay of stripes
and phonons.


\subsubsection{Anisotropic magnetic spectra and nematic order}
\label{sec:exp_nematic}

A state with broken rotational symmetry causes a neutron scattering intensity
$\chi_s''(\vec{q},\w)$ with directional $\vec{q}$ space anisotropy, and vice versa.
However, such a spectral asymmetry is wiped out if domains of different orientation
co-exist and are probed by the neutron beam.
This complication can be avoided by a small symmetry-breaking field.
Among the cuprates, de-twinned crystals of YBCO have precisely this property:
For dopings $\delta\geq0.4$ the presence of the CuO chains induces a structural
orthorhombic distortion, which should be able to align nematic domains.

In a remarkable experimental effort,
the spin-fluctuation spectrum of de-twinned superconducting YBCO crystals has been
studied in detail by Hinkov {\em et al.} \cite{hinkov04,hinkov07,hinkov08a}.
In YBCO, one expects bilayer splitting of all magnetic modes; all
the following information apply to magnetic excitations which are odd under bilayer
exchange unless otherwise noted.

In moderately underdoped YBCO-6.85 and YBCO-6.6 the spin excitations are gapped.
Below $\Tc$, the spectrum is consistent with the hour-glass shape
described above. While its high-energy part above the resonance energy approximately
obeys the square-lattice symmetry, the low-energy part is significantly anisotropic
in both compounds \cite{hinkov04,hinkov07}:
In YBCO-6.6, the intensity at 33\,meV displays strong incommensurate peaks along the a axis,
whereas the corresponding peaks along the b axis are weak \cite{hinkov07} -- this
anisotropy may be consistent with nematic behavior.
However, as function of temperature, the spectra change smoothly except at the
superconducting $\Tc$, i.e., no signature of a nematic ordering transition has been
detected.
The results of Hinkov {\em et al.} \cite{hinkov04,hinkov07} are consistent with earlier reports of anisotropic
spin fluctuations at 24\,meV on a partially de-twinned YBCO-6.6 crystal \cite{mook00}.

\begin{figure*}
\ifhires
\includegraphics[width=2.4in]{nema_cuts1.eps}
\includegraphics[width=3.0in]{nema_pd1.eps}
\else
\includegraphics[width=2.4in]{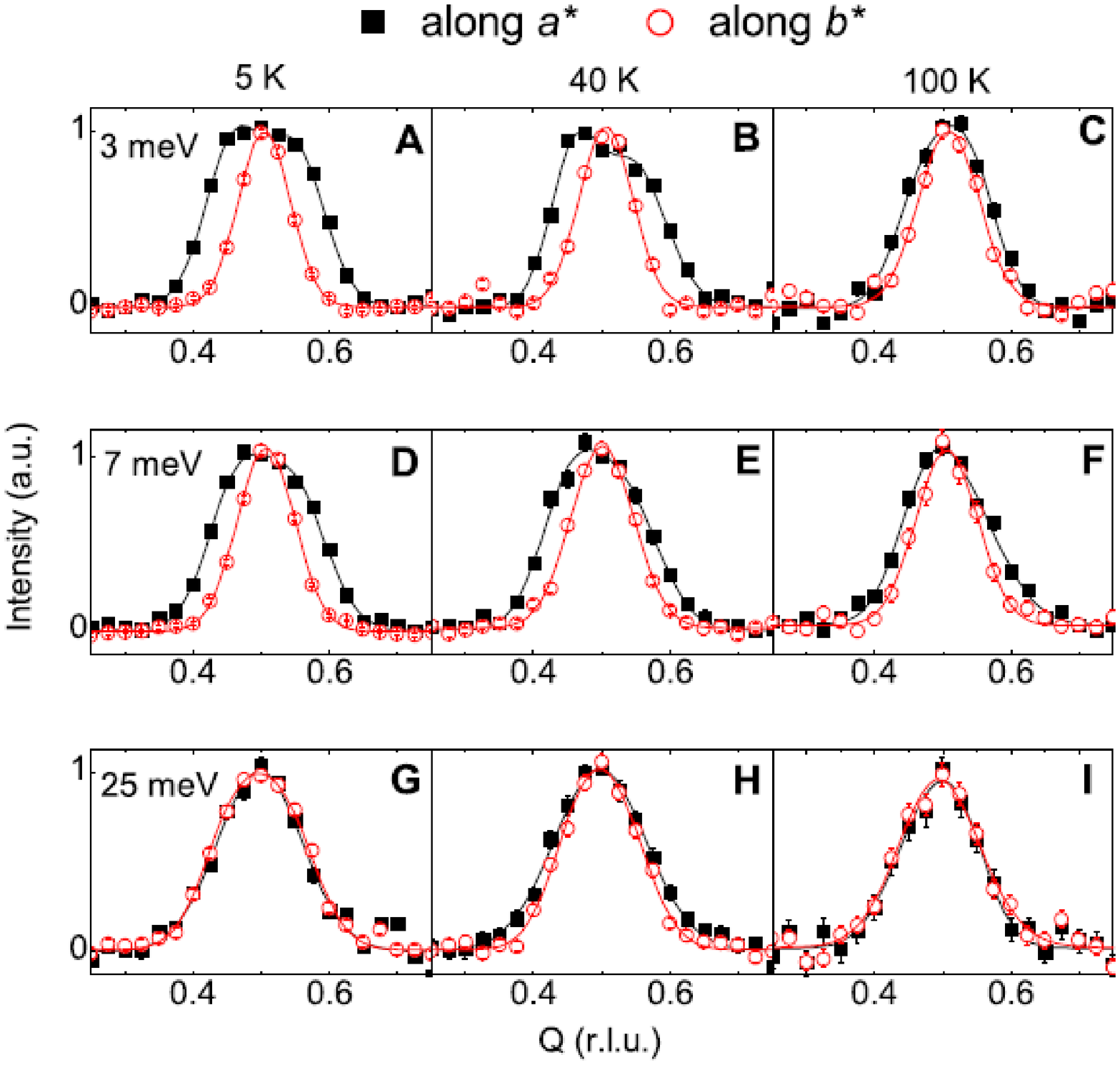}
\includegraphics[width=3.0in]{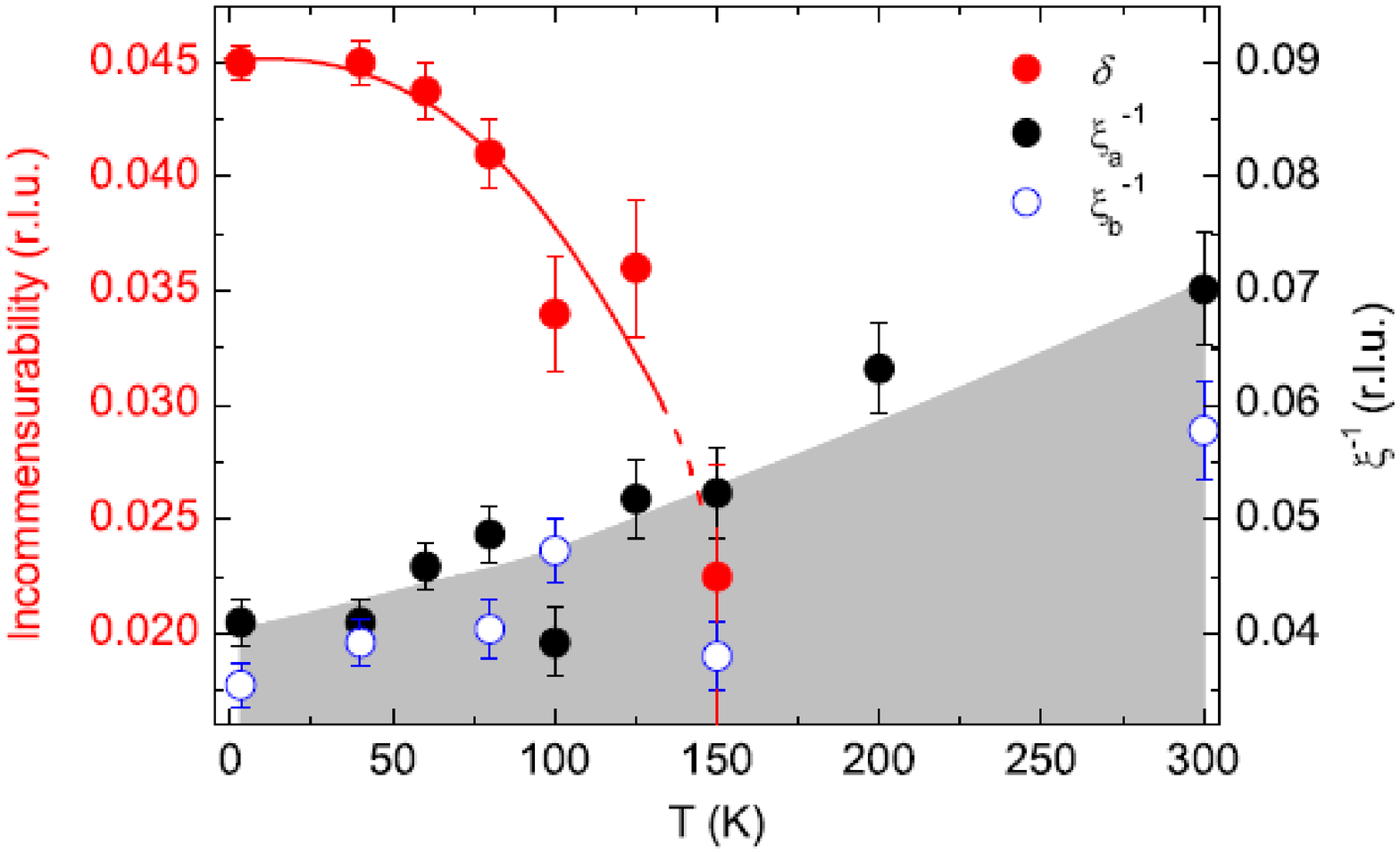}
\fi
\caption{
Signatures of nematic order in YBCO-6.45
(from Ref.~\cite{hinkov08a}, Science {\bf 319}, 597 (2008), reprinted with permission from AAAS).
Left: Energy and temperature evolution of
the a-b anisotropy of the spin correlations.
Full squares and empty circles represent data points
measured at fixed K along a and at
fixed H along b, respectively.
Scans are normalized and background-subtracted.
Solid lines represent the results of
fits with one or two Gaussians.
Right:
Temperature and energy evolution of parameters characterizing
the spin excitation spectrum. The parameters are the results of fits to the
raw data.
Shown are incommensurability $\delta\equiv\epsilon_s$ (red symbols),
half-width-at-half-maximum of the incommensurate peaks along a
($\xi_a^{-1}$, black symbols) and along b ($\xi_b^{-1}$, open blue
symbols) in reciprocal lattice units, measured at 3\,meV.
}
\label{fig:ins_nematic}
\end{figure*}

More intriguing is the behavior in strongly underdoped YBCO-6.45 \cite{hinkov08a}
with $\Tc\!=\!35\,K$:
At this doping concentration, the neutron spectrum is essentially gapless at 40\,K
and below.
At low energies of e.g. 3 and 7\,meV, the intensity distribution in $\chi_s''(\vec{q},w)$
takes the form of an ellipse in momentum space around $(\pi,\pi)$,
i.e., the intensity is broadly distributed
along the a axis while it is less broad along the b axis, Fig.~\ref{fig:ins_nematic}.
At energies of 3\,meV and below, the intensity along the a axis is well be fitted by the
sum of two Lorentzians at incommensurate wavevectors $2\pi(0.5\pm\epsilon_s,0.5)$.
Plotting the incommensurability $\epsilon_s$ at 3\,meV as function of temperature shows an
order-parameter-like behavior, i.e., it decreases with increasing temperature and
vanishes at around 150\,K, i.e., far above the superconducting $\Tc$,
Fig.~\ref{fig:ins_nematic}.
Although the orthorhombic structural distortion is expected to smear out a sharp nematic phase
transition, a reasonable interpretation of the data is in terms of a spontaneous onset of {\em both}
magnetic incommensurability and magnetic anisotropy of the electronic system at
around 150\,K.
This is suggestive of a nematic phase transition.
The strong increase of the in-plane transport anisotropy below 200\,K \cite{ando02}
of a YBCO sample of similar doping level appears consistent with this
interpretation, see Sec.~\ref{sec:exp_trans}.
(Whether the NMR/NQR work on YBCO of Ref.~\cite{haase03}, showing two inequivalent
planar O sites but only a single type of Cu site, is related to nematic order is not
known.)

The quasi-static neutron signal shows a significant upturn below 30\,K,
but, according to $\mu$SR, static magnetic order sets in only below 2\,K,
with an ordered moment of about 0.05 $\mu_B$. This is consistent with very slow, perhaps
glassy, spin dynamics at low temperatures \cite{hinkov08a}.

Interpreting the incommensurate magnetism in YBCO-6.45 in terms of stripes, then the
stripes run along the b direction, which appears consistent with the resistivity
anisotropy, $\rho_a>\rho_b$.
However, as noted above, reports on static charge order in underdoped YBCO are
controversial.

The anisotropies of the spin fluctuation spectra in YBCO suggest the existence of a
nematic QCP around $\delta_c=0.5$ \cite{hinkov08a}
(although one cannot exclude the possibility that nematic transitions also
exist at larger $\delta$, but with a strong smearing due to the large orthorhombicity).
A magnetic QCP may exist at the same or a slightly
smaller doping, but static magnetic order for $\delta<\delta_c$ and zero field
\cite{haug_mf} appears to be restricted to extremely low temperatures,
while nematic order extends well into the pseudogap regime.


\subsection{Scanning tunneling microscopy}
\label{sec:exp_stm}

The techniques of scanning tunneling microscopy and spectroscopy (STM/STS) have contributed
enormously to the exciting progress in the field of cuprates over the past
decade \cite{stm_rmp}.
With the caveat of being only sensitive to the physics of the sample surface,
STM has provided real-space images of apparently intrinsic inhomogeneities in BSCCO,
allowed for a detailed analysis of local impurity physics, and unraveled the tendency
towards ordering phenomena accompanied by lattice symmetry breaking -- the latter shall
be summarized here.
As STM is essentially a static measurement, ``fluctuating'' stripes can only be detected
if pinned by impurities.

High-resolution STM experiments require high-quality sample surfaces, which currently
restricts the application to BSCCO and CCOC. Although STM data have been reported for
YBCO and 214 compounds as well, atomic resolution is often not achieved, and the quality
of the surface layer can be problematic.

STM experiments measure the current $I$ between tip and sample as function of voltage $V$ and
position $\vec r$. Assuming an energy-independent electronic density of states in the tip,
the measured $dI/dV$ is equivalent to the spatially resolved local density of states (LDOS),
$\rho(\vec{r},E)$, up to an $\vec{r}$-dependent tunnel matrix element which depends on
the set-point conditions \cite{kohsaka07,hanaguri07}.
For cuprates, it is commonly assumed that the measured LDOS at low energies reflects the
properties of the CuO$_2$ layers, as the additional layers between crystal surface and
topmost CuO$_2$ plane are insulating. However, the tunneling path through those layers may
non-trivial, as discussed in Sec.~\ref{sec:th_stm}.

\subsubsection{Quasiparticle interference vs. charge order}

In order to extract possibly periodic signals from LDOS maps $\rho(\vec{r},E)$, a Fourier
transformation to momentum space is routinely used. The resulting quantity
$\rho(\vec{k},E)$, dubbed FT-LDOS, can show well-defined structures for various
reasons.\footnote{
The analysis of the FT-LDOS is usually restricted to its power spectrum
$|\rho(\vec{k},E)|^2$.
}
(i) If a modulation in the charge sector with wavevector $\vec Q$ is present,
then $\rho(\vec{k},E)$ will show ``Bragg'' peaks at $\vec k\!=\!\vec Q$ at all energies $E$.
(ii) Friedel oscillations caused by impurities will contribute to
momentum-space structures in $\rho(\vec{k},E)$ as well.
Importantly, these structures will be energy-dependent due to the energy--momentum
dispersion of the single-particle excitations \cite{capriotti03}.

The first observation of such generalized Friedel oscillations,
or ``quasiparticle interference'' (QPI), made by Hoffman {\em et al.} \cite{hoffman02b}
in the superconducting state of optimally doped BSCCO,
was subsequently confirmed and extended \cite{mcelroy03,mcelroy05,mcelroy06}.
In interpreting the experimental data, a simple recipe, dubbed ``octet model'', was
used to extract information about the single-particle dispersion from the peak locations in
the FT-LDOS.
The extracted Fermi surface and superconducting gap show good agreement with results from
ARPES \cite{mcelroy06}, which is remarkable, as STM is usually not a probe with
momentum-space resolution.
There are, however, a number of problems with the octet-model interpretation of QPI,
and we will give a theoretical discussion in Sec.~\ref{sec:th_stm}.

The presence of QPI phenomena poses a serious problem in the search for charge order:
As doped cuprates are intrinsically dirty, the QPI signals are not weak and therefore are
not easy to disentangle from modulations due to collective charge ordering,
in particular if the charge order is only short-ranged due to strong disorder pinning.
Therefore, the experimental search for charge order concentrated on identifying
non-dispersive peaks in the FT-LDOS $\rho(\vec{k},E)$, in particular near the wavevectors
$(\pi/2,0)$, $(0,\pi/2)$ corresponding to period-4 charge order as known
from 214 cuprates.

\subsubsection{LDOS modulations}

Spatial modulations in the $dI/dV$ signal, suggestive of charge order,
were first detected near the vortex cores in slightly overdoped BSCCO in a applied
field of 7\,T \cite{hoffman02a}. The spatial pattern resembled
a checkerboard, for more details see Sec.~\ref{sec:fieldtune} below.

\ifrmp
\begin{figure}[!b]
\else
\begin{figure}
\fi
\begin{center}
\includegraphics[width=3.2in]{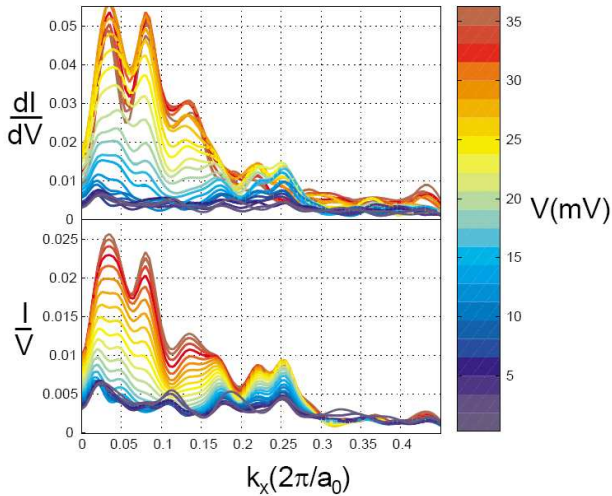}
\caption{
FT-LDOS data, $\rho(\vec{k},E)$, of optimally BSCCO, taken at 8\,K
(reprinted with permission from Ref.~\cite{kapi03a}, copyright 2003 by the American Physical Society).
Shown are line scans as a
function of $k_x$ along the $(1,0)$ direction,
and as a function of energy (color scale).
Top: LDOS ($dI/dV$).
Bottom: LDOS integrated up to the given energy ($I/V$).
The peak at $2\pi(0.25,0)$ displays little dispersion.
}
\label{fig:kapi}
\end{center}
\end{figure}

Subsequently, different measurements of the FT-LDOS in zero magnetic field
in optimally doped and underdoped BSCCO lead to some controversy:
While Howald {\em et al.} \cite{kapi03a,kapi03b} and Fang {\em et al.} \cite{kapi04}
interpreted their data as evidence for an underlying charge-density modulation
(co-existing with QPI features),
Hoffman {\em et al.} \cite{hoffman02b} and McElroy {\em et al.} \cite{mcelroy03}
asserted that their data are consistent with QPI, not showing signs of a CDW.
In this debate, Howald {\em et al.} have pointed out that the peaks in the FT-LDOS
which were associated with QPI in Refs.~\cite{hoffman02b,mcelroy03} display a much
weaker dispersion below 15\,meV compared to what is expected from the octet
model (Fig.~\ref{fig:kapi}).
Whether this is simply a failure of the octet model or indeed evidence for charge
ordering is difficult to decide, see Sec.~\ref{sec:th_stm}.
Moreover, Fang {\em et al.} \cite{kapi04} show that the height of the superconducting coherence
peaks display a modulation very similar to that of the low-energy LDOS, with a period of
$4.5\,a_0$.

\begin{figure*}
\begin{center}
\ifhires
\includegraphics[width=4in]{yazdani1.eps}
\else
\includegraphics[width=4in]{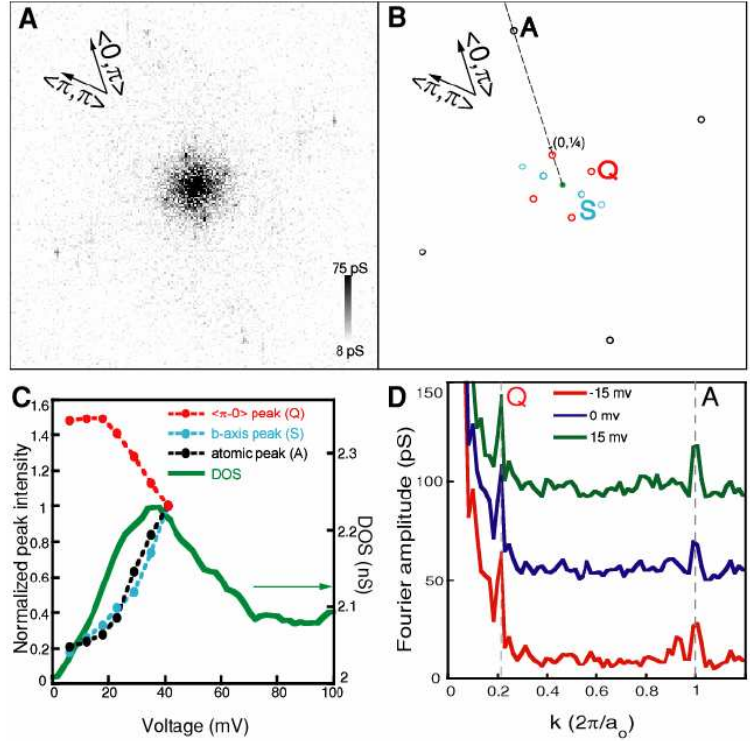}
\fi
\caption{
Fourier analysis of DOS modulations measured in slightly underdoped BSCCO ($\Tc=80$\,K)
in the pseudogap regime at 100\,K
(from Ref.~\cite{yazdani04}, Science {\bf 303}, 1995 (2004), reprinted with permission from AAAS).
(A) Fast Fourier transform (FFT) of an unprocessed conductance map acquired over a 380 \AA\ by 380 \AA\
field of view at 15 mV.
(B) The FFT has peaks corresponding to atomic sites (colored black and labeled A), primary (at $2\pi/6.8a_0$) and secondary
peaks corresponding to the b-axis supermodulation (colored cyan and labeled S), and peaks
at $\approx 2\pi/4.7a_0$ along the $(H,0)$ and $(0,H)$ directions (colored red and labeled Q).
(C) The energy evolution of the peaks in (B), scaled by their respective
magnitudes at 41 mV.
(D) Two-pixel-averaged FFT profiles taken along the dashed line in (B) for the DOS measurement at 15 mV shown in (A) and
measurements acquired simultaneously at 0 mV and $-15$ mV. The positions of key peaks are shown
by dashed lines and labeled according to their location in (B).
}
\label{fig:yaz}
\end{center}
\end{figure*}

Unambiguous evidence for charge oder (albeit with a short correlation length)
came from LDOS measurements in the pseudogap regime of BSCCO by Vershinin {\em et al.}
\cite{yazdani04}. An underdoped BSCCO crystal with a $\Tc$ of 80\,K, measured at 100\,K,
displayed non-dispersive peaks in the FT-LDOS below 40\,meV, corresponding to a modulation
period of $4.7\,a_0$, Fig.~\ref{fig:yaz}.
The observed FT-LDOS signal was shown to be incompatible with QPI \cite{yazdani04,misra04},
essentially because the octet wavevectors $q_1$ and $q_5$ did not disperse at all.
(The measured data below $\Tc$ displayed dispersing peaks, compatible with
Ref.~\cite{hoffman02b}.)
A plausible interpretation is that QPI signatures become progressively weaker with
increasing temperature, as quasiparticles are ill-defined in the pseudogap regime over a
large fraction of the Brillouin zone. As a result, it becomes easier to identify
signatures of CDW formation. Whether, in addition, charge order is enhanced in the
pseudogap regime is unclear.

More recent STM measurements of various underdoped BSCCO samples have established that
non-dispersive FT-LDOS peaks also occur in the superconducting state, namely at elevated
energies above $30\ldots40$\,meV. While those peaks were originally associated with
pseudogap patches of the inhomogeneous sample \cite{mcelroy05}, they have been recently
argued to be generic features of the approach to the Mott insulator \cite{kohsaka08}
(see also next subsection).
A connection between strong gap inhomogeneities and the presence of non-dispersive
FT-LDOS peaks has been pointed out in Refs.~\cite{hashi06,liu07}:
In both the superconducting and the pseudogap regimes, non-dispersive peaks occurred predominantly
in strongly inhomogeneous samples or sample regions. This appears plausible under the
assumption that pinning due to disorder plays a dominant role in inducing the static
modulation.

A charge-order signal much stronger than that in BSCCO was identified in the low-temperature
LDOS of underdoped samples of CCOC with dopings $x\!=\!0.08\ldots 0.12$ \cite{hanaguri04}.
Here, the LDOS at all energies below the pseudogap energy of 100\,meV displayed a clearly
visible checkerboard modulation with spatial period 4.\footnote{
Early STM data taken on CCOC at a bias voltage of 200\,meV have been interpreted in
terms of nanoscale inhomogeneities \cite{kohsaka04}. The more recent high-quality data
of Ref.~\cite{kohsaka07} show that those inhomogeneities correspond to uni-directional
bond-centered modulation patterns.
}
ARPES experiments on similar samples of CCOC detected nearly nested, but incoherent, antinodal Fermi
surface pieces with almost doping-independent nesting wavevectors close to $(\pi/2,0)$,
$(0,\pi/2)$. The authors proposed that these antinodal regions are responsible for
charge ordering in the spirit of a weak-coupling scenario \cite{shen05}.

Very recently, non-dispersive FT-LDOS peaks at energies below 30\,meV were reported in single-layer
BSCCO-2201 \cite{wise08}. The modulations displayed a doping-dependent spatial period between
$4.5\,a_0$ (underdoped) and $6.2\,a_0$ (optimally doped).
This tendency is opposite from the one of the CDW wavevector in 214 compounds (where the modulation
period becomes smaller with increasing doping), and consequently a weak-coupling scenario has been
suggested as the origin of the modulations \cite{wise08}.

Field-induced LDOS signals near vortex cores have been investigated in more detail
in BSCCO \cite{levy05,matsuba07,takagi08} and also observed in YBCO \cite{yeh08},
and will be described in Sec.~\ref{sec:fieldtune} below.


In the experiments described so far, clear-cut evidence for (locally) broken rotation symmetry
was lacking (although the data of both Hoffman {\em et al.} \cite{hoffman02a} and
Howald {\em et al.} \cite{kapi03a} indicate a weak local breaking of $C_4$ symmetry).
Superficially, this might be more consistent with checkerboard than stripe order.
However, as discussed in Sec.~\ref{sec:dis}, pinning of stripes by impurities can result in
checkerboard (instead of stripe) patterns, in particular if the clean system is on the
disordered side of a stripe-ordering transition \cite{maestro,robertson}.

\subsubsection{Tunneling asymmetry}

A critique which has been voiced against the analysis of modulations in $dI/dV$,
as described in the last subsection,
is related to the possibly position-dependent tunneling matrix element
in the STM experiment.
The standard measurement protocol is to adjust the tip height at each position such
that a constant current flows at some (high) set-point voltage.
In the presence of significant charge inhomogeneities, the tip height will then
be modulated as a function of $\vec r$ as well, rendering $dI/dV$ inequivalent to
the LDOS. Moreover, this effect will depend on the set-point conditions.

To separate physical modulations from set-point effects,
it was proposed to study the LDOS ratios
\begin{equation}
Z(\vec{r},E) = \frac{\rho(\vec{r},E)}{\rho(\vec{r},-E)},~
R(\vec{r},E) = \frac{\int_0^E d\omega\rho(\vec{r},\omega)}{\int_{-E}^0 d\omega
\rho(\vec{r},\omega)}.
\end{equation}
The physical content of these ratios, which measure spectral particle--hole asymmetry,
is non-trivial.
For one-band models of weakly doped Mott insulators, both $Z$ and $R$
have been argued to be proportional to the hole density \cite{pwa_as,mohit_as}.

\begin{figure*}
\begin{center}
\ifhires
\includegraphics[width=4in]{kohsaka_vbc1.eps}
\else
\includegraphics[width=4in]{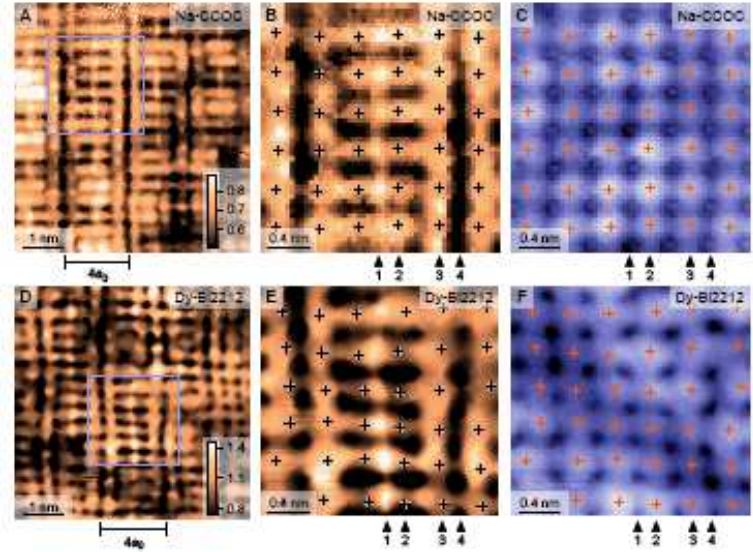}
\fi
\caption{
Valence-bond glass as seen in STM
(from Ref.~\cite{kohsaka07}, Science {\bf 315}, 1380 (2007), reprinted with permission from AAAS).
A) and D): R maps of Na-CCOC and Dy-Bi2212, respectively (taken at 150 mV).
The blue boxes in A), D) indicate areas shown in panels B), C) and E), F).
B and E: Higher-resolution R map within equivalent domains from Na-CCOC and Dy-
Bi2212, respectively. The locations of the Cu atoms are shown as
black crosses.
C) and F) Constant-current topographic images simultaneously taken with panels B)
and E), respectively. The markers show atomic locations, used also in B) and E).
}
\label{fig:vbglass}
\end{center}
\end{figure*}

Maps of $R(\vec{r},E)$, taken in underdoped BSCCO and CCOC, have been analyzed by Kohsaka
{\em et al.} \cite{kohsaka07}. The tunneling asymmetry at 150\,mV was found to be strongly
inhomogeneous, with a spatial pattern interpreted as ``electronic cluster glass''.
A number of properties of the observed $R(\vec{r},E)$ are indeed remarkable:
(i) The modulations are centered on the Cu-O-Cu bonds and are strongest on
the O (instead of Cu) sites, with a contrast of up to a factor of two.
(ii) The modulation pattern locally breaks to $C_4$ symmetry down to $C_2$, i.e.,
uni-directional domains, suggestive of stripe segments, are clearly visible.
(iii) The modulations in the $R$ map of BSCCO and CCOC are essentially indistinguishable
(although differences exist in the energy-resolved LDOS spectra).
In both cases, the spatial correlation length of the modulation pattern is of order 10
lattice spacings.
(iv) In both cases, the order co-exists with well-established superconductivity,
with a $\Tc$ of 45\,K and 21\,K for BSCCO and CCOC, respectively.

Subsequently, maps of $Z(\vec{r},E)$ were used to analyze QPI in slightly underdoped
CCOC \cite{hanaguri07} and strongly underdoped BSCCO \cite{kohsaka08}.
In CCOC, the spatial period-4 modulations are so strong that QPI features are hard to
detect in the LDOS \cite{hanaguri04}. In contrast, QPI peaks were observed in the LDOS
ratio $Z(\vec{k},E)$ below 20\,meV \cite{hanaguri07}.
The fact that $Z$ is more sensitive to QPI features than the LDOS itself was rationalized
arguing that charge order causes modulations in the LDOS which are approximately in-phase
between positive and negative energies, whereas QPI modulations occur approximately
anti-phase.
In BSCCO, dispersing QPI peaks in $Z(\vec{k},E)$ were observed at low energies as well \cite{kohsaka08}.
However, above a doping-dependent crossover energy the $Z(\vec{k},E)$ spectra change:
Some QPI peaks disappear, while others cease to disperse.
Interpreting the dispersive peaks within the octet model suggests that the QPI signal
exists for quasiparticles inside the antiferromagnetic Brillouin zone only
\cite{kohsaka08}.
The non-dispersive features at elevated energies correspond to a modulation period of
approximately $4\,a_0$, with little doping dependence.
Interestingly, the modulations in $Z(\vec{r},E)$ appear strongest (as function of energy)
at the local pseudogap energy $\Delta_1$ which is of order 100\,meV.

Although a comprehensive understanding of the asymmetry maps $Z(\vec{r},E)$ and
$R(\vec{r},E)$ is difficult, as it necessarily involves Mott physics,
the described results strongly suggest that a tendency toward bond-centered stripe
order is present at the surface of both underdoped BSCCO and CCOC.
This stripe-like order, primarily visible at elevated energies, co-exists with well-defined
low-energy quasiparticles and bulk superconductivity.\footnote{
It should be noted that a precise relation between the observed modulations in
the tunneling asymmetry and those in the LDOS has not been established experimentally.
}
In both BSCCO and CCOC, the charge order exists without long-range magnetic order,
although spin-glass-like magnetism has been reported in CCOC \cite{ohishi05}.
The result in Fig.~\ref{fig:vbglass}, showing that the main modulation is on oxygen,
suggests that a quantitative description requires a three-band instead of a one-band
model for the CuO$_2$ planes.


\subsection{Other probes}

Besides neutron scattering and STM, a broad variety of other experimental techniques have been
employed to detect and investigate ordering phenomena accompanied by lattice symmetry
breaking. Here we shall give a quick overview, without pretense of completeness.

\subsubsection{Transport}
\label{sec:exp_trans}

Transport measurements have been used to search for tendencies towards order.
Two possible signatures are obvious: (i) If the ordered phase is less conducting
than the disordered phase, then the resistivity will show an upturn below the ordering
temperature. (ii) If the ordered phase globally breaks rotation symmetry, then the
resistivity will develop an in-plane anisotropy.

With the exception of the small-doping insulating regime, i.e., $x<0.055$ in LSCO,
stripe phases appear generally conducting, i.e., are not of Wigner crystal type.
Temperature-dependent resistivity measurements often show a small upturn feature upon
cooling below the stripe ordering temperature $\Tch$.
Although $d\rho/dT<0$ at low $T$ in stripe-ordered cuprates (above $\Tc$),
an exponential rise of the resistivity signaling insulating behavior is never seen.

Optical conductivity measurements in LBCO-1/8 \cite{homes06} find a residual Drude peak
even in the charge-ordered regime at low $T$, which has been interpreted as a ``nodal metal''
state, i.e. ungapped nodal quasiparticles coexist with stripe order.
This finding is consistent with the absence of a gap in earlier optical-conductivity data
taken on LNSCO-1/8 \cite{dumm02}.
The presence of nodal quasiparticles in stripe phases
is also compatible with results from photoemission and STM experiments
\cite{valla06,hanaguri07,borisenko08}.
Note that, on the theory side, arguments against \cite{castellani00,lee04}
and in favor \cite{zaanen_coin,granath01,berg08a,mvor08} of nodal quasiparticles in stripe
phases have been put forward.
For small stripe amplitude, the survival of nodal quasiparticles immediately
follows from the fact that the ordering wavevector $\vec Q$ does {\em not} connect
the nodal points \cite{vzs00a,vzs00b,berg08a}, while for larger stripe amplitude a substantial
$d$-wave component of the charge order can protect nodal quasiparticles
\cite{mvor08,mv08}.

The optical conductivity data in LBCO-1/8 \cite{homes06} show a rapid loss of spectral
weight below 40\,meV which occurs below about $60\,K \approx \Tch$, consistent with the development of
an anisotropic gap.
However, the existence of strong far-infrared peaks (between 20 and 100 cm$^{-1}$)
at low temperatures, present in earlier data on LSCO and LNSCO \cite{dumm02,luca03}
and interpreted in terms of stripe pinning \cite{benfatto03},
could {\em not} be confirmed. This disagreement has been tentatively attributed to
surface problems \cite{homes06}.

A detailed resistivity study of LSCO at finely spaced dopings \cite{komiya05} identified
specific doping levels, defined by peaks of the resistivity at fixed temperatures of 50
and 100\,K as function of doping. These peaks, albeit being weak, have been interpreted
as magic doping fractions 1/8, 1/16, 3/16, 3/32, 5/32, and suggested to be a signature of
a hierarchy of checkerboard charge-ordered states. While this idea is interesting, other
measurements in 214 materials are more consistent with stripe instead of checkerboard
order. The presence of special commensurate doping levels with enhanced ordering is also
expected for lattice pinning of stripes, although the set of special doping fractions may
be different.

Let me now turn to more direct ordering signatures.
States with broken in-plane 90-degree rotation symmetry will display
anisotropic transport properties, e.g., a d.c. resistivity tensor with $\rho_{xx} \neq
\rho_{yy}$. This applies both to nematic and to stripe phases, but, as discussed
in Sec.~\ref{sec:symfield}, requires a single-domain sample. Hence, de-twinned YBCO is
again the prime candidate.
The magnitude of the anisotropy will depend on details of the order and the electronic
scattering mechanisms: A metallic nematic state has a full Fermi surface (the same
applies to most small-amplitude stripe states), hence velocity and scattering anisotropies
become important.

In-plane resistivity anisotropies in de-twinned YBCO with dopings $\delta=0.35 \ldots 1$ have
been investigated by Ando {\em et al.} \cite{ando02}. While an anisotropy is present for
all temperatures and dopings due to the presence of the CuO chains in this material, the
anisotropy decreases with cooling below 200\,K for $\delta>0.6$, but increases for
$\delta<0.6$. This effect is particularly significant for $\delta=0.35$ and 0.45, where
the ratio $\rho_a/\rho_b$ increases from 1.3 at room temperature to $2\ldots 2.5$ at low $T$.
This result is consistent with the onset of nematic order inferred from neutron
scattering in YBCO-6.45 \cite{hinkov08a}, see Sec.~\ref{sec:exp_nematic}.
Note that the presence of a {\em sharp} onset temperature cannot be expected, because
rotation symmetry is broken by the orthorhombic distortion of the crystal from the
outset.

In stripe-ordered 214 compounds, global resistivity anisotropies cannot be observed, due
to the plane-to-plane alternation in the stripe direction in the LTT phase. Noda {\em et al.}
\cite{noda99} instead measured magnetotransport in stripe-ordered LNSCO at various doping
levels. For doping $x<1/8$, the data show a distinct drop of the Hall coefficient at a
temperature of order 80\,K where stripe charge order is believed to set in. This has been
interpreted as evidence for one-dimensional charge transport deep in the stripe-ordered
phase of LNSCO for $x<1/8$, with a crossover to more two-dimensional transport for
$x>1/8$.
However, subsequent theoretical work indicated that this picture is too simple:
A nearly vanishing Hall coefficient can be obtained in a quarter-filled stripe phase
independent of the total doping level, not as a result of one-dimensional transport,
but of an approximate particle--hole symmetry at this particular stripe
filling \cite{emery00,prelovsek01}.\footnote{
An approximate particle-hole symmetry was also inferred from thermopower
measurements in the antiferromagnetic low-doping state of YBCO \cite{ong01}.
Its relation to stripe physics is unclear.
A distinctly doping-dependent behavior of the thermopower was measured in
stripe-ordered LNSCO, which was interpreted as evidence for a Fermi-surface
change near optimal doping \cite{daou09}.
}
A rapid drop in the Hall coefficient upon cooling has also been observed in
LBCO-0.11 \cite{adachi01}, with the Hall coefficient becoming negative at lowest
temperatures. A simplistic interpretation would be in terms of electron pockets arising
from band backfolding \cite{leboeuf07}, but correlation effects may again change the picture
\cite{prelovsek01}.


Measurements of the Nernst effect in both LNSCO and LESCO  showed a distinct
low-temperature enhancement of the Nernst signal for dopings between 0.12 and 0.20,
with the signal onset occurring at a temperature $T_\nu$ which tracks the
pseudogap temperature \cite{cyr09}. (Interestingly, the $T_\nu$ of LSCO-0.12 and LESCO-0.12
are essentially equal, around 140\,K.)
As $T_\nu$ roughly follows the onset of stripe order, the Nernst signal was interpreted
as evidence for a stripe-induced Fermi-surface reconstruction.
However, the present data appear as well consistent with a pseudogap-induced
Nernst signal of non-stripe origin. Measurements at lower dopings might help
to disentangle these possibilities.

\subsubsection{Photoemission}
\label{sec:exp_arpes}

Angle-resolved photoemission (ARPES) is the method of choice to obtain momentum- and
energy-resolved information on the single-electron spectral
function.\footnote{
Complications arising from non-trivial final states and from a
dispersion perpendicular to the CuO$_2$ planes shall be ignored here.
}
Compared to neutron scattering, ARPES is only sensitive to a surface
layer of the sample with a thickness of order $5 \ldots 20$\,\AA, depending on
the photon energy.
Quality and resolution of ARPES data have improved tremendously over the
past decades, although the current ARPES energy resolution, being typically
2\ldots 10\,meV, is not as good as that of state-of-the-art neutron scattering.
An experimental problem is that ARPES requires samples with high-quality surfaces;
those are routinely available for BSCCO and CCOC. Extensive ARPES studies have also
been performed on a number of 214 cuprates, with by now comparable data quality.
In contrast, YBCO suffers from a charge imbalance at the surface, rendering
the interpretation of corresponding ARPES difficult.

I start with a description of a few general features of ARPES spectra on underdoped
cuprates, primarily obtained of BSCCO crystals.
A common observation \cite{zxshen_rmp,yoshida_rev,zhou04,kanigel,shen05}
is that low-energy electronic states along the diagonal (nodal)
direction in momentum space appear to be rather well defined, i.e. produce sharp
peaks in energy-distribution curves (EDC). In contrast, states near $(\pm\pi,0)$
and $(0,\pm\pi)$ (antinodal points) are broad and appear gapped even above $\Tc$.
The disparate behavior of the two regions in momentum space, also dubbed
``nodal--antinodal dichotomy'', is also reflected in the temperature dependence
of the linewidths: Upon lowering $T$, nodals become sharper, while antinodals tend to
be broader.

Below $\Tc$, experiments find a gap consistent with $d$-wave symmetry, sometimes mixed with higher
harmonics such that the gap near the nodes is smaller compared to a pure $d$-wave shape.
Moving into the pseudogap regime above $\Tc$, antinodals remain gapped, whereas
near-nodal states are essentially gapless, leading to apparent segments of Fermi surface.
These observation lead to the concept of ``Fermi arcs'', with a temperature and doping
dependent arc length \cite{kanigel}.
From theoretical studies, it has been proposed that the arcs may
in fact be Fermi pockets centered along the nodal direction, with the outer part of the
pockets being nearly invisible to ARPES due to matrix-element effects \cite{sudip03,imada08}.
The distinct temperature and doping dependence of the antinodal and near-nodal gaps
has also prompted proposals of a two-gap scenario, with the gaps in the two regions in
momentum space being caused by different underlying physics \cite{huefner08}.

The absence of well-defined antinodal quasiparticle has sometimes been interpreted as (indirect)
evidence for ordering tendencies (e.g. of CDW type) primarily carried by antinodals \cite{shen05},
which could be consistent with an RPA picture of CDW formation.
However, opposite experimental views have also been put forward \cite{chatterjee06}.

I now turn to the question how ARPES spectra are directly affected by symmetry-breaking
order: Broken rotation symmetry should be visible directly in momentum-resolved data of
properly aligned crystals, and broken translation symmetry should lead to Bragg gaps and
multiple bands via band backfolding.

To my knowledge, unambiguous evidence for spontaneously broken rotation symmetry
has not been detected in cuprate ARPES to date.
As discussed above, the simultaneous presence of domains with different
preferred directions -- either in a single plane or in adjacent planes -- makes rotation
symmetry breaking difficult to observe (unless the experiment would be sensitive to the
topmost CuO$_2$ layer only, which in addition had to be single-domain).
The exception would again be YBCO, with complications arising from the presence of CuO
chains; however, high-quality photoemission data for underdoped YBCO, showing the full
temperature dependence of the spectrum, is not available to my knowledge.

With broken translation symmetry (or other stripe signatures) in ARPES, the situation is
more involved.
An early experiment on stripe-ordered LNSCO-0.12 \cite{zhou99} at 20\,K has found distinct
signatures near the Fermi energy, namely a cross-shaped intensity distribution in
the Brillouin zone and no evidence for nodal quasiparticles.
These features were attributed to stripes with strong charge modulation;
in particular, the straight horizontal and vertical Fermi-surface pieces near the
antinodal points were interpreted as evidence for prominent one-dimensional
behavior.\footnote{The conclusions in Ref.~\cite{zhou99} were mainly drawn from ARPES data
which had been integrated over an energy window of 0.5~eV around the Fermi level.
A theoretical consideration on energy-integration effects in ARPES in the presence of
fluctuating charge order is in Ref.~\cite{grilli09}.
}
This view was subsequently supported by model calculations which started from
weakly coupled chains or ladders and {\em assumed} a strong 1d-like modulation (Sec.~\ref{sec:th_arpes}).
However, the results of \cite{zhou99} remained controversial \cite{zhou01,claesson07,chang08b},
and later experiments with improved samples and energy resolution could neither verify the strong
cross-shaped intensity pattern nor the absence of nodal quasiparticles \cite{claesson07};
instead, Fermi arcs were found at 15\,K above $\Tc=7$\,K \cite{chang08b}.
However, clear-cut stripe signatures were not detected.

For stripe-ordered LBCO-1/8, an ARPES study \cite{valla06} at 16\,K, much below the
charge-ordering temperature,
also failed to detect signatures of stripe-like superstructures, but instead reported
an anisotropic gap along the Fermi surface, with angle dependence being consistent with a
$d$-wave form factor. A recent higher-resolution experiment \cite{he09} mainly confirmed
this data, but also indicated that the gap slope changes rather abruptly along the Fermi
surface, suggestive of a two-component gap. While the large antinodal gap may be
interpreted in terms of pseudogap physics, the origin of the smaller near-nodal gap
is unclear at present, with fluctuating $d$-wave pairing \cite{berg07}
or $d$-wave stripe order \cite{mvor08} being two candidates.

Other recent ARPES measurements are worth mentioning:
Data obtained on stripe-ordered LESCO-1/8 \cite{borisenko08}
displayed weak signatures of both rotation symmetry breaking and band backfolding,
which may be consistent with the expected period-4 charge order.
However, not all features are consistent with model calculations (Sec.~\ref{sec:th_arpes}).
For stripe-ordered LNSCO-1/8, Ref.~\cite{chang08b} verified the existence of Fermi arcs
and, in addition, observed a second Fermi surface crossing near the Brillouin zone diagonal,
suggestive of band backfolding leading to Fermi surface pocket.
The location of this second branch appears consistent with $(\pi,\pi)$ order,
i.e., a unit cell doubling. The branch is visible up to 110\,K, i.e., far above the
charge ordering and LTT transitions, and a similar branch is seen in LSCO-0.12
(but not at other LSCO doping levels).
We note that hints for the existence Fermi surface pockets have also been
detected in ARPES in the pseudogap regime of underdoped BSCCO \cite{valla08},
however, with the pocket {\em not} being centered around $(\pi/2,\pi/2)$.
At present, the explanation of these observations is open.

In summary, a satisfactory experimental verification of the theoretical expectation
for electronic spectra of stripe phases is still missing.
Clearly, more experiments are called for, with candidate materials being the stripe-ordered
214 materials, studied systematically as function of doping, temperature, and field,
and underdoped YBCO, where signatures of rotation symmetry breaking have already been
detected in neutron scattering.

\subsubsection{$\mu$SR, NMR, and NQR}

Techniques sensitive to local properties have been very useful in characterizing ordering
tendencies in cuprates. $\mu$SR has been used to detect magnetic order via static (or
slowly fluctuating) hyperfine fields; magnetic ordering as determined by $\mu$SR
typically gives lower ordering temperatures $\Tsp$ as compared to elastic neutron
scattering (e.g. Fig.~\ref{fig:pd3}), due to the different frequency windows of the two
techniques.\footnote{
Further differences in $\Tsp$ may originate from the different behavior of polycrystals and
single crystals, compare e.g. Refs.~\cite{arai03} and \cite{savici05}.}
$\mu$SR also allows to determine the magnitude of the magnetic moment relative
to some known reference (e.g. the undoped parent compound) as well as the volume fraction
of the order. For striped 214 compounds, the moment size is typically found to be
half of that in the undoped compound, but with reduced volume fraction \cite{nachumi98,savici02}.
$\mu$SR has been used to map out the magnetic phase diagram of LESCO,
where spin stripe order replaces superconductivity over a large doping range \cite{klauss00,klauss04},
as well as of YBCO, LSCO, and LBCO \cite{nieder98,arai03}.

NMR and NQR provide related information and have been used extensively in the context of
stripes. The broadening of NMR lines at low temperature in LSCO \cite{julien99,haase00}
provides evidence for local spatial inhomogeneities.
NQR measurements have detected a so-called ``wipeout effect'', i.e. the gradual loss of the Cu
NQR signal below some temperature. While this was originally interpreted as direct measure of
charge stripe order \cite{imai99}, subsequent work has established that the wipeout is
due to a slowing down of spin fluctuations which accompany the tendency toward stripe order
\cite{teitel00,julien01,simovic03}.

Subsequent NMR/NQR studies provided details about the spatial distribution of doped holes
\cite{singer02,haase03,haase04} in LSCO and YBCO, giving evidence for spatial
inhomogeneitities at low doping.
In underdoped YBCO, with oxygen content below 6.4 and located close to
the transition to the insulator, $\mu$SR has detected static magnetic order of glassy
character which co-exists with superconductivity \cite{hinkov08a,sanna04}.
For YBCO at larger doping, it has been argued from NQR measurements that charge
inhomogeneities arise only from the presence of local oxygen vacancies \cite{ofer06};
this would be consistent an SDW and/or CDW critical point being located at very small
doping.
A particular remarkable finding is that of two inequivalent planar O sites, with
only one type of Cu site, in YBCO over essentially the entire doping range \cite{haase03}.
This interesting result may be related to the tendency toward electronic nematic order.
Unfortunately, a detailed experimental study of the temperature dependence of
oxygen NMR/NQR has not been performed to date. Together with an in-depth theoretical analysis,
this would be highly revealing about nematic ordering.
NMR measurements have also exploited the field dependence of magnetism in YBCO
\cite{curro00,mitro01}, see Sec.~\ref{sec:fieldtune} below.

NMR has been as well applied to LESCO \cite{curro00b,grafe06} which is known to display {\em static}
stripes, with charge order setting in around 75\,K at $x=1/8$.
The results show a distinct change in the O NMR spectra below 80\,K, and
have been interpreted \cite{grafe06} in terms of a correlation between local hole doping and domain
walls in the spin modulation, as expected from stripe phases.
Also, the static stripe order and/or the LTT distortion characteristic of LESCO
apparently act to suppress the very slow spin fluctuations,
which are present in LSCO, in favor of ordered magnetism \cite{curro00b,grafe06}.


\subsection{Relation to superconductivity}
\label{sec:exp_stripesc}

The experiments described so far have established the existence of stripe-like order
in certain underdoped compounds, and the data suggest that the phenomenon is in fact
common to a variety of cuprate families.
Then, a central question is about the relation between stripes and superconductivity.
A number of relevant experiments have been performed that will be summarized in the
following. A detailed theoretical discussion will be given in Sec.~\ref{sec:th_stripesc}.

\subsubsection{$\Tc$ suppression and 1/8 anomaly}
\label{sec:anomaly}
\label{sec:fluctsc}

Do stripes promote or inhibit superconductivity?
First information can be obtained by looking at the transition temperatures of both
ordering phenomena. The bulk superconducting $\Tc$ as function of doping follows the
well-known parabolic dome shape -- this applies in particular to compounds with
little stripe signatures, e.g. multilayer BSCCO.
In contrast, in the single-layer 214 materials which tend to stripe formation, the situation is
different:
The superconducting materials LSCO and LBCO display a pronounced $\Tc$ suppression at
approximately 1/8 doping, first observed in LBCO \cite{mooden88} (which was later found to
display static stripe order).
This phenomenon has been termed ``1/8 anomaly''.
If the LTT distortion is stabilized by co-doping as in LNSCO and LESCO,
then superconductivity is replaced by non-superconducting stripe order over a
large part of the doping range.
Notably, also bilayer YBCO displays features which may be of similar origin as the
1/8 anomaly in LSCO:
The doping dependence of $\Tc$ in YBCO shows a shoulder near oxygen content 6.4,
and the penetration depth is found to have a distinct maximum \cite{sonier07}.

As stripe order is known to be particularly strong near 1/8 doping, the 1/8 anomaly is
commonly attributed to stripes {\em competing} with superconductivity.
In line with this interpretation, most cuprate materials display
{\em either} quasi-long-range stripe order (like LESCO and LNSCO for $x\!\leq\!1/8$)
{\em or} well-developed bulk superconductivity.
Exceptions, with static stripes apparently co-existing with superconductivity at
lowest $T$, are LBCO near $x\!=\!1/8$ as well as LNSCO and LESCO at larger hole doping,
$x\!\gtrsim\!1/8$.\footnote{
For LNSCO and LESCO, it has been debated whether the superconductivity
co-existing with stripes is of true bulk nature \cite{buechner94}.
}
In LSCO with $x\leq 1/8$ superconductivity co-exists with quasi-static incommensurate spin
order, but here no direct evidence of charge order has been reported.
In superconducting LSCO, spin order can be induced for larger $x$ by applying a magnetic
field, see Sec.~\ref{sec:fieldtune}.
In underdoped BSCCO and CCOC, the signatures of charge order seen by STM co-exist
with well-developed superconductivity.

For LBCO with $0.11\!\leq\!x\!\leq\!0.15$, the stripe ordering temperature is above the
superconducting $\Tc$.
The most extensive data set is available at doping $x\!=\!1/8$, displaying a
remarkable hierarchy of temperature scales \cite{li07,jt08} which demonstrates that physics is
far more complicated than a simple competition of two ordering phenomena would suggest.
The structural transition to the LTT phase is located at 55\,K, immediately followed by
the charge ordering transition around 54\,K. Spin order sets in around 42\,K,
accompanied by a large drop in the in-plane resistivity $\rho_{ab}$.
However, $\rho_{ab}$ only vanishes below 16\,K, whereas $\rho_c$ vanishes below 10\,K.
Finally, a bulk Meissner effect is seen below 4\,K.
This set of highly non-trivial data is puzzling:
In addition to competing orders, it likely involves dimensional crossover and disorder effects.
The temperature regime above 10\,K has been interpreted in terms of fluctuating 2d
superconductivity, and we will come back to this in Sec.~\ref{sec:antiphase}.
A detailed experimental discussion is in Ref.~\cite{jt08}.

The competition between superconductivity and stripes, which are stabilized by the LTT
distortion of some of the 214 compounds, is visible in a number of other effects:
In LESCO, where stripes exist for $x<0.2$, the oxygen isotope effect on the
superconducting $\Tc$ was studied in in Ref.~\cite{takagi_iso}. For a non-stripe-order
$x\!=\!0.24$ sample essentially no isotope effect on $\Tc$ was seen, whereas $\Tc$ decreased
significantly by 1.5\,K in an $x\!=\!0.16$ sample upon substituting $^{16}$O by $^{18}$O.
A plausible interpretation is that the isotope substitution primarily strengthens stripe
order, thereby suppressing superconductivity.
This would then also be a nice confirmation of stripe--lattice coupling, although a detailed
modeling is not available.

Related are the pressure studies of 214 compounds in
Refs.~\cite{takagi_strain,yamada92b,arumugam02,simovic04,crawford05,huecker09}.
Hydrostatic pressure applied to stripe-ordered LNSCO-0.12 leads to a large
increase of the superconducting $\Tc$ \cite{arumugam02,crawford05},
similar data have been obtained for LBCO \cite{yamada92b,huecker09} and
LESCO \cite{simovic04}.
Uni-axial in-plane pressure increases the superconducting $\Tc$ in both
non-stripe-ordered LSCO and stripe-ordered LESCO and LNSCO.
However, while the pressure directions [1\,0\,0] and [1\,1\,0] are equivalent
for LSCO, there is a significant anisotropy for stripe-ordered LESCO-0.16
\cite{takagi_strain}, with the $\Tc$ increase occurring for pressure along the
[1\,1\,0] direction.
From these experiments, it is believed that pressure reduces the tilt angle of the
CuO$_6$ octahedra, i.e. tends to stabilize the orthorhombic symmetry of the
lattice, thereby suppressing stripe order and hence enhancing superconductivity.
The maximum $\Tc$ is typically reached at high pressures in the HTT phase
\cite{crawford05} when the CuO$_2$ planes are flat.
Interestingly, the pressure-induced increase in $\Tc$ is rather slow for $x\!=\!1/8$
as compared to $x\!=\!0.12$ or 0.13 \cite{yamada92b}, and signatures of stripe order have
been found to survive in the HTT phase of LBCO-1/8 \cite{huecker09}.

\subsubsection{Temperature dependence of stripe signatures}

Insight into the interplay of ordering phenomena can be gained by studying their
temperature dependence. Distinct behavior is found depending on whether stripes
or superconductivity dominate.

I begin with the materials showing well-developed stripe order according to
scattering probes.
For LBCO with $0.11\leq x\leq 0.15$, LNSCO, and LESCO
charge order sets in below the LTT transition
followed by quasi-static spin order at lower $T$,
and superconductivity with a very small (or zero) $\Tc$.
While $\Tch \lesssim T_{\rm LTT}$ in LBCO and LNSCO, the two transition are well
separated with $\Tch<T_{\rm LTT}$ in LESCO.
All compounds have $\Tsp<\Tch$, i.e., charge order exists without spin order.\footnote{
Note that $\Tsp$ depends on the employed probe, suggesting a wide regime of slow
fluctuations.
}
No appreciable changes of charge and spin order have been reported upon crossing $\Tc$.

The materials LSCO-0.12 \cite{waki01a} and La$_2$CuO$_{4.12}$ \cite{oxy},
displaying SDW order (without reported
signatures of charge order), share an interesting feature:
Within the experimental accuracy, the superconducting $\Tc$ equals $\Tsp$.
This may be related to a property of LBCO-1/8 noted above:
The spin-ordering transition at $\Tsp$ is accompanied by a large drop in the in-plane
resistivity (but bulk superconductivity is not reached here).
Whether these findings are coincidental or rather signify a positive correlation
between magnetic order and pair formation is unclear at present.
Note, however, that a magnetic field separates $\Tsp$ from $\Tc$
in La$_2$CuO$_{4.12}$ \cite{khaykovich02} and $\Tsp$ from the resistivity drop
in LBCO-1/8 \cite{li07}.
Phase separation has been under discussion for La$_2$CuO$_{4+\delta}$ \cite{moho06};
interestingly, a theoretical scenario based on phase coexistence of {\em competing} SDW and
superconducting orders has been shown to allow for $\Tc\!=\!\Tsp$ for a range of
parameters \cite{KAE01}.

I now turn to the temperature dependence of incommensurate spin excitations.
For stripe-ordered LBCO-1/8, the evolution of the low-$T$ hour-glass spectrum
was studied in Ref.~\cite{xu07}. The low-energy incommensurate signal is well visible at
65\,K above $\Tch\approx55$\,K, but deteriorates at 300\,K. However, the upper excitation
branch above 50\,meV changes little from 5\,K to 300\,K.
This was interpreted as evidence for a stripe liquid in Ref.~\cite{xu07}.
In LSCO, the temperature dependence of the excitation spectrum is similar:
While excitations at high energies display only a weak temperature dependence,
the low-energy excitations become much more structured at low $T$.
While in optimally doped LSCO this crossover is tied to $\Tc$ \cite{jt04b},
it occurs at a higher temperature in underdoped LSCO-0.085, which has been attributed
to the pseudogap \cite{lips08}.
Remarkably, incommensurate excitations below $\Tc$ are even seen in overdoped
LSCO \cite{waki04}, but disappear for doping $x\!=\!0.30$ outside the superconducting
dome.

In optimally and slightly underdoped YBCO, a clear hour-glass dispersion with a sharp
resonance at $(\pi,\pi)$ is observed at low $T$.
In YBCO-6.95, the lower branch disappears completely above $\Tc$, with damping being
rather strong \cite{reznik08c}.
In YBCO-6.6 with $\Tc=61$\,K,
the low-$T$ hour glass is replaced by a ``Y''-shaped dispersion at 70\,K,
and the resonance intensity is strongly reduced \cite{hinkov07}.
Turning to strongly underdoped YBCO, a ``Y''-shaped dispersion is seen in
YBCO-6.45 ($\Tc=35$\,K) both above and below $\Tc$, with the low-energy intensity
being strongly enhanced at low $T$, however, this increase does not appear tied to
$\Tc$ \cite{hinkov08a}.

It is worth noting an interesting aspect of the temperature dependence of spin
fluctuations: While in optimally and overdoped compounds, cooling and in particular
the onset of superconductivity usually reduces the magnetic low-energy spectral weight
(i.e. often a clear spin gap opens below $\Tc$),
this is not the case in strongly underdoped materials, where instead magnetism appears
enhanced.

Finally, I briefly discuss the CDW signatures as detected with STM.
Here, only a few studies at elevated temperatures are available and restricted to
BSCCO \cite{yazdani04,liu07}. The data indicate robust period-4 CDW features
in the pseudogap regime, likely pinned by sample inhomogeneities.
The comparison with data obtained at low $T$ is complicated by the
signatures of quasiparticle interference at low $T$ which are weak or absent
at elevated $T$:
At low $T$, it is difficult to assign peaks in the FT-LDOS uniquely to either
quasiparticle interference or charge order (while this ambiguity does not exist above
$\Tc$), see Sec.~\ref{sec:exp_stm}.
Whether this implies that CDW tendencies are enhanced in the pseudogap regime as compared
to the low-$T$ superconducting phase is not known.

While a one-line summary is difficult, a common feature seems that in underdoped
compounds the ordering tendencies develop below the pseudogap temperature (although
long-range order may set in only at very low $T$).
Moving towards optimal doping, ordering tendencies weaken, and
the influence of the superconducting $\Tc$ becomes more significant,
with strongly damped behavior above $\Tc$.

\subsubsection{Magnetic field tuning of order}
\label{sec:fieldtune}

The application of a c-axis magnetic field to superconducting cuprates has been found to
enhance the tendency towards spin and charge order in a variety of cuprates.
These experiments provide strong evidence for a competition of stripes and
superconductivity, as will be discussed below.
The concrete modeling of this competition is described in Sec.~\ref{sec:comp_op}.

A series of experiments is concerned with enhanced magnetism in an applied field.
Particularly striking are low-temperature neutron scattering experiments on 214 compounds.
Where well-developed superconductivity and quasi-static incommensurate spin order
coexist, the latter is found to be significantly enhanced by a field of order 10\,T.
This applies to LSCO with $x\lesssim 1/8$ \cite{lake02,lake05,chang07,chang08} and to
La$_2$CuO$_{4+\delta}$ \cite{khaykovich02,khaykovich03}.
For LSCO-0.10, the ordered moment increases from $0.03\,\mu_B$
at zero field to $0.15\,\mu_B$ per Cu at 14.5 T \cite{lake02}.
Inelastic measurements show that the spectral-weight transfer is restricted to energies
below 1.5\,meV, albeit with a non-trivial temperature dependence \cite{chang07}.
In superconducting compounds without quasi-static order, but located close to a magnetic
instability, an applied field can reduce the gap to spin excitations and eventually
drive the system into a magnetically ordered state. This is again nicely seen in
LSCO \cite{lake01,jt04b,khaykovich05,chang08,chang09}, see Fig.~\ref{fig:fieldtune} for
data from LSCO-0.145 \cite{chang08}.
For this composition, with $\Tc=36$\,K, inelastic neutron scattering show that the low-temperature
spin gap decreases continuously as function of the applied field and closes at about 7\,T,
consistent with a field-induced quantum critical point.
%
In $\mu$SR experiments on 214 compounds field-enhanced magnetism
is observed as well \cite{savici05,chang08}.

\begin{figure*}
\begin{center}
\includegraphics[width=2.2in]{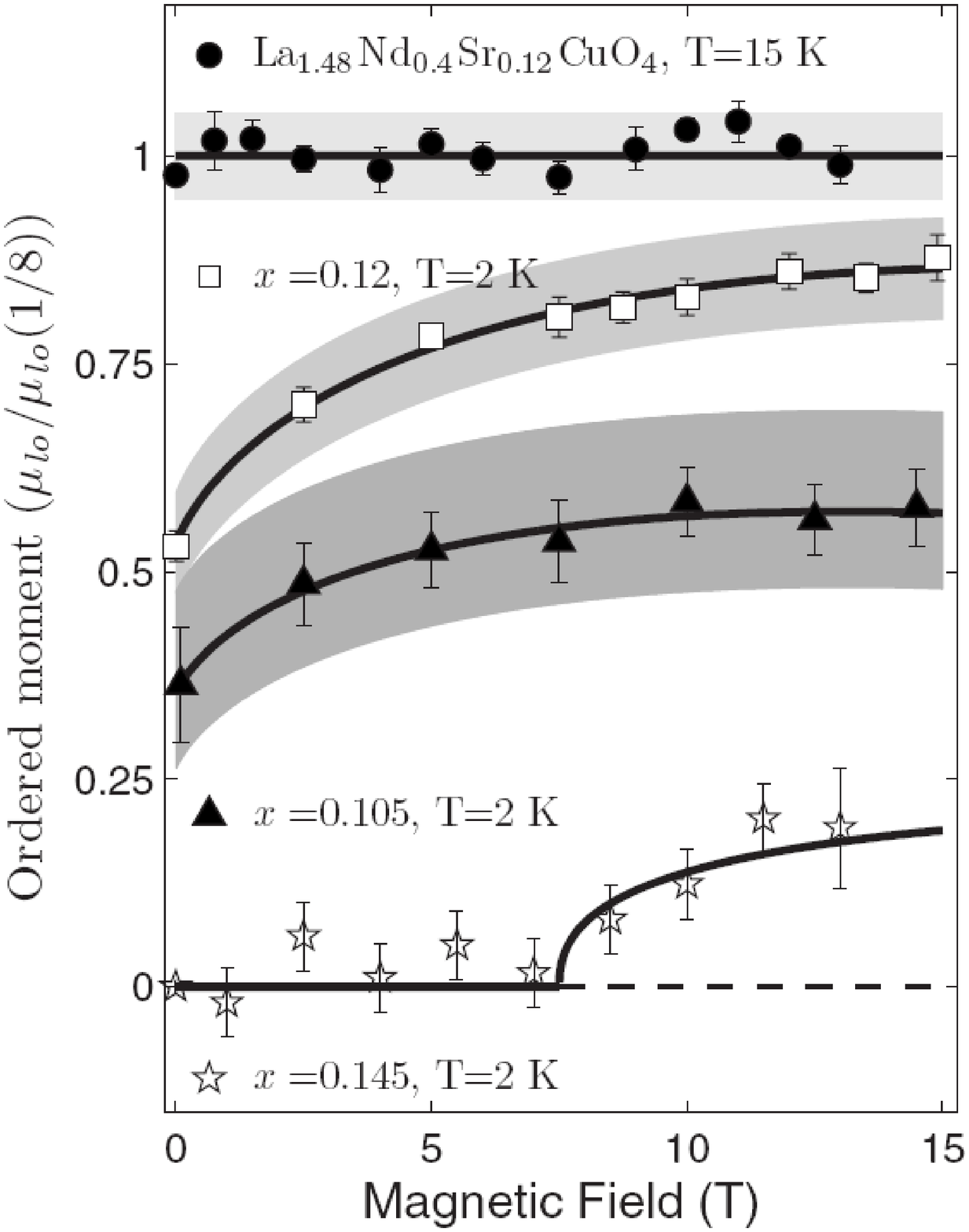}
\hspace*{20pt}
\ifhires
\includegraphics[width=1.9in]{vortex_stm1.eps}
\else
\includegraphics[width=1.9in]{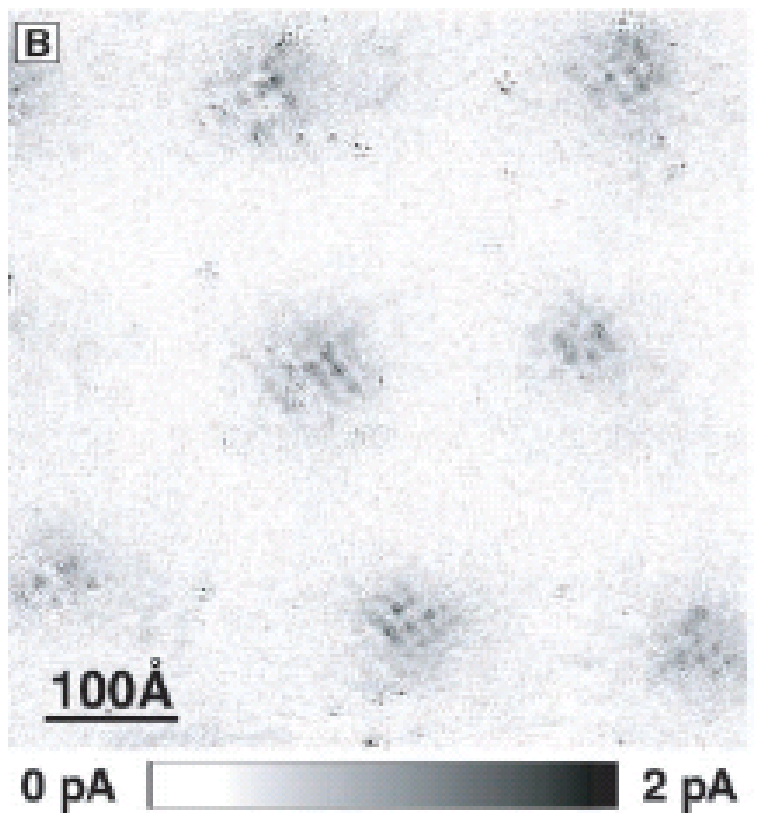}
\fi
\caption{
Left:
Field tuning of magnetic order in LSCO
(reprinted with permission from Ref.~\cite{chang08}, copyright 2008 by the American Physical Society).
Shown are the field dependences of the elastic neutron scattering
intensity at $\Qsp$ for LSCO with $x\!=\!0.105$, $x\!=\!0.12$, $x\!=\!0.145$,
and LNSCO-0.12.
The solid lines are fits to the theory of Demler {\em et al.} \cite{demler1},
see Eq.~\eqref{field_intensity} in Sec.~\ref{sec:comp_op}.
Moment sizes are plotted relative to LNSCO-0.12; the moments at zero field were
estimated from $\mu$SR and the gray colors indicate the corresponding error.
The LSCO-0.145 data are presented in arbitrary units.
Right:
Low-energy LDOS, integrated between 1 and 12\,meV, of slightly overdoped BSCCO
in a field of 5\,T, as measured by STM
(from Ref.~\cite{hoffman02a}, Science {\bf 295}, 466 (2002), reprinted with permission from AAAS).
A checkerboard-like modulation around each vortex is clearly visible.
}
\label{fig:fieldtune}
\end{center}
\end{figure*}

In contrast to these observations of field-enhanced order,
in compounds with well established stripes the field effect in neutron scattering
is either weak, as in LBCO-1/8 \cite{wen08}, or absent, as in LBCO-0.095
\cite{duns08}, LNSCO-0.12 \cite{chang08} and LNSCO-0.15 \cite{waki03}.
High-resolution x-ray scattering in LBCO-1/8 found a slight field enhancement
of the correlation length of the charge order, but no measurable increase of the
intensity \cite{kim08b}.
Note, however, that transport experiments on LBCO and LNSCO have been interpreted
in terms of field-enhanced charge order \cite{adachi05,adachi06}.

Oxygen NMR experiments in fields up to 42 T, mainly performed on YBCO, found two components with
distinct temperature dependencies in the spin-lattice relaxation rate \cite{curro00,mitro01}.
The results were interpreted as a enhanced antiferromagnetic fluctuations in the vortex
cores, and have been linked to the enhanced magnetism in neutron scattering,
described above.

A clear-cut neutron scattering observation of field-enhanced static order in YBCO
is that of Haug {\em et al.} \cite{haug_mf} in de-twinned YBCO-6.45.
At 2\,K, the quasieleastic intensity at incommensurate wavevectors was found to increase by a
factor of 2 from zero field to 15\,T, with the intensity increase $I(B)$ being approximately
linear. In turn, the inelastic response between 2.5 and 4\,meV decreased with field.
From $\mu$SR, the zero-field ordered moment in YBCO-6.45 is rather small,
about 0.05\,$\mu_B$ at 2\,K, indicating that YBCO-6.45 is closely located to a magnetic QCP.

A search for field-induced ordering with the additional advantage of spatial resolution
has been performed using scanning tunneling microscopy.
Hoffman {\em et al.} \cite{hoffman02a} observed a checkerboard modulation in the local density
of states in slightly overdoped BSCCO in an applied field of 5\,T,
Fig.~\ref{fig:fieldtune}.
The LDOS, integrated over an energy window between 1 and 12\,meV, showed a
modulation with period $4.3\,a_0$ around each vortex,
with a decay length of approximately 8\,$a_0$, significantly larger than the vortex
core size.

As with the zero-field modulation, it is interesting to check whether the spatial
periodicity of the field-induced modulation is energy-dependent or not.
For BSCCO, Levy {\em et al.} \cite{levy05} reported an energy-independent modulation
with period $(4.3\pm0.3)\,a_0$ and fourfold symmetry.
Matsuba {\em et al.} \cite{matsuba07} detected a small breaking of the $C_4$ symmetry
towards $C_2$, and furthermore emphasized that the modulations at positive and negative
bias are anti-phase.
In optimally doped YBCO, STM experiments in a field of 5 T also found
signatures of charge order with a period of $(4.25\pm0.25)\,a_0$ \cite{yeh08}.
While these experiments suggest field-induced CDW order near the vortices,
a recent detailed analysis of quasiparticle interference in a field
in CCOC \cite{takagi08} was interpreted differently:
The field enhancement of the QPI vectors $q_1$ and $q_5$ may also induce
a checkerboard-like structure in the LDOS.

At this point, a crucial difference between the spin and charge channels in an applied field
has to be emphasized: While static order in the spin channel is necessarily accompanied by
spontaneous symmetry breaking, charge fluctuations are pinned by the vortices which break
the lattice translation symmetry and couple linearly to the CDW order parameter,
similar to impurities.
Hence, incommensurate fluctuations (in either the spin or the charge channel),
which would be gapped in the absence of vortices or impurities,
cause weak inhomogeneous modulations in the {\em charge} channel (i.e. of both the density and
the LDOS) due to pinning.
In such a case, the LDOS modulations are {\em not} expected to be energy-independent
\cite{tolya03,podol03}, see Sec.~\ref{sec:th_stm}.

Summarizing, in superconducting compounds where spin or charge order in zero field is
weak or absent (but ``almost'' ordered), an applied magnetic field operates to stabilize the order,
whereas fully developed stripe order displays little field dependence.
In principle, the physical origin of these phenomena
could be either a {\em direct} coupling of the field
to the spin/charge order parameter, or an {\em indirect} coupling where the field-induced
vortices in the superconducting order parameter shift the balance between
superconductivity and stripe order towards stripes.
The strength of the field effect and the nearly linear increase or the order
parameter with field point towards the second, {\em indirect}, possibility,
because effects of a direct coupling between field and CDW/SDW would be quadratic
in the field.
Indeed, a phenomenological theory based on competing order parameters \cite{demler1,demler2}
account for a large part of the data, see Sec.~\ref{sec:comp_op} and Fig.~\ref{fig:fieldtune}.

Finally, I note that, for electron-doped cuprates, reports both
pro \cite{kang03,sonier03} and contra \cite{matsuda02b}
field-enhanced magnetism are in the literature.
However, in all cases the spin correlations were found at the commensurate wavevector
$(\pi,\pi)$.


\subsection{Glassy behavior}
\label{sec:exp_glass}

Glassy behavior, usually arising from a manifold of low-lying many-body states
which are separated by large energy barriers,
has been seen in a variety of underdoped cuprates at low
temperatures.
While canonical spin-glass physics arises in spin systems under the influence of both
disorder and geometric frustration, the situation in cuprates is more involved due to the
presence of charge carriers.

A spin-glass phase has been identified early on in LSCO in the insulating doping range
$0.02<x<0.055$ \cite{cho92,chou95,waki00b}.
Irreversibility and glassy freezing, as evidenced e.g. in magnetization and ac susceptibility,
are seen below temperatures of $5\ldots10$\,K.
Remarkably, similar glassy behavior is also seen for larger doping, where
spin-glass order appears to coexist with superconductivity \cite{nieder98,julien99,mitro08}.
Taking together NMR and NQR results with those of neutron scattering,
one concludes that this phase is not a ``simple'' spin glass with little spatial
correlations, but instead displays well-defined incommensurate antiferromagnetic
correlations.
This has been dubbed ``cluster spin glass'' \cite{cho92,ek93}.

In fact, the incommensurate magnetic order in the 214 compounds as well as in
underdoped YBCO is characterized by a wide regime of very slow fluctuations.
The best indication is that the apparent ordering temperatures extracted from different
probes are significantly different:
In elastic neutron scattering, the onset of quasi-static order is often seen at
temperatures by a factor of two (or more) higher than the onset of order in $\mu$SR.
Moreover, the ordered state is often characterized \cite{jt99b,buyers05,buyers08}
by only medium-range magnetic
correlations and sometimes even by almost isotropic spin orientations
(dubbed ``central mode'' behavior by Stock {\em et al.} \cite{buyers05} for YBCO).
All these properties point towards the glassy nature of the incommensurate order.
The slowing down of spin fluctuations in the low-temperature limit has been
argued to persist essentially up to optimal doping \cite{pana02,pana05,mitro08}.
Spin-glass-like behavior was also reported in CCOC in the doping range
$0.05\!<\!x\!<\!0.12$; for $x\!=\!0.12$ it co-exists with superconductivity
\cite{ohishi05}.

A number of observations have also been attributed to glassy behavior in the {\em charge}
sector. In the insulating regime of LSCO, resistivity noise measurements suggest the existence of a
charge glass, but only below 0.3\,K, i.e., far below the spin freezing temperature
\cite{popovic08}. This is supported by detailed impedance spectroscopy in LSCO-0.03
\cite{pana08}.
For stripe-ordered LESCO, NMR experiments \cite{simovic03} revealed a distinct slowing down of the spin
dynamics below 30\,K (whereas glassy spin freezing only appears at 5\,K) -- this was
attributed to a glassy ``stripe liquid'' forming at 30\,K, which in turn slows down the
spins. In addition, resistivity measurements in LSCO in both the insulating and
superconducting regimes have been interpreted in terms of glassy behavior in the charge
sector, see Ref.~\cite{pana05} for a discussion.

Finally, the STM results of Kohsaka {\em et al.} \cite{kohsaka07} suggest the existence
of a period-4 valence-bond glass in BSCCO and CCOC, based on the fact that the observed order is static,
but short-ranged. However, dynamical aspects of glassy behavior were not studied there.

On the theory side, relatively little work has been done to investigate these collective
glass states.
Glassy behavior in the charge sector is not unexpected, given the presence of strong
collective effects in the spin-singlet channel, i.e., valence bonds and stripes,
which couple linearly to disorder of potential-scattering type.
The vast amount of literature on the random-field Ising model is relevant in this
context, see Ref.~\cite{natter} for an overview. However, the dynamics in the quantum
regime is not well understood.
The interplay of singlet formation and disorder has recently been investigated
in a large-$N$ framework and shown to lead to a distinct valence-bond glass
phase \cite{biroli08}.

The origin of glassy behavior in the spin sector is clearly related to carrier doping,
as the underlying square lattice structure of cuprates is unfrustrated.
In the insulating regime at low doping, a picture \cite{sushkov06}
of holes that are localized at random positions and induce local spiral distortions
of the spin background \cite{shraiman,ss94} appears appropriate.
At higher doping, the situation is less clear. However, it is conceivable that
ordering tendencies in the charge sector, which are short-range due to quenched disorder,
plus competing interactions cause frustration in the spin sector, driving it glassy.

I note that behavior reminiscent of randomness-induced glassiness can even arise
in the absence of quenched disorder, i.e., purely from interactions.
This remarkable phenomenon has been theoretically demonstrated in Ref.~\cite{schmalian00}
in a simple stripe model with competing (i.e. frustrated) interactions, but is poorly understood
in general.
The NMR results of Ref.~\cite{mitro08} for LSCO-0.12, showing spin-glass features
which depend only weakly on the level of chemical disorder, may point in this direction
as well.

In summary, both the spin and charge order observed in underdoped cuprates display glassy
features, which, however, strongly mix with collective effects. More theoretical work is
needed, e.g., in order to understand how the glassy behavior influences the
low-temperature magnetic excitation spectrum measured in neutron scattering.



\section{Microscopic mechanisms}
\label{sec:mic}

After the survey of experimental evidence for lattice symmetry breaking in cuprates,
we now turn to the theory side.
The most important questions appear to be:
(i) Which microscopic ingredients (on the level of model Hamiltonians) are required to
obtain symmetry-broken phases as observed in experiment?
(ii) Which general principles can be identified as driving forces of such
symmetry breaking?
(iii) Beyond symmetries, concrete theoretical results for observables (e.g. neutron or STM spectra)
have to be contrasted with experimental data.
In this section, I shall address (i) and (ii),
covering mainly stripes (but briefly also spiral and nematic states),
while (iii) is subject of Sec.~\ref{sec:eff}.

Remarkably, early mean-field studies of the Hubbard model {\em predicted}
the formation of inhomogeneous states at small doping, with stripes being
one variant.
These theory papers, published in 1989--1990, preceded the 1995 experimental reports on
evidence for stripes, with an enormous subsequent growth of theory activities.
Naturally, this section can only give a partial coverage,
and I refer the reader to previous review articles for alternative expositions
\cite{antonio_rev,brom_rev,oles_rev}.


\subsection{Microscopic models: Hubbard and $t$--$J$}

Based on band structure calculations, the low-energy electronic properties of
the cuprate superconductors are commonly assumed to be dominated
by the Cu $3d_{x^2-y^2}$ and O $2p_{x,y}$ orbitals of the CuO$_2$ planes,
Fig.~\ref{fig:pd1}.
Supplementing the one-particle terms by local Coulomb repulsion
results in a three-band Hubbard or ``Emery'' model \cite{emerymodel}.
At half filling, the system is a charge-transfer insulator, with
one hole per Cu orbital and filled O orbitals.

It is commonly assumed\footnote{
Ref.~\cite{reiter88} has shown that an {\em exact} mapping from the three-band model
to a one-band Hubbard or $t$--$J$ model is {\em not} possible even in suitable parameter
limits. This result, however, does not imply that the low-energy physics of three-band
and one-band models is necessarily different.
}
that under certain conditions, namely the
hopping matrix elements being smaller than both the charge transfer energy
and the on-site Coulomb repulsion, the low-energy physics of such a
model can be mapped onto a simpler one-band Hubbard model \cite{Hubbard},
\begin{equation}
H\, =\, - \sum_{\langle ij\rangle \sigma} t_{ij}
      (c^\dagger_{i\sigma} c_{j\sigma} +
       c^\dagger_{j\sigma} c_{i\sigma})
    + U \sum_i  n_{i\uparrow} n_{i\downarrow}
\,,
\label{H_HUBBARD}
\end{equation}
defined on the square lattice formed by the copper sites.
The Hubbard repulsion $U$ is roughly given by the charge-transfer energy
of the original CuO$_2$ system;
for the cuprates $U=8t$ is a common choice, where $t$ is the nearest-neighbor hopping.

For large Coulomb repulsion, $U\gg t$, the one-band Hubbard model can in turn be mapped
onto a $t$--$J$ model \cite{pwa87,tJ}, with $J_{ij} = 4\,t_{ij}^2/U$.
In this mapping, doubly occupied sites are excluded from the Hilbert space,
and virtual hopping processes between neighboring sites transform into
an antiferromagnetic Heisenberg exchange, leading to
\begin{equation}
H\, =\, - \sum_{\langle ij\rangle \sigma} t_{ij}
      (\hat c^\dagger_{i\sigma} \hat c_{j\sigma} + h.c.)
    + \sum_{\langle ij\rangle} J_{ij} \ ({\vec S}_i \cdot {\vec S}_j - {{n_i n_j} \over 4} )
\label{H_TJ}
\end{equation}
where the electron operators $\hat c_{i \sigma}^{\dagger}$ exclude double occupancies,
$\hat c^{\dagger}_{i\sigma} = c^{\dagger}_{i\sigma} (1-n_{i,-\sigma})$.
At half filling, all sites are singly occupied, hopping processes are
suppressed, and the $t$--$J$ model reduces to the well-known square lattice
spin-$\frac{1}{2}$ Heisenberg model.
The mapping from the Hubbard model to the $t$--$J$ model can be
understood as expansion in $t/U$; we note that higher orders in this expansion
generate additional terms in Eq.~\eqref{H_TJ} like ring exchange and spin-dependent
three-site hopping which are in general not negligible,
see e.g. Refs.~\cite{cherny04,hub_tj}.
Very accurate perturbative mappings up to high order can be obtained using
the method of continuous unitary transformations; this has been done for
the Hubbard $\leftrightarrow$ $t$--$J$ mapping in Ref.~\cite{hub_tj}.
At half filling, a direct mapping from the three-band Hubbard model to the Heisenberg
model has also been discussed \cite{emery_tj}.
It is worth emphasizing that the discussed mappings imply transformations for all
operators, i.e., care has to be taken when calculating observables with one of the
effective models.

In general, models like \eqref{H_HUBBARD} and \eqref{H_TJ} have to be supplemented by
longer-range Coulomb interactions, as metallic screening can be expected to be poor in
particular close to the Mott insulating phase. However, such terms are often neglected
for simplicity.

At present, an open question is how much of the cuprate physics is captured by one-band
models of Hubbard or $t$--$J$ type. This concerns, e.g., the fundamental issues
of superconductivity and of the strong asymmetry between hole-doped and electron-doped
materials.
Numerical methods applied to the one-band models have reproduced a number of
salient features of the cuprate phase diagram (see e.g. Ref.~\cite{jarrell06}),
including superconductivity, but the results also indicate that
other properties are {\em not} properly reproduced, see Sec.~\ref{sec:num} below.
However, other workers have raised doubts regarding the existence of superconductivity
in the one-band models, based on numerical results \cite{imada_pair} not showing
sufficiently strong pairing.\footnote{
Based on variational Monte-Carlo studies showing superconductivity in the Hubbard model,
Ref.~\cite{baeriswyl} suggested that the absence of superconductivity in the simulations
in Ref.~\cite{imada_pair} is in fact related to the neglect of second-neighbor hopping
$t'$ and the large doping values of $x\geq0.18$ studied in Ref.~\cite{imada_pair}:
In Ref.~\cite{baeriswyl}, superconductivity disappears for $x\geq0.18$ for $t'\!=\!0$,
whereas it survives up to $x\!=\!0.25$ for $t'/t\!=\!-0.3$ relevant for cuprates.
}
While this debate is open, physics of the three-band model will eventually be important
to fully understand cuprate properties.\footnote{
The importance of three-band physics has also been emphasized in the context of
more phenomenological approaches, see e.g. Refs.~\cite{varma99,newns}.
}


\subsection{Stripes: Concepts and mean-field theories}
\label{sec:mic_mf}

Nearly in parallel, Zaanen and Gunnarsson \cite{za89},
Poilblanc and Rice \cite{poil89}, Schulz \cite{schulz89},
and Machida \cite{machida89,machida90} employed mean-field approximations to
the one-band and three-band Hubbard models to study the physics slightly off
half-filling.
Their real-space Hartree-Fock approaches allowed for inhomogeneous charge and spin
distributions, with mean fields representing on-site densities.
The results showed a clear tendency of holes preferring to agglomerate in the presence of
an antiferromagnetic background. The most favorable configurations were such that the
holes formed parallel one-dimensional lines, later on dubbed ``stripes''.
The Hartree-Fock stripes have a number of interesting properties:
(i) The stripes form $\pi$ (or``antiphase'') domain walls of the antiferromagnetic order
(Fig.~\ref{fig:stripe1}).
(ii) The number of holes per unit length of stripe, $\rho_l$, is unity,
such that a stripe distance of $n$ lattice spacings gives a total doping of $1/n$.
Those stripes are often called ``filled stripes'' (or ``empty'' in an electron picture),
and result in insulating behavior.

The formation of such stripes reflects a tendency of the antiferromagnetic background to
expel holes. Indeed, in the $t$--$J$ model at low doping, phase separation into hole-rich
and hole-poor regions was found \cite{ekl}, and hole droplets also appeared in numerical
studies of the three-band Hubbard model \cite{horsch89}.
For the mean-field Hubbard stripes, it was argued that quantum fluctuations lead to a
bunching of stripes \cite{ek93}.
In all situations, the presence of additional long-range Coulomb interactions is expected to
disfavor phase separation, and stripe-ordered states may result.
Such {\em frustrated phase separation} \cite{ek93} has become an important concept in
correlation physics.
Subsequent phenomenological theory work supported this notion:
The presence of interactions with two distinct length scales
was shown to give rise to stripe-like structures \cite{loew94,sk1}.

The experiments of Tranquada {\em et al.} \cite{jt95,jt96},
establishing stripe order in LBCO and LNSCO,
uncovered deficiencies of the early Hartree-Fock stripes.
Experimentally, stripes were found to be conducting instead of insulating and, consistent with this,
the ordering wavevectors were more compatible with half-filled instead of filled stripes.
At lowest temperatures, stripes were even found to co-exist with bulk superconductivity.

As a result, analytical and numerical activities were directed to find and study
metallic and superconducting stripe phases.
While the numerical works will be summarized in the following subsections,
I shall briefly discuss alternative mean-field approaches here, most of which are based
on slave-boson formulations of the Hubbard or $t$--$J$ models.

Metallic stripes with a filling close to 1/2 were found in a Gutzwiller variational
treatment of the one-band Hubbard model \cite{seibold98}. Metallic vertical stripes
compete with insulating diagonal stripes, and long-range Coulomb interaction is required
to stabilize the former.
Later, a systematic slave-boson analysis of the three-band Hubbard model \cite{lorenzana02}
found nearly half-filled metallic stripes to be stable without long-range Coulomb
interactions. These stripes were bond-centered, with a doping evolution of the stripe
distance reminiscent of what is found experimentally, i.e., the Yamada plot of
Fig.~\ref{fig:yamada}.
Similar metallic stripes were also obtained in mean-field treatments of one-band Hubbard
models including second-neighbor hopping \cite{machida99,seibold04}.
Lattice anisotropies, as exist in the LTT structure of 214 materials, were taken into account
in the real-space Hartree-Fock study of Ref.~\cite{normand01}. As expected, both hopping
and exchange anisotropies are quite effective in stabilizing stripe structures.
In all cases, the stripe order co-exists with well-developed
antiferromagnetism in the hole-poor regions.

\begin{figure*}
\includegraphics[width=2.4in]{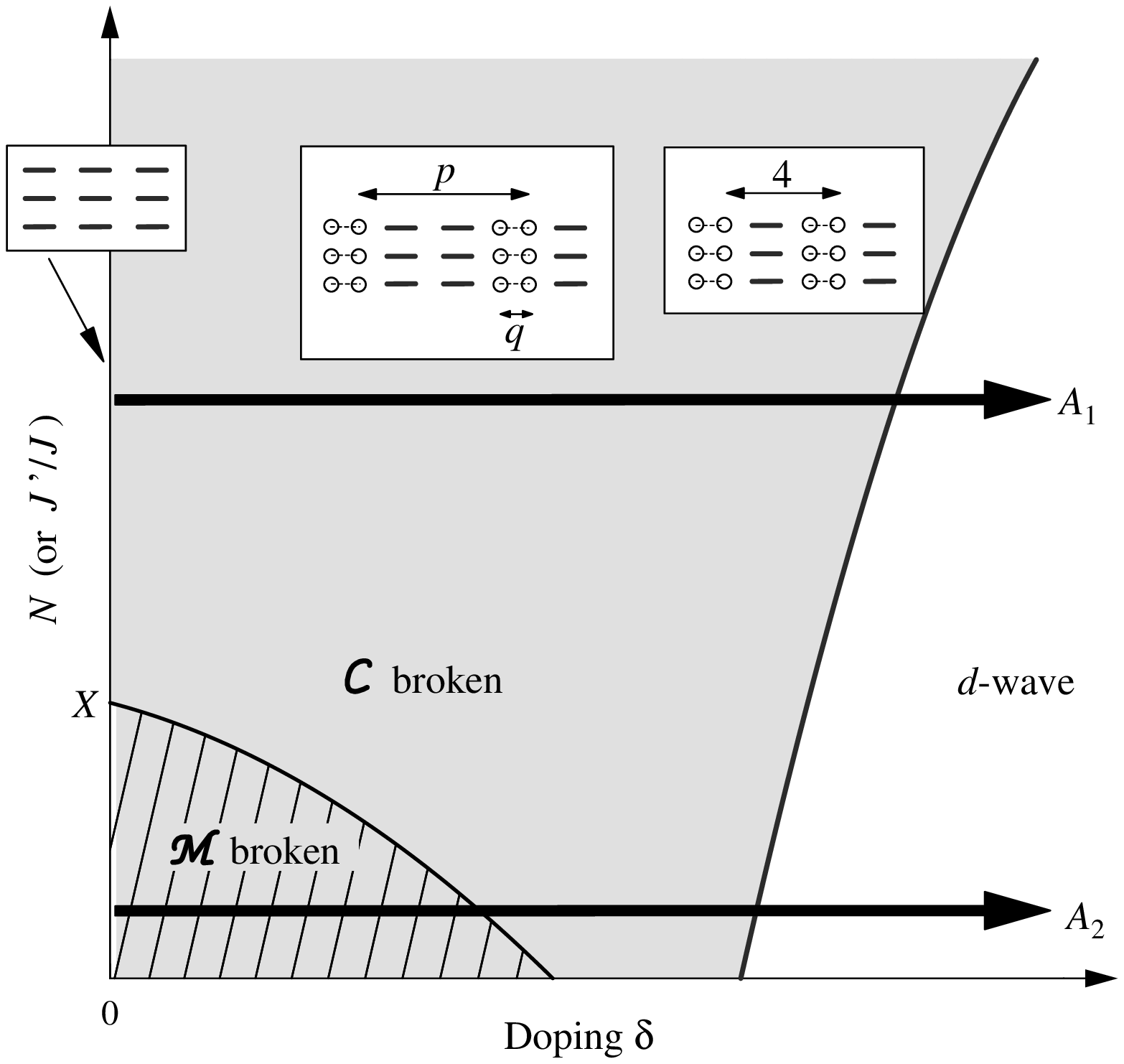}
\hspace*{10pt}
\includegraphics[width=2.7in]{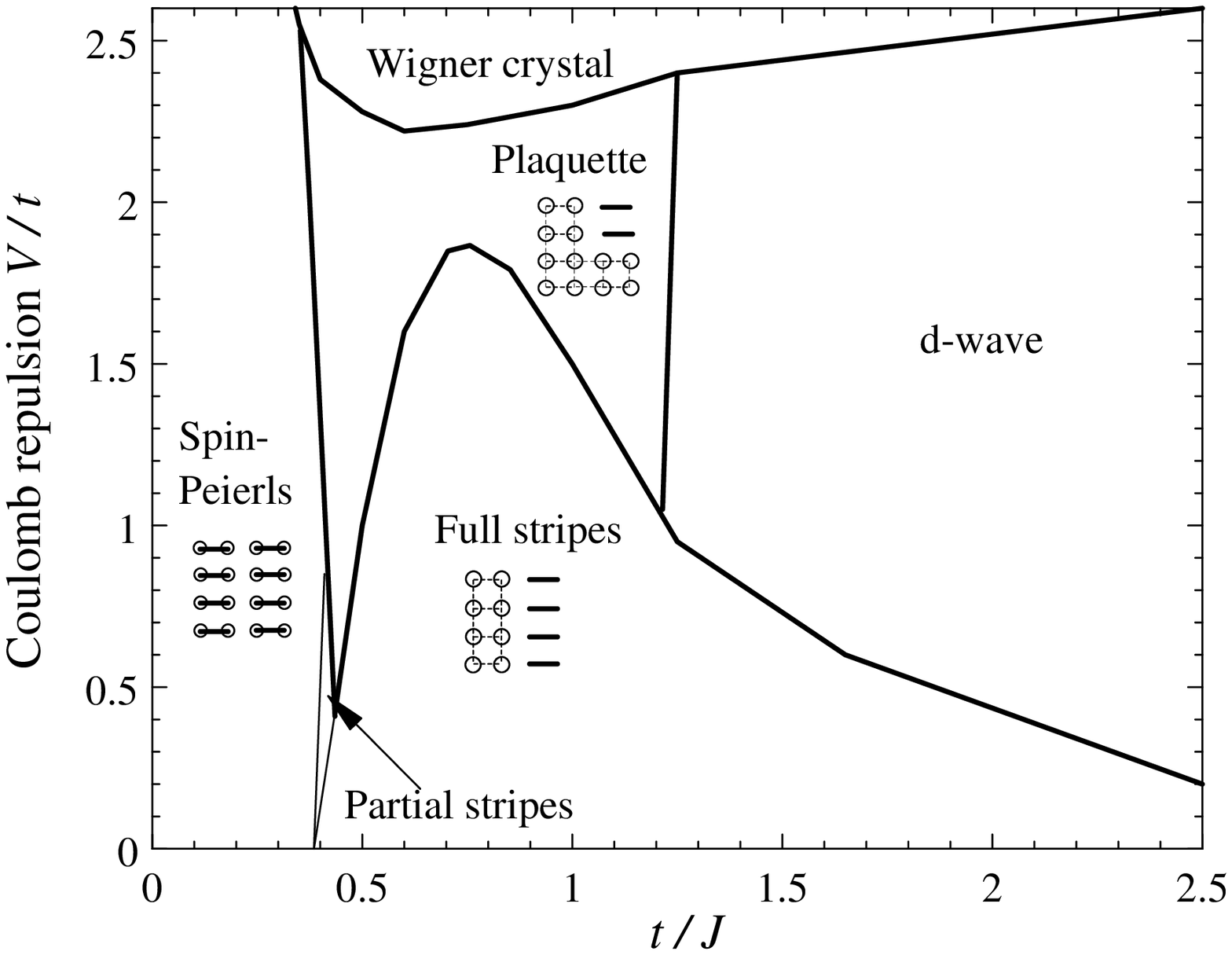}
\caption{
Left: Global ground-state phase diagram for doped Mott insulators on the square lattice,
proposed in Ref.~\cite{vs99}
(reprinted with permission from Ref.~\cite{vs99}, copyright 1999 by the American Physical
Society),
as function of doping and the amount of quantum fluctuations in the spin sector.
The latter is tuned by $N$ in an  $Sp(2N)$ generalization of the spin symmetry
or by the strength $J'/J$ of a competing frustrating interaction.
$\mathcal{C}$ ($\mathcal{M}$) denote the translation (magnetic) symmetries
which can be spontaneously broken.
All states at finite doping are superconducting;
insulating Wigner crystal states will occur at very small doping (not shown).
The insets illustrate the type of lattice symmetry breaking, with
circle sizes denoting the amount of hole doping per site,
and line widths denoting the strengths of the square-lattice bonds.
Superconducting stripe states with periodicity $p$, with $p$ even,
occur naturally as a result of hole doping into a paramagnetic dimerized Mott
insulator.
At large doping, the state becomes a $d$-wave superconductor without additional
symmetry breaking.
Right: Phase diagram emerging from the mean-field treatment of an extended $Sp(2N)$ $t$-$J$ model
(reprinted with permission from Ref.~\cite{mv02}, copyright 2002 by the American Physical Society),
as function of $t/J$ and the strength of the long-range Coulomb interaction $V/t$.
at a fixed doping of 0.20.
All states with the exception of the Wigner crystal are superconducting.
``Full stripes'' refers to states where the charge modulation in the
large-$N$ limit is maximal, i.e., the hole density is zero in the
hole-poor regions, whereas the ``partial stripe'' states have a finite
hole density there.
Plaquette (or checkerboard) states generically compete with stripes.
}
\label{fig:pd_vs}
\end{figure*}

While the mean-field theories sketched so far used decoupling fields on the
lattice {\em sites}, conceptually different mean-field descriptions employ {\em bond}
variables. Those appear in the resonating-valence-bond (RVB) \cite{rvbmf}
and large-$N$ mean-field theories \cite{affmar,sr_spn} of the $t$--$J$ model.
The $Sp(2N)$ large-$N$ limit of the $t$--$J$ model \cite{sr_spn},
with mean fields being anomalous bond variables, was used to describe
stripes in Refs.~\cite{vs99,vzs00a,mv02}.
In the $Sp(2N)$ approach,
fermionic pairing is naturally implemented, while magnetic interactions lead
to singlet dimerization, i.e., spin-Peierls-like bond order \cite{ssrmp}, instead
of antiferromagnetic long-range order.
In this mean-field theory, states with finite doping are superconducting at $T\!=\!0$.
The dimerization of the undoped paramagnetic Mott insulator survives for a finite doping
range, while at large doping a homogeneous $d$-wave superconductors emerges.
At small doping, the system is unstable towards phase separation. Upon including
long-range Coulomb interactions, superconducting stripe states are found,
with bond order inherited from the undoped system.
Thus, the width $q$ and periodicity $p$ of the stripes are even,
with $p$ and $q$ depending on doping and Coulomb repulsion.
For $t/J$ values relevant to cuprates and $q=2$, the stripe filling is close to 1/2,
and the doping evolution of $1/p$ similar to the Yamada plot.
On the basis of these results, a global phase diagram for doped Mott insulators on the
square lattice was proposed in Ref.~\cite{vs99}, see Fig.~\ref{fig:pd_vs}.
Superconducting checkerboard states were investigated as well \cite{mv02},
which turned out to closely compete with stripes.
Noteworthy, indications for paramagnetic stripes being favored
over antiferromagnetic stripes were found before in numerical
calculations for coupled $t$--$J$ ladders \cite{tsu95}.

The evolution from a bond-ordered Mott insulator to a $d$-wave superconductor
was investigated in more detail using a bond-operator approach \cite{kwon}.
Other models for striped superconductors have been proposed \cite{martin01},
and corresponding mean-field theories have been used extensively to describe
concrete experiments, see Sec.~\ref{sec:eff}.
More recently, renormalized mean-field and variational Monte-Carlo calculations
have been used extensively to study in detail the real-space structure of paramagnetic
superconducting stripe states \cite{himeda02,poil07,raczk08,rice08}.

A first-principles view on stripes was obtained from an LDA+U calculation
of a CuO$_2$ plane, focused on a period-4 stripe state (in the charge sector) at 1/8
doping \cite{anisimov04}. The self-consistent stable stripe solution was found to
be bond-centered and strongly antiferromagnetic with period 8 in the spin sector
(note that singlet formation cannot be described by LDA+U).
The calculated photoemission intensity, integrated over a large energy window,
was found to be roughly consistent with the early ARPES result of Ref.~\cite{zhou99}.

A common deficiency of most mean-field-like approaches to stripes is that the charge
modulation within a unit cell is very strong, i.e., the hole-poor regions are essentially
undoped.
Quantum fluctuations can be expected to reduce the modulation, see
Fig.~\ref{fig:dmrg_stripes} for a result beyond mean-field.
Further, it should be emphasized that the filling of stripes, $\rho_l$, cannot
be sharply locked to a particular doping-independent value (e.g. $\rho_l\!=\!1/2$):
If this were the case, then it could only be the consequence of the existence
of an incompressible state at this filling, which, however, then would be insulating
instead of metallic.

From a one-dimensional perspective, metallic stripe phases may display further
instabilities: Apart from superconductivity, density waves with $2k_F$ or $4k_F$ along
the stripe direction may occur. Those have been seen in DMRG \cite{dmrg00},
but not experimentally to date.

Some theory papers have also discussed Wigner-crystal states, e.g. Wigner crystals of
Cooper pairs \cite{pdw1}, in particular in connection with checkerboard structures observed
in STM.
However, considering the metallic character of stripes in 214 cuprates, this type
of approach may be too far on the strong-coupling side.


\subsection{Numerical results}
\label{sec:num}

Numerically exact solutions of microscopic models for the CuO$_2$ planes
play an essential role in the cuprate high-$\Tc$ research.
Naturally, stripe (and other symmetry-broken) states have been searched for,
with limited success.

Of the available methods, exact diagonalization is limited to system sizes
below 40 sites, being too small to detect clear-cut stripe signatures.
Quantum Monte Carlo (QMC) and density matrix renormalization group (DMRG) techniques
can treat larger systems, with the former usually being unable to access low
temperatures and the latter being restricted to the ground state and to quasi-1d systems.
Dynamical mean-field theory (DMFT) \cite{metzner89,georges96}, extended to a small number of
inequivalent sites, may be used to search for stripes.
However, such a method does not capture inter-site correlations beyond the
single-particle level.
This problem can be overcome using modern cluster extensions \cite{dca_review} of DMFT,
provided that the clusters are sufficiently large to host stripes.
However, these computationally demanding method share the QMC problem of being
restricted to elevated temperatures.
In this section, we shall discuss results, obtained with the listed techniques,
relevant for lattice symmetry breaking in cuprate superconductors -- the focus will be on
the existence of and conditions for symmetry-broken phases.

Due to the limited system or cluster sizes, currently all numerical techniques
have severe difficulties in providing detailed momentum- and energy-resolved
spectral information, which could be compared to experimental data.
Therefore, simplified effective models are commonly employed for this purpose,
with an overview given in Sec.~\ref{sec:eff} below.

\subsubsection{Stripes in DMRG}

In an effort to search for stripe states, White and Scalapino have
applied DMRG to the $t$--$J$ model on clusters of sizes up to $19\times8$
sites \cite{dmrg98,dmrg99,dmrg00}.
DMRG, being most suitable for the investigation of 1d systems, was applied
with periodic boundary conditions in $y$ direction, but open boundaries in $x$
direction.

The initial calculations \cite{dmrg98} showed stripe states for $J/t=0.35$ and doping 1/8,
with a periodicities in the charge and spin sector of 4 and 8, respectively,
i.e., very similar to the experimental data on 214 compounds.
Depending on details of the boundary conditions, the stripes were either site-centered or
bond-centered.
In all cases, the amplitude of the charge density modulation was about $\pm 40 \ldots 50\%$.
Such half-filled stripes, i.e. with $\rho_l\approx1/2$ hole per unit length of stripe,
were found to be present for all dopings $x\leq 1/8$.
In contrast, for dopings $0.17<x<0.3$ stripes with $\rho_l\approx1$ were found,
and the region with $1/8<x<0.17$ displayed phase separation between $\rho_l\approx1/2$ and
$\rho_l\approx1$ stripes.

In Ref.~\cite{dmrg00}, the calculations were extended to include a next-neighbor hopping
$t'$ and to study pairing along with stripe formation.
\footnote{
The experimentally measured Fermi-surface shapes of hole-doped cuprates
are reproduced if one assumes $t'/t<0$, with $t'/t=-0.3$ being a typical value.
}
In general, $t'/t>0$ was found to enhance $d$-wave-like pairing while $t'/t<0$ had the opposite effect.
$t'$ with both signs suppressed the tendency toward stripe formation, such that stable
stripes only occurred for $|t'|\lesssim 0.2$.
A plausible interpretation is that stripes compete with superconductivity,
but the two may co-exist, with superconductivity here being two-dimensional rather
than one-dimensional \cite{dmrg00,dmrg01}.
Also, the dependence on boundary conditions was examined in more detail.
For stripes along the $x$ direction, the open boundaries at the stripe ends induce
a $2k_F$ density wave in stripe direction.

Motivated by the observation of checkerboard-like structures in STM \cite{hanaguri04},
the authors also investigated the possible occurrence of checkerboard modulations,
i.e., coexisting vertical and horizontal CDW.
Somewhat surprisingly, true checkerboards (i.e. with $C_4$ symmetry intact) were not found as
low-energy states. However, approximate checkerboard patterns can arise from stripes with additional
CDW formation along the stripes \cite{dmrg04}.

\begin{figure*}
\begin{center}
\includegraphics[clip,width=1.7in]{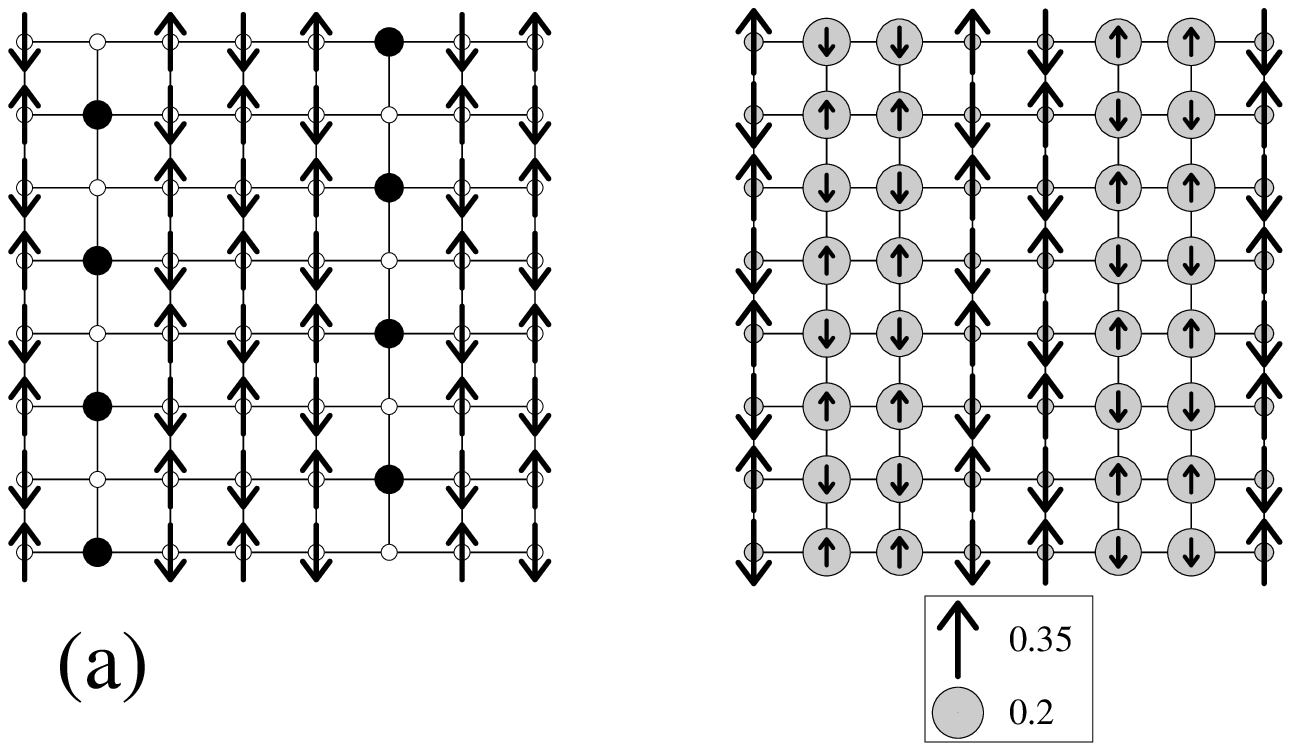}
\hspace*{10pt}
\includegraphics[width=3.1in]{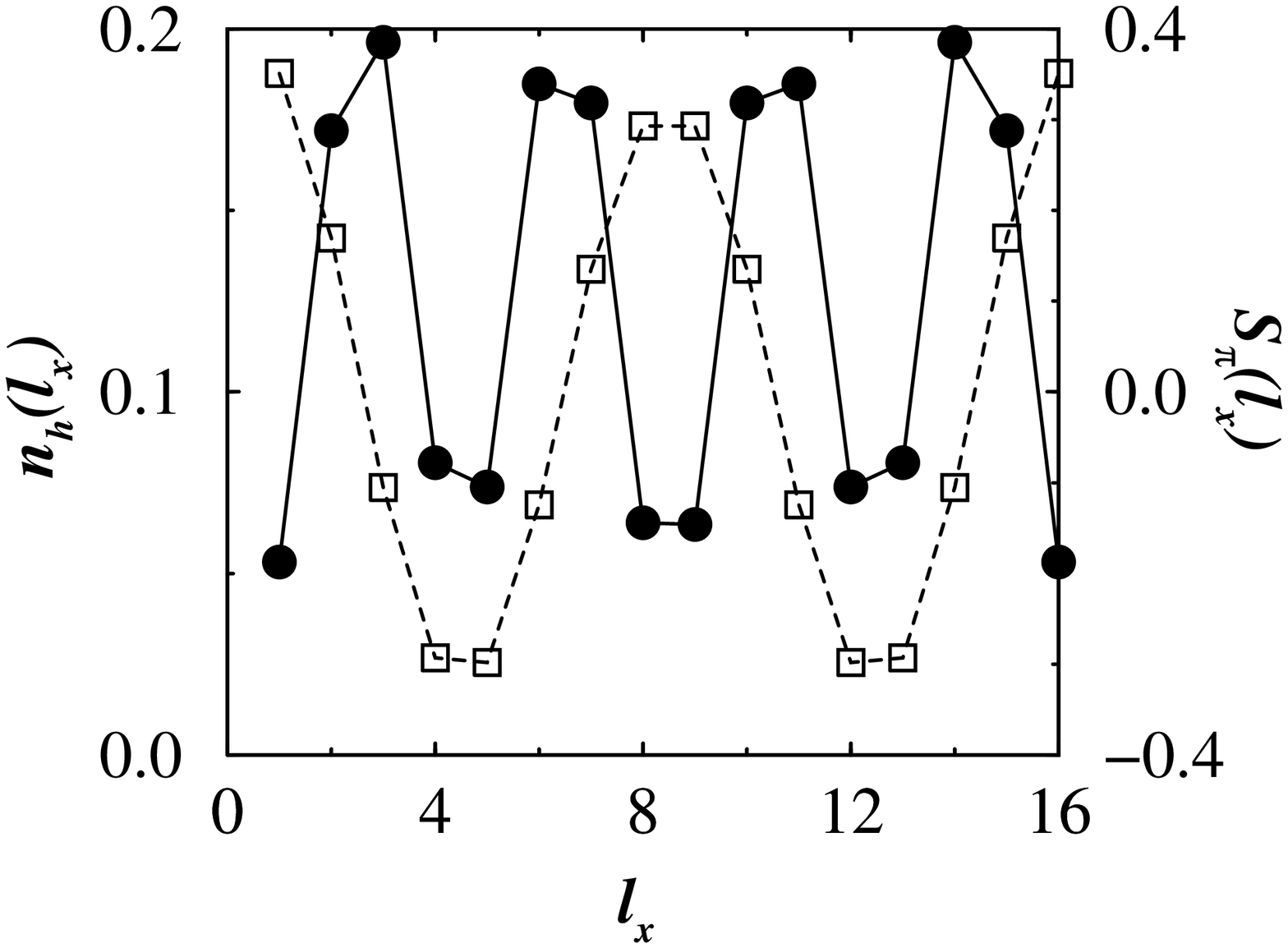}
\caption{
Bond-centered stripe state in the $t$--$J$ model as obtained from DMRG
(reprinted with permission from Ref.~\cite{dmrg99}, copyright 1999 by the American Physical Society).
Left: Hole density and spin moments for the central $8\times 8$ region of a $16\times 8$
$t$--$J$ system. The diameter of the gray circles is proportional to the hole density $n_h$,
and the length of the arrows is proportional to $\langle S_z\rangle$,
according to the scales shown.
Right: Hole density $n_h$ (solid circles) and spin structure function $S_\pi$ (open
squares), averaged over each column on the $16\times 8$ system.
}
\label{fig:dmrg_stripes}
\end{center}
\end{figure*}

The DMRG results have been criticized for their dependence of boundary conditions.
On the basis of exact-diagonalization results, Hellberg and Manousakis \cite{hellberg99}
concluded that the ground state of the $t$--$J$ model is {\em not} striped.
Instead, stripe states may appear as excited states, which then in turn could be favored
by suitable boundary conditions or lattice anisotropies; the effect of the latter was
explicitly studied in DMRG in Ref.~\cite{kampf01}.

While this debate \cite{hellberg99,dmrg00} has, to my knowledge, not been completely
settled, the results indicate that tendencies toward stripe formation are part of the
$t$--$J$ model physics.
Note that long-range Coulomb interaction was not included in the cited calculations.
Hence, the scenario of frustrated phase separation (see
Sec.~\ref{sec:frust_ps}) does not literally apply.

A disturbing feature of the DMRG results is that no robust pairing is obtained
for the physical parameter regime of $t'/t=-0.3$. This may cast doubts on whether the
one-band $t$--$J$ model contains the essential ingredients required to describe
cuprate superconductivity.

Subsequently, DMRG has been applied to study stripes in the Hubbard model.
A large-scale study of six-leg Hubbard ladders \cite{jeckel} investigated stripe
states, modulated along the leg direction, for a doping of $x=9.5\%$ at $U/t=3$ and 12.
Finite-size scaling indicated that stripes are {\em stable} in the infinite-length
limit for $U/t=12$, whereas those at $U/t=3$ are artifacts of DMRG boundary conditions.
For $U/t=12$, this result suggests a small value of the Luttinger parameter $K_\rho$
and hence only weak superconducting correlations.

\subsubsection{DMFT and quantum cluster methods}

While single-site DMFT is designed to treat homogeneous systems, it can be easily
generalized to inhomogeneous or modulated situations, by assuming a local, but site-dependent
electronic self-energy. Then, the lattice problem maps onto a set of single-impurity
problems supplemented by self-consistency conditions.
The inhomogeneous DMFT approach treats correlation effects only locally, and modulations
are captured in mean-field fashion. This implies that valence-bond physics is {\em not} part
of this method, and $d$-wave superconductivity cannot be described beyond mean-field
(in contrast to cluster DMFT to be discussed below).

Fleck {\em et al.}  \cite{fleck00} have applied an inhomogeneous DMFT method,
using supercells up to $36\times8$ sites, to the single-band Hubbard model in the absence
of superconductivity. They found metallic stripes for doping levels
$0.03<x<0.2$, with a number of remarkable properties:
The stripe orientation changed from diagonal to horizontal/vertical around $x=0.05$,
and the stripe separation followed closely the Yamada plot. For $x<0.17$, the stripes
were site-centered -- this is not surprising given the absence of bond singlet physics in
the approach. Also, nodal quasiparticles were gapped by the stripe order;
while this was consistent with some early ARPES results \cite{zhou99},
it is inconsistent with more recent data \cite{valla06,claesson07,chang08b,hanaguri07,borisenko08}.

The problem of missing non-local correlations is overcome in powerful cluster
extensions \cite{dca_review} of DMFT, with the dynamic cluster
approximation (DCA) and the cluster DMFT (CDMFT) being the most prominent variants.
DCA is based on coarse-graining the momentum dependence of electronic self-energy,
and the lattice problem is now mapped onto a correlated {\em cluster} embedded
in a dynamic bath in the single-particle sector.
In CDMFT, the cluster is formed in real space instead of momentum space.
Solving the cluster impurity problems, typically using
QMC or exact diagonalization methods, is numerically much more demanding than treating a
single impurity, which limits the applicability of the method.
Most practical calculations use $N_c=4$ cluster sites -- the minimum required for $d$-wave
superconductivity -- but calculations up to $N_c=64$ (albeit restricted to high
temperatures) have also been reported.
Although the results should become increasingly reliable with growing cluster size,
the mean-field character of the method has to be kept in mind, i.e.,
the self-consistency loop will converge to dynamic saddle points of the
problem.

DCA and CDMFT have been quite successfully applied
in the study of cuprate superconductivity \cite{jarrell06}.
Starting from the standard 2d Hubbard model, supplemented by a nearest-neighbor hopping
$t'$, these methods have reproduced key features of the cuprate phase diagram as function
of doping and temperature: Antiferromagnetism and superconductivity occur in the roughly
correct doping and temperature ranges \cite{maier05},
a pseudogap occurs in the underdoped regime \cite{macridin06,stanescu06,jarrell08a},
and the normal state around optimal doping is characterized by ill-defined quasiparticles
and a large scattering rate \cite{haule07,jarrell08b}.
The dichotomy between nodal and antinodal quasiparticles in the underdoped regime,
i.e., well-defined nodals and broad incoherent antinodals,
has been discussed in detail \cite{civelli08}.

So far, states with broken lattice symmetries have not been unambiguously identified in
DCA or CDMFT. Clearly, the smallest cluster size of $2\times2$ is insufficient for
stripes, as non-magnetic stripes require at least a $4\times4$ cluster (and low
temperatures).
What has been found are strong tendencies towards bond order in the Hubbard model: The
parent Mott insulator with additional magnetic frustration displays a dimerized phase in
CDMFT on a $2\times2$ cluster, but also the neighboring phases are characterized by soft
singlet-to-singlet fluctuations \cite{domenge08}. Using DCA, the doped Hubbard model has
been shown to display a bond-order susceptibility which diverges as $T\to0$ in the
underdoped regime of doping $x<0.22$ \cite{jarrell08a}. This has been taken as
evidence for a quantum critical point at optimal doping between a Fermi liquid and a
bond-ordered phase.
While the precise interpretation of these numerical results may be problematic due to finite-size
and finite-temperature effects, they support the notion that valence-bond physics
is a relevant player in underdoped cuprates.

The self-energy-functional theory of Potthoff \cite{pott03} allows to construct
more general dynamical cluster theories.
Here, correlated clusters coupled to uncorrelated bath sites,
all with variationally determined parameters, are solved exactly.
DMFT and CDMFT arise in the limit of an infinite number of bath sites.
In contrast, without bath sites one obtains the so-called variational cluster
approach (VCA) \cite{vca1}.
Compared to CDMFT, larger clusters can be used and access to low temperatures
is simplified.
This method has been employed recently to the 2d Hubbard model \cite{vca1,vca2,vca3},
and the general structure of the phase diagram has been obtained in overall agreement
with DCA and CDMFT.
So far, stripe states have not been systematically searched for.
A recent extension allows for the calculation of two-particle correlations
beyond the bubble approximations \cite{brehm08},
and results will be discussed below.

Interestingly, the present results from quantum cluster approaches
indicate that the difference between electron and hole
doping in the cuprates is not fully captured by the single-band $t$--$t'$ Hubbard model:
Although antiferromagnetism is found to be more stable for electron doping, in agreement with
experiment, the low-temperature amplitude of the superconducting $d$-wave order parameter
is roughly equal for electron and hole doping \cite{jarrell06,vca3},
in disagreement with experiment.
An initial DCA study of the three-band Hubbard model \cite{kent08},
with microscopic parameters extracted from LDA
bandstructure calculations and limited to small cluster of four Cu sites,
has indeed shown that the superconducting $\Tc$ depends
sensitively on microscopic parameters.


\subsection{Weak-coupling approaches}

For small band filling, or equivalently large doping, it is natural
to expect weakly correlated Fermi liquid behavior in the Hubbard and $t$--$J$ models.
Assuming that this persists up to band fillings of $\sim 0.7$, weak-coupling
many-body techniques may be used to access the physics of high-$\Tc$ cuprates
from the overdoped side.

\subsubsection{Random-phase approximation}

The simplest way to capture collective instabilities of a Fermi liquid
is the random-phase approximation (RPA).
This has been primarily employed to describe the spin excitations in hole-doped cuprates
and their interplay with superconductivity.
RPA is an infinite resummation of bubble diagrams for the susceptibility,
leading to
\begin{equation}
\chi_{\rm RPA}(\vec q,\w) = \frac{\chi_0(\vec q,\w)} {1-g(\vec q) \chi_0(\vec q,\w)}
\label{rpa}
\end{equation}
where $\chi_0$ is the Lindhard susceptibility of the non-interacting systems
and $g$ is the fermionic four-point vertex in the singlet (triplet) channel
for the charge (spin) susceptibility.

RPA can predict instabilities towards ordered phases:
$\chi_{\rm RPA}(\vec q,\w\!=\!0)$ diverges when $g(\vec q) \chi_0(\vec q,\w\!=\!0)$
reaches unity at some wavevector $\vec q=\vec Q$.
In addition, RPA is able to describe collective modes: Sharp modes are reflected in
poles of $\chi_{\rm RPA}$ at some finite frequency, which then have to lie outside the
particle--hole continuum of $\chi_0$, whereas damped modes can exist inside the
continuum.
In fact, RPA has been initially used to model the emergence of the so-called spin
resonance mode in the superconducting state of cuprates: This mode at $(\pi,\pi)$
is overdamped in the normal state, but pulled below the particle--hole continuum by the
onset of superconductivity, see Sec.~\ref{sec:eff_spin}.

For an interaction vertex $g$ with little structure in momentum space,
RPA predicts the instability wavevector $\vec Q$ to be given by the maximum of
$\chi_0(\vec q,\w\!=\!0)$, i.e., $\vec Q$ is determined by Fermi-surface properties.
From the measured Fermi surfaces of underdoped cuprates,
there are two candidate wavevectors $\vec Q$ where $\chi_0$ displays peaks:
(i) a wavevector close to $(\pi,\pi)$, connecting regions near the nodal points -- this
has frequently been related to incommensurate SDW order, and
(ii) a short wavevector in $(1,0)$ or $(0,1)$ direction which connects the often nearly
straight antinodal segments of the Fermi surface and which may be related to CDW order.
With increasing hole doping, the magnitude of both vectors decreases.
For the SDW nodal nesting wavevector this means that the difference to $(\pi,\pi)$ increases,
i.e., the modulation period w.r.t. commensurate antiferromagnetism decreases,
which is in qualitative agreement with spin fluctuations observed in various
cuprates.
In fact, RPA has been used quite successfully to describe features
of the cuprate spin fluctuation spectrum, with a brief overview given in
Sec.~\ref{sec:eff_spin}.
In contrast, the possible CDW period deduced from the antinodal nesting wavevector increases with
hole doping, in sharp contrast to ordered stripes in 214 cuprates.
As charge order sets in at a higher temperature than spin order (see
Sec.~\ref{sec:exp_elscatt}), this indicates the failure of RPA in the underdoped regime,
at least for the 214 compounds.
(Note, however, that a CDW wavevector decreasing with doping has been extracted from
STM measurements in BSCCO-2201 \cite{wise08}.)
A failure of RPA has also been pointed out for the dynamic spin excitations
in optimally doped YBCO \cite{reznik08c} and BSCCO \cite{xu09} as well as
in electron-doped Pr$_{0.88}$LaCe$_{0.12}$CuO$_4$ \cite{krueger07}, based on strong
mismatch between RPA and experiment regarding the spin fluctuation intensities and the
structure of the signal above $\Tc$.
A more detailed critical discussion of the applicability of RPA is in Sec.~\ref{sec:wkstr}.

In applications to the Hubbard model a common choice is $g=U$, whereas for the square-lattice
$t$--$J$ model $g=J(\cos k_x + \cos k_y)$. In practical applications, the amplitude of
$g$ is often treated as a fit parameter to adjust the position of the instability,
diminishing the predictive power of RPA.

Although it is frequently assumed that RPA is a controlled approximation for weakly
interacting systems, this belief is incorrect:
In situations with competing instabilities, RPA is unreliable {\em in principle}
even at weak coupling.
This can be easily seen for the half-filled Hubbard model with perfect nesting,
where the $U$ dependence of the antiferromagnetic transition temperature is predicted
incorrectly by RPA.
As one of the main features of cuprates is the presence of multiple instabilities,
approaches beyond RPA are required.

\subsubsection{Functional renormalization group}
\label{sec:frg}

A systematic and consistent treatment of competing weak-coupling instabilities is
provided by the functional renormalization group (fRG) treatment of the underlying microscopic
model.
The practical applications of the fRG method to the Hubbard model have been summarized
in detail by Honerkamp \cite{honer_rev}.

The starting point is an exact formulation of the renormalization
group in terms of flow equations for $N$-particle irreducible vertex functions, which are
truncated such that only the flow of the one-particle vertex functions is kept.
To solve the flow equations numerically, the frequency dependence of the vertex functions
is neglected, and the momentum dependence is discretized via suitable patches in momentum
space.
Different flow schemes have been developed, with either the momentum-space cutoff, the
temperature, or the interaction as flow parameter \cite{honer_rev}.
In the temperature-flow scheme, one starts at high temperatures $T$ and calculates the
temperature dependence of the momentum-resolved interaction vertices in the particle--particle
and particle--hole channels.
Some of these vertices will grow upon reduction of $T$, with a divergence indicating an
instability towards an ordered state.
Usually, this stops the applicability of the fRG scheme.\footnote{
A proposal to circumvent this problem is the application of a small symmetry-breaking
field which is kept as additional flow parameter \cite{frg_ord}.
}

Importantly, the fRG method is able to deal with competing ordering phenomena in an
unbiased manner, in contrast to RPA.
The perturbative character of the fRG limits its applicability to small and
moderate values of $U$, practically values of $U/t$ up to 4 have been used.
It furthermore implies that the physics of the Mott
insulator cannot be described, although precursors of Mott physics driven by strong
Umklapp scattering have been discussed \cite{honer_rev}.

The published fRG calculations for the 2d Hubbard model \cite{HM00,GKW02,honer01a,honer01b,metzner05}
consistently show antiferromagnetism (at small doping) and $d$-wave pairing (at larger doping)
for both signs of the second-neighbor hopping $t'$, corresponding to electron and hole
doping, respectively.
The hole-doped regime with $t'/t<0$ is addition characterized by a so-called saddle-point
regime where many instabilities compete and re-enforce each other,
in addition to antiferrmagnetism and $d$-wave pairing these are a $d$-wave Pomeranchuk (nematic)
instability and, to a weaker extent, $d$-density wave order \cite{honer01b}.
In the overdoped regime for $0.15\leq x\leq0.30$, a recent fRG calculation \cite{ossadnik08}
found strong angle-dependent scattering at $U/t=4$.
This leads to an apparent violation of Fermi-liquid behavior, consistent with
transport experiments on overdoped Tl$_2$Ba$_2$CuO$_{6+x}$ \cite{abdel},
but does not exclude that the underlying low-temperature state is a Fermi liquid.

In present fRG studies, instabilities toward charge order are found to subleading.
Interestingly, a recent study \cite{fu05} of a Hubbard-like model supplemented by electron--phonon
interactions found that phonons strengthen the tendency towards a CDW with a $d$-wave form
factor and a wavevector given by the nested antinodal Fermi-surface pieces. However, also
this instability appears to be magnetism-driven.
In summary, genuine CDW instabilities do not seem to occur in the Hubbard model at weak
coupling.


\subsection{Origin of stripe formation}
\label{sec:origin}
\label{sec:frust_ps}

After having reviewed results from microscopic calculations,
I will try to summarize the current phenomenological ideas on the origin of
stripes.
This is particularly important, as collected experimental and theoretical
results described in Secs.~\ref{sec:exp} and \ref{sec:mic} leave us with
what looks like a conflict:
On the one hand, stripes have been unambiguously identified in experiment, but, on the
other hand, stripes do not readily appear in theoretical treatments (beyond mean-field)
of the popular microscopic models. An incomplete list of possible explanations is:
(i) Without the LTT distortion of some 214 compounds, stripes are never the lowest-energy
state of CuO$_2$ planes.
While this is certainly a possibility, tendencies toward stripes have also been identified
in cuprates with tetragonal CuO$_2$ planes, and hence should be detectable in numerics.
(ii) The present limitations inherent to the numerical methods, i.e. cluster size,
temperature etc., do not yet allow to detect stripes.
(iii) The investigated single-band models do not contain all ingredients required for stripe
formation -- this appears to be a likely explanation.

About the origin of stripes:
Phenomenologically, popular lines of thought -- all from a strong-coupling perspective and
not being mutually exclusive -- have been
(A) frustrated phase separation (or micro phase separation),
(B) spin-charge ``topological'' properties, and
(C) valence-bond solid formation,
all of which have been invoked to rationalize the formation of conducting stripes
in cuprates. While (A) is general, (B) and (C) refer to more microscopic aspects.

Frustrated phase separation \cite{ek93,loew94,sk1} requires little explanation:
A system of particles on a lattice, moving under the influence of short-range attractions
and long-range repulsions, minimizes its energy by forming linear domains of
enhanced and reduced density.\footnote{
Ref.~\cite{pryadko} has argued, based on a Ginzburg-Landau analysis,
that long-range physics {\em alone} is insufficient to produce the experimentally
observed antiphase magnetic stripe order.
A more formal reasoning on frustrated phase separation can be found in
Ref.~\cite{chayes96}.
}
This argument invokes long-range Coulomb repulsion as
crucial, and leaves open the issue of how the short-range physics exactly looks like.
Being classical, it generically results in conducting instead of insulating stripes.

Microscopically, frustrated phase separation in cuprates is related to the fact that an
undoped Mott insulator (partially) expels holes.\footnote{
This is most easily seen by the fact that hole motion in an antiferromagnetic
background is frustrated due to the creation of spin defects.}\footnote{
In the phenomenological SO(5) theory, a tendency towards phase separation is found
as well \cite{so5stripe}, here with superconducting and antiferromagnetic domains.
}
A resulting stripe state then not only
balances magnetic (i.e. short-range attractive) and Coulomb (i.e. long-range repulsive)
energies, but also the hole kinetic (i.e. quantum) energy:
Holes can gain kinetic energy by moving along the hole-rich stripes, but at the same
time leave the hole-poor domains essentially unaffected.
Due to the presence of quantum effects, long-range Coulomb interaction may not be
absolutely necessary for stripe formation (although it is in some mean-field approaches
which underestimate quantum effects).

Inspired by the physics of one-dimensional systems, spin-charge ``topological'' properties
have been discussed as driving force of stripe formation.
Hole-rich stripes form charged domain walls of the background antiferromagnetic order,
which may be interpreted as a two-dimensional generalization of holons \cite{zaanen96}.
Within this picture, stripes form elastic strings on the lattice,
which may be pinned by disorder \cite{KFE98,hassel98,hassel99},
and the destruction of a stripe states is driven by transverse stripe fluctuations
or topological defects.
The term ``topological'' is somewhat misleading here, as domain walls are topological
defects of Ising antiferromagnets, but {\em not} of ones with Heisenberg symmetry.
Hence, there is no topological protection for the strings.
Microscopic considerations of the motion of holes along AF domain walls within the $t$--$J$ model
show that motion both along the domain wall and into the AF domains (which leads to some
spin-polaron-like dressing) is required to compensate for the cost of the domain wall
\cite{cherny02}.
A general problem with this set of ideas is that it is intimately tied to magnetic order.
Experimentally, charge stripes exist without long-range magnetic order
(Sec.~\ref{sec:exp}), although some short-range order is certainly present.

In contrast, valence-bond solid formation is a concept relying on physics in the singlet
sector. It starts from the observation that the destruction of antiferromagnetic order in
an undoped Heisenberg magnet on the square lattice is accompanied by lattice symmetry
breaking, most likely in the form of a columnar valence-bond solid \cite{rs0,ssrmp}.
It has been argued that this type of symmetry breaking survives with hole doping
\cite{vs99,jpsj}, leading to a VBS metal or superconductor.
Moreover, the resulting system, effectively consisting of coupled two-leg ladders,
may display a stripe (or frustrated phase separation) instability,
because weakly doped ladders have strong rungs,
whereas kinetic energy is better gained by motion along the legs.
The STM data of Ref.~\cite{kohsaka07}, showing bond-centered stripes with strong bond
modulations, appear to support this concept.
It should be pointed out that valence-bond stripes, although not relying on magnetic
long-range order, are primarily driven by magnetic exchange interactions, as those set
the scale for VBS formation.

In both microscopic pictures, the doping dependence of the stripe ordering wavevector is
explained by the hole-rich stripes having a filling close to half a hole per unit length of
stripe which weakly depends on the overall doping $x$ for small $x$.
As a result, $\Qch \approx (4\pi x,0)$. For larger $x$ and
short stripe distances, stripes strongly repel each other, resulting in a plateau in
$\Qch(x)$, before stripes eventually disappear.

While stripes may in principle arise from weak-coupling Fermi-liquid instabilities,
many cuprate experiments point toward a strong-coupling picture being more appropriate,
as detailed in Sec.~\ref{sec:wkstr}.
However, it is well conceivable that nearly nested antinodal Fermi surface pieces
have a share in driving charge order.
This speculation is particularly appealing as an explanation of
increased stability of stripes at $x\!=\!1/8$. This is commonly attributed to lattice
commensuration effects, may be further enforced by Fermi surface nesting.
In this scenario, $x\!=\!1/8$ is the doping concentration where the antinodal
nesting wavevector and the strong-coupling stripe wavevector coincide.


\subsection{Spiral magnetism}
\label{sec:spiral}

Holes doped into a quantum antiferromagnet have been argued to induce
an instability towards a spiral state \cite{shraiman,ss94}.
It has been proposed that the physics of spin spirals dominates
the insulating regime of cuprates at low doping \cite{sushkov06}.
In this scenario, holes are pinned at random positions by the electrostatic
potential of the dopant atoms. The local spiral distortions
of the classical spin background lead to elastic peaks in neutron scattering
around wavevectors $\Qsp = 2\pi(0.5\pm\epsilon_s',0.5\pm\epsilon_s')$,
with a doping evolution of $\epsilon_s'$ as observed experimentally.
Although a doped spiral state is unstable w.r.t. phase separation,
this may be circumvented here by strong pinning.
A distinct but related scenario was proposed in Ref.~\cite{juricic06}:
Magnetic anisotropies were argued to stabilize a canted N\'eel state at
doping $x \lesssim 2\%$, whereas they lead to a helicoidal magnetic phase at
larger doping (but still in the insulating phase), with a small out-of-plane
magnetization component. This proposal appears more consistent with susceptibility
measurements in weakly doped LSCO \cite{lavrov01}.

In the superconducting state at higher doping, mobile holes and quantum effects become
more important and drastically change the underlying physics. Therefore, the
physics of spin spirals is commonly assumed to be irrelevant to cuprate superconductivity.
However, the smoking-gun polarized neutron scattering experiment has not been performed
to date, see Sec.~\ref{sec:exp_elscatt}.


\subsection{Nematic order}
\label{sec:mic_nematic}

As with stripes, both weak-coupling and strong-coupling approaches have
been employed to search for nematic phases in correlated electron models.
A common feature of the studies described below,
which make {\em no} reference to fluctuating stripes,
is that a nematic instability
(equivalent to a spin-symmetric $d$-wave, $l\!=\!2$, Pomeranchuk instability)
is driven by strong forward scattering.

Strong-coupling mean-field calculations in the standard RVB slave-boson formalism
find a nematic phase in the 2d $t$--$J$ model, which competes with homogeneous $d$-wave
superconductivity \cite{YK00,yamase06}.
For the 2d Hubbard model, weak-coupling functional RG has been applied \cite{HM00,GKW02,NM03},
with qualitatively similar conclusions:
In particular near the van-Hove filling, there is a tendency towards a $d$-wave Fermi
surface deformation. Subsequently, antiferromagnetism or $d$-wave superconductivity
may develop at low temperatures inside the nematic state. If, however, antiferromagnetism or
superconductivity set in before the nematic order, then the forward scattering
interactions stop to grow, and a nematic instability does not develop.
The overall picture of nematic order competing with $d$-wave pairing is supported
by exact diagonalization \cite{miyanaga06} and variational Monte-Carlo \cite{gros06}
studies of the $t$--$J$ model.
A nematic phase has also been shown to occur in a special strong coupling limit
of the three-band Hubbard model \cite{nema_emery}.

In 3d lattice models, the Pomeranchuk instability for even $l$ is generically of first
order due to the presence of cubic terms in the Ginzburg-Landau theory. It has been
argued that even in $d=2$ strong fluctuations may drive the transition
first order \cite{KKC03,KCOK04}. Full quantum critical behavior is
restored in these 2d models if a sufficiently strong repulsive term is
added to the forward-scattering interaction \cite{YOM05}.

Mean-field treatments of nematic ordering often employ effective models with a
quadrupole--quadrupole interaction explicitly designed to produce an $l\!=\!2$
instability \cite{OKF01,KKC03,KCOK04,YOM05}.
Recently, it has been shown that generic central interactions in 2d can
produce Pomeranchuk instabilities as well \cite{schofield08}.

Notably, nematic order in an {\em isotropic} Fermi liquid (i.e. without lattice)
leads to non-Fermi liquid behavior due to the overdamped Goldstone fluctuations
of the director order parameter \cite{OKF01}.

A natural expectation is that the formation of stripes should be favored in a $d$-wave
nematic phase due to the uni-axial anisotropy \cite{KFE98}.
However, to my knowledge, a comprehensive study of a microscopic model (using mean-field
or more elaborate techniques) with a sequence of disordered, nematic, and stripe phases
is not available.


\section{Effective models: Linking theory and experiment}
\label{sec:eff}

In this section, we describe theoretical activities aimed on a detailed modeling
of experimental data which have been obtained in (or close to) phases with broken
lattice symmetries.
Given the difficulties with the numerically exact treatment of strongly correlated
Hubbard or $t$--$J$ models, most efforts are constrained to simplified
effective models representing spin or renormalized single-particle degrees of
freedom.
Most of the section will again be devoted to stripe order, but nematic and loop-current
order will also be covered.


\subsection{Weak modulations vs. coupled chains and ladders}

Often, effective models {\em assume} the presence of static stripe-like modulations
in both the charge and spin sectors.
Two seemingly distinct viewpoints can then be employed for model building:
Either
(A) one starts from a two-dimensional system and adds (weak) modulations, or
(B) one starts from one-dimensional chains or ladders and adds a (weak) transverse coupling.
The states which result from these two approaches are usually equivalent regarding
symmetries, but for a given experimental situation one of the two may be more
appropriate.
Intuitively, (A) leads to a weakly modulated state,
whereas (B) corresponds to strong modulations.
Experimentally, the amplitude of the modulations in the charge sector is not precisely
known, and extracting modulation amplitudes from actual data is model-dependent.
From x-ray scattering, the variation in the oxygen hole concentration in LBCO-1/8
has been estimated to be about a factor 4 \cite{abbamonte05}; the STM tunneling asymmetry
maps show a contrast of a factor 1.5--2 \cite{kohsaka07}. This suggests that the
modulations are not weak, but also not close to the maximum limit.

However, this is not the full story:
1d building blocks often feature fractionalized excitations described by a Luttinger
liquid, whereas excitations in 2d are conventional.
Then, the elementary excitations of the starting points of (A) and (B) are different.
Nevertheless, upon including the perturbations, the low-energy physics is the same in both cases,
as inter-chain coupling is a relevant perturbation of the Luttinger-liquid fixed point
(except for certain frustrated cases),
and even {\em weakly} coupled metallic chains form a 2d Fermi liquid in the low-energy limit.
Consequently, the answer as to whether approach (A) or (B) leads to a better description
of experimental data may depend on the observable and the energy range to be considered.
In principle, 1d spinons and holons could be a good description of the
excitations at elevated energies provided that the modulations are strong --
this has been proposed \cite{orgad01}
in interpreting certain photoemission data, but is controversial.

For some observables, the situation is simpler:
For instance, the spin excitation spectrum of bond-ordered stripes can be described
both by coupled spin ladders and by a combined theory of CDW and SDW order parameters
in a 2d system, with essentially identical results,
see Sec.~\ref{sec:eff_stripemag}.
One reason for this equivalence is that the spin excitations of a two-leg ladder are
conventional spin-1 triplons.


\subsection{Static, fluctuating, and disorder-pinned stripes}
\label{sec:pin}

The picture of long-ranged fluctuationless stripe order in the charge sector
is certainly idealized: Even at lowest temperatures (i.e. without thermal fluctuations),
both quantum fluctuations and quenched disorder (arising e.g. from dopant atoms)
will induce deviations from ideal order.
Deep in the charge-ordered phase, such deviations are often negligible
(however, e.g. photoemission spectra are strongly
influenced even by small amounts of stripe disorder).

The most interesting regime is close to the charge ordering transition,
where -- in the absence of quenched disorder -- collective degrees of freedom
associated with the CDW order are slowly fluctuating in space and time.
Here, the CDW physics on short scales is quantum critical,
and a comprehensive theoretical treatment becomes difficult.

In building approximations, the question arises which type of stripe fluctuations
dominates. In the language of the CDW order parameter $\phi_c$, one can distinguish
amplitude and phase fluctuations. The latter imply that stripes remain well-defined
objects, but fluctuate in the transverse direction -- here the picture of elastic
strings on a lattice \cite{hassel98,hassel99} has been frequently used.
However, this is not the full story: The phase field can display topological
defects, corresponding to stripe end points. Moreover, in a tetragonal environment,
there will be generalized phase fluctuations between horizontal and vertical stripes.
In high dimensions, it is known that transitions are driven by amplitude fluctuations,
in 2d the situation is less clear, and the importance of certain fluctuations depends on
whether the system is more in the weak-coupling or strong-coupling limit.

Including the effects of quenched disorder complicates matters,
but may also imply some simplification (Sec.~\ref{sec:dis}).
Most importantly, disorder is of random-field type in the charge sector
and tends to pin fluctuations.
Then, on the one hand, short-range ordered CDW configurations are rendered static
(and hence visible e.g. in STM).
Pinning will act on both horizontal and vertical stripes and thus tends to
smear the distinction between stripe and checkerboard order \cite{maestro,robertson}.
On the other hand, the finite-frequency CDW dynamics becomes glassy,
and the charge-ordering transition is smeared.

In the literature, various routes have been followed to tackle the situation of
fluctuating stripes:
(i) Ignoring subtleties of critical physics and disorder, one may treat the
CDW modes perturbatively, e.g., by calculating the lowest-order diagrams describing the
interaction of electrons with CDW modes \cite{castellani95,caprara99}.
This approach is suitable if amplitude fluctuations of the CDW order are dominant.
(ii) Theories of quantum strings on a lattice have been invoked to discuss the phase
diagram and the dynamics of fluctuating stripes in the presence of both disorder and
lattice pinning \cite{KFE98,hassel98,hassel99,zachar00}.
In particular, disorder pinning may lead to a Bragg-glass-type state \cite{KFE98}.
Based on a calculation of the ratio between spin and charge correlation lengths in scenarios
with and without topological string defects, it was argued that stripes in 214 cuprates
are dominated by non-topological fluctuations \cite{zachar00}.
(iii) Numerical simulations of order parameter theories can in principle account for all possible
fluctuations. As the full dynamical treatment of coupled modes is rather complicated,
a Born-Oppenheimer approach to the coupled SDW--CDW problem was proposed in Ref.~\cite{vvk}:
It is {\em assumed} that the collective charge modes are slow compared to the spin modes.
The CDW modes are approximated as static on the time scale of the spin
fluctuations, and the spin sector is solved for fixed charge configurations.
This approach neglects charge dynamics, i.e., inelastic processes
involving CDW modes, but is otherwise non-perturbative and allows to access the full
spin dynamics, as discussed in Refs.~\cite{vvk,mv08}.
The assumption of slow charge dynamics appears particularly good in the presence of
strong pinning by quenched disorder. Then, the Born-Oppenheimer approximation may even be used
down to zero energy, e.g. to calculate photoemission spectra.
(iv) A strongly simplified version of the Born-Opperheimer approach consists of treating
static parallel stripes with random spacings, which only accounts for a very special type of
CDW phase fluctuations.

While these approaches provide some insight into different limits of the
fluctuating-stripe problem, a concise theoretical treatment of the random-field
glassy dynamics in the quantum regime is not available to date.


\subsection{Spin excitations}
\label{sec:eff_spin}

Magnetic excitations in doped Mott insulators such
as cuprates can either be calculated numerically from one of the relevant microscopic
models (with the limitations discussed in Sec.~\ref{sec:num}) or can be obtained from simplified
effective descriptions: Here, magnetism can be modeled in an ``itinerant'' or in a ``localized'' concept.
In an itinerant description, one starts from band electrons and captures
interaction-generated collective dynamics via RPA or more sophisticated methods.
Alternatively, in a localized picture, the modeling is centered around collective modes,
typically obtained from Heisenberg models of localized spins, and the physics of mobile
carriers is added perturbatively or ignored entirely. While the latter option implies a
quite drastic simplification, it may be sufficient to describe collective modes (seen as
peaks in $\chi_s''$) which reside on top of a broad continuum of particle--hole excitations
-- this continuum is then simply not part of the model.

The itinerant description is obviously more appropriate at small band filling or large
doping, where correlation effects are expected to be weak, whereas local-moment
collective magnetism should prevail in the small-doping regime. In the absence of
symmetry breaking, the two situations may or may not be adiabatically connected --
theoretical scenarios for both cases are known: In a one-band Hubbard model at
half-filling and close to nesting, the small-$U$ itinerant antiferromagnet is
continuously connected to the large-$U$ local-moment antiferromagnet \cite{hubb_af}. On
the other hand, in two-band (Kondo-lattice) models of interacting electrons, metallic
paramagnetic phases with and without local moments can exist, which are {\em not}
continuously connected: While the latter is a conventional Fermi liquid, the former is
an exotic state dubbed ``fractionalized Fermi liquid'' \cite{flst}.
For the cuprates, it is not known whether the itinerant and localized descriptions of
magnetism are adiabatically connected. As detailed below, various experiments in doped
compounds can in principle be described within both concepts, however, the standard
RPA (or ``fermiology'') models fail in some important cases.

\subsubsection{Spin excitations of ordered stripes}
\label{sec:eff_stripemag}

Static stripe order in both the spin and charge sectors, which has to be accompanied by
linearly dispersing low-energy magnetic excitations due to Goldstone's theorem,\footnote{
Magnetic anisotropies will open a gap in the spin-wave spectrum,
but are often ignored.
}
lends itself to relatively simple theoretical descriptions.

\begin{figure*}
\begin{center}
\ifhires
\includegraphics[width=2in]{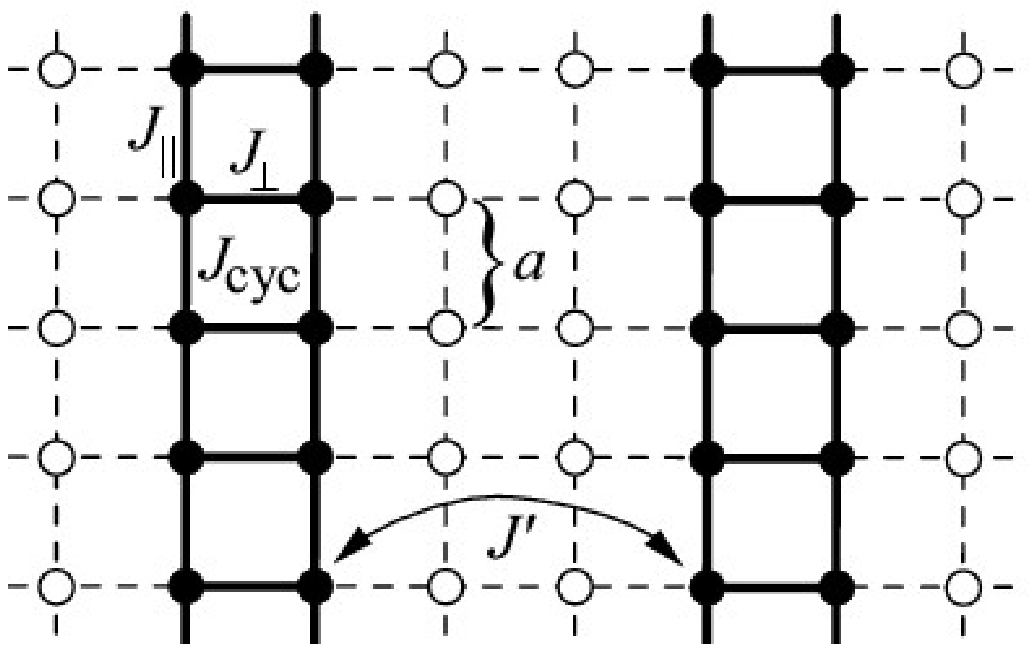}
\hspace*{20pt}
\includegraphics[width=2in]{uhrig1.eps}
\else
\includegraphics[width=2in]{gsu_ladd1.eps}
\hspace*{20pt}
\includegraphics[width=2in]{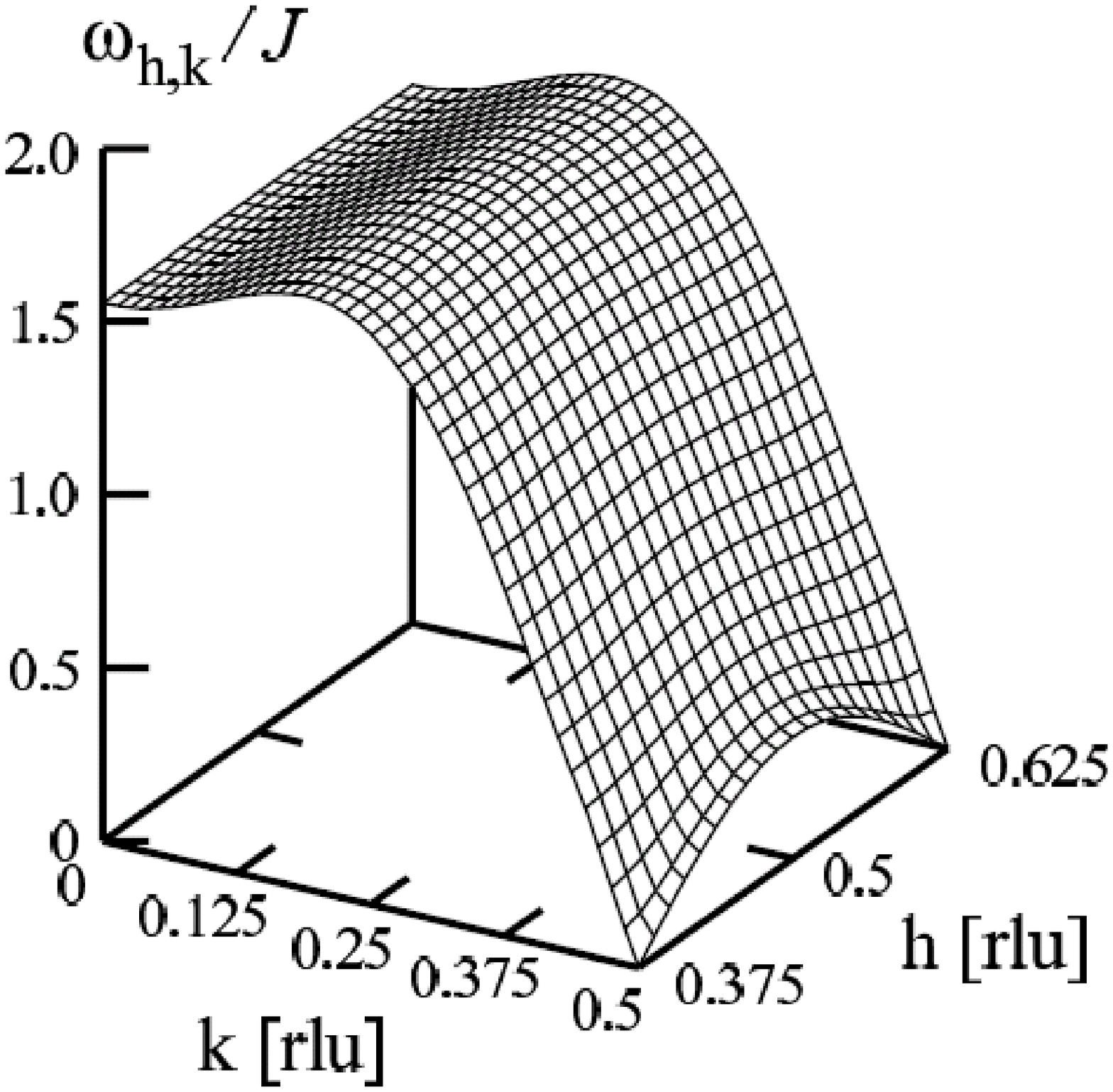}
\fi
\caption{
Left:
2D square lattice split into two-leg $S=1/2$ ladders (full dots) coupled by
bond-centered charge stripes (open dots). $J_\parallel = J_\perp = J$ is the intra-ladder
exchange, $J_{\rm cyc}$ is an additional cyclic exchange.
Magnetic properties are calculated from the two-leg ladders coupled by $J'$.
Right:
Excitation spectrum $\w_{h,k}$ of coupled two-leg spin ladders as model for
period-4 stripes as function of the momenta perpendicular (h) and along (k)
the ladders.
The inter-ladder coupling is ferromagnetic and chosen such that the excitation
gap vanishes.
The spectrum displays low-energy ``spin waves'' and high-energy ``triplons'' which meet
at a ``resonance peak'' at $(\pi,\pi)$.
(Reprinted with permission from Ref.~\cite{gsu04}, copyright 2004 by the American Physical
Society.)
}
\label{fig:gsu}
\end{center}
\end{figure*}

In particular, in a picture of spin moments localized in the hole-poor regions of the
square lattice, one can write down a Heisenberg model for coupled spin ladders.
\begin{equation}
\mathcal{H} = J \sum_{\langle ij\rangle} {\vec S}_i \cdot {\vec S}_j +
J' \sum_{\langle ik\rangle} {\vec S}_i \cdot {\vec S}_k
\end{equation}
with the $\sum_{\langle ij\rangle}$ running over intra-ladder bonds,
while the sum $\sum_{\langle ik\rangle}$ connects neighboring sites from different
ladders. The model may be supplemented by longer-range couplings or ring-exchange terms.
Microscopically, $J$ is the usual antiferromagnetic exchange between
neighboring Cu spins, whereas $J'$ is an {\em effective} exchange which is thought to be
mediated via the hole-rich stripes, Fig.~\ref{fig:gsu}.
Note that the sign of $J'$ determines whether ordered
antiferromagnetism will be in-phase or anti-phase.
The hole-rich stripes themselves are not part of the model; they are assumed to
only contribute a broad continuum to the spin excitations, and their damping effect is
ignored.

For site-centered stripes with a period 4 in the charge sector, the canonical choice are
three-leg ladders  \cite{scheidl03,carlson06}.
``Antiphase'' magnetic order, with an ordering wavevector
of $\Qsp=(3\pi/4,\pi)$, is obtained from antiferromagnetic inter-ladder coupling, $J'>0$.
For bond-centered period-4 stripes, two-leg ladders appear natural \cite{mvtu04,gsu04}, but four-leg
ladders have also been used \cite{carlson06}; anti-phase magnetism requires $J'<0$ in both cases.
For odd-leg ladders, an infinitesimal $J'$ is sufficient to drive the coupled system into a
magnetically ordered ground state, while a finite $J'$ is required in the even-leg case
in order to close the spin gap of the isolated ladder.
The resulting spin model can be treated by semiclassical spin-wave theory (assuming a
magnetically ordered state) or by more elaborate methods, to be described below.

Most of the published calculations were triggered by the inelastic neutron scattering
experiments of Tranquada {\em et al.} \cite{jt04} on \lbcoo, which showed an upper dispersion
branch well described by the spectrum of an isolated two-leg spin ladder in a situation
with period-4 charge order.
Consequently, coupled two-leg ladders were used in Refs.~\cite{mvtu04,gsu04}.
Ref.~\cite{mvtu04} employed a linearized bond-operator description for a situation inside
the magnetically ordered phase close to the QCP, i.e., where weak magnetic order
co-exists with strong dimerization, and the magnetism is far from the semiclassical
limit. Ref.~\cite{gsu04} used a more sophisticated approach of continuous unitary
transformations (which, however, cannot be easily applied to the ordered phase)
right at the QCP.

The two approaches \cite{mvtu04,gsu04} yield dispersions of effectively non-interacting bosonic excitations,
with rather similar results. They nicely show a dual character of the excitations (Fig.~\ref{fig:gsu}):
At low energies, the excitations resemble spinwave-like Goldstone modes, whereas at higher
energies, the character of the triplon dispersion of the single ladder is reproduced. The
low-energy and high-energy branches meet in a saddle point at wavevector $(\pi,\pi)$. If
$|J'|$ is not too large, the energy of this saddle point is essentially given by the spin
gap of an isolated ladder. In fact, the idea of low-energy incommensurate excitations and
the resonance peak being part of a unified picture was proposed earlier by Batista {\em
et al.} \cite{batista01}.

The excitation spectra of the coupled-ladder model compare favorably
to the experimental data of Ref.~\cite{jt04}, after taking an average of horizontal and vertical
stripes (assuming stripes running orthogonal to each other in adjacent planes),
see Fig.~\ref{fig:ecuts} in Sec.~\ref{sec:exp_inelscatt}.
Once matrix elements are properly taken into account, the low-energy signal essentially
consists of four spots near the ordering wavevectors, forming a square with corners
along the (1,0) and (0,1) directions. Upon increasing the energy, the four spots meet
at $(\pi,\pi)$ at what one might call a ``resonance peak''. Above this energy
the response move away from $(\pi,\pi)$ and forms a square which is now rotated by 45
degrees. Within this description, $J$ is taken to be $130 \ldots 160$\,meV \cite{gsu04}
which is the canonical value for 214 compounds). $J'$ is a fit parameter chosen to
place the system near criticality; within the ordered phase, the spectrum is only weakly
sensitive to the value of $J'$.

Subsequently, Ref.~\cite{carlson06} employed spin-wave theory to systems of coupled three-leg
and four-leg ladders. For both situations, reasonable agreement with the experimental
data of \cite{jt04} could be obtained for suitable values of $J$ and $J'$.
(For three-leg ladders, the results are qualitatively similar to Ref.~\cite{scheidl03}.)
The remarkable fact that semiclassical spin-wave theory and strongly quantum mechanical
bond-operator theory produce rather similar results can be related to the fact that
the anomalous dimension $\eta$ of the magnetic order parameter, characterizing the
magnetic QCP, is small ($\eta=0.07$ for the 2d $O(3)$ model).

Coupled-ladder models can be employed as well to model the magnetism of stripes at
smaller doping. Taking the doping dependence of the real-space stripe period as input,
CDW periods of $2M$ sites are naturally described using
coupled even-leg ladders with $(2M-2)$ legs; other periods can be mimicked by an
alternating arrangement of even-leg ladders of different width. Such models reproduce
essential features of the experimental spectra, for instance the relation
$E_{\rm cross} \propto x$ where $E_{\rm cross}$ is the saddle-point energy at
$(\pi,\pi)$ \cite{mvtu_unpub}.

Calculations of stripe magnetism beyond localized spin models have also been performed.
Mean-field plus RPA calculations of spin and charge-ordered stripes (i.e. with broken
symmetries) can be found in Refs. \cite{kaneshita01,andersen05}.
The obtained spin excitations reasonably describe the low-energy part of the experimental
data. However, as seen in Ref.~\cite{andersen05}, the upper branch does not emerge from the
resonance peak and is moreover heavily damped, which is not in agreement with experiment.
The problem of not correctly describing the upper branch of the hour-glass excitation
spectrum is shared by essentially all RPA calculations, see next subsection.

A more elaborate time-dependent Gutzwiller approach to the Hubbard model was put forward
by Seibold and Lorenzana \cite{seibold05}, which places the system somewhere in between
the itinerant and localized regimes. Good agreement with the data of Ref.~\cite{jt04} was found for
bond-centered period-4 stripes.

Finally, the collective magnetism of static stripes can also be described in a lattice
order-parameter theory: Starting from a $\phi^4$ theory for 2d commensurate magnetism,
stripe physics can be implemented via uni-directional modulations in mass and gradient
terms \cite{mvss06}.
The resulting spin fluctuation spectrum is essentially identical to that of the
coupled-ladder two-leg model \cite{mvtu04,gsu04}, with the advantage that the
order-parameter theory can be generalized to fluctuating or disordered stripes \cite{vvk},
see next subsection.

What can be learned from these theory excursions?
(i) For ordered stripes, local-moment approaches work, whereas the simplest RPA fails.
(ii) While bond-centered stripes work marginally better in comparison to experiment,
site-centered stripes cannot be ruled out (on the basis of the neutron scattering data).
Note that the STM result of \cite{kohsaka07} on BSCCO and CCOC gives very clear
evidence for bond-centered charge order.
(iii) Scenarios of checkerboard (instead of stripe) charge order can be ruled out
(at least for LBCO):
A spin-wave calculation for checkerboard order \cite{carlson08} gives results in
disagreement with the data, and this conclusion is consistent with the calculations in
Ref.~\cite{vvk} on fluctuating charge order.

\subsubsection{Gapped incommensurate spin excitations: RPA vs. fluctuating stripes}
\label{sec:eff_gappedmag}

The modeling of gapped spin excitations in superconducting cuprates goes back to the
so-called ``resonance mode'', seen in neutron scattering at wavevector $(\pi,\pi)$
and energies 30\ldots50\,meV \cite{respeak1,respeak1b,respeak1c,respeak2,respeak3}.
RPA and related Fermi-liquid-based methods, like slave-bosons plus RPA \cite{brilee99} or
the so-called fluctuation-exchange approximation \cite{rpa1},
have been very successful in capturing the essential features of the early experimental
observations:
The resonance mode only appears as a sharp mode in the superconducting state, where it is
pulled below the particle--hole continuum (i.e. exists below $2\Delta$),
whereas only overdamped response exists in the normal state.
The energy of the resonance roughly scales with the superconducting $\Tc$.
RPA also describes the development of incommensurate excitations below the resonance
energy, albeit with a small intensity \cite{brilee99,rpa1}.

\begin{figure}
\begin{center}
\includegraphics[width=2.8in]{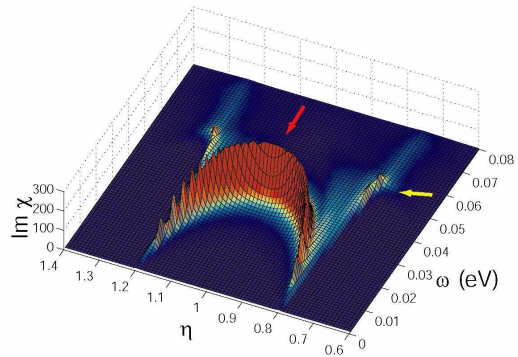}
\caption{
RPA result for $\chi_s''$ of a $d$-wave superconductor
(reprinted with permission from Ref.~\cite{eremin05b}, copyright 2005 by the American Physical Society),
as a function of frequency and momentum along ${\vec q}=\eta (\pi,\pi)$ (note that
this is not the direction in which the incommensurate low-energy peaks are commonly observed).
The RPA interaction function was chosen as
$g({\vec q})=g_0[1-0.1(\cos q_x +\cos q_y)]$,
with $g_0=0.573$\,eV.
The arrows show the resonance mode around $(\pi,\pi)$ and the weak excitation above the
resonance.
}
\label{fig:eremin}
\end{center}
\end{figure}

The neutron scattering experiments of Refs.~\cite{hinkov04,buyers04,buyers05,hayden04,face}
changed the overall picture. In addition to the resonance, both upward and downward dispersing
excitation branches (the so-called ``hourglass'') were detected in underdoped YBCO.
Moreover, sharp features in $\chi_s''$ were also detected above $\Tc$.
Subsequently, modifications and refinements of RPA were proposed,
mainly consisting in adjusting the interaction function $g(\vec q)$ in Eq.~\eqref{rpa},
including its momentum dependence.
As a result, weak excitations above the resonance energy could be described
\cite{eremin05b}, see Fig.~\ref{fig:eremin}.
RPA was also used to describe the anisotropic low-energy excitations in de-twinned YBCO
\cite{eremin05a,schnyder06}, see Sec.~\ref{sec:th_nematic}.
However, a common feature of all RPA calculations is that a full excitation branch
above the resonance energy is never obtained, and the intensity in this energy range
is always weak.

While Fermi-liquid-based methods like RPA are not expected to be applicable
for underdoped cuprates, the failure of RPA is already apparent in optimally
doped YBCO-6.95, as discussed in detail in Ref.~\cite{reznik08c}:
Neutron scattering observes dispersing collective modes above the resonance energy,
both below and above $\Tc$ (the latter fact is different from the early experimental
results, possibly due to better resolution and crystal quality).
In contrast, there is little structure in the RPA spin excitations.
A similar inapplicability of RPA was recently reported for optimally doped
BSCCO \cite{xu09}.

Consequently, strong-coupling theories are called for -- those should account both for
collective spin and charge modes.
An explicit proposal for a phenomenological order-parameter theory of coupled spin and charge
fluctuations was made in Ref.~\cite{mvss06}, with two crucial ingredients:
(i) Both spin and charge sectors are defined on the microscopic lattice; hence the theory
accounts for short-wavelength effects and lattice pinning.
(ii) The bare spin fluctuations live at wavevector $(\pi,\pi)$, and any
incommensurabilities in the spin sector are driven by modulations in the charge sector,
via couplings of the type
${\rm Re}[\exp(i\Qch\cdot\vec x) \phi_c(x)] |\phi_{s\alpha}(x)|^2$:
Microscopically, this reflects that both the spin density and the magnetic couplings get
modulated along with the charge. Importantly, the magnetic couplings can even change
sign, switching the spin ordering wavevector from $(\pi,\pi)$ to one dictated by the
charge order -- this renders the effect of the coupling non-perturbative.
For perfect charge order, the results of this approach are essentially identical to that
of a microscopic theory of coupled spin ladders \cite{mvss06}.

\begin{figure*}
\begin{center}
\includegraphics[width=5in]{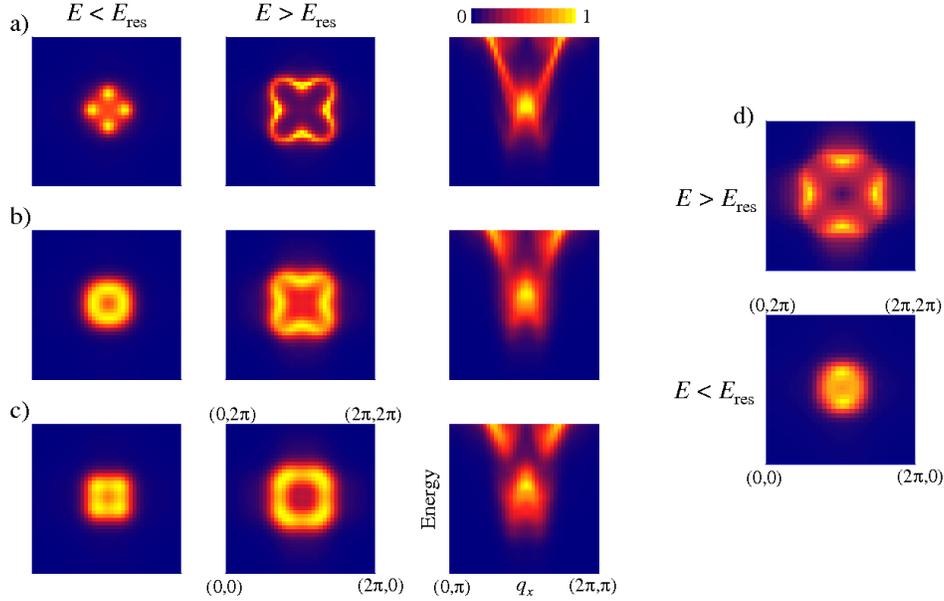}
\caption{
Dynamic spin susceptibility $\chi_s''({\vec q},\omega)$
for bond-centered fluctuating/disordered stripes
(reprinted with permission from Ref.~\cite{vvk}, copyright 2006 by the American Physical Society).
Left/Middle: cuts at a constant energy, at roughly $60\%$ and $150\%$ of the
resonance energy, $E_{\rm res}$, as function of momentum.
Right: cuts along $(q_x,\pi)$ as function of $q_x$ and energy,
showing the universal ``hour-glass'' spectrum.
a) Strong repulsion between $\psi_x$ and $\psi_y$, correlation length $\xi\approx 30$.
b) Weak repulsion,  $\xi\approx 20$.
c) Weak attraction, $\xi\approx 20$.
d) As in panel b), but in the presence of an in-plane anisotropy preferring
horizontal stripes. The ratio of the charge gradients in ${\cal S}_\psi$ is
1.005. While the anisotropy is significant at low energies,
$\w \approx 60\% E_{\rm res}$ (lower panel),
it is much less pronounced at $200\% E_{\rm res}$ (upper panel).
}
\label{fig:flstr}
\end{center}
\end{figure*}

Ref.~\cite{vvk} proposed an adiabatic approximation for the coupled order-parameter
theory of Ref.~\cite{mvss06}, being appropriate for slowly fluctuating or disorder-pinned
stripes.
By combining lattice Monte Carlo simulations for the charge sector with exact
diagonalization of the spin sector, it was possible to determine the spin
excitations of fluctuating stripes. The full numerical treatment of the charge sector
ensures that positional stripe fluctuations as well as fluctuations between horizontal
and vertical stripes including proper domain walls are correctly described. Also, the
crossover from stripe to checkerboard order is part of the theory.
Remarkably, the results display an hour-glass excitation spectrum over a wide range of
parameters, see Fig.~\ref{fig:flstr}. For strongly repulsive horizontal and vertical
stripes, Fig.~\ref{fig:flstr}a,
the constant-energy cuts display distinct high-intensity spots as in the ordered-stripe
calculation, whereas a situation with weakly repulsive stripes including checkerboard
domain walls, Fig.~\ref{fig:flstr}b, results in a rather isotropic intensity distribution
in $\vec q$ space, despite the underlying stripe physics.
The lower branch of the hour glass becomes smeared with decreasing charge correlation
length $\xich$; for $\xich<10$, only a broad vertical feature in $\w$-$\vec q$ space
is left at low energies (i.e. a ``Y''-shaped dispersion).
Including a small in-plane anisotropy in the charge sector has a strong
effect on the magnetic excitations at low energies, whereas those at high energies are
changed rather little, Fig.~\ref{fig:flstr}d. This energy dependence of the anisotropy
is very similar to what is experimentally observed in YBCO \cite{hinkov04,hinkov07}.
Theoretically, this behavior can be rationalized considering that the charge sector is
close to symmetry breaking, and hence is very sensitive to anisotropies. However,
only the lower branch of the spin excitations is determined by the properties of the charge
sector, whereas the upper branch essentially reflects the excitations of a commensurate
gapped antiferromagnet.
Note that the adiabatic approximation for the CDW modes neglects inelastic processes.
As discussed in Sec.~\ref{sec:pin},
this might be particularly appropriate for stripes pinned by quenched disorder,
which then are static, but only short-range ordered, as also seen in STM \cite{kohsaka07}.
While the results of Ref.~\cite{vvk} are encouraging, linking them to microscopic control
parameters like doping or temperature is difficult.

A remark on fermionic damping of collective modes is in order: Here the RPA
and the purely collective-mode description constitute two extreme cases. In the former,
little signal is left when the resonance moves into the continuum, whereas damping is
absent e.g. in the calculation of Ref.~\cite{vvk} (but could in principle be included).
The experiments give evidence for collective modes also at elevated energies and above
$\Tc$, albeit with weaker intensity, implying that the truth is in the middle.

In this context, I mention that a recent extension of the variational cluster
approach (Sec.~\ref{sec:num}) allows to calculate two-particle quantities of the
one-band Hubbard model. This strong-coupling approach has been shown to reproduce
an hour-glass shaped excitation spectrum \cite{brehm08}.
However, a more detailed comparison with experiments has not yet been performed.

Finally, I note that the failure of RPA appears even more drastic in
electron-doped cuprates.
For Pr$_{0.88}$LaCe$_{0.12}$CuO$_4$, the intensity of the resonance peak as calculated by
RPA is by a factor of 10 too small as compared to experiment.
Furthermore, RPA predicts downward dispersing ``wings'' of the resonance which are not
seen in experiment \cite{krueger07}.


\subsection{Photoemission spectra of stripe phases}
\label{sec:th_arpes}

As with neutron scattering, angle-resolved photoemission (ARPES) experiments are performed
routinely on cuprates.
Theoretical calculations of ARPES spectra of stripe phases have appeared in a vast number of
papers over the last decade.
Consequently, I will only mention a few important results, and their possible connection to
experiments.

On general grounds, translational symmetry breaking induces Bragg scattering and band
backfolding. As a result, the Fermi surface breaks up into an arrangement of pockets of
both electron and hole type as well as some open pieces, see Ref.~\cite{millis_pockets}.
However, such a structure is not observable in ARPES due to matrix-element effects,
as concrete model calculations show.

Mean-field approaches readily provide access to the single-particle spectrum.
As sketched in Sec.~\ref{sec:mic_mf}, charge order may be implemented by modulated site or bond
variables, which are either calculated self-consistently within a Hubbard or $t$--$J$ model
or imposed ad-hoc onto a single-particle Hamiltonian.
Explicit results for ARPES spectra were reported e.g. in
Refs.~\cite{salkola96,seibold00,granath02,granath04,granath08,orgad08,wollny08}.
Common properties of the results are:
(i) ARPES appears little sensitive to details of the spatial structure of stripe order,
like site vs. bond modulations and site vs. bond centering.
(ii) Bragg scattering opens gaps on quasiparticle branches separated by wavevectors
$\Qch$ or $\Qsp$.
(iii) Shadow features, e.g. Fermi-surface pieces shifted by $\Qch$ or $\Qsp$,
are present, but suppressed by matrix-element effects.
(iv) Stripes induce almost straight Fermi-surface segments near the antinodal points
located in $\Qch$ direction, i.e., near $(\pi,0)$ for vertical stripes.
(v) For small to intermediate stripe amplitudes, the Fermi surface crossings along the
momentum space diagonals are preserved, which also implies a state with co-existing
stripes and $d$-wave-like pairing possesses zero-energy nodal quasiparticles.
For small stripe amplitudes, this follows from the ordering wavevector $\vec Q$ not
connecting the nodal points of the pure $d$-wave superconductor \cite{vzs00a,vzs00b,berg08a}.
In contrast, for large stripe amplitudes, the results resemble those expected from
quasi-1d systems, and additional pairing eliminates low-energy quasiparticles.

In fact, the early photoemission experiments of Zhou et al. \cite{zhou99} on LNSCO-0.12,
displaying a cross-shaped low-energy intensity distribution and no evidence for nodal
quasiparticles, were interpreted in terms of 1d behavior.
The concept of effectively weakly coupled chains/ladders was followed in the cluster-perturbation theory
treatment of the Hubbard model \cite{zacher00}: Here, ladders were treated by
exact diagonalization, whereas the inter-ladder coupling was included on the one-particle
level in an RPA-like fashion. For half-filled stripes with maximal charge
modulation, good agreement with the data of Ref.~\cite{zhou99} were obtained.
However, as those ARPES results were not confirmed by later experiments,
see Sec.~\ref{sec:exp_arpes},
the concept of strong 1d behavior of Ref.~\cite{zacher00} appears too drastic.

Other theoretical approaches to stripe ARPES spectra beyond mean field
have also been put forward,
namely supercell DMFT of the Hubbard model \cite{fleck00},
a spin-polaron approach to the $t$--$J$ model \cite{wrobel06},
and exact diagonalization of a $t$--$J$ Hamiltonian in the presence of a stripe
potential \cite{tohyama99,eder04}.
The results are broadly consistent with those from mean-field studies,
the latter calculations, however, suffered from a poor momentum resolution.

While many theory works only account for perfect charge order,
static spatial disorder of the stripe pattern can be treated at least in simple mean-field theories.
The physical picture here is that of impurity-pinned static charge order.
A simple approach is using a random spacing of uni-directional stripes
\cite{salkola96,seibold00,granath02,granath08}, alternatively short-range-ordered charge configurations
can be generated from a full Monte-Carlo simulation of a stripe order-parameter theory
\cite{mv08,wollny08}.
From the results, it is apparent that stripe signatures in ARPES are quickly smeared
by a combination of spatial disorder and superposition of horizontal and vertical
stripes. For instance, at stripe correlation lengths smaller than 20 lattice spacings,
clear-cut stripe signatures become essentially invisible \cite{wollny08}.
This implies that short-range-ordered stripes, while nicely visible in STM,
are difficult to detect using probes without spatial resolution.
The influence of purely dynamic CDW fluctuations on ARPES lineshapes has also been discussed
in a few papers. In particular, in has been invoked as explanation for broad lineshapes and
dispersion kinks \cite{seibold01}.

Interestingly, none of the ARPES experiments on stripe-ordered 214 cuprates
(after Ref.~\cite{zhou99})
displayed clear-cut stripe signatures,
perhaps with the exception of recent data from LESCO-1/8 \cite{borisenko08}.
Here, various features of the data have been argued to be consistent with
stripes of moderate amplitude. However, the behavior near the antinodal points
is not understood, and the limited experimental resolution renders a detailed
comparison to theory difficult.


\subsection{STM spectra}
\label{sec:th_stm}

Measurements of the local density of states (LDOS) using scanning tunneling spectroscopy
in a number of cuprates have triggered enormous theoretical activities. While calculating
the LDOS of a homogeneous BCS $d$-wave superconductor is a textbook exercise, the most
important complications arise from intrinsic inhomogeneities and from Mott physics.
In the following, we shall only discuss theory work on periodic (or
quasi-periodic) modulations in the STM signal, possibly related to
symmetry-broken states.
In contrast, spatially irregular inhomogeneities, reflected e.g.
in a broad distribution of local gaps in the superconducting state and possibly related
to the influence of oxygen dopants \cite{stm_rmp,alloul_rev,mcelroy05b},
are outside the scope of this review.

A standard assumption in modeling STM data is that the measured signal at low energies locally
reflects properties of the topmost CuO$_2$ layer.
This is not trivial: Commonly used crystals of BSCCO
cleave such that insulating BiO layers are exposed on the surface. This has the advantage
of charge neutrality, but implies that electrons may follow a non-trivial tunneling path
between the STM tip and the topmost CuO$_2$ layer, with the result of a
momentum-dependent tunneling matrix element (which could also differ between the various
cuprate families).
Theoretical work on this issue has proposed such a ``filter effect'', with the matrix
element being of $d$-wave type for BSCCO \cite{filter1,filter2}.
However, experimental results are most consistent with direct, i.e., momentum-independent
tunneling -- this conclusion can be drawn from the general energy dependence of the LDOS,
the spatial shape of impurity resonances, and the fact that STM results of BSCCO and
CCOC are grossly similar, despite the different surface layers.
Recent LDA calculations for impurity states \cite{filter_lda} also argued against
a general filter effect, but instead proposed impurity-specific features of the
wavefunctions to explain details of the impurity resonances.

\subsubsection{Quasiparticle interference}

Isolated elastic scatterers in an otherwise homogeneous metallic system are known to cause
Friedel oscillations in the local density. An energy-resolved version of this effect,
dubbed ``generalized Friedel oscillations'' or ``quasiparticle interference'' (QPI), can
possibly be observed in the LDOS. Then, energy-dependent spatial oscillations in
$\rho(\vec{r},E)$ will show up in the Fourier-transformed LDOS (FT-LDOS) $\rho(\vec{k},E)$
(or its power spectrum).
These features typically takes the shape of arcs or ridges in momentum space
\cite{capriotti03}.
The existence, intensity, and dispersion of such structures depend strongly on the shape of
the Fermi surface, the presence and shape of a superconducting gap, and the nature of the
scatterers. Understanding QPI is important in order to be
able to distinguish it from genuine charge order which also leads to real-space
oscillations in the LDOS.

Remarkable, the initial experiments revealing QPI in BSCCO \cite{hoffman02b} showed
relatively clear {\em peaks} in momentum space. Those were interpreted using
the ``joint quasiparticle density of states'' of a BCS $d$-wave superconductor:
Peaks in $\rho(\vec{k},E)$ were assumed to occur at wavevectors $\vec k$ separating those points
on iso-energy contours of energy $E$ which have a minimal velocity -- for a $d$-wave
superconductor with a typical cuprate Fermi surface, these are eight points at
the tips of the ``banana-shaped'' iso-energy contours.
This idea leads to the so-called ``octet model'': The eight points define a set of seven
wavevectors $\vec q_1 \ldots \vec q_7$ (Fig.~\ref{fig:octet}),
whose positions appear to match the
experimental locations of QPI peaks, and agreement with Fermi surfaces as measured by ARPES has been
pointed out \cite{mcelroy03,mcelroy06}.

\begin{figure}
\begin{center}
\ifhires
\includegraphics[width=2.7in]{octet1.eps}
\else
\includegraphics[width=2.7in]{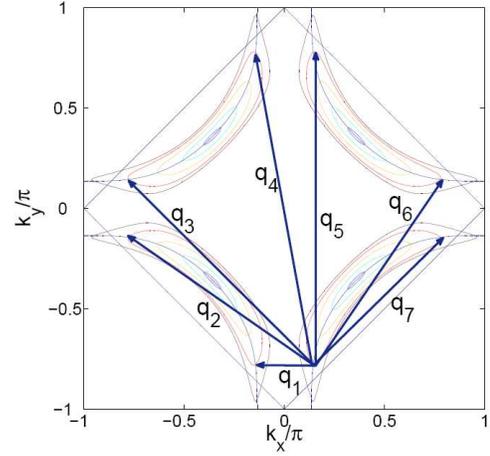}
\fi
\caption{
Contours of constant quasiparticle energy for a
$d$-wave superconductor with $\Delta=0.6t$ at energies
$\omega/t=0.0, 0.075, 0.225, 0.375, 0.5, 0.57$. Also
shown are the underlying normal state Fermi surface, the AF
zone boundary, and the seven distinct wavevectors ${\vec q}_i$
of the octet model which connect the banana tips.
(From Ref.~\cite{hirschfeld08})
}
\label{fig:octet}
\end{center}
\end{figure}

In subsequent theory work, the problem of scattering from impurities in BCS $d$-wave
superconductors was studied in detail \cite{qpith1,qpith2,qpith3,qpith4}.
For a single point-like impurity at site ${\vec r}_0$, the formal result for the impurity-induced
change in the LDOS is
\begin{equation}
\delta\rho(\vec{r},E) = -\frac 1 \pi {\rm Im} \big[
G(\vec{r}-\vec{r}_0,E) T(E) G(\vec{r}_0-\vec{r},E)
\big]
\label{eq:qpi}
\end{equation}
where $G(\vec{r},E)$ is the single-particle Green's function of the clean
translational-invariant system, and $T(E)$ is the T matrix of the scatterer. In the
superconducting state, both quantities are matrices in Nambu space.
This formula clarifies an important point: $\delta\rho$ is not only determined by
the quasiparticle DOS, ${\rm Im} G$, but also by real parts of Green's functions.
Therefore, the octet model is certainly oversimplified.

Numerical calculations for a single impurity in a $d$-wave BCS state found structures in
the FT-LDOS reminiscent of what was seen experimentally.
However, the theoretical FT-LDOS landscape contained both arcs and peaks, whereas the
experimental data show mainly peaks, and also the intensity distribution and precise peak locations
did not agree \cite{qpith1,qpith2}.
Interference phenomena between multiple impurities were taken into account in
Ref.~\cite{qpith3,qpith4}, without a significant improvement.
It became clear that the momentum-space intensity distribution in $\rho(\vec{k},E)$
depends strongly on the momentum-space structure of the scatterers.
An extensive theoretical study \cite{nunner06} of QPI caused by various possible scattering sources
revealed that smooth scatterers in both the particle--hole and the particle--particle channel, possibly
arising from out-of-plane defects, are crucial in modeling the experimental data.
Combined with the effect of a small concentration of point-like unitary scatterers,
Ref.~\cite{nunner06} was able to reproduce the gross intensity distribution in the
experimental $\rho(\vec{k},E)$.
Nevertheless, a complete understanding of the observed FT-LDOS patterns is lacking,
although the experimental existence of dispersing peaks can be associated with QPI
phenomena with reasonable certainty.

A few remarks are in order:
(i) The octet model describes the data remarkably well, with a few caveats:
The energy dispersion of the experimentally observed $\vec q_1$ peaks is significantly
weaker, and both $\vec q_1$ and $\vec q_5$ have a higher intensity, than predicted
by the octet model when using the ARPES dispersions as input.
(ii) Why does the octet model work so well? This is unclear at present.
One ingredient is the strong velocity anisotropy in the superconducting state,
$v_F \gg v_\Delta$, which renders the ``banana tips'' very sharp,
but this is not sufficient.
(iii) Essentially all existing theory work is based on non-interacting Bogoliubov
quasiparticles of a BCS superconductor. The influence of strong correlation
effects on QPI has not been investigated.
Among other things, the quasiparticle weights of partially incoherent quasiparticles
should enter the strength of the QPI signatures. This also implies a strong temperature
dependence of QPI phenomena. Experimentally, no QPI has been detected in the pseudogap
regime \cite{yazdani04}.
(iv) The experimentally observed partial disappearance of QPI signatures upon approaching
the Mott insulator \cite{kohsaka08} is not understood. It is worth pointing out that the
absence of QPI peaks does {\em not} imply the absence of quasiparticles, but may simply be
related to a change in the structure of iso-energy contours.
It has been proposed that magnetic ordering tendencies could play a role here
\cite{hirschfeld08}, but other phases with pocket-like Fermi surfaces can lead
to a similar effect.

An exception to (iii) is a recent consideration of QPI near a nematic quantum critical
point \cite{lawler08}: Using a one-loop self-energy, it was shown that both the velocity
anisotropy and the dominance of the ``banana tips'' for the QPI signal are enhanced due
to scattering off nematic critical fluctuations. However, a more detailed study is needed to see
whether this proposal solves the problems noted above.

\subsubsection{Stripes and checkerboards}

Atomically resolved STM spectra from charge-ordered states contain a wealth of
information on the nature of the charge order. The direct real-space picture reveals the
spatial symmetries (e.g. site centering vs. bond centering), and the
quantitative information can be used to determine the nature of the modulation
(e.g. primarily on sites or bonds).

Within mean-field treatments for both charge order and superconductivity,
the LDOS spectra of stripe states can be easily calculated.
A comprehensive analysis for weak charge order on top of a BCS $d$-wave superconducor
has been presented by Podolsky {\em et al.} \cite{podol03}.
In particular, these authors calculated the energy dependence of the Fourier component
of $\rho(\vec{k},E)$ at the charge-ordering wavevector $\vec k = \Qch$
which was measured in Refs.~\cite{kapi03a,kapi03b}.
Based on their comparison of the spectra of different modulation types,
the authors concluded that an on-{\em site} modulation alone cannot explain the STM data,
but a sizeable modulation on the {\em bonds}, either of kinetic energy or pairing strength,
is required.
Subsequently, the LDOS has been calculated for a variety of mean-field theories
of charge-ordered states, both for stripes \cite{machida02,mv02,orgad08}
and for checkerboards \cite{mv02,dwave_checker2,pdw1}

The clear-cut observation of stripe order on the surface of both BSCCO and CCOC
\cite{kohsaka07} showed that the static order is only short-ranged, likely due to
strong impurity pinning, is moreover dominated by bond (instead of site) modulations,
and apparently co-exists with well-defined low-energy quasiparticles
\cite{hanaguri07,kohsaka08}.
The STM signatures of short-range valence-bond stripe order, together with the interplay
with impurity scattering, were theoretically investigated in Ref.~\cite{mv08}.
Here, an electronic mean-field theory was combined with an order-parameter description of
short-range charge order, which was used before to model the spin excitations of
disordered stripes \cite{vvk}.
The results show that stripe order is strongly visible in the LDOS and $Z$ map
at elevated energies, in particular near the superconducting gap energy,
whereas QPI signatures dominate at low energies, Fig.~\ref{fig:vbsstr},
in good agreement with experiment.
As pointed out in Sec.~\ref{sec:ph_cdw},
this dichotomy is related to the fact that valence-bond stripes
have an approximate $d$-wave form factor and hence display little coupling to nodal
quasiparticles.

\begin{figure}
\begin{center}
\ifhires
\includegraphics[width=3.5in]{qpi1b.eps}
\else
\includegraphics[width=3.5in]{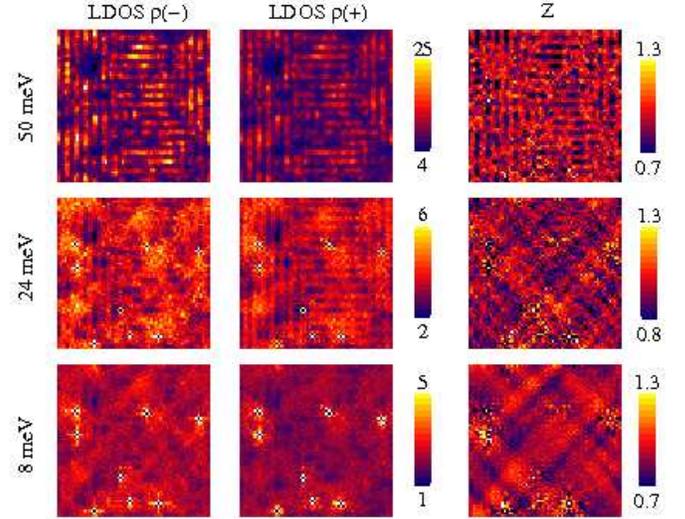}
\fi
\caption{
Theoretical results for the LDOS $\rho(\vec{r},E)$ at negative (left) and positive (middle) bias,
together with $Z(\vec{r},E)$ (right),
for pinned short-range ordered stripes
with additional impurities on a $64^2$ lattice,
at energies 50, 24, 8\,meV (from top to bottom).
While stripe signatures dominate at elevated energies,
the low-energy data, in particular $Z(\vec{r},E)$, display QPI features.
(Reprinted with permission from Ref.~\cite{mv08}, copyright 2008 by the American Physical
Society.)
}
\label{fig:vbsstr}
\end{center}
\end{figure}

I now come to field-induced order as seen in STM.
The initially observed checkerboard patterns around vortices in BSCCO \cite{hoffman02a}
triggered a variety of theory works. One class of explanations assumed that static
collinear SDW order is induced in or near the vortex cores by the applied
field \cite{zhu02,ting02,chen02,scz02}.
Then, a static charge-density modulation is automatically induced.
Another class of theories is based on pinning of otherwise fluctuating order
by a vortex core (or another type of impurity) \cite{franz02,tolya02}.
While pinning of a CDW trivially induces a charge modulation, the pinning effect on a collinear SDW
will also lead to a static charge modulation, however, static spin order is {\em not} required \cite{tolya02}.
The reason is that the pinning potential at position $\vec x_0$ couples as $\zeta \phi_c(\vec x_0)$ or
$\zeta \phi_s^2(\vec x_0)$ to CDW and SDW order parameters, respectively.
In the case of pinning of SDW fluctuations, the resulting peaks in the FT-LDOS
are {\em not} energy-independent: The collective modes entering e.g. a one-loop
self-energy are fully dispersive, which causes a dispersion
in the FT-LDOS as well. This implies a subtle interplay of induced charge order
and QPI \cite{tolya03}.


It should be noted that a distinct explanation of the vortex checkerboard in terms of
field-enhanced quasiparticle interference has been proposed \cite{takagi08,tami08}, without reference
to collective effects. Indeed, a calculation assuming the existence of QPI peaks at the octet
model spots in the FT-LDOS shows that both $q_1$ and $q_5$ are field-enhanced \cite{tami08}.
This effect is related to the coherence factors of the superconducting state and may indeed
enhance a checkerboard-like signal in the LDOS.
While it is at present not unambiguously established which of the interpretations of the
vortex checkerboard is correct, the collected experimental evidence for field-induced ordering
tendencies in cuprates is overwhelming, see Sec.~\ref{sec:fieldtune}. Therefore, it seems
very likely that the vortex STM signal involves an intrinsic ordering phenomenon as
well, in particular because stripe-like ordering is known \cite{kohsaka07} to
occur at the surface of BSCCO and CCOC, the materials where the vortex checkerboard has been studied.

Finally, let me point out that essentially all model calculations for cuprate STM spectra are done for one-band
models of the CuO$_2$ plane. Considering that the stripe-like modulation observed in
Ref.~\cite{kohsaka07} is strongest on the oxygen orbitals, this may not be justified:
A separate treatment of Cu and O orbitals within a three-band model could be required to
fully understand the physics of bond order in cuprates.\footnote{
While bond order can in principle be described in a one-band model,
the physics of the three-band model is certainly richer.
}


\subsection{Stripes and superconductivity}
\label{sec:th_stripesc}

Considering that superconductivity and stripes are two prominent ordering phenomena in
the cuprates, obvious questions are:
(i) Do stripes and superconductivity co-exist?
(ii) Do stripes and superconductivity compete or co-operate?
(iii) Can external tuning parameters employed to tune this interplay?

Some relevant experiments are described in Sec.~\ref{sec:exp} above, in particular
Sec.~\ref{sec:exp_stripesc}.
The answer to (i) is yes, most clearly seen in LBCO for $0.1\leq x\leq 0.15$, where scattering
experiments have established stripe order and thermodynamic measurements show bulk
superconductivity (albeit with a strongly suppressed $\Tc$).
As for (ii), the suppression of the superconducting $\Tc$ is usually taken as evidence
for a {\em competition} of static stripes and superconductivity. This is consistent with
the possibility (iii) of magnetic field tuning of incommensurate spin order,
and the fact that (static) stripe signatures have not been detected in the multi-layer cuprates
with the highest $\Tc$.

I note that a few other observations in 214 compounds have prompted speculations about a
co-operative interplay of stripes and superconductivity.
For LSCO, the incommensurate low-energy spin excitations have been found to survive
into the overdoped regime and to disappear concomitantly with superconductivity
\cite{waki04}.
However, whether the incommensurate fluctuations are the cause or the result of
pairing cannot be deduced from the data.
For LBCO-1/8, the experimental data include an
unusual gap in the in-plane optical conductivity appearing together with charge order
\cite{homes06}, the apparent $d$-wave gap in the charge-ordered state above $\Tc$
\cite{valla06,he09}, and the resistivity drop at the spin-ordering temperature \cite{li07,berg07,jt08}.
The latter finding has been interpreted in terms of ``antiphase superconductivity'',
to be described in more detail in Sec.~\ref{sec:antiphase} below,
but alternative explanations have been proposed as well \cite{tsvelik07}.
At present, a concise theoretical picture has not emerged, which certainly is related
to the challenge in understanding the pseudogap regime.

\subsubsection{Competing order parameters}
\label{sec:comp_op}

The competition of antiferromagnetism and superconductivity has played a central
role in the cuprate phenomenology early on. It appeared in various theoretical flavors,
including the SO(5) theory of Zhang \cite{zhang97,so5rmp,so5vortex}.
Also without appealing to a higher underlying symmetry, the interplay of stripes and
superconductivity can be modeled using a coupled order-parameter field theory -- such
an approach assumes the presence of the two ordering phenomena without making
reference to their microscopic origin.

A concrete theory for the competition of SDW order and superconductivity
in the presence of an external magnetic field has been worked out by Demler
{\em et al.} \cite{demler1,demler2}, with the focus on the SDW transition inside
the superconducting state.
The ingredients are a classical Ginzburg-Landau free energy for the
superconducting condensate $\psi(\vec x)$ in the presence of an external field $H$,
and a quantum $\phi^4$ theory for the SDW order parameter $\phi_{s\alpha}(\vec x,\tau)$,
with a density-density coupling $v |\psi|^2 |\phi_{s\alpha}|^2$.
The primary effect of a small applied field is to induce vortices in the superconducting
order parameter. Vortices are accompanied by a suppression of $\psi$ in a region around the
vortex core, such that the balance between SC and SDW is locally changed.
Importantly, the periodic ``potential'' for the $\phi$ order parameter,
resulting from an Abrikosov vortex lattice, enhances the magnetic fluctuations not only in
the vortex cores, but over the entire sample -- this effect eventually causes a field-induced
transition from a SC to a SC+SDW state.

The schematic phase diagram from this consideration is in Fig.~\ref{fig:fieldpd};
it has been verified by a full numerical analysis of the coupled field theory.
Among the important results is the behavior of the phase boundary $H(s)$,
where $s$ is the tuning parameter of the SDW order,
near the zero-field quantum phase transition at $s_c$:
\begin{equation}
H/H_{c2}^0 = 2 (s - s_c)/[v \ln (1/(s-s_c))]\,.
\label{field_boundary}
\end{equation}
Remarkably, the phase boundary cannot be obtained from an analytic expansion
in $H$, the reason being the infinite diamagnetic susceptibility of the superconductor.
A related result is the behavior of the staggered moment,
experimentally measured as Bragg peak intensity,
as function of the applied field:
\begin{eqnarray}
\label{field_intensity}
\langle | \phi_{s\alpha} | \rangle ^2 &\propto& s_c - s(H), \\
s(H) &\equiv& s - (v H/(2 H_{c2}^0)) \ln (\vartheta H_{c2}^0/H) \nonumber
\end{eqnarray}
with $s(H)$ the renormalized $\phi_s$ mass, $H_{c2}^0$ the upper critical field of
the bare superconductor, and $\vartheta$ a number of order unity.
Both results \eqref{field_boundary} and \eqref{field_intensity}
have been found to be quantitatively obeyed in the
neutron scattering experiments of Lake {\em et al.} \cite{lake01,lake02}
and of Chang {\em et al} \cite{chang08}, see Fig.~\ref{fig:fieldtune} in
Sec.~\ref{sec:fieldtune}.
In particular, the field-induced intensity  $\langle |\phi_{s\alpha}|\rangle^2$
increases almost linearly with field near the zero-field SDW transition \cite{lake02}.
The results of Haug {\em et al.} on YBCO-6.45 \cite{haug_mf} also appear consistent
with Eq.~\eqref{field_intensity}.

The theory of Demler {\em et al.} \cite{demler1,demler2},
originally restricted to $T\!=\!0$ and two space dimensions,
was subsequently refined and extended to include inter-layer coupling \cite{kiv_field}.
For low order parameter symmetry (Ising or XY), the 3d situation admits an interesting
quasi-1d ordered phase, where ordering occurs within each vortex line individually.
Then, the $B\to 0$ limit of the order--disorder phase boundary does not match the $B\!=\!0$
critical point \cite{kiv_field}.

\begin{figure}
\begin{center}
\includegraphics[width=2.8in]{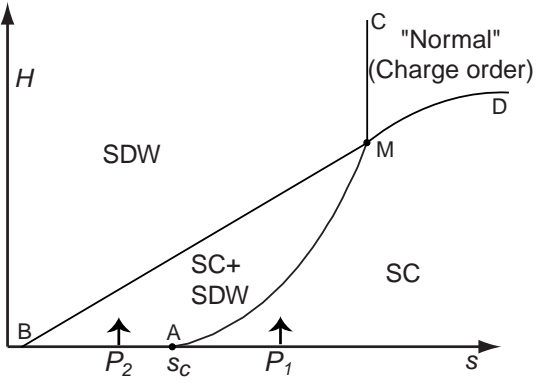}
\caption{
Zero-temperature phase diagram for the field tuning of SDW order in the
superconducting state
(reprinted with permission from Ref.~\cite{demler2}, copyright 2002 by the American Physical Society).
$H$ is the external field, and $s$ is the SDW tuning parameter, with zero-field order for
$s<s_c$.
The line A--M is the field-tuned SDW transition, whereas superconductivity disappears
above the line B--M--D. P$_1$ and P$_2$ denote paths taken by experiments
which start outside \cite{lake01,jt04b,khaykovich05,chang08,chang09} or
inside \cite{lake02,lake05,chang07,chang08,khaykovich02,khaykovich03}
the SDW phase, respectively.
}
\label{fig:fieldpd}
\end{center}
\end{figure}

A few remarks on the theoretical analysis and its consequences
are in order:
(i) The Zeeman effect of the applied field has been neglected. Its effect only
appears at order $H^2$ and is small for $H\ll J$, with $J$ being a typical magnetic
exchange energy.
(ii) The theory implies that the field-enhanced magnetism does {\em not} arise primarily
from the vortex cores, but -- via the superflow surrounding each vortex -- from the entire
sample.
In fact, if the magnetism would only come from the vortex core region, then
the observed neutron scattering intensity would imply large local moments of
size 0.6 $\mu_B$, as in undoped cuprates, which appears unlikely.
However, other theories have proposed vortices with antiferromagnetic cores
\cite{so5vortex}.
(Note that the nature of the vortex cores in cuprates is still not understood,
but the continuum theory of Refs.~\cite{demler1,demler2} gives a reliable description of
the physics outside the vortex cores.)
(iii) Although the theory focuses on the SDW order parameter competing with
superconductivity, a similar approach could be applied to other field-induced orders as
well.

Indeed, an open issue is whether the field directly enhances the SDW component
of the stripe order or whether the field enhances the CDW component.
In the CDW is primarily enhanced, one may expect separate transitions associated with
the CDW and SDW order, with an intermediate SC+CDW phase.
However, due to pinning of the CDW fluctuations by the vortex cores the CDW transition
will be smeared.
Nevertheless, a simultaneous study of field-induced SDW and CDW intensities in one sample
might be able to distinguish between the two scenarios.

Under the assumption that stripes and superconductivity compete, the 1/8 anomaly in $\Tc$
finds a natural explanation in terms of the enhanced stripe stability at doping
$x\!=\!1/8$ (which in turn is commonly attributed to lattice commensuration effects, but may
be enhanced by Fermi surface nesting).
I note that, in addition, the $2k_F$ CDW instability of one-dimensional stripes may
contribute to the $\Tc$ suppression: For nearly half-filled stripes, $2k_F$ corresponds
to a CDW period of four lattice spacings along the stripes,
which coincides with the modulation period perpendicular to the stripes near $x\!=\!1/8$.
Hence, coupling between the planes could be efficient in stabilizing the $2k_F$ CDW
which would then suppress pairing \cite{dmrg98}.

Last but not least it is worth mentioning that the observation of $\Tc\approx\Tsp$,
as observed e.g. in La$_2$CuO$_{4+\delta}$,
is {\em not} necessarily inconsistent with a concept of competing SDW and superconducting
orders. Within a phenomenological Landau theory, which admits a region of phase
coexistence of the two competing orders at low $T$, there is a regime of parameters with
simultaneous onset of superconductivity and magnetism \cite{KAE01}.

\subsubsection{Fluctuating pairing and antiphase superconductivity}
\label{sec:antiphase}

A remarkable hierarchy of temperature scales was found in stripe-ordered LBCO-1/8
\cite{li07,jt08}, with a sharp resistivity drop around 42\,K, but no corresponding signature
in $\rho_c$ and no bulk Meissner effect.
This has triggered an interpretation in terms of 2d fluctuating superconductivity without
3d phase coherence \cite{li07,berg07}.
While the resistivity drop seems to coincide with the spin ordering temperature
in zero field, a magnetic field separates the two.
The data raise a number of questions:
(i) What is the nature of the fluctuating pairing state and its relation to magnetism?
(ii) Why is bulk superconductivity, i.e., a Meissner signal, only established around 5\,K?

A particular scenario, proposed in Refs.~\cite{berg07,berg08b}, is that of ``antiphase''
superconductivity in a stripe state, also dubbed ``pair density wave''.
Here, the superconducting condensate is modulated, i.e., it is $d$-wave-like
within each stripe, but undergoes a $\pi$ phase shift between neighboring stripes.
If adjacent planes have alternating stripe directions, then it is easy to show
that the leading-order Josephson coupling between neighboring planes vanishes
by symmetry, and bulk superconductivity will be strongly suppressed.
(Higher-order couplings, however, will be finite.)

In the order-parameter language, antiphase superconductivity is very similar to
an FFLO state, i.e., it is characterized by a finite-momentum condensate,
$\langle c_{\vec{k}+\Qp\uparrow} c_{-\vec{k}\downarrow}\rangle\neq 0$,
with $\Qp$ being the pairing modulation wavevector.
In the context of stripe phases, such a state was first proposed by
Himeda {\em et al.} \cite{himeda02} and then independently by Berg {\em et al.} \cite{berg07}.
A condensate at wavevector $\Qp$ naturally couples to the charge density at wavevector
$2\Qp$, see Sec.~\ref{sec:coupling}, hence the pairing modulation in
a period-4 charge stripe state is of period 8.
Spectral properties of antiphase-condensate states have been discussed in
Ref.~\cite{orgad08}.
In contrast to superconductors with zero-momentum condensate,
the state has a full Fermi surface, which is visible mainly as arcs near the
nodal direction after matrix elements have been taken into account.

While the idea of vanishing Josephson coupling by symmetry is appealing, a few issues are
open:
(i) The antiphase-condensate state alone does not explain the full hierarchy of temperature
scales seen experimentally. For instance, in the simplest model of weakly coupled planes there is
still a single ordering temperature (which approaches the Kosterlitz-Thouless temperature in
the limit of vanishing inter-plane coupling).
A plausible assumption, also consistent with the global cuprate phase diagram,
would be that bulk superconductivity at low $T$ emerges
from a zero-momentum condensate, i.e., the transition into the Meissner state involves a
{\em different} condensate. If this co-exists with the condensate at finite $\Qp$,
then charge order at wavevector $\Qp$ is induced, which is {\em not}
a higher harmonic of the charge order at $2\Qp$. This is testable experimentally.
In addition, defects in the stripe order and glass-like behavior are likely required
to explain the complicated hierarchy of scales in LBCO \cite{berg08b}.
(ii) The full Fermi surface of the antiphase-condensate state \cite{orgad08}
is not easily compatible with the $d$-wave-like gap observed in ARPES and STM experiments
on LBCO-1/8 \cite{valla06,he09}.
(iii) It is not obvious that the antiphase-condensate state is favored over a more
conventional superconducting state \cite{berg08b}.
Whereas variational Monte Carlo calculations for the $t$--$J$ model \cite{himeda02}
found an interval of $t'/t$ where the antiphase-condensate state is slightly favored,
results from renormalized mean-field theory \cite{poil07,raczk08,rice08}
indicate that the antiphase-condensate state,
although competitive, is always higher in energy than a stripe state with
zero-momentum condensate.

An alternative model for the resistivity drop in LBCO invokes spin-liquid
physics emerging from nearly straight Fermi-surface segments near the antinodal
points \cite{tsvelik07}, but its relation to the established stripe order in LBCO is
open.


\subsection{Stripes in other experiments}

In this section, we turn to other experimental findings in cuprates,
which have been interpreted as to provide indirect evidence for translational
symmetry breaking.

\subsubsection{Quantum oscillations and Fermi-surface reconstruction}

Recently, quantum oscillations have been observed for the first time
in underdoped cuprates, both in de-Haas-van-Alphen and Shubnikov-de-Haas measurements.
The compounds used for these experiments are YBCO-6.5 \cite{doiron07,sebastian08}
and YBCO-124 \cite{yelland08}, both having exceptionally low disorder due to ordered oxygen
dopants.
The findings came as a surprise,
as some theoretical scenarios assumed the complete absence of
coherent electronic quasiparticles in the underdoped regime.
The observed quantum oscillations are instead interpreted as evidence for
the presence of a Fermi surface of quasiparticles, at least in the regime
of the large magnetic fields of order 50 T applied in the experiment.

A follow-up question is concerned with momentum-space area of the oscillation orbit
which is a direct measure of the enclosed Fermi volume in two dimensions.
A standard analysis of the oscillation frequencies gives a Fermi volume somewhat smaller
than the nominal {\em hole} doping:
The oscillation with frequency (530\,T)$^{-1}$ in YBCO-6.5 translates
into a Fermi pocket of size 0.075 per Cu, while the doping level is about $10\%$.
A similar mismatch is found for YBCO-124.
(Note that oscillations with a frequency corresponding to the {\em electron}
concentration, i.e., to a large Fermi surface, are not observed.)
Parallel magnetotransport measurements revealed a negative Hall resistance
in both materials, which was then used to argue in favor of electron (instead of hole)
pockets \cite{leboeuf07}.
A later de-Haas-van-Alphen experiment in YBCO-6.5 \cite{sebastian08} identified a second
oscillation period of (1650\,T)$^{-1}$, associated with an additional smaller pocket.

One line of interpretation is in terms of a conventional metallic state where
(perhaps field-induced) translational symmetry breaking and associated Fermi-surface
reconstruction (via band backfolding) induces small Fermi surface pockets.
A detailed analysis of such pockets in various candidate ordered states shows
that $(\pi,\pi)$ antiferromagnetism is insufficient,
but incommensurate (or long-period commensurate) SDW and/or CDW order could in principle
lead to hole and electron pockets which would cause oscillations consistent with the experimental
observations \cite{sebastian08,millis_pockets}.

While this interpretation may support the notion that static stripe-like order is crucial
for the phenomenology of underdoped high-$\Tc$ cuprates, care is required.
One complication is that the experiments are performed in the mixed state.
This situation was analyzed by Stephen \cite{stephen92} in a regime where quasiparticle
scattering on vortex lines can be treated perturbatively.
Considering that the effective Fermi energy for the relevant orbits may be small (i.e. of the
same order of magnitude as the superconducting gap), it is unclear at present whether
Stephen's analysis is applicable \cite{palee08}.

\subsubsection{Stripes and Raman scattering}

Inelastic light scattering has revealed rich information on various correlated electron
systems, for a review see Ref.~\cite{hackl_rmp}. Being a $\vec q=0$ probe, it couples to
both SDW and CDW modes at finite $\vec q$ only via two-particle processes.
Nevertheless, Raman scattering was used to reveal both charge ordering at low temperature
and collective CDW motion at high temperatures in the ladder system Sr$_{12}$Cu$_{24}$O$_{41}$
\cite{blumberg02}.

For the cuprates, a number of Raman scattering studies have been published on the 214
materials \cite{hackl02,hackl05}.
In LSCO the difference between B$_{1g}$ and B$_{2g}$ spectra  has been
used to argue in favor of tendencies toward stripe formation \cite{hackl05}:
At low temperature an anomalous low-energy peak shows up in the B$_{2g}$ channel for
$x\!=\!0.02$ and in the B$_{1g}$ channel for $x\!=\!0.10$ -- note that the selection rule of the two
geometries are equivalent up to a 45$^\circ$ in-plane rotation.
This interpretation is supported by a theoretical calculation of the Raman response in
the presence of soft CDW collective modes \cite{caprara05}.
One should, however, note that details of the Raman response are not understood
\cite{hackl_rmp}.

\subsubsection{Doping evolution of the chemical potential}
\label{sec:chempot}

Core-level photoemission experiments have been utilized to determine the doping dependence of
the chemical potential $\mu$ in cuprates. Particularly interesting is the behavior in 214
materials: While electron-doped Nd$_{2-x}$Ce$_x$CuO$_4$ displays a roughly constant
chemical-potential slope $d\mu/dx$, $\mu$ is very weakly doping dependent in hole-doped
LSCO for $x<0.15$, while the slope for larger $x$ is similar to the one on the electron-doped
side \cite{ino97,harima01}.
The anomalous small-doping behavior has been attributed to stripe physics:
In a picture with doping-independent stripe filling $\rho_l$ (which also results in a
stripe incommensurability $\epsilon \propto x$), the chemical potential $\mu(x)$ will be constant,
whereas a deviation from $\epsilon \propto x$ will induce a slope in $\mu(x)$.
Hence, the behavior of the stripe incommensurability as shown in the Yamada plot,
Fig.~\ref{fig:yamada}, naturally ties in with the behavior of $\mu(x)$.
Of course, this picture is too simplistic, as a constant $\mu(x)$ implies phase
separation, and a fixed $\rho_l(x)$ (which could only arise from locking due
to an incompressibility of the underlying state) implies insulating stripes.
Nevertheless, a weakly doping-dependent $\rho_l$ for $x\leq 1/8$ appears consistent with
the experimental observations. This is also borne out from mean-field studies of
metallic stripes, see e.g. Ref.~\cite{lorenzana02}.

A few caveats of this interpretation should be noted:
(i) No static charge stripes have been reported in LSCO (without Nd or Eu co-doping).
Whether fluctuating stripes would be consistent with the experimental data is unclear.
(ii) The chemical potential in CCOC has been found to display a much larger slope
over the entire doping range as compared to LSCO \cite{yagi06}.
This has been attributed to the periodicity of the charge order in CCOC being
weakly doping dependent.

Numerical studies of $t$--$J$ models found a rather strong influence of the longer-range
hopping terms, $t'$ and $t''$, on the slope of $\mu(x)$ \cite{tohyama03}.
This might explain the experimentally detected differences in $d\mu/dx$
between the cuprate families \cite{yagi06}.
Obviously, a more detailed study of the chemical potential in other cuprates,
including those with static charge order, would be desirable.

\subsubsection{Possible signatures of fluctuating stripes}

If the tendency to stripe order is common to many underdoped cuprates,
but static long-range order is restricted to certain 214 compounds,
then an obvious question is:
``How to detect fluctuating stripes?'' \footnote{
Incidentally, this was the title of the 2003 review article by Kivelson {\em et al.} \cite{kiv_rmp}.
}
A rough estimate of a relevant fluctuation frequency, using
the STM charge correlation length $\xi_c \approx 10\ldots20\,a_0$ and a
characteristic UV cut-off scale of 100\,meV,
results in mode frequencies $\w_f$ of order of a few meV or, equivalently, one THz.

The observation strategies for fluctuating orders are different for the spin and charge sectors,
due to the limitations in experimental probes and due to the different effects of
quenched disorder on both sectors.
In the following, we only discuss the charge sector;
in the spin sector, the low-energy incommensurate fluctuations seen by inelastic
neutron scattering are a well-established precursor of order.

A direct observation of stripe fluctuations would be via low-energy collective modes in
the charge sector, by measuring the dynamic charge susceptibility $\chi_c''(\vec q,\w)$,
or the dielectric function $\epsilon(\vec q,\w)$,
at wavevectors $\vec q\sim \Qch$ and $\w\sim\w_f$.
Alternatively, it was proposed that a superconductor with fluctuating stripe should display a
shear photon mode \cite{cvet07}. This mode shows up in the dielectric response as well,
now at $q \sim 1/\xi_c$ and $\w\sim\w_f$.
In both cases, no present-day experimental technique has the required energy resolution,
as electron energy-loss spectroscopy (EELS) is currently limited to resolutions $>0.1$ eV.

Less direct signatures of fluctuating charge order may be found in the phonon and spin
excitation spectra, in the optical and Raman response, as well as in single-particle spectra.
However, the interpretation of these probes is rarely unambiguous, with examples given
throughout this article.

If an experimental parameter (like pressure or magnetic field) is available to
continuously tune the system through a quantum phase transition associated with stripe
order, then the system close to quantum criticality can be expected to display fluctuating
stripes. However, given the large intrinsic energy scales of cuprates, such tuning is
only possible if the material at zero pressure/field is already close to criticality.
With the exception of field-induced SDW order in LSCO and underdoped YBCO
(Sec.~\ref{sec:fieldtune}), no conclusive experiments in this direction have been
reported.

Taking into account impurity pinning, ``fluctuating'' stripes do no longer fluctuate.
This simplifies matters, i.e., allows for a detection using static probes,
but at the same time complicates matters, because the distinction between ordering
tendencies and impurity effects becomes subtle,
see Secs.~\ref{sec:dis} and \ref{sec:exp_stm}.


\subsection{Weak-coupling vs. strong-coupling description of incommensurate order}
\label{sec:wkstr}

As became clear from the theoretical approaches sketched in Sec.~\ref{sec:mic},
symmetry-breaking order can emerge in conceptually different ways.
In metals, weak interactions of low-energy quasiparticles can lead to Fermi-surface
instabilities. In the presence of strong interactions, the possibilities are richer
and dominated by collective effects instead of Fermi-surface properties.

From a symmetry point of view, weak-coupling and strong-coupling limits of density-wave
order may be continuously connected; exceptions are insulating or non-Fermi liquid
ordered states which cannot be obtained from weak coupling.\footnote{
There are a few interesting cases where insulating or non-Fermi liquid behavior can be
obtained from weak interactions:
Perfect Fermi-surface nesting,
Luttinger liquids in strictly one-dimensional systems,
or the nematic Fermi fluid in a continuum system described in Ref.~\cite{OKF01}.
}
An obvious requirement for a meaningful weak-coupling treatment is the existence of
well-defined fermionic quasiparticles in the relevant temperature range.
A feature of weak-coupling approaches to cuprate models is that ordering is generically
spin-driven, and charge order is parasitic to a collinear SDW state.

A number of quantitative, but possibly significant, differences between weak-coupling and
strong-coupling approaches can be identified from the available theories \cite{kiv_rmp}.
The energy dependence of the dynamic spin susceptibility $\chi_s''$ provides one criterion:
In weak coupling, the intensity will be broadly distributed over a wide energy range up
to the Fermi energy $E_F$, whereas $\chi_s''$ of the local-moment antiferromagnet
is dominated by energies of order of or less than the exchange energy
$J$. Furthermore, the strong-coupling $\chi_s''$ is typically much more structured in momentum
space. This is also reflected in the size of the ordered moment in an SDW state,
which is of order unity and of order $\Tsp/E_F$ (in units of $\mu_B$) in the strong and
weak-coupling limits, respectively.
The physics in the vicinity of the ordering transition may provide a second criterion:
In weak coupling, the order is carried by low-energy quasiparticles, implying spectral
weight transfer over small energy scales only. Moreover, the fluctuation regime of
a weak-coupling transition is usually very narrow. In both respects, the opposite is true
for strong coupling.
A third criterion is provided by the response to quenched disorder in a nearly ordered situation:
In the weak-coupling limit, disorder will primarily scatter quasiparticles and thus broaden
Fermi-surface-related features, whereas pinning of low-energy collective modes is dominant in the
strong-coupling case. Thus, the low-energy parts of both $\chi_s''$ and $\chi_c''$
will be suppressed (enhanced) by quenched disorder in weak (strong) coupling.

Before coming to experimental data, it should be emphasized that experiments may well be
in an intermediate-coupling regime, where both approaches describe at least part of the
data.
Moreover, falsifying a weak-coupling RPA calculation is much simpler than falsifying the
statement that ``strong-coupling physics is involved'':
RPA can provide numbers, e.g., for spin-fluctuation weights which can be directly compared to
experiments, whereas strong-coupling approaches are diverse and often more phenomenological.

Having said this, the collected experimental data for stripe order in 214 cuprates
speak in favor of a strong-coupling description:
(i) The ordering temperatures obey $\Tch>\Tsp$, i.e., magnetic order is not a
prerequisite for charge order.
(ii) The doping dependence of the CDW wavevector is opposite to what is expected from a
nesting scenario.
(iii) The ordered moments in the SDW phases are not small.
The energy scale of magnetic excitations appears to be set by $J$.
A large fraction of the spectral weight is found at energies below $J$ and near
wavevector $(\pi,\pi)$.
(iv) The fluctuation regime of magnetism is generically wide.
(v) Substituting Zn for Cu pins stripes in LSCO.
(vi) Although cuprate stripes are not insulating, a well-defined full Fermi surface
is not a property of stripe compounds: Stripe order emerges from the pseudogap
regime, not from a well-developed Fermi liquid.
I note that the existence of low-energy (nodal) quasiparticles near the momentum-space
diagonals is not necessarily in contradiction with a strong-coupling perspective:
For instance, valence-bond stripes display nodal quasiparticles even for sizeable
modulation amplitudes because of the approximate $d$-wave form factor of the charge order
\cite{mvor08}.

The STM observations in BSCCO-2212, BSCCO-2201, and CCOC are less clear-cut:
Low-energy LDOS modulations have been reported which display a doping dependence
of the CDW wavevector compatible with a nesting scenario.
As these materials are good superconductors (with the exception of underdoped CCOC),
this may not be surprising: It is known that the onset of superconductivity in
the cuprates renders quasiparticles much more coherent as compared to the
normal state, opening an avenue for Fermi-surface driven ordering.
However, the tunnel-asymmetry maps show a rather robust period-4 signal
which is particularly strong a elevated energies.

At present, it is unclear whether the data constitute a true contradiction.
A possibility is that strong-coupling and nesting effects co-operate
in producing ordered phases.


\subsection{Consequences of nematic order}
\label{sec:th_nematic}

Static nematic order of $d_{x^2-y^2}$ type may arise, e.g., from a Pomeranchuk
instability of the Fermi surface or as a precursor to stripe order,
see Secs.~\ref{sec:ph_nematic} and \ref{sec:mic_nematic}.
Nematic order breaks the rotation symmetry in the CuO$_2$ plane
from $C_4$ down to $C_2$.
This has a number of consequences, among them
(i) locally anisotropic single-particle properties, visible in STM,
(ii) globally anisotropic single-particle properties, e.g. a distorted Fermi surface,
visible in ARPES,
(iii) anisotropic spin fluctuation spectra, and
(iv) anisotropic transport.
With the exception of (i), these signatures require an experimental system in a
single-domain nematic state, as expected in de-twinned YBCO.

About anisotropic STM spectra: It has been proposed \cite{kiv_rmp} to employ spatial derivatives
of LDOS maps to detect nematic order. The simplest quantities are
$Q_{xx}(\vec r,E) = (\partial_x^2-\partial_y^2) \rho(\vec r,E)$ and
$Q_{xy}(\vec r,E) = 2 \partial_x\partial_y \rho(\vec r,E)$,
which then should be integrated over some energy interval to remove noise and QPI
features to only retain long-wavelength information.
A practical problem might be that gap modulations inherent to BSCCO \cite{stm_rmp}
hamper the procedure.
To my knowledge, a detailed analysis along these lines has not been performed.
However, the STM data of Kohsaka {\em et al.} \cite{kohsaka07} show stripe-like
patterns in the tunnel-asymmetry map which obviously break the $C_4$ rotation symmetry
locally.

I now turn to features of single-domain broken rotation symmetry.
Assuming a Pomeranchuk instability without underlying charge order,
issues (ii) and (iii) have been investigated theoretically on the basis of a RVB-type slave-boson
approach \cite{yamase06,yamase08}, with an eye towards the experiments on underdoped YBCO.
The starting point is a $t$--$J$ model with a small built-in hopping anisotropy of
$(\delta t/t)_0 = 5\%$. The mean-field solution then leads to an effective, i.e.
correlation-enhanced, hopping anisotropy of up to $(\delta t/t)_{\rm eff} \sim 20\%$
at low temperature and doping.
The renormalized band structure is then used to calculate the spin-fluctuation spectrum
using standard RPA, with a downward-renormalized interaction adjusted such that the system
at doping $0.07$ is close to the magnetic QCP.

In Ref.~\cite{yamase06}, the authors presented a detailed study of the RPA spin
fluctuation spectrum, as function of temperature both above and below the
pairing temperature (denoted $T_{\rm RVB}$ in Ref.~\cite{yamase06}).
The theory reproduces a number of features found in neutron scattering on
de-twinned YBCO \cite{hinkov04,hinkov07}, i.e., a ``resonance peak''
and anisotropic incommensurate correlations developing below the pairing temperature.
A number of differences should also be noted:
The anisotropy of the intensity distribution obtained from RPA is somewhat smaller than
obtained experimentally in YBCO-6.6, the RPA energy dependence of the anisotropy is too
weak, and the RPA resonance peak is somewhat too sharp.
The subsequent Ref.~\cite{yamase08} studied a non-superconducting state at low doping.
The obtained momentum-space profile of the low-energy spin fluctuations
shows some agreement with the neutron scattering data from YBCO-6.45 of Hinkov {\em et al.}
\cite{hinkov08a}. However, the spin correlations from RPA are {\em not} incommensurate,
in contrast to the experimental data, which may be due to the neglect of pairing.
(Pairing, however, leads to a distinct resonance peak in RPA which is not observed
experimentally.)

The neutron scattering data on less underdoped YBCO-6.85
\cite{hinkov04} have also been modeled using RPA for a fixed
anisotropic hopping with $\delta t/t = 6\%$ \cite{eremin05a,schnyder06}.
Although the calculations differ in details of the band structure and RPA interaction
functions, reasonable agreement with experiment was found in both cases.
In addition, Ref.~\cite{schnyder06} also accounted for a sub-dominant $s$-wave
pairing component which is generically present in YBCO -- this was found to lead to a
90$^\circ$ rotation of the anisotropy pattern as function of energy.

Possible consequences of a Pomeranchuk scenario have also been investigated for
214 compounds \cite{yamase01,yamase07}. Slave-boson plus RPA calculations
reproduce an approximate hour-glass spin excitation spectrum. However, the experimentally
observed doping evolution of the incommensurability is not easily recovered in this
approach.

A distinct viewpoint onto nematic phases, e.g. in YBCO, is based on underlying fluctuating
stripes. Then, spin incommensurabilities are primarily driven by fluctuating charge
modulations instead of Fermi surface distortions.
As shown in Sec.~\ref{sec:eff_gappedmag}, such calculations, albeit based on
phenomenological input, can describe salient features of the experimental data on YBCO
as well.

An important property of all nematic models based on Fermi-surface distortions only
is that the effective Fermi surface anisotropies, which are required to fit the neutron
data, are {\em large}, with the Fermi surface
topology changed compared to the undistorted case, see Fig. 4 of Ref.~\cite{yamase06}.
This should be easily detectable in ARPES experiments and hence provides a clear-cut
distinction to a scenario of a nematic phase originating from fluctuating stripes,
where the Fermi surface distortion for plausible parameters is much smaller \cite{wollny08}.


\subsection{Consequences of loop-current order}
\label{sec:th_varma}

Two types of loop-current order have been proposed to occur in the pseudogap regime of
the cuprates, namely the Cu-O loop currents within a unit cell with $\vec Q\!=\!0$
of Varma \cite{varma99,varma02}, and the $d$-density wave state with
$\vec Q\!=\!(\pi,\pi)$ of Chakravarty {\em et al.} \cite{ddw}.
In both cases, the resultant orbital antiferromagnetic order should be visible in
elastic neutron scattering, and the relevant experiments were summarized in
Sec.~\ref{sec:exp_elscatt}. Interestingly, the phase transitions can be related
to those of vertex models in classical statistical mechanics \cite{ddw_6vertex,varma06}.
We shall discuss a few theoretical aspects and additional properties of these states
in the following.

The Cu-O loop-current order of Varma has been originally derived from a mean-field theory of the
three-band Hubbard model. It has a number of interesting properties, which have been
reviewed in Ref.~\cite{varma06}.
For instance, the fermionic spectrum (at the mean-field level) displays a $d$-wave-like
gap tied to the Fermi level, with the quasiparticle energies given by
\begin{equation}
E^{\gtrless}_{\vec k} = \epsilon_{\vec k} \pm D({\vec k})
~~\textrm{for}~E_{\vec k} \gtrless \mu,
\end{equation}
where $\epsilon_{\vec k}$ is the bare dispersion,
$D({\vec k}) \propto \cos^2(2\phi)/[1+(\epsilon_{\vec k}/\epsilon_c)^2]$
the mean-field order parameter including form factor,
$\phi$ is the angle of $\vec k$, and $\epsilon_c$ is a band cutoff energy.
Thus, the $T\!=\!0$ spectrum consists of four Fermi points along the Brillouin zone
diagonals, which have been shown to broaden into arcs at finite temperatures
\cite{varma07a}.
The orbital moments of the loop-current state should be oriented perpendicular to the
CuO$_2$ planes. However, the experimentally detected magnetic order at $\vec Q = 0$
is characterized by moment directions canted by roughly 45$^\circ$ \cite{fauque06}.
Spin-orbit coupling in the low-symmetry YBCO structure has been proposed as a
source of canting \cite{varma07b}.
However, the fact that moment directions are very similar in HgBa$_2$CuO$_{4+\delta}$,
with this type of spin-orbit coupling being absent due to the tetragonal symmetry,
suggest that currents involve oxygen orbitals outside the CuO$_2$ planes
\cite{fauque08b}.

The transition into the Varma loop-current state is believed\footnote{
Not all symmetries of the Ashkin-Teller model are shared
by the original loop-current model (at the microscopic level).
} to be described by a
variant of the Ashkin-Teller model \cite{varma06,sudbo08a}, with weak thermodynamic
signatures at the finite temperature transition \cite{sudbo08b}.
The quantum critical fluctuations associated with the breakup
of the loop-current order have been shown \cite{varma06} to be of the
scale-invariant form hypothesized to lead to a Marginal Fermi
Liquid \cite{mfl89,mfl_rev}.
In particular, the fluctuations have been proposed to mediate $d$-wave pairing \cite{varma08}
as well as to cause a linear-in-$T$ resistivity \cite{varma06} and
a linear-in-$T$ single-particle lifetime \cite{varma07c}.

Currently, the microscopic conditions for the appearance of the Varma loop-current state
(e.g. its doping dependence) are not well investigated.
The reliability of the original mean-field approach is unclear,
considering that numerical investigations  \cite{greiter07,weber09} of three-band models
did not provide evidence for loop-current order of sufficient strength to
explain the pseudogap physics
(unless strong hybridizations with apical oxygen atoms are included \cite{weber09}).

Let me now come to $d$-density wave order \cite{ddw}.
This type of order appears frequently in RVB-type mean-field theories
of Hubbard and $t$--$J$ model \cite{affmar},
although the proposal in Ref.~\cite{ddw} was of phenomenological nature and did
not rely on RVB physics.
Within a one-band model, $d$-density wave order can be characterized by an order parameter of the type
$\sum_k i d_{\vec k} \langle c_{\vec k+\vec Q\sigma}^\dagger c_{\vec k\sigma}\rangle$
with wavevector $\vec Q=(\pi,\pi)$ and form factor $d_{\vec k} = \cos k_x\!-\!\cos k_y$.
(Also incommensurate variants of $d$-density wave order have been discussed
\cite{incommddw} in connection to the quantum oscillation experiments
\cite{sebastian08}.)
For $\vec Q=(\pi,\pi)$, the fermionic spectrum displays again a $d$-wave gap,
\begin{equation}
E^{\pm}_{\vec k} =
\frac{\epsilon_{\vec k} + \epsilon_{\vec k+\vec Q}}{2}
\pm
\frac{\sqrt{({\epsilon_{\vec k} - \epsilon_{\vec k+\vec Q})^2 + 4 W_{\vec k}^2}}}{2},
\end{equation}
with $W_{\vec k} \propto d_{\vec k}$ being the order parameter.
Here, the resultant gap between the two quasiparticle bands is {\em not} tied to the
Fermi level, i.e., in the presence of doping
and particle--hole asymmetry the state displays Fermi pockets (of both electron and hole
type) and a dip in the density of states at some finite energy away from the Fermi level.
This gap behavior has been one reason for criticism towards the
$d$-density wave state being an explanation for the pseudogap regime:
The gap reported in STM measurements appears to be centered at the Fermi level
also above $\Tc$ \cite{valla06,yazdani04,liu07}.
Also, ARPES experiments did not detect clear-cut evidence for electron pockets near
$(\pi,0)$, expected in a $d$-density wave state.
It has been proposed \cite{jia08} that long-range correlated disorder in a $d$-density-wave state
may lead to ARPES spectra with Fermi arcs, but no pockets near $(\pi,0)$.

The transition into the $d$-density wave state has been argued to be the same as that of
the 6-vertex model \cite{ddw_6vertex}. As above, the thermodynamic singularity at the
finite-temperature transition is weak. To my knowledge, the quantum critical properties
have not been worked out in detail.


\section{Implications for the cuprate phase diagram}
\label{sec:impl}

Understanding the physics of cuprates is still a major challenge in condensed matter
physics. Although the superconducting state appears reasonably well described by
BCS-type $d$-wave pairing, issues of ongoing debate are the nature of the pairing
mechanism, the non-Fermi liquid normal-state properties, and the pseudogap regime at
small doping above $\Tc$ \cite{np_feat}.
Ordering phenomena beyond superconductivity and associated
quantum phase transitions play a central role in this debate.

This final section is therefore devoted to a critical discussion of broader aspects of lattice
symmetry breaking in the cuprates. Among other things, it was proposed
(i) that stripes or stripe fluctuations are a central ingredient to the cuprate pairing mechanism,
(ii) that stripe quantum criticality is responsible for the non-Fermi liquid behavior around optimal doping, and
(iii) that stripe order is the cause of the pseudogap in underdoped samples.
Similar proposals have also been made in the context of other ordering phenomena.


\subsection{Lattice symmetry breaking: Universality?}
\label{sec:exp_univ}

Given the experimental evidence for lattice symmetry breaking in a variety of cuprates,
a crucial question is that of universality:
Which of the described features are special to a particular family of compounds,
and which may be common to all high-$\Tc$ cuprates?
A problem is that not all probes are available for all cuprate families (e.g. due to
surface problems or the lack of large single crystals),
making a direct comparison difficult.

Static stripe order with large spatial correlation length is only present in the
single-layer 214 cuprates. The STM data of Ref.~\cite{kohsaka07}, showing a period-4
valence-bond glass in underdoped BSCCO and CCOC, suggest that these materials have a
tendency towards stripe order as well -- this moderate conclusion remains true
even if the {\em static} order seen in STM is a surface effect only.
The idea of universal stripe physics is indirectly supported by the hour-glass
magnetic excitation spectrum, observed in 214 compounds as well as in YBCO and BSCCO,
which is consistent with a concept of fluctuating or spatially disordered stripes, see
Sec.~\ref{sec:eff_gappedmag}.
However, in some cuprates with particularly high $\Tc$, e.g. multilayer Bi and Hg compounds,
no stripe signatures have been detected to my knowledge.

The experimental data may be summarized in the following hypothesis:
The tendency toward bond/stripe order appears common to doped cuprates,
but competes with $d$-wave superconductivity.
Two trends are suggested by the data:
(i) The LTT phase of 214 compounds is most effective in hosting stripes.
(ii) An increasing number of CuO$_2$ layers per unit cell shifts the balance
between stripes and superconductivity towards homogeneous superconductivity.

Trend (i) is straightforward to explain: The LTT distortion pattern induces an electronic
in-plane anisotropy which is favorable for stripe order. Hence, stripes are stabilized
by electron--phonon coupling.

About trend (ii): Why would multilayer cuprates be less stripy?
The phase diagram of multilayer Hg compounds \cite{mukuda08} indicates that,
with increasing number of CuO$_2$ layers,
superconductivity is stabilized together with commensurate antiferromagnetism.
While the presence of different hole doping levels in the different layers
certainly plays a role here, the data prompts a speculation:
An increasing number of CuO$_2$ layers shifts the spin-sector competition between
commensurate antiferromagnetism and valence-bond order towards antiferromagnetism,\footnote{
The competition between antiferromagnetism and VBS order is well known
and studied in undoped Mott insulators on the square lattice \cite{ssrmp}.
}
possibly due to an increasing effective inter-layer coupling.
Assuming that valence-bond order is the driving force for stripe formation,
then stripes are suppressed together with valence bonds.

Notably, {\em no} signatures of stripe or nematic physics have been identified in any of
the electron-doped compounds. All neutron scattering experiments show that magnetic order
or magnetic fluctuations are peaked at the antiferromagnetic wavevector $\qaf$, rather
than away from it, see e.g. Refs.~\cite{wilson06,motoyama07,krueger07,fujita08}.
Similarly, no evidence for charge inhomogeneities has been reported,
but here little experimental data is available.
In line with the above speculation, the increased stability of commensurate
antiferromagnetism for electron doping may be connected to the absence of stripes.

Signatures of loop-current order with $\vec Q\!=\!0$
have been detected by now in YBCO and HgBa$_2$CuO$_{4+\delta}$, but not elsewhere.
Therefore, it is too early for a judgement regarding universality here.


\subsection{Phase transitions and quantum criticality}

All ordered phases discussed in this article admit zero-temperature phase transitions
\cite{ssbook} which could be associated with interesting quantum critical behavior.
In the cuprates, quantum criticality has been discussed widely, on both the experimental and
theoretical side.

In the normal state, the most puzzling piece of data is the linear temperature
dependence of the normal-state resistivity around optimal doping \cite{gurvitch87,takagi92}.
This and other experimental results appear well described in the framework of the marginal Fermi
liquid phenomenology \cite{mfl89,mfl_rev}
which is based on the assumed behavior of the electronic self-energy
\begin{equation}
\Sigma(k_F,\w) \simeq \lambda \left(
\w \ln \frac{\w_c}{\w} + i|\w|
\right)
\end{equation}
where $\w_c$ is a ultraviolet cutoff energy.
The linear-in-$T$ resistivity has been attributed to quantum criticality early on,
although reliable transport calculations are scarce.
Other normal-state observations also point towards a quantum critical point near
optimal doping, associated with the disappearance of local-moment magnetism
\cite{pana02}, a change in the Fermi surface \cite{daou08,daou09,cyr09},
and a distinct crossover behavior in the resistivity \cite{hussey09}.
In contrast, signatures of quantum criticality {\em inside} the superconducting phase
have not been conclusively identified.

The next two subsections contain a summary on what is known theoretically about quantum phase
transitions into states with broken lattice symmetry.
The quantum critical behavior depends on whether or not a transition takes place in the
presence of background superconductivity. Hence, we shall distinguish critical theories
in the $d$-wave superconducting state and in the metallic normal state, with the former (latter)
being appropriate for temperatures below (above) the superconducting $\Tc$.
We shall focus our discussion to criticality in two space dimensions.

\subsubsection{Nematic transition}

The first transition to be considered is between a disordered state, without broken
lattice symmetries, and a nematic state, with focus on the $d$-wave nematic.
The continuum and lattice situations need to be distinguished.
The continuum case, with the order parameter being a director field, has been
studied in Refs.~\cite{OKF01,lawler07}.
In the following, we concentrate on the square-lattice case, where the order
parameter is a real scalar, i.e. of Ising type.

Importantly, in the presence of a full Fermi surface, there is Landau damping of the
order parameter, resulting in a dynamic exponent of $z\!=\!3$. Hence, the normal-state phase
transition is above its upper-critical dimension. Critical behavior can be calculated by
a perturbative expansion about the Gaussian fixed point, as is standard in the
Landau-Ginzburg-Wilson (LGW) (or Hertz-Millis) approach to metallic
criticality \cite{hertz76,millis93,hvl_rmp}.
However, the effect of critical
fluctuations on the fermions is strong due to $\vec Q\!=\!0$. Physically, the Fermi surface
becomes soft at the transition \cite{MRA03}. The dynamical Fermi-surface fluctuations
at criticality in $d\!=\!2$ have been analyzed recently \cite{DM06}: the electronic
self-energy scales as $\w^{2/3}$, thus destroying the Fermi liquid at all wavevectors
except for the momentum-space diagonals (i.e. at ``cold spots'').\footnote{
The Landau damping of the nematic
order parameter is not qualitatively changed by the $\w^{2/3}$ self energy as compared to
the Fermi-liquid case \cite{DM06}.
}
The quantum-critical transport scattering rate was found to be linear in $T$ except
at the cold spots, leading to a resistivity varying as $\rho(T) \propto T^{3/2}$ in
the clean limit.
In contrast, in an impurity-dominated regime, $\rho(T) - \rho_0 \propto T$ \cite{DM07}.

Within higher-dimensional bosonization it has been demonstrated that the fermionic
correlation functions display ``local'' behavior in the non-Fermi liquid quantum
critical regime, i.e., spatial correlations remain short-ranged \cite{lawler07}.
Note that this result has been derived in the continuum limit.
While it has been suggested that the quantum critical properties of continuum and lattice
cases are similar, this is not entirely correct in $d\!=\!2$:
In the continuum situation, a second type of critical mode with $z\!=\!2$ appears due to the
combination of rotational invariance, $d$-wave nature of
the order parameter, and Landau damping. This leads to a complicated interplay of two
time scales near the transition, resulting in non-trivial logarithmic corrections \cite{garst09}.

In the $d$-wave superconducting state, fermionic excitations only exist at the nodal
points along the momentum space diagonals. In the present $\vec{Q}\!=\!0$ case,
the order-parameter fluctuations acquire a relevant coupling to the nodal fermions.
The critical theory has been focus of recent work and turns out to be subtle.
In Ref.~\cite{vzs00b,vzs00c}, an expansion in $\epsilon=3-d$ was employed, with
the result that no stable critical fixed point was found
(for cases D and E in \cite{vzs00b,vzs00c}).
This suggested a fluctuation-induced first-order transition.
However, a recent approach based on a $1/N_f$ expansion, where $N_f$ is the number of fermion flavors,
uncovered that a continuous transition is possible in $d\!=\!2$ \cite{kivss08,huhss08}.
Interestingly, the velocity anisotropy of the nodal quasiparticles becomes large near
criticality, in contrast to the theories of QPT to $d+is$ or $d+id$ states,
where the ordered state preserves the $C_4$ lattice symmetry and the quantum critical
theory takes a Lorentz-invariant form \cite{vzs00b,vzs00c}.
The strong velocity anisotropy may be important for understanding ARPES experiments
and quasiparticle interference as measured by STM \cite{lawler08}.

\subsubsection{CDW and SDW transitions}

Next I turn to transitions into a state with broken translational symmetry,
which could either display a charge modulation only, or both CDW and SDW.
The transition into a uni-directional density-wave state can occur as direct transition
from a disordered state, or from a state with nematic order;
the latter can then can be understood as a precursor of density-wave order.

The order parameter $\phi$ of the density wave is a complex scalar (CDW) or vector (SDW)
and carries a lattice momentum $\vec Q\neq 0$.
Commensurate lattice pinning reduces the $U(1)$ symmetry of the complex phase of $\phi$
to $Z_N$.
Multiple copies of $\phi$ are needed for the inequivalent directions of
$\vec Q$, e.g., a cuprate CDW or SDW transition in a tetragonal environment requires two
fields for horizontal and vertical modulations.

As above, the presence of a full Fermi surface leads to Landau damping, here with $z\!=\!2$,
i.e., the damping arises from hot spots (lines) on the Fermi surface connected by $\vec Q$
in 2d (3d).
In 2d, the normal-state phase transitions are at their upper-critical dimension.
This can be treated by the standard LGW approach, but requires a re-summation of
perturbation theory, with the result of mean-field behavior supplemented by
logarithmic corrections.\footnote{
It has been argued that the LGW approach breaks down for the 2d metallic
antiferromagnet, as the low-energy modes of the Fermi liquid induce an infinite
number of marginal operators. As a result, a continuous transition with non-trivial
exponents emerges \cite{abanov04}.
}
A calculation of the electronic self-energy shows that the quasiparticle picture
breaks down in $d<3$, but only along the hot lines.
A reliable transport theory becomes difficult even for the simplest antiferromagnetic
transition:
In contrast to $d\!=\!3$ where a Boltzmann description is possible \cite{rosch00},
in $d\!=\!2$ more elaborate methods are required.
Ref.~\cite{kontani99} studied transport near a 2d
antiferromagnetic critical point, taking into account vertex corrections, but neglecting
impurity effects. The result for the resistivity shows approximately $T$-linear behavior
over an intermediate range of temperatures. However, from the 3d results \cite{rosch00} a
complex interplay of magnetic and impurity scattering can be expected also in 2d which has
not been studied to date.

Concrete applications to cuprates of criticality associated with an incommensurate metallic CDW
have been worked out by the Rome group \cite{castellani95,castellani96,castellani97,
castellani98,caprara99}.
In particular, singular scattering near CDW-induced hot spots of the Fermi surface has
been invoked to explain the non-Fermi liquid characteristics in both the single-particle
and transport properties. However, a full transport calculation was not presented.
From an experimental perspective, a strain-controlled CDW QCP has been suggested,
based on EXAFS measurements of the bond-length distributions in various cuprates
\cite{strain_qcp}.

A non-trivial interplay of nematic and density-wave fluctuations occurs for
a transition from a nematic to a CDW (or ``smectic'') state in the continuum
(i.e. without underlying lattice) in $d\!=\!2$ \cite{sun08}.
The reason is the Goldstone (director) mode of the nematic phase which causes non-Fermi liquid
behavior inside the nematic phase \cite{OKF01} and strongly influences the critical
properties of the CDW transition \cite{sun08}.

In the $d$-wave superconducting state, the fate of the critical theory depends on whether
or not the ordering wavevector $\vec Q$ connects two nodal points.
This is not the case without fine tuning, and such fine tuning seems to be absent experimentally.
Then the critical theory is that of an insulator, with dynamical exponent $z\!=\!1$.
While non-trivial exponents will occur in order-parameter correlations, the fermions are
bystanders only.
However, this is not the full truth: As discussed in Ref.~\cite{ssafm08} for the case of
a SDW transition in a $d$-wave superconductor on the square lattice, the coupling between
the nodal fermions and an Ising nematic field, constructed from the two SDW fields,
is irrelevant, but with a tiny scaling dimension. Hence, although fermions and
order-parameter fluctuations formally decouple at the QCP, there will be strong damping
of the nodal quasiparticles, with a nearly $T$-linear scattering rate, from nematic
fluctuations.
If, instead $\vec Q$ is fine-tuned to connect nodal points, then non-trivial critical
behavior the coupled system of order parameter and fermions can be expected,
in analogy to Refs.~\cite{kivss08,huhss08}.

Finally, I mention an interesting route to exotic phases arising from the tendency toward
SDW order. As discussed in Ref.~\cite{demler2}, the order parameter for a collinear
SDW can be written as $\phi_{s\alpha} = e^{i \Theta} n_\alpha$. This description involves
a $Z_2$ degeneracy associated with a simultaneous change $\Theta\rightarrow\Theta+\pi$ and $n_\alpha\rightarrow-n_\alpha$,
which may be implemented into a field theory via a $Z_2$ gauge field. Consequently, it is
conceivable to have a deconfined phase with ``excitation fractionalization'': This is a
phase without broken symmetries, but with topological order, where the $\Theta$ and
$n_\alpha$ degrees of freedom form separate excitations \cite{demler2}.
Similar ideas of ``stripe fractionalization'' were put forward in the context of fluctuating
stripes in Ref.~\cite{zaanen01}.
Experimental signatures of such fractionalized phases (which are very different from
spin-charge separated states in RVB-like theories) have not been identified to date.

\subsubsection{Cuprate quantum criticality?}
\label{sec:resis}

Various cuprate experiments suggest the existence of a quantum critical point near optimal
doping, with quantum critical signatures above $\Tc$, i.e., in the so-called strange-metal regime.
The theoretically most challenging observation is the linear-in-$T$ resistivity
which extends up to high temperatures, namely 600\,K in YBCO and almost 1000\,K in LSCO
\cite{gurvitch87,takagi92}.
In the following, I try to summarize the theoretical status on this issue.

In general, quantum critical points can be interacting or mean-field-like, depending on
whether the theory is below or above its upper critical dimensions.
Interacting critical fixed points display single-parameter scaling in thermodynamics
and strong hyperscaling properties \cite{ssbook}.
The hypothesis of single-parameter scaling in the cuprates was explored in
Ref.~\cite{zaanen04}, where a number of scaling laws for thermodynamic data were
proposed. Tests require detailed measurements of the doping dependence of the
chemical potential and the electronic compressibility, which are not available to date.

Proposals which relate cuprate quantum criticality to transitions into states
with lattice symmetry breaking face a number of objections:
(i) Order-parameter theories of Hertz-Millis type \cite{hertz76,millis93,hvl_rmp}
have a dynamical exponent $z\!=\!2$ or 3, and are not below their upper critical
dimension in $d\!=\!2$. Consequently, hyperscaling is violated,
and it is unclear whether a robust linear $\rho(T)$ can be expected.
This applies e.g. to the charge-density wave criticality of
Ref.~\cite{castellani98}.
(ii) An order parameter carrying a finite momentum $\vec Q$ (like stripes) will primarily affect
low-energy fermions which can be connected by $\vec Q$ in momentum space. However,
ARPES experiments appear to be characterized by the absence of quasiparticles at all
wavevectors in the strange-metal regime.
(iii) Quantum criticality associated with a conventional order parameter is unlikely to
yield quantum critical behavior up to 600\,K.

Let me discuss the last objection in somewhat more detail:
Usually, quantum critical behavior with well-defined power laws
requires that the order-parameter correlation length is large compared to microscopic scales,
and that no other energy scales intervene.
For simple quantum magnets, characterized by an exchange scale $J$, it has been shown that the
quantum critical regime can extend up to $J/2$ \cite{chakravarty05}.
However, it appears unlikely that long-ranged correlations
exist in optimally doped cuprates at temperatures of several hundred Kelvin.
On the one hand, such correlations, if existing in the spin or charge channels, should
have been detected experimentally, but e.g. neutron scattering has found the magnetic
correlation length at optimal doping to be around two lattice constants at low $T$.
On the other hand, numerical investigations of relevant microscopic models using, e.g.,
quantum Monte-Carlo techniques, have not found appreciable correlations
at optimal doping and elevated temperatures.
These arguments could be invalid if the order parameter is difficult to detect,
like nematic order (under the influence of quenched disorder)
or the circulating current patterns proposed by Varma \cite{varma99,varma02};
here also objections (i) and (ii) may not apply.
It remains to explain why phonons, existing in the temperature range up to 600\,K,
do not affect the resistivity.

It is fair to say that, at present, the robustness of the linear-in-$T$ resistivity is a
puzzle. Even under the hypothesis of an interacting critical fixed point, a
linear $\rho(T)$ in the quantum critical regime does {\em not} automatically follow.
In fact, explicit candidates for a linear $\rho(T)$ are scarce.
Perhaps with the exceptions of early gauge-theory descriptions of Mott physics \cite{gauge_lin}
and a scenario of a doped disordered spin-liquid Mott insulator \cite{page99},
no robust theoretical explanation seems available.
This leaves open the options of
(A) some unknown form of strong quantum criticality or
(B) the resistivity not being truly linear, in which case the search for a quantum
critical mechanism could be pointless.

Interestingly, recent numerical studies of the 2d Hubbard \cite{jarrell08b} and $t$--$J$
models \cite{haule07}, using cluster extensions of DMFT, have provided some hints toward
criticality around optimal doping.
However, the results of Refs.~\cite{jarrell08b,haule07} are at best indicative
and provide little phenomenological understanding on the physics of the critical point,
moreover they disagree with respect to the nature of the phase below optimal doping.
Clearly, more investigations in this direction are required.


\subsection{Pseudogap}

The suppression of low-energy fluctuations in underdoped cuprates significantly
above $\Tc$, dubbed ``pseudogap'', is central to the cuprate phenomenology \cite{timusk}.
The pseudogap temperature, $T^\ast$, monotonically decreases with doping, in
striking difference to the superconducting $\Tc$. $T^\ast$ has been suggested to
extrapolate to the scale $J$ of the magnetic exchange in the limit of zero doping, and to
vanish either around optimal doping or at the overdoped end of the superconducting dome.
To my knowledge, pseudogap signatures have been unambiguously identified in all
hole-doped cuprates, while in electron-doped materials the issue is controversial.

The list of proposed explanations for the pseudogap is long and
ranges from genuine Mott-gap physics over preformed, phase-fluctuating Cooper pairs to
ordered states competing with superconductivity. While an extensive discussion of
pseudogap physics is beyond the scope of this review, I will briefly summarize a few
important aspects.

Genuine Mott physics is difficult to describe on a phenomenological level.
In single-site DMFT \cite{georges96}, a Mott gap occurs in the large-$U$ insulating phase
of the single-band Hubbard model, however, this phase suffers from an artificial spin
degeneracy. With spatial correlations included, short-range singlet formation may
be responsible for a partial gap formation \cite{macridin06,stanescu06,jarrell08a}.
In the framework of Hubbard-model field theories, attempts have been made to identify
the Mott gap with the dynamics of a charge-$2e$ boson which connects the
low-energy sector to the upper Hubbard band \cite{phillips}.
The verification of theoretical ideas in this direction is open.

Signatures of preformed pairs above $\Tc$ \cite{preformed} have been identified
in a number of experiments, most notably Nernst effect measurements
\cite{nernst_exp1,nernst_exp2}, photoemission \cite{valla08,kanigel08},
and STM studies \cite{yazdani07}.
Theoretical calculations, relating the Nernst signal to phase fluctuating
superconductivity \cite{nernst_theory}, give a plausible description of most of the data.
This interpretation is supported by the observation of fluctuating diamagnetism
which often varies in proportion to the Nernst coefficient \cite{nernst_diamag}.
In general, however, care is required, as there are three sources of a sizeable Nernst
signal \cite{nernst_exp2}: quasiparticles with a small Fermi energy, vortices (i.e. phase
fluctuations), and short-lived Cooper pairs (i.e. amplitude fluctuations), and it remains
to be seen which is most important.\footnote{
Nernst effect measurements in LESCO and LNSCO have observed a
distinct enhancement of the Nernst signal far above $\Tc$,
which was interpreted in terms of a stripe-induced reconstruction of the
Fermi surface \cite{cyr09}.
}
In any case, the characteristic onset temperature of pairing fluctuations,
as determined from Nernst effect or STM,
is significantly below the established pseudogap temperature in underdoped cuprates.
This casts doubts on the assumption of preformed pairs being the exclusive source
of the pseudogap.

A third class of proposals links the pseudogap to an ordering phenomenon in the
particle--hole channel which is assumed to compete with superconductivity. This would
also offer a natural explanation for the suppression of $\Tc$ in the underdoped regime.
Among the concrete proposals for competing phases are spin and charge density waves, e.g.,
stripes, and various forms of circulating-current orders.
An common objection against these proposals is that the pseudogap line at $T^\ast$
does not appear to be associated with a thermodynamic phase transition.
However, this can be circumvented either by invoking quenched disorder which
tends to smear the transition (see Sec.~\ref{sec:dis}) or by postulating
a special form of phase transition with weak thermodynamic singularities,
e.g. of Kosterlitz-Thouless or Ashkin-Teller type \cite{sudbo08b}.
In fact, a weak, but rather sharp, signature in the uniform susceptibility has recently
been detected in YBCO samples of different doping, tracking the pseudogap
temperature \cite{monod08}. This may represent a distinct thermodynamic phase
transition, and is possibly connected to the broken time reversal as detected in
polar Kerr effect measurements \cite{kerr08}.

In the context of the present article's topic, I will briefly discuss the
hypothesis that stripes are the cause of the pseudogap, which has been voiced
on the basis of both phenomenological theory \cite{castellani98,benfatto00} and
experimental data \cite{wise08}.
In my view, this hypothesis is problematic on several grounds:
(i) Order in the particle-hole channel with a finite wavevector $\vec Q$ will cause
distinct signatures in the quasiparticle band structure, in particular gaps at momenta
separated by $\vec Q$. Those have not been observed.
(ii) The pseudogap appears to be a universal phenomenon in hole-doped cuprates,
with very similar properties in the different families. This suggests that the pseudogap
has a common origin in all hole-doped cuprates.
In contrast, stripes are strongest in single-layer compounds of the 214 family,
but are weak or absent e.g. in materials with more than two CuO$_2$ layers per unit cell.

On the theory side, microscopic calculations using cluster extensions of DMFT have
established the existence of a pseudogap in the 2d Hubbard model at small doping above
the superconducting instability \cite{macridin06,stanescu06,jarrell08a}.
Unfortunately, the numerical data provide limited insight into the origin of gap
formation:
The pseudogap occurs in the {\em absence} of long-range order and is apparently
related to strong short-range correlations (which may eventually become long-ranged at
low temperatures).
Whether this pseudogap should be attributed to genuine Mott physics is unclear;
alternatively, both antiferromagnetic \cite{macridin06} and bond-order \cite{jarrell08a}
correlations have been made responsible for the pseudogap.

In summary, experimental and theoretical results consistently show that forms of
non-superconducting order are enhanced at small doping, i.e., in the pseudogap regime.
This includes spin and charge density waves, nematic order, and possibly loop-current
order.
Then, such order can either be
(A) the cause of the pseudogap, in which case the phenomenon should be common to all cuprates, or
(B) a secondary effect of some other phenomenon causing the pseudogap.
For stripes, experimental evidence points towards scenario (B).
For other forms of order, more experiments are required to check or verify their
universal occurrence.


\subsection{Pairing mechanism}

While it appears well established that the superconducting state in the cuprates
is characterized by Cooper pairs of $d$-wave symmetry \cite{uni_dwave},
the pairing mechanism is controversial.
Weak-coupling calculations in the Hubbard-model framework,
using RPA \cite{miyake86,scalapino86}
or the more sophisticated functional renormalization group (fRG, see Sec.~\ref{sec:frg}) \cite{honer_rev},
show that $d$-wave pairing can in principle be mediated by antiferromagnetic
fluctuations.
Electron-phonon coupling, various experimental signatures for which have been identified,
has also been suggested as pairing force -- this, however, is not easily
compatible with $d$-wave symmetry.
The distinct maximum in $\Tc$ as function of doping has triggered alternative proposals,
with strong pairing driven by the presence of a quantum critical point near optimal
doping.

Stripe physics as as a source of pairing was suggested by different groups,
either via quantum critical CDW fluctuations \cite{castellani96,castellani97},
or via topological stripe fluctuations \cite{zaanen01},
or via the interplay of co-existing Luttinger-liquid and hole-pair excitations
\cite{antonio01}.
At present, it is fair to say that none of these suggestions has been worked
out into a testable theory of $d$-wave superconductivity.

A different but related proposal is that of Kivelson and co-workers
\cite{inhom_sc03,inhom_sc05,inhom_sc08} who argue that the
presence of stripes in a superconducting system enhances the transition temperature $\Tc$.
This idea starts from the assumption of a pairing scenario where the superconducting gap
and stiffness are anti-correlated as a function of doping,
as is the case in cuprates below optimal doping.
Then, one can show that an inhomogeneous structure, combining regions of large gap with
those of large stiffness, leads to an increased $\Tc$ as compared to the homogeneous system,
at the expense of having a smaller superfluid stiffness.
This concept of ``optimal inhomogeneity'' was proposed to be relevant for understanding
the high transition temperature in cuprates.

While at present it is difficult to judge these proposals, they meet the same problem
as was noted above: The empirical anticorrelation between ``stripyness'' and high $\Tc$
points towards stripes being a competitor to superconductivity.\footnote{
The resistivity drop in LBCO-1/8, occurring at $\Tsp$ and interpreted as fluctuating
superconductivity \cite{li07,jt08}, together with the fact that pairing sets in at a
lower $T$ for all other doping levels of LBCO, has been invoked as evidence for a positive
correlation between stripes and pairing. However, a concise picture for the data of
Refs.~\cite{li07,jt08} has not yet emerged, see Sec.~\ref{sec:antiphase}.
}
The possibility that stripe {\em fluctuations} at elevated energies mediate pairing
cannot be ruled out, however. 


\section{Conclusions}

Ordered phases in strongly correlated electronic systems offer a fascinating variety
of phenomena. In this article, I aimed to give a balanced and critical account on a
particularly intensively studied class of ordered phases, namely those associated with
lattice symmetry breaking in cuprate high-temperature superconductors.
I tried to cover exciting experimental developments, which include
the observation of charge order via resonant soft x-ray scattering and scanning
tunneling microscopy and the identification of a seemingly universal spin excitation
spectrum. I also discussed theoretical works, in particular those dealing with the microscopic
origins of symmetry breaking as well as those providing a more phenomenological modeling of
experimental data.
Considering the wealth of published papers and the existence of previous review articles,
the main emphasis was on recent works which appeared during the last five years.

A quick physics summary
may be given as follows:
The tendency towards states with modulated spin and charge densities (stripes)
appears to be common to hole-doped cuprates -- a view which has only been established
recently --
although the strength of the phenomenon varies from family to family.
While general scenarios of stripe formation have been developed,
the underlying microscopic physics is not completely understood.
In particular, oxygen orbitals appear to play an important role, implying that the
three-band Hubbard model has to be considered to gain a quantitative picture.
Although tendencies to stripes are strongest in the pseudogap regime, stripes may well
be a {\em result} rather than a {\em cause} of the pseudogap.
The same applies to other experimentally identified symmetry-broken states, e.g.,
nematic and loop-current order.
Indications for these ordering phenomena being crucial ingredients to the cuprate
pairing mechanism are weak at present.

Independent of their actual role for cuprate superconductivity,
stripes and nematics appear to be common to a variety of correlated oxides,
ranging from cuprates to nickelates \cite{cheong97,jt98}, manganites \cite{mori98},
and perhaps also ruthenates \cite{meta_pom}.
While this suggests a common driving mechanism, differences are apparent:
While ruthenates are good metals, stripe phases in nickelates and manganites are
insulating and display rather robust charge order, rendering them more ``classical''.
Thus, cuprates are indeed special, as they offer a unique combination of
the proximity to a Mott insulator and very strong quantum effects.
Understanding their puzzles remains a challenge.


\section*{Acknowledgments}

It is my pleasure to thank
S. Borisenko, B. B\"uchner, J. C. Davis, J. Fink, R. Hackl, W. Hanke, V. Hinkov, P. Hirschfeld,
C. Honerkamp, M. H\"ucker, R. K. Kaul, B. Keimer, S. Kivelson, P. A. Lee, W. Metzner,
M. R. Norman, C. Panagopoulos, A. Polkovnikov, A. Rosch, O. R\"osch, S. Sachdev, Q. Si,
L. Taillefer, J. Tranquada, T. Ulbricht, T. Vojta, P. W\"olfle, A. Wollny, H. Yamase, A. Yazdani,
J. Zaanen, and Y. Zhang
for illuminating conversations and collaborations over the last years.
I thank C. Vojta for moral and technical support and
Y. Kodak for proofreading the references.
This work was supported by the Deutsche Forschungsgemeinschaft through
FOR 538, FOR 960, SFB 608, and SFB/TR-12.


\label{lastpage}


\begin{thebibliography}{999}



\bibitem{bednorz}
G.~Bednorz and K.~A.~M\"uller, Z. Phys. B {\bf 64}, 189 (1986).

\bibitem{bcs}
J. Bardeen, L. N. Cooper, and J. R. Schrieffer,
Phys. Rev. {\bf 108}, 1175 (1957).

\bibitem{ssrmp}
S. Sachdev, Rev. Mod. Phys. {\bf 75}, 913 (2003).

\bibitem{leermp}
P. A. Lee, N. Nagaosa, and X.-G. Wen,
Rev. Mod. Phys. {\bf 78}, 17 (2006).

\bibitem{timusk}
T. Timusk and B. W. Statt,
Rep. Prog. Phys. {\bf 62}, 61 (1999).

\bibitem{carlson_rev}
E. W. Carlson, V. J. Emery, S. A. Kivelson, and D. Orgad,
in: The Physics of Conventional and Unconventional Superconductors, K. H. Bennemann and J. B. Ketterson, eds.
(Springer, 2003).


\bibitem{norman_ap}
M. R. Norman, D. Pines, and C. Kallin,
Adv. Phys. {\bf 54}, 715 (2005).

\bibitem{lee_rev}
P. A. Lee,
Rep. Prog. Phys. {\bf 71}, 012501 (2008).


\bibitem{pnas}
V. J. Emery, S. A. Kivelson, and J. M. Tranquada,
Proc. Natl. Acad. Sci. USA {\bf 96}, 8814 (1999).

\bibitem{jan}
J. Zaanen,
Physica C {\bf 317}, 217 (1999).

\bibitem{kiv_rmp}
S. A. Kivelson, I. P. Bindloss, E. Fradkin, V. Oganesyan, J. M. Tranquada,
A. Kapitulnik, and C. Howald,
Rev. Mod. Phys. {\bf 75}, 1201 (2003).

\bibitem{brom_rev}
H. B. Brom and J. Zaanen,
in: Handbook of Magnetic Materials, Vol. xx, K. H. J. Buschow, ed. (Elsevier, 2003).

\bibitem{antonio_rev}
A. H. Castro Neto and C. Morais Smith,
in: Strong Interactions in Low Dimensions, D. Baeriswyl and L. Degiorgi, eds. (Kluwer, 2004),
pg. 277.

\bibitem{oles_rev}
M. Raczkowski, A. M. Ole\'{s}, and R. Fr\'{e}sard,
Low. Temp. Phys. {\bf 32}, 305 (2006).

\bibitem{za89}
J. Zaanen and O. Gunnarsson,
Phys. Rev. B {\bf 40}, 7391 (1989).

\bibitem{poil89}
D. Poilblanc and T. M. Rice,
Phys. Rev. B {\bf 39}, 9749 (1989).

\bibitem{schulz89}
H. J. Schulz, J. de Physique {\bf 50}, 2833 (1989).

\bibitem{machida89}
K.~Machida, Physica C {\bf 158}, 192 (1989).


\bibitem{jt95}
J. M. Tranquada, B. J. Sternlieb, J. D. Axe, Y. Nakamura, and S. Uchida,
Nature {\bf 375}, 561 (1995).

\bibitem{jt96}
J. M. Tranquada, J. D. Axe, N. Ichikawa, Y. Nakamura, S. Uchida, and B. Nachumi,
Phys. Rev. B {\bf 54}, 7489 (1996).

\bibitem{KFE98}
S. A. Kivelson, E. Fradkin, and V. J. Emery,
Nature {\bf 393}, 550 (1998).

\bibitem{cheong97}
S.-H. Lee and S-W. Cheong,
Phys. Rev. Lett. {\bf 79}, 2514 (1997).

\bibitem{jt98}
J. M. Tranquada,
J. Phys. Chem. Solids {\bf 59}, 2150 (1998).

\bibitem{mori98}
S. Mori, C. H. Chen, and S.-W. Cheong,
Nature {\bf 392}, 473 (1998).

\bibitem{zxshen_rmp}
A. Damascelli, Z. Hussain, and Z. X. Shen,
Rev. Mod. Phys. {\bf 75}, 473 (2003).

\bibitem{yoshida_rev}
T. Yoshida {\em et al.}, 
J. Phys.: Cond. Matter {\bf 19}, 125209 (2007).

\bibitem{stm_rmp}
{\O}. Fischer, M. Kugler, I. Maggio-Aprile, C. Berthod, and C. Renner,
Rev. Mod. Phys. {\bf 79}, 353 (2007).

\bibitem{so5rmp}
E. Demler, W. Hanke, and S.-C. Zhang,
Rev. Mod. Phys. {\bf 76}, 909 (2004).

\bibitem{ogata_rop}
M. Ogata and H. Fukuyama,
Rep. Prog. Phys. {\bf 71}, 036501 (2008).

\bibitem{ddw}
S. Chakravarty, R. B. Laughlin, D. K. Morr, and C. Nayak,
Phys. Rev. B {\bf 63}, 094503 (2001).

\bibitem{varma99}
C.~M.~Varma,
Phys. Rev. Lett. {\bf 83}, 3538 (1999).

\bibitem{varma02}
M. E. Simon and C. M. Varma,
Phys. Rev. Lett. {\bf 89}, 247003 (2002).



\bibitem{thio88}
T. Thio, T. R. Thurston, N. W. Preyer, P. J. Picone, M. A. Kastner, H. P. Jenssen,
D. R. Gabbe, C. Y. Chen, R. J. Birgeneau, and A. Aharony,
Phys. Rev. B {\bf 38}, 905 (1988).


\bibitem{dwave_checker1}
J.-X. Li, C.-Q. Wu, and D.-H. Lee,
Phys. Rev. B {\bf 74}, 184515 (2006).

\bibitem{dwave_checker2}
K. Seo, H.-D. Chen, and J. Hu,
Phys. Rev. B {\bf 76}, 020511(R) (2007).

\bibitem{mvor08}
M. Vojta and O. R\"osch,
Phys. Rev. B {\bf 77}, 094504 (2008).

\bibitem{kohsaka07}
Y. Kohsaka, C. Taylor, K. Fujita, A. Schmidt, C. Lupien, T. Hanaguri, M. Azuma, M. Takano,
H. Eisaki, H. Takagi, S. Uchida, and J. C. Davis,
Science {\bf 315}, 1380 (2007).

\bibitem{uni_dwave}
C. C. Tsuei, J. R. Kirtley, G. Hammerl, J. Mannhart, H. Raffy, and Z. Z. Li,
Phys. Rev. Lett. {\bf 93}, 187004 (2004).

\bibitem{FF}
P. Fulde and R. A. Ferrell, Phys. Rev. {\bf 135}, A550 (1964).

\bibitem{LO}
A. I. Larkin and Y. N. Ovchinnikov, Zh. Eksp. Teor. Fiz. {\bf 47}, 1136 (1964).

\bibitem{wosnitza07}
R. Lortz, Y. Wang, A. Demuer, P. H. M. B\"ottger, B. Bergk, G. Zwicknagl, Y. Nakazawa, and J. Wosnitza,
Phys. Rev. Lett. {\bf 99}, 187002 (2007).

\bibitem{himeda02}
A. Himeda, T. Kato, and M. Ogata,
Phys. Rev. Lett. {\bf 88}, 117001 (2002).

\bibitem{berg07}
E. Berg, E. Fradkin, E.-A. Kim, S. A. Kivelson, V. Oganesyan, J. M. Tranquada, and S. C. Zhang,
Phys. Rev. Lett. {\bf 99}, 127003 (2007).

\bibitem{pomer}
I. J. Pomeranchuk, Sov. Phys. JETP {\bf 8}, 361 (1958).

\bibitem{VZ05}
C. M. Varma and L. Zhu, Phys. Rev. Lett. {\bf 96}, 036405 (2006).

\bibitem{meta_pom}
R. A. Borzi, S. A. Grigera, J. Farrell, R. S. Perry, S. J. S. Lister, S. L. Lee, D. A. Tennant,
Y. Maeno, and A. P. Mackenzie,
Science {\bf 315}, 214 (2007).

\bibitem{vzs00b}
M. Vojta, Y. Zhang, and S. Sachdev,
Phys. Rev. Lett. {\bf 85}, 4940  (2000);
{\bf 100}, 089904 (2008).

\bibitem{ddw1}
D. A. Ivanov, P. A. Lee, and X. G. Wen,
Phys. Rev. Lett. {\bf 84}, 3958 (2000).

\bibitem{ddw2}
C. Nayak,
Phys. Rev. B {\bf 62}, 4880 (2000).

\bibitem{affmar}
I.~Affleck and J.~B.~Marston,
Phys. Rev. B {\bf 37}, 3774 (1988).

\bibitem{zachar98}
O.~Zachar, S.~A.~Kivelson, and V.~J.~Emery,
Phys. Rev. B {\bf 57}, 1422 (1998).

\bibitem{ek93}
V. J. Emery and S. A. Kivelson,
Physica C {\bf 209}, 597 (1993).

\bibitem{berg08b}
E. Berg, E. Fradkin, and S. A. Kivelson,
preprint arXiv:0810.1564.



\bibitem{supermod}
Y. He, S. Graser, P. J. Hirschfeld, and H.-P. Cheng,
Phys. Rev. B {\bf 77}, 220507(R) (2008).

\bibitem{klauss00}
H.-H. Klauss, W. Wagener, M.~Hillberg, W. Kopmann, H. Walf, F. J. Litterst, M. H\"ucker, and B. B\"uchner,
Phys. Rev. Lett {\bf 85}, 4590 (2000).

\bibitem{kampf01}
A. P. Kampf, D. J. Scalapino, and S. R. White,
Phys. Rev. B {\bf 64}, 052509 (2001).

\bibitem{andersen08}
O. K. Andersen and T. Saha-Dasgupta,
unpublished.

\bibitem{takagi_strain}
N. Takeshita, T. Sasagawa, T. Sugioka, Y. Tokura, and H. Takagi,
J. Phys. Soc. Jpn. {\bf 73}, 1123 (2004).

\bibitem{alloul_rev}
H. Alloul, J. Bobroff, M. Gabay, and P. J. Hirschfeld,
Rev. Mod. Phys. {\bf 81}, 45 (2009).

\bibitem{harris}
A. B. Harris,
J. Phys. C {\bf 7}, 1671 (1974).

\bibitem{imryma}
Y. Imry and S.-k. Ma,
Phys. Rev. Lett. {\bf 35}, 1399 (1975).

\bibitem{natter}
T. Nattermann,
in: Spin Glasses and Random Fields, A. P. Young, ed.
(World Scientific, Singapore, 1998).


\bibitem{tvojta}
T. Vojta,
J. Phys. A {\bf 39}, R143 (2006).

\bibitem{schmalian00}
J. Schmalian and P. G. Wolynes,
Phys. Rev. Lett. {\bf 85}, 836 (2000).


\bibitem{maestro}
A. Del Maestro, B. Rosenow, and S. Sachdev,
Phys. Rev. B {\bf 74}, 024520 (2006).

\bibitem{robertson}
J. A. Robertson, S. A. Kivelson, E. Fradkin, A. C. Fang, and A. Kapitulnik,
Phys. Rev. B {\bf 74}, 134507 (2006).

\bibitem{metlitski08}
R. K. Kaul, R. G. Melko, M. A. Metlitski, and S. Sachdev,
Phys. Rev. Lett. {\bf 101}, 187206 (2008).



\bibitem{takagi_iso}
Suryadijaya, T. Sasagawa, and H. Takagi,
Physica C {\bf 426}, 402 (2005).

\bibitem{fink08}
J. Fink {\em et al.},
Phys. Rev. B {\bf 79}, 100502(R) (2009).


\bibitem{huecker07}
M. H\"ucker {\em et al.},
Physica C {\bf 460}, 170 (2007).

\bibitem{mooden88}
A. R. Moodenbaugh, Y. Xu, M. Suenaga, T. J. Folkerts, and R. N. Shelton,
Phys. Rev. B {\bf 38}, 4596 (1988).

\bibitem{huecker05}
M. H\"ucker, G. D. Gu, and J. M. Tranquada,
Phys. Rev. B {\bf 78}, 214507 (2008).

\bibitem{jt08}
J. M. Tranquada {\em et al.},
Phys. Rev. B {\bf 78}, 174529 (2008).

\bibitem{duns08}
S. R. Dunsiger, Y. Zhao, Z. Yamani, W. J. L. Buyers, H. Dabkowska, and B. D. Gaulin,
Phys. Rev. B {\bf 77}, 224410 (2008).

\bibitem{arai03}
J. Arai, T. Ishiguro, T. Goko, S. Iigaya, K. Nishiyama, I. Watanabe, and K. Nagamine,
J. Low Temp. Phys. {\bf 131}, 375 (2003).

\bibitem{savici05}
A. T. Savici {\em et al.},  
Phys. Rev. Lett. {\bf 95}, 157001 (2005).


\bibitem{jt97}
J. M. Tranquada, J.~D.~Axe, N.~Ichikawa, A.~R.~Moodenbaugh, Y.~Nakamura, and S.~Uchida,
Phys. Rev. Lett. {\bf 78}, 338 (1997).

\bibitem{ichikawa00}
N. Ichikawa, S. Uchida, J.~M. Tranquada, T. Niem\"oller, P.~M. Gehring, S.-H. Lee, and J.~R. Schneider,
Phys. Rev. Lett. {\bf 85}, 1738 (2000).

\bibitem{yamada98}
K. Yamada {\em et al.},
Phys. Rev. B {\bf 57}, 6165 (1998).

\bibitem{waki99}
S. Wakimoto {\em et al.},
Phys. Rev. B {\bf 60}, R769 (1999).

\bibitem{waki00}
S. Wakimoto {\em et al.},
Phys. Rev. B {\bf 61}, 3699 (2000).

\bibitem{matsuda00}
M. Matsuda {\em et al.},
Phys. Rev. B {\bf 61}, 4326 (2000).

\bibitem{matsuda00a}
M. Matsuda, M. Fujita, K. Yamada, R.~J. Birgeneau, M.~A. Kastner, H. Hiraka, Y. Endoh, S. Wakimoto, and
G. Shirane,
Phys. Rev. B {\bf 62}, 9148 (2000).

\bibitem{matsuda02a}
M. Matsuda, M. Fujita, K. Yamada, R.~J. Birgeneau, Y. Endoh, and G. Shirane,
Phys. Rev. B {\bf 65}, 134515 (2002).

\bibitem{wells}
B.~O. Wells, Y.~S. Lee, M.~A. Kastner, R.~J. Christanson, R.~J. Birgeneau, K. Yamada, Y. Endoh, and G. Shirane,
Science {\bf 277}, 1067 (1997).

\bibitem{oxy}
Y. S. Lee, R. J. Birgeneau, M. A. Kastner, Y. Endoh,
S. Wakimoto, K. Yamada, R. W. Erwin, S.-H. Lee, and G. Shirane,
Phys. Rev. B {\bf 60}, 3643 (1999).

\bibitem{kim99}
H. Kimura {\em et al.},
Phys. Rev. B {\bf 59}, 6517 (1999).

\bibitem{jt99a}
J. M. Tranquada, N. Ichikawa, K. Kakurai, and S. Uchida,
J. Phys. Chem. Solids {\bf 60}, 1019 (1999).

\bibitem{dai01}
P. Dai, H.~A. Mook, R.~D. Hunt, and F. Do\v{g}an,
Phys. Rev. B {\bf 63}, 054525 (2001).

\bibitem{mook02}
H.~A.~Mook, P.~Dai, and F.~Do\v{g}an,
Phys. Rev. Lett. {\bf 88}, 097004 (2002).

\bibitem{hinkov08a}
V. Hinkov, D. Haug, B. Fauqu\'e, P. Bourges, Y. Sidis, A. Ivanov, C. Bernhard, C. T. Lin, and B. Keimer,
Science {\bf 319}, 597 (2008).

\bibitem{cheong91}
S.-W. Cheong, G. Aeppli, T. E. Mason, H. Mook, S. M. Hayden, P. C. Canfield, Z. Fisk, K. N. Clausen, and J. L.
Martinez,
Phys. Rev. Lett. {\bf 67}, 1791 (1991).

\bibitem{yamada92}
T. R. Thurston, P. M. Gehring, G. Shirane, R. J. Birgeneau, M. A. Kastner,
Y. Endoh, M. Matsuda, K. Yamada, H. Kojima, and I. Tanaka,
Phys. Rev. B {\bf 46}, 9128 (1992).

\bibitem{waki01a}
S. Wakimoto, R. J. Birgeneau, Y. S. Lee, and G. Shirane,
Phys. Rev. B {\bf 63}, 172501 (2001).

\bibitem{aeppli97}
G. Aeppli, T. E. Mason, S. M. Hayden, H. A. Mook, and J. Kulda,
Science {\bf 278}, 1432 (1997).

\bibitem{haug_635}
D. Haug {\em et al.},
to be published.

\bibitem{buyers06}
C. Stock, W. J. L. Buyers, Z. Yamani, C. L. Broholm, J.-H. Chung, Z. Tun, R. Liang, D. Bonn, W. N. Hardy, and R. J.
Birgeneau,
Phys. Rev. B {\bf 73}, 100504 (2006).

\bibitem{buyers08}
C. Stock, W. J. L. Buyers, Z. Yamani, Z. Tun, R. J. Birgeneau, R. Liang, D. Bonn,
and W. N. Hardy,
Phys. Rev. B {\bf 77}, 104513 (2008).

\bibitem{waki01b}
S. Wakimoto, J. M. Tranquada, T. Ono, K. M. Kojima, S. Uchida, S.-H. Lee, P. M. Gehring,
and R. J. Birgeneau,
Phys. Rev. B {\bf 64}, 174505 (2001).


\bibitem{hirota01}
K. Hirota,
Physica C {\bf 357-360}, 61 (2001).

\bibitem{fink90}
A.~M.~Finkelstein, V.~E.~Kataev, E.~F.~Kukovitskii, and G.~B.~Teitelbaum,
Physica C {\bf 168}, 370 (1990).

\bibitem{alloul91}
H.~Alloul, P.~Mendels, H.~Casalta, J.-F.~Marucco, and J.~Arabski,
Phys. Rev. Lett. {\bf 67}, 3140 (1991).

\bibitem{julien00}
M.-H.~Julien, T.~Feher, M.~Horvatic, C.~Berthier O.~N.~Bakharev, P.~Segransan, G.~Collin, and J.-F.~Marucco,
Phys. Rev. Lett. {\bf 84}, 3422 (2000).

\bibitem{fujita04}
M. Fujita, H. Goka, K. Yamada, J. M. Tranquada, and L. P. Regnault,
Phys. Rev. B {\bf 70}, 104517 (2004).

\bibitem{moho06}
H. E. Mohottala, B. O. Wells, J. I. Budnick, W. A. Hines, C. Niedermayer, L. Udby, C. Bernhard,
A. R. Moodenbaugh, and F.-C. Chou,
Nature Mater. {\bf 5}, 377 (2006).

\bibitem{KAE01}
S. A. Kivelson, G. Aeppli, and V. J. Emery,
Proc. Natl. Acad. Sci. USA {\bf 98}, 11903 (2001).

\bibitem{kim00}
H. Kimura, H. Matsushita, K. Hirota, Y. Endoh, K. Yamada, G. Shirane, Y. S. Lee, M. A. Kastner, and R. J. Birgeneau,
Phys. Rev. B {\bf 61}, 14366 (2000).

\bibitem{nachumi98}
B. Nachumi {\em et al.}, 
Phys. Rev. B {\bf 58}, 8760 (1998).

\bibitem{savici02}
A. T. Savici {\em et al.}, 
Phys. Rev. B {\bf 66}, 014524 (2002).

\bibitem{klauss04}
H.-H. Klauss,
J. Phys. Cond. Matter {\bf 16}, S4457 (2004).

\bibitem{christensen07}
N. B. Christensen, H. M. R{\o}nnow, J. Mesot, R. A. Ewings, N. Momono, M. Oda, M. Ido, M. Enderle,
D. F. McMorrow, and A. T. Boothroyd,
Phys. Rev. Lett. {\bf 98}, 197003 (2007).

\bibitem{jt99b}
J. M. Tranquada, N. Ichikawa, and S. Uchida,
Phys. Rev. B {\bf 59}, 14712 (1999).

\bibitem{lake05}
B.~Lake {\em et al.}, 
Nature Mater. {\bf 4}, 658 (2005).

\bibitem{fujita02}
M. Fujita, H. Goka, K. Yamada, and M. Matsuda,
Phys. Rev. Lett. {\bf 88}, 167008 (2002).

\bibitem{kimura04}
H. Kimura, Y. Noda, H. Goka, M. Fujita, K. Yamada, M. Mizumaki, N. Ikeda, and H. Ohsumi,
Phys. Rev. B {\bf 70}, 134512 (2004).

\bibitem{huecker09}
M. H\"ucker, private communication;
M. H\"ucker, M. v. Zimmermann, M. Debessai, J. S. Schilling, and G. D. Gu,
to be published.

\bibitem{shraiman}
B.~I.~Shraiman and E.~D.~Siggia,
Phys. Rev. B {\bf 42}, 2485 (1990).

\bibitem{sushkov06}
A. L\"uscher, G. Misguich, A. I. Milstein, and O. P. Sushkov,
Phys. Rev. B {\bf 73}, 085122 (2006).

\bibitem{juricic06}
V. Juricic, M. B. Silva Neto, and C. Morais Smith,
Phys. Rev. Lett. {\bf 96}, 077004 (2006).



\bibitem{sidis01}
Y. Sidis, C. Ulrich, P. Bourges, C. Bernhard, C. Niedermayer, L. P. Regnault, N. H. Andersen,
and B. Keimer,
Phys. Rev. Lett. {\bf 86}, 4100 (2001).

\bibitem{mook01}
H. A. Mook, P. Dai, and F. Do\v{g}an,
Phys. Rev. B {\bf 64}, 012502 (2001).

\bibitem{mook02b}
H. A. Mook, P. Dai, S. M. Hayden, A. Hiess, J. W. Lynn, S.-H. Lee, and F. Do\v{g}an,
Phys. Rev. B 66, 144513 (2002).

\bibitem{buyers02}
C. Stock, W. J. L. Buyers, Z. Tun, R. Liang, D. Peets, D. Bonn, W. N. Hardy, and L. Taillefer,
Phys. Rev. B {\bf 66}, 024505 (2002).

\bibitem{hodges02}
J. A. Hodges, Y. Sidis, P. Bourges, I. Mirebeau, M. Hennion, and X. Chaud,
Phys. Rev. B {\bf 66}, 020501 (2002).

\bibitem{mook04}
H. A. Mook, P. Dai, S. M. Hayden, A. Hiess, S.-H. Lee, and F. Do\v{g}an,
Phys. Rev. B {\bf 69}, 134509 (2004).


\bibitem{niemoller99}
T. Niem\"oller, N. Ichikawa, T. Frello, H. H\"unnefeld, N. H. Andersen, S. Uchida,
J. R. Schneider, and J. M. Tranquada,
Eur. Phys. J. B {\bf 12}, 509 (1999).

\bibitem{kim08a}
Y.-J. Kim, G. D. Gu, T. Gog, and D. Casa,
Phys. Rev. B {\bf 77}, 064520 (2008).

\bibitem{bianconi01}
N. L. Saini, H. Oyanagi, A. Lanzara, D. Di Castro, S. Agrestini, A. Bianconi, F. Nakamura,
and T. Fujita,
Phys. Rev. B {\bf 64}, 132510 (2001).

\bibitem{bianconi96}
A. Bianconi, N. L. Saini, A. Lanzara, M. Missori, T. Rossetti, H. Oyanagi, H. Yamaguchi,
K. Oka, and T. Ito,
Phys. Rev. Lett. {\bf 76}, 3412 (1996).

\bibitem{zimmermann98}
M. von Zimmermann {\em et~al.},  
Europhys. Lett. {\bf 41}, 629 (1998).

\bibitem{abbamonte02}
P. Abbamonte {\em et al.},
Science {\bf 297}, 581 (2002).

\bibitem{abbamonte05}
P. Abbamonte, A. Rusydi, S. Smadici, G. D. Gu, G. A. Sawatzky, and D. L. Feng,
Nature Phys. {\bf 1}, 155 (2005).

\bibitem{lorenzana02}
J. Lorenzana and G. Seibold,
Phys. Rev. Lett. {\bf 89}, 136401 (2002).

\bibitem{smadici06}
S. Smadici, P. Abbamonte, M. Taguchi, Y. Kohsaka, T. Sasagawa, M. Azuma, M. Takano,
and H. Takagi,
Phys. Rev. B {\bf 75}, 075104 (2007).

\bibitem{hanaguri04}
T. Hanaguri, C. Lupien, Y. Kohsaka, D.-H. Lee, M. Azuma, M. Takano, H. Takagi, and J. C. Davis,
Nature {\bf 430}, 1001 (2004).


\bibitem{valla06}
T. Valla, A. V. Fedorov, J. Lee, J. C. Davis, and G. D. Gu,
Science {\bf 314}, 1914 (2006).


\bibitem{fauque06}
B. Fauqu\'e, Y. Sidis, V. Hinkov, S. Pailh\`es, C.T. Lin, X. Chaud, and P. Bourges
Phys. Rev. Lett. {\bf 96}, 197001 (2006).

\bibitem{fauque08a}
H. A. Mook, Y. Sidis, B. Fauqu\'e, V. Bal\'edent, and P. Bourges,
Phys. Rev. B {\bf 78}, 020506 (2008).

\bibitem{fauque08b}
Y. Li, V. Bal\'edent, N. Barisic, Y. Cho, B. Fauqu\'e, Y. Sidis, G. Yu, X. Zhao, P. Bourges, and M. Greven,
Nature {\bf 455}, 372 (2008).

\bibitem{kaminski02}
A. Kaminski {\em et al.},
Nature {\bf 416}, 610 (2002).

\bibitem{kerr08}
J. Xia, E. Schemm, G. Deutscher, S. A. Kivelson, D. A. Bonn, W. N. Hardy, R. Liang, W. Siemons,
G. Koster, M. M. Fejer, and A. Kapitulnik,
Phys. Rev. Lett. {\bf 100}, 127002 (2008).

\bibitem{luke08}
G. J. MacDougall, A. A. Aczel, J.P. Carlo, T. Ito, J. Rodriguez, P. L. Russo, Y. J. Uemura,
S. Wakimoto, and G. M. Luke,
Phys. Rev. Lett. {\bf 101}, 017001 (2008).

\bibitem{shekhter08}
A. Shekhter, L. Shu, V. Aji, D. E. MacLaughlin, and C. M. Varma,
Phys. Rev. Lett. {\bf 101}, 227004 (2008).

\bibitem{straessle08}
S. Str\"assle, J. Roos, M. Mali, H. Keller, and T. Ohno,
Phys. Rev. Lett. {\bf 101}, 237001 (2008).

\bibitem{monod08}
B. Leridon, P. Monod, and D. Colson,
preprint arXiv:0806.2128.


\bibitem{jt04}
J. M. Tranquada, H. Woo, T. G. Perring, H. Goka, G. D. Gu, G. Xu, M. Fujita, and K. Yamada,
Nature {\bf 429}, 534 (2004).

\bibitem{mvtu04}
M. Vojta and T. Ulbricht,
Phys. Rev. Lett. {\bf 93}, 127002 (2004).

\bibitem{gsu04}
G. S. Uhrig, K. P. Schmidt, and M. Gr\"uninger,
Phys. Rev. Lett. {\bf 93}, 267003 (2004).

\bibitem{seibold05}
G. Seibold and J. Lorenzana,
Phys. Rev. Lett. {\bf 94}, 107006 (2005).

\bibitem{carlson06}
D. X. Yao, E. W. Carlson, and D. K. Campbell,
Phys. Rev. Lett. {\bf 97}, 017003 (2006).

\bibitem{vvk}
M. Vojta, T. Vojta, and R. K. Kaul,
Phys. Rev. Lett. {\bf 97}, 097001 (2006).

\bibitem{respeak1}
J.~Rossat-Mignod, L.~P.~Regnault, C.~Vettier, P.~Bourges, P.~Burlet, J.~Bossy, J.~Y.~Henry, and G.~Lapertot,
Physica C {\bf 185-189}, 86 (1991).

\bibitem{respeak1b}
H.~A.~Mook, M.~Yehiraj, G.~Aeppli, T.~E.~Mason, and T.~Armstrong,
Phys. Rev. Lett. {\bf 70}, 3490 (1993).

\bibitem{respeak1c}
H. F. Fong, P. Bourges, Y. Sidis, L. P. Regnault, J. Bossy, A. Ivanov, D. L. Milius, I. A. Aksay, and B. Keimer,
Phys. Rev. B {\bf 61}, 14773 (2000).

\bibitem{respeak2}
H. F. Fong, P. Bourges, Y. Sidis, L. P. Regnault, A. Ivanov, G. D. Gu, N. Koshizuka, and B. Keimer,
Nature {\bf 398}, 588 (1999).

\bibitem{respeak3}
H. F. He, P. Bourges, Y. Sidis, C. Ulrich, L. P. Regnault, S. Pailh\`es, N. S. Berzigiarova,
N. N. Kolesnikov, and B. Keimer,
Science {\bf 295}, 1045 (2002).

\bibitem{hinkov04}
V. Hinkov, S. Pailh\`es, P. Bourges, Y. Sidis, A. Ivanov, A. Kulakov, C. T. Lin, D. P. Chen, C. Bernhard, and B. Keimer,
Nature {\bf 430}, 650 (2004).

\bibitem{hayden04}
S. M Hayden, H. A. Mook, P. Dai, T. G. Perring, and F. Do\v{g}an,
Nature {\bf 429}, 531 (2004).

\bibitem{christensen04}
N. B. Christensen, D. F. McMorrow, H. M. R{\o}nnow, B. Lake, S. M. Hayden, G. Aeppli,
T. G. Perring, M. Mangkorntong, M. Nohara, and H. Takagi,
Phys. Rev. Lett. {\bf 93}, 147002 (2004).

\bibitem{vignolle07}
B. Vignolle, S. M. Hayden, D. F. McMorrow, H. M. R{\o}nnow, B. Lake, C. D. Frost, and T. G. Perring,
Nature Phys. {\bf 3}, 167 (2007).

\bibitem{face}
S. Pailh\`es, Y. Sidis, P. Bourges, V. Hinkov, A. Ivanov, C. Ulrich, L. P. Regnault, and B. Keimer,
Phys. Rev. Lett. {\bf 93}, 167001 (2004).

\bibitem{buyers04}
C. Stock, W. J. L. Buyers, R. Liang, D. Peets, Z. Tun, D. Bonn, W. N. Hardy, and R. J. Birgeneau,
Phys. Rev. B {\bf 69}, 014502 (2004).

\bibitem{buyers05}
C. Stock {\em et al.},
Phys. Rev. B {\bf 71}, 024522 (2005).



\bibitem{lips08}
O. J. Lipscombe, B. Vignolle, T. G. Perring, C. D. Frost, and S. M. Hayden,
Phys. Rev. Lett. {\bf 102}, 167002 (2009).

\bibitem{xu09}
G. Xu, G. D. Gu, M. H\"ucker, B. Fauqu\'e, T. G. Perring, L. P. Regnault, and J. M. Tranquada,
preprint arXiv:0902.2802.

\bibitem{jt06}
J. M. Tranquada,
in: Handbook of High-Temperature Superconductivity: Theory and Experiment,
J. R. Schrieffer, J. S. Brooks, eds.
(Springer, Amsterdam, 2007).

\bibitem{mook00}
H. A. Mook, P. Dai, F. Do\v{g}an, and R. D. Hunt,
Nature {\bf 404}, 729 (2000).

\bibitem{jt05}
J. M. Tranquada, H. Woo, T. G. Perring, H. Goka, G. D. Gu, G. Xu, M. Fujita, and K. Yamada,
J. Phys. Chem. Solids {\bf 67}, 511 (2006).

\bibitem{hinkov07}
V. Hinkov, P. Bourges, S. Pailh\`es, Y. Sidis, A. Ivanov, C. D. Frost, T. G. Perring, C. T. Lin, D. P. Chen, and B. Keimer,
Nature Phys. {\bf 3}, 780 (2007).

\bibitem{reznik08c}
D. Reznik, J.-P. Ismer, I. Eremin, L. Pintschovius, T. Wolf, M. Arai, Y. Endoh, T. Masui,
and S. Tajima,
Phys. Rev. B {\bf 78}, 132503 (2008).

\bibitem{chang07}
J. Chang {\em et al.}, 
Phys. Rev. Lett. {\bf 98}, 077004 (2007).

\bibitem{jt04b}
J. M. Tranquada, C. H. Lee, K. Yamada, Y. S. Lee, L. P. Regnault, and H. M. R{\o}nnow,
Phys. Rev. B {\bf 69}, 174507 (2004).

\bibitem{matsuda08}
M. Matsuda, M. Fujita, S. Wakimoto, J. A. Fernandez-Baca, J. M. Tranquada, and K. Yamada,
Phys. Rev. Lett. {\bf 101}, 197001 (2008).

\bibitem{waki04}
S. Wakimoto, H. Zhang, K. Yamada, I. Swainson, H. Kim, and R. J. Birgeneau,
Phys. Rev. Lett. {\bf 92}, 217004 (2004).

\bibitem{waki07}
S. Wakimoto, K. Yamada, J. M. Tranquada, C. D. Frost, R. J. Birgeneau, and H. Zhang,
Phys. Rev. Lett. {\bf 98}, 247003 (2007).


\bibitem{reznik06}
D. Reznik, L. Pintschovius, M. Ito, S. Iikubo, M. Sato, H. Goka, M. Fujita, K. Yamada, G. D. Gu,
and J. M. Tranquada,
Nature {\bf 440}, 1170 (2006).

\bibitem{reznik07}
D. Reznik, L. Pintschovius, M. Fujita, K. Yamada, G. D. Gu, and J. M. Tranquada,
J. Low Temp. Phys. {\bf 147}, 353 (2007).

\bibitem{reznik08a}
D. Reznik, T. Fukuda, D. Lamago, A. Q. R. Baron, S. Tsutsui, M. Fujita, and K. Yamada,
J. Phys. Chem. Solids {\bf 69}, 3103 (2008).

\bibitem{mcq99}
R. J. McQueeney, Y. Petrov, T. Egami, M. Yethiraj, G. Shirane, and Y. Endoh,
Phys. Rev. Lett. {\bf 82}, 628 (1999).

\bibitem{braden99}
L. Pintschovius and M. Braden,
Phys. Rev. B {\bf 60}, R15039 (1999).

\bibitem{fukuda05}
T. Fukuda {\em et al.},
Phys. Rev. B {\bf 71}, 060501(R) (2005).

\bibitem{pintsch02}
L. Pintschovius, W. Reichardt, M. Kl\"aser, T. Wolf, and H. v. L\"ohneysen,
Phys. Rev. Lett. {\bf 89}, 037001 (2002).

\bibitem{kaneshita02}
E. Kaneshita, M. Ichioka, and K. Machida,
Phys. Rev. Lett. {\bf 88}, 115501 (2002).

\bibitem{reznik08b}
D. Reznik, L. Pintschovius, J. M. Tranquada, M. Arai, Y. Endoh, T. Masui, and S. Tajima,
Phys. Rev. B {\bf 78}, 094507 (2008).

\bibitem{mukhin07}
S. I. Mukhin, A. Mesaros, J. Zaanen, and F. V. Kusmartsev,
Phys. Rev. B {\bf 76}, 174521 (2007).

\bibitem{heat98}
O. Baberski, A. Lang, O. Maldonado, M. H\"ucker, B. B\"uchner, and A. Freimuth,
Europhys. Lett. {\bf 44}, 335 (1998)

\bibitem{heat03}
C. Hess, B. B\"uchner, U. Ammerahl, and A. Revcolevschi,
Phys. Rev. B {\bf 68}, 184517 (2003).



\bibitem{ando02}
Y. Ando, K. Segawa, S. Komiya, and A. N. Lavrov,
Phys. Rev. Lett. {\bf 88}, 137005 (2002).

\bibitem{haase03}
J. Haase and C. P. Slichter,
J. Supercond. {\bf 16}, 473 (2003).


\bibitem{hanaguri07}
T. Hanaguri, Y. Kohsaka, J. C. Davis, C. Lupien, I. Yamada, M. Azuma, M. Takano, K. Ohishi,
M. Ono, and H. Takagi,
Nature Phys. {\bf 3}, 865 (2007).

\bibitem{capriotti03}
L. Capriotti, D. J. Scalapino, and R. D. Sedgewick,
Phys. Rev. B {\bf 68}, 014508 (2003).

\bibitem{hoffman02b}
J. E. Hoffman, K. McElroy, D. H. Lee, K. M. Lang, H. Eisaki, S. Uchida, and J. C. Davis,
Science {\bf 297}, 1148 (2002).

\bibitem{mcelroy03}
K. McElroy, R. W. Simmonds, J. E. Hoffman, D.-H. Lee, J. Orenstein, H. Eisaki, S. Uchida, and J. C. Davis,
Nature {\bf 422}, 592 (2003).

\bibitem{mcelroy05}
K. McElroy, D.-H. Lee, J. E. Hoffman, K. M. Lang, J. Lee, E. W. Hudson, H. Eisaki, S. Uchida, and J. C. Davis,
Phys. Rev. Lett. {\bf 94}, 197005 (2005).

\bibitem{mcelroy06}
K. McElroy, G.-H. Gweon, S. Y. Zhou, J. Graf, S. Uchida, H. Eisaki, H. Takagi, T. Sasagawa,
D.-H. Lee, and A. Lanzara,
Phys. Rev. Lett. {\bf 96}, 067005 (2006).

\bibitem{hoffman02a}
J.~E.~Hoffman, E. W. Hudson, K. M. Lang, V. Madhavan, H. Eisaki, S. Uchida, and J. C. Davis,
Science {\bf 295}, 466 (2002).

\bibitem{kapi03a}
C. Howald, H. Eisaki, N. Kaneko, M. Greven, and A. Kapitulnik,
Phys. Rev. B {\bf 67}, 014533 (2003).

\bibitem{kapi03b}
C. Howald, H. Eisaki, N. Kaneko, and A. Kapitulnik,
Proc. Natl. Acad. Sci. USA {\bf 100}, 9705 (2003).

\bibitem{kapi04}
A. Fang, C. Howald, N. Kaneko, M. Greven, and A. Kapitulnik,
Phys. Rev. B {\bf 70}, 214514 (2004).


\bibitem{yazdani04}
M. Vershinin, S. Misra, S. Ono, Y. Abe, Y. Ando, and A. Yazdani,
Science {\bf 303}, 1995 (2004).

\bibitem{misra04}
S. Misra, M. Vershinin, P. Phillips, and A. Yazdani,
Phys. Rev. B {\bf 70}, 220503 (2004).


\bibitem{kohsaka08}
Y. Kohsaka {\em et al.},
Nature {\bf 454}, 1072 (2008).


\bibitem{hashi06}
A. Hashimoto, N. Momono, M. Oda, and M. Ido,
Phys. Rev. B {\bf 74}, 064508 (2006).

\bibitem{liu07}
Y. H. Liu, K. Takeyama, T. Kurosawa, N. Momono, M. Oda, and M. Ido,
Phys. Rev. B {\bf 75}, 212507 (2007).

\bibitem{kohsaka04}
Y. Kohsaka, K. Iwaya, S. Satow, T. Hanaguri, M. Azuma, M. Takano, and H. Takagi,
Phys. Rev. Lett. {\bf 93}, 097004 (2004).


\bibitem{shen05}
K. M. Shen {\em et al.},
Science {\bf 307}, 901 (2005).

\bibitem{wise08}
W. D. Wise, M. C. Boyer, K. Chatterjee, T. Kondo, T. Takeuchi, H. Ikuta, Y. Wang, and E. W. Hudson,
Nature Phys. {\bf 4}, 696 (2008).

\bibitem{levy05}
G. Levy, M. Kugler, A. A. Manuel, and {\O}. Fischer,
Phys. Rev. Lett. {\bf 95}, 257005 (2005).

\bibitem{matsuba07}
K. Matsuba, S. Yoshizawa, Y. Mochizuki, T. Mochiku, K. Hirata, and N. Nishida,
J. Phys. Soc. Jpn. {\bf 76}, 063704 (2007).

\bibitem{takagi08}
T. Hanaguri, Y. Kohsaka, M. Ono, M. Maltseva, P. Coleman, I. Yamada, M. Azuma, M. Takano, K. Ohishi, and H. Takagi,
Science {\bf 323}, 923 (2009).

\bibitem{yeh08}
A. D. Beyer, M. S. Grinolds, M. L. Teague, S. Tajima, and N.-C. Yeh,
preprint arXiv:0808.3016.

\bibitem{pwa_as}
P. W. Anderson and N. P. Ong,
J. Phys. Chem. Solids {\bf 67}, 1 (2006).

\bibitem{mohit_as}
M. Randeria, R. Sensarma, N. Trivedi, and F.-C. Zhang,
Phys. Rev. Lett. {\bf 95}, 137001 (2005).

\bibitem{ohishi05}
K. Ohishi, I. Yamada, A. Koda, W. Higemoto, S. R. Saha, R. Kadono, K. M. Kojima, M. Azuma, and M. Takano,
J. Phys. Soc. Jpn. {\bf 74}, 2408 (2005).


\bibitem{homes06}
C. C. Homes, S. V. Dordevic, G. D. Gu, Q. Li, T. Valla, and J. M. Tranquada,
Phys. Rev. Lett. {\bf 96}, 257002 (2006).

\bibitem{dumm02}
M. Dumm, D. N. Basov, S. Komiya, Y. Abe, and Y. Ando,
Phys. Rev. Lett. {\bf 88}, 147003 (2002).

\bibitem{borisenko08}
V. B. Zabolotnyy {\em et al.},
preprint arXiv:0809.2237.

\bibitem{castellani00}
C. Castellani, C. Di Castro, M. Grilli, and A. Perali,
Physica C {\bf 341-348}, 1739 (2000).

\bibitem{lee04}
P. A. Lee, Physica C {\bf 408}, 5 (2004).

\bibitem{zaanen_coin}
J. Zaanen and Z. Nussinov,
in: Open Problems in Strongly Correlated
Electron Systems, Nato Science Series II/15,
J. Bonca, P. Prelovsek, A. Ramsak and S. Sarkar, eds.
(Kluwer Ac. Pub., Dordrecht, 2001)

\bibitem{granath01}
M. Granath, V. Oganesyan, S. A. Kivelson, E. Fradkin, and V. J. Emery,
Phys. Rev. Lett. {\bf 87}, 167011 (2001).

\bibitem{berg08a}
E. Berg, C.-C. Chen, and S. A. Kivelson,
Phys. Rev. Lett. {\bf 100}, 027003 (2008).

\bibitem{vzs00a}
M. Vojta, Y. Zhang, and S. Sachdev,
Phys. Rev. B {\bf 62}, 6721 (2000).

\bibitem{mv08}
M. Vojta,
Phys. Rev. B {\bf 78}, 144508 (2008).

\bibitem{luca03}
A. Lucarelli {\em et al.}, 
Phys. Rev. Lett. {\bf 90}, 037002 (2003).

\bibitem{benfatto03}
L. Benfatto and C. Morais Smith,
Phys. Rev. B {\bf 68}, 184513 (2003).

\bibitem{komiya05}
S. Komiya, H.-D. Chen, S.-C. Zhang, and Y. Ando,
Phys. Rev. Lett. {\bf 94}, 207004 (2005).

\bibitem{noda99}
T. Noda, H. Eisaki, and S. Uchida,
Science {\bf 286}, 265 (1999).

\bibitem{emery00}
V. J. Emery, E. Fradkin, S. A. Kivelson, and T. C. Lubensky,
Phys. Rev. Lett. {\bf 85}, 2160 (2000).

\bibitem{prelovsek01}
P. Prelovsek, T. Tohyama, and S. Maekawa,
Phys. Rev. B {\bf 64}, 052512 (2001).

\bibitem{ong01}
Y. Wang and N. P. Ong,
Proc. Natl. Acad. Sci. USA {\bf 98}, 11091 (2001).


\bibitem{daou09}
R. Daou, O. Cyr-Choini\`ere, F. Lalibert\'e, D. LeBoeuf, N. Doiron-Leyraud, J.-Q. Yan, J.-S. Zhou, J. B. Goodenough,
and L. Taillefer,
preprint arXiv:0810.4280.

\bibitem{adachi01}
T. Adachi, T. Noji, and Y. Koike,
Phys. Rev. B {\bf 64}, 144524 (2001).

\bibitem{leboeuf07}
D. LeBoeuf {\em et al.},
Nature {\bf 450}, 533 (2007).

\bibitem{cyr09}
O. Cyr-Choini\`ere {\em et al.},
Nature {\bf 458}, 743 (2009).




\bibitem{zhou04}
X. J. Zhou {\em et al.},
Phys. Rev. Lett. {\bf 92}, 187001 (2004).

\bibitem{kanigel}
A. Kanigel {\em et al.},
Nature Phys. {\bf 2}, 447 (2006).

\bibitem{sudip03}
S. Chakravarty, C. Nayak, and S. Tewari,
Phys. Rev B {\bf 68}, 100504(R) (2003).

\bibitem{imada08}
S. Sakai, Y. Motome, and M. Imada,
Phys. Rev. Lett. {\bf 102}, 056404 (2009).

\bibitem{huefner08}
S. H\"ufner, M. A. Hossain, A. Damascelli, and G. A. Sawatzky,
Rep. Prog. Phys. {\bf 71}, 062501 (2008).

\bibitem{chatterjee06}
U. Chatterjee {\em et al.},
Phys. Rev. Lett. {\bf 96}, 107006 (2006).


\bibitem{zhou99}
X. J. Zhou, P. Bogdanov, S. A. Kellar, T. Noda, H. Eisaki, S. Uchida, Z. Hussain,
and Z. X. Shen,
Science {\bf 286}, 268 (1999).

\bibitem{grilli09}
M. Grilli, G. Seibold, A. Di Ciolo, and J. Lorenzana,
preprint arXiv:0809.2197.

\bibitem{zhou01}
X. J. Zhou {\em et al.},
Phys. Rev. Lett. {\bf 86}, 5578 (2001).

\bibitem{claesson07}
T. Claesson {\em et al.},
preprint arXiv:0712.4039.

\bibitem{chang08b}
J. Chang {\em et al.},
New J. Phys. {\bf 10}, 103016 (2008).

\bibitem{he09}
R.-H. He {\em et al.},
Nature Phys. {\bf 5}, 119 (2009).

\bibitem{valla08}
H.-B. Yang, J. D. Rameau, P. D. Johnson, T. Valla, A. Tsvelik, and G. D. Gu,
Nature {\bf 456}, 77 (2008).


\bibitem{nieder98}
C. Niedermayer, C. Bernhard, T. Blasius, A. Golnik, A. Moodenbaugh, and J. I. Budnick,
Phys. Rev. Lett. {\bf 80}, 3843 (1998).

\bibitem{julien99}
M.-H. Julien, F. Borsa, P. Carretta, M. Horvatic, C. Berthier, and C. T. Lin,
Phys. Rev. Lett. {\bf 83}, 604 (1999).

\bibitem{haase00}
J. Haase, C. P. Slichter, R. Stern, C. T. Milling, and D. G. Hinks,
Physica C {\bf 341-348}, 1727 (2000).

\bibitem{imai99}
A. W. Hunt, P. M. Singer, K. R. Thurber, and T. Imai,
Phys. Rev. Lett. {\bf 82}, 4300 (1999).

\bibitem{teitel00}
G. B. Teitel'baum, I. M. Abu-Shiekah, O. Bakharev, H. B. Brom, and J. Zaanen,
Phys. Rev. B {\bf 63}, 020507(R) (2000).

\bibitem{julien01}
M.-H. Julien {\em et al.},
Phys. Rev. B {\bf 63}, 144508 (2001).

\bibitem{simovic03}
B. Simovic, P. C. Hammel, M. H\"ucker, B. B\"uchner, and A. Revcolevschi,
Phys. Rev. B {\bf 68}, 012415 (2003).

\bibitem{singer02}
P. M. Singer, A. W. Hunt, and T. Imai,
Phys. Rev. Lett. {\bf 88}, 047602 (2002).

\bibitem{haase04}
J. Haase, O. P. Sushkov, P. Horsch, and G. V. M. Williams,
Phys. Rev. B {\bf 69}, 094504 (2004).

\bibitem{sanna04}
S. Sanna, G. Allodi, G. Concas, A. D. Hillier, and R. De Renzi,
Phys. Rev. Lett. {\bf 93}, 207001 (2004).

\bibitem{ofer06}
R. Ofer, S. Levy, A. Kanigel, and A. Keren,
Phys. Rev. B {\bf 73}, 012503 (2006).

\bibitem{curro00}
N. J. Curro, C. Milling, J. Haase, and C. P. Slichter,
Phys. Rev. B {\bf 62}, 3473 (2000).

\bibitem{mitro01}
V. F. Mitrovi\'c, E. E. Sigmund, M. Eschrig, H. N. Bachman, W. P. Halperin, A. P. Reyes, P. Kuhns,
and W. G. Moulton,
Nature {\bf 413}, 501 (2001).

\bibitem{curro00b}
N. J. Curro, P. C. Hammel, B. J. Suh, M. H\"ucker, B. B\"uchner, U. Ammerahl, and A. Revcolevschi,
Phys. Rev. Lett. {\bf 85}, 642 (2000).


\bibitem{grafe06}
H.-J. Grafe, N. J. Curro, M. H\"ucker, and B. B\"uchner,
Phys. Rev. Lett. {\bf 96}, 017002 (2006).


\bibitem{sonier07}
J. E. Sonier {\em et al.},
Phys. Rev. B {\bf 76}, 134518 (2007).

\bibitem{buechner94}
B. B\"uchner, M. Breuer, A. Freimuth, and A. P. Kampf,
Phys. Rev. Lett. {\bf 73}, 1841 (1994).

\bibitem{li07}
Q. Li, M. H\"ucker, G. D. Gu, A. M. Tsvelik, and J. M. Tranquada,
Phys. Rev. Lett. {\bf 99}, 067001 (2007).

\bibitem{yamada92b}
N. Yamada and M. Ido,
Physica {\bf 203}, 240 (1992).

\bibitem{arumugam02}
S. Arumugam, N. Mori, N. Takeshita, H. Takashima, T. Noda, H. Eisaki, and S. Uchida,
Phys. Rev. Lett. {\bf 88}, 247001 (2002).

\bibitem{simovic04}
B. Simovic, M. Nicklas, P. C. Hammel, M. H\"ucker, B. B\"uchner and J. D. Thompson,
Europhys. Lett. {\bf 66}, 722 (2004).

\bibitem{crawford05}
M. K. Crawford {\em et al.},
Phys. Rev. B {\bf 71}, 104513 (2005).



\bibitem{khaykovich02}
B. Khaykovich, Y. S. Lee, R. W. Erwin, S.-H. Lee, S. Wakimoto, K. J. Thomas, M. A. Kastner, and R. J. Birgeneau,
Phys. Rev. B {\bf 66}, 014528 (2002).

\bibitem{xu07}
G. Xu, J. M. Tranquada, T. G. Perring, G. D. Gu, M. Fujita, and K. Yamada,
Phys. Rev. B {\bf 76}, 014508 (2007).


\bibitem{lake02}
B.~Lake {\em et al.},
Nature {\bf 415}, 299 (2002).

\bibitem{chang08}
J. Chang {\em et al.}, 
Phys. Rev. B {\bf 78}, 104525 (2008).

\bibitem{khaykovich03}
B. Khaykovich {\em et al.},
Phys. Rev. B {\bf 67}, 054501 (2003).

\bibitem{lake01}
B.~Lake {\em et al.},
Science {\bf 291}, 1759 (2001).

\bibitem{khaykovich05}
B. Khaykovich {\em et al.},
Phys. Rev. B {\bf 71}, 220508(R) (2005).

\bibitem{chang09}
J. Chang {\em et al.}, 
Phys. Rev. Lett. {\bf 102}, 177006 (2009).


\bibitem{wen08}
J. Wen, Z. Xu, G. Xu, J. M. Tranquada, G. Gu, S. Chang, and H. J. Kang,
Phys. Rev. B {\bf 78}, 212506 (2008).

\bibitem{waki03}
S. Wakimoto {\em et al.},
Phys. Rev. B {\bf 67}, 184419 (2003).

\bibitem{kim08b}
J. Kim, A. Kagedan, G. D. Gu, C. S. Nelson, and Y.-J. Kim,
Phys. Rev. B {\bf 77}, 180513(R) (2008).



\bibitem{adachi05}
T. Adachi, N. Kitajima, T. Manabe, Y. Koike, K. Kudo, T. Sasaki, and N. Kobayashi,
Phys. Rev. B {\bf 71}, 104516 (2005).

\bibitem{adachi06}
T. Adachi, K. Omori, T. Kawamata, K. Kudo, T. Sasaki, N. Kobayashi, and Y. Koike,
J. Phys.: Conf. Series {\bf 51}, 259 (2006).

\bibitem{haug_mf}
D. Haug {\em et al.}, 
preprint arXiv:0902.3335.

\bibitem{tolya03}
A. Polkovnikov, S. Sachdev, and M. Vojta,
Physica C {\bf 388-389}, 19 (2003).

\bibitem{podol03}
D. Podolsky, E. Demler, K. Damle, and B. I. Halperin,
Phys. Rev. B {\bf 67}, 094514 (2003).

\bibitem{kang03}
H. J. Kang {\em et al.}, 
Nature {\bf 423}, 522 (2003).

\bibitem{sonier03}
J. E. Sonier, K. F. Poon, G. M. Luke, P. Kyriakou, R. I. Miller, R. Liang, C. R. Wiebe, P. Fournier,
and R. L. Greene,
Phys. Rev. Lett. {\bf 91}, 147002 (2003).

\bibitem{matsuda02b}
M. Matsuda, S. Katano, T. Uefuji, M. Fujita, and K. Yamada,
Phys. Rev. B {\bf 66}, 172509 (2002).



\bibitem{cho92}
J. H. Cho, F. Borsa, D. C. Johnston, and D. R. Torgeson,
Phys. Rev. B {\bf 46}, 3179 (1992).

\bibitem{chou95}
F. C. Chou, N. R. Belk, M. A. Kastner, R. J. Birgeneau, and A. Aharony,
Phys. Rev. Lett. {\bf 75}, 2204 (1995).

\bibitem{waki00b}
S. Wakimoto, S. Ueki, Y. Endoh, and K. Yamada,
Phys. Rev. B {\bf 62}, 3547 (2000).

\bibitem{mitro08}
V. F. Mitrovic, M.-H. Julien, C. de Vaulx, M. Horvatic, C. Berthier, T. Suzuki, and K. Yamada,
Phys. Rev. B {\bf 78}, 014504 (2008).

\bibitem{pana02}
C. Panagopoulos, J. L. Tallon, B. D. Rainford, T. Xiang, J. R. Cooper, and C. A. Scott,
Phys. Rev. B {\bf 66}, 064501 (2002).

\bibitem{pana05}
C. Panagopoulos and V. Dobrosavljevic,
Phys. Rev. B {\bf 72}, 014536 (2005).

\bibitem{popovic08}
I. Raicevic, J. Jaroszynski, D. Popovic, C. Panagopoulos, and T. Sasagawa,
Phys. Rev. Lett. {\bf 101}, 177004 (2008).

\bibitem{pana08}
G. R. Jelbert, T. Sasagawa, J. D. Fletcher, T. Park, J. D. Thompson, and C. Panagopoulos,
Phys. Rev. B {\bf 78}, 132513 (2008).

\bibitem{biroli08}
M. Tarzia and G. Biroli,
Europhys. Lett. {\bf 82}, 67008 (2008).

\bibitem{ss94}
S.~Sachdev,
Phys. Rev. B {\bf 49}, 6770 (1994).



\bibitem{emerymodel}
V. J. Emery,
Phys. Rev. Lett. {\bf 58}, 2794 (1987).

\bibitem{reiter88}
V. J. Emery and G. Reiter,
Phys. Rev. B {\bf 38}, 11938 (1988).

\bibitem{Hubbard}
J. Hubbard, Proc. R. Soc. London A {\bf 276}, 238 (1963);
M.~C.~Gutzwiller, Phys. Rev. Lett. {\bf 10}, 159 (1963);
J. Kanamori, Prog. Theor. Phys. {\bf 30}, 275 (1963).

\bibitem{pwa87}
P.~W.~Anderson, Science {\bf 235}, 1196 (1987).

\bibitem{tJ}
F.~C.~Zhang and T.~M.~Rice, Phys. Rev. B {\bf 37}, 3759 (1988).

\bibitem{cherny04}
A. L. Chernyshev, D. Galanakis, P. Phillips, A. V. Rozhkov, and A.-M. S. Tremblay,
Phys. Rev. B {\bf 70}, 235111 (2004).

\bibitem{hub_tj}
A. Reischl, E. M\"uller-Hartmann, and G. S. Uhrig,
Phys. Rev. B {\bf 70}, 245124 (2004).

\bibitem{emery_tj}
E. M\"uller-Hartmann and A. Reischl,
Eur. Phys. J. B {\bf 28}, 173 (2002).

\bibitem{jarrell06}
T. A. Maier, M. S. Jarrell, and D. J. Scalapino,
Physica C {\bf 460-462}, 13 (2007).

\bibitem{imada_pair}
T. Aimi and M. Imada,
J. Phys. Soc. Jpn. {\bf 76}, 113708 (2007).

\bibitem{baeriswyl}
D. Eichenberger and D. Baeriswyl,
preprint arXiv:0808.0433.

\bibitem{newns}
D. M. Newns and C. C. Tsuei,
Nature Phys. {\bf 3}, 184 (2007).



\bibitem{machida90}
M.~Kato, K.~Machida, H.~Nakanishi, and M.~Fujita,
J. Phys. Soc. Jpn. {\bf 59}, 1047 (1990).

\bibitem{ekl}
V.~J.~Emery, S.~A.~Kivelson, and H.~Q.~Lin,
Phys. Rev. Lett. {\bf 64}, 475 (1990);
Phys. Rev. B {\bf 42}, 6523 (1990).

\bibitem{horsch89}
W. H. Stephan, W. v. d. Linden, and P. Horsch,
Phys. Rev. B {\bf 39}, 2924 (1989).

\bibitem{sk1}
S.~A.~Kivelson and V.~J.~Emery, in:
Strongly Correlated Electronic Materials: The Los Alamos Symposium 1993,
K. S. Bedell, Z. Wang, D.~E. Meltzer, A.~V. Balatsky, and E. Abrahams, eds.
(Addison-Wesley, Reading, Massachusetts, 1994) p. 619.

\bibitem{loew94}
U. L\"ow, V. J. Emery, K. Fabricius, and S. A. Kivelson,
Phys. Rev. Lett. {\bf 72}, 1918 (1994).

\bibitem{sr_spn}
S.~Sachdev and N.~Read,
Int. J. Mod. Phys. B {\bf 5}, 219 (1991).

\bibitem{vs99}
M.~Vojta and S.~Sachdev,
Phys. Rev. Lett. {\bf 83}, 3916 (1999).

\bibitem{mv02}
M. Vojta,
Phys. Rev. B {\bf 66}, 104505 (2002).

\bibitem{seibold98}
G. Seibold, C. Castellani, C. Di Castro, and M. Grilli,
Phys. Rev. B {\bf 58}, 13506 (1998).

\bibitem{machida99}
M.~Ichioka and K.~Machida,
J. Phys. Soc. Jpn. {\bf 68}, 4020 (1999).

\bibitem{seibold04}
G. Seibold and J. Lorenzana,
Phys. Rev. B {\bf 69}, 134513 (2004).

\bibitem{normand01}
B. Normand and A. P. Kampf,
Phys. Rev. B {\bf 64}, 024521 (2001).


\bibitem{rvbmf}
F. C. Zhang, C. Gros, T. M. Rice, and H. Shiba,
Supercond. Sci. Technol. {\bf 1}, 36 (1988).

\bibitem{tsu95}
H. Tsunetsugu, M. Troyer, and T. M. Rice,
Phys. Rev. B {\bf 51}, 16456 (1995).

\bibitem{kwon}
K.~Park and S.~Sachdev,
Phys. Rev. B {\bf 64}, 184510 (2001).


\bibitem{martin01}
I. Martin, G. Ortiz, A. V. Balatsky, and A. R. Bishop,
Europhys. Lett. {\bf 56}, 849 (2001).

\bibitem{poil07}
M. Raczkowski, M. Capello, D. Poilblanc, R. Fr\'esard, and A. M. Ole\'s,
Phys. Rev. B {\bf 76}, 140505(R) (2007).

\bibitem{raczk08}
M. Capello, M. Raczkowski, and D. Poilblanc,
Phys. Rev. B {\bf 77}, 224502 (2008).

\bibitem{rice08}
K.-Y. Yang, W.-Q. Chen, T. M. Rice, M. Sigrist, and F.-C. Zhang,
preprint arXiv:0807.3789.

\bibitem{anisimov04}
V. I. Anisimov, M. A. Korotin, A. S. Mylnikova, A. V. Kozhevnikov, D. M. Korotin, and J. Lorenzana,
Phys. Rev. B {\bf 70}, 172501 (2004).

\bibitem{dmrg00}
S. R. White and D. J. Scalapino,
Phys. Rev. B {\bf 61}, 6320 (2000).

\bibitem{pdw1}
H.-D. Chen, O. Vafek, A. Yazdani, and S.-C. Zhang,
Phys. Rev. Lett. {\bf 93}, 187002 (2004).


\bibitem{dmrg98}
S. R. White and D. J. Scalapino,
Phys. Rev. Lett. {\bf 80}, 1272 (1998);
{\em ibid} {\bf 81}, 3227 (1998).

\bibitem{dmrg99}
S. R. White and D. J. Scalapino,
Phys. Rev. B {\bf 60}, R753 (1999).

\bibitem{hellberg99}
C. S. Hellberg and E. Manousakis,
Phys. Rev. Lett. {\bf 83}, 132 (1999);
{\em ibid.} {\bf 84}, 3022 (2000);
S. R. White and D. J. Scalapino,
{\em ibid.} {\bf 84}, 3021 (2000).

\bibitem{dmrg04}
S. R. White and D. J. Scalapino,
Phys. Rev. B {\bf 70}, 220506(R) (2004).

\bibitem{dmrg01}
D. J. Scalapino and S. R. White,
Found. Phys. {\bf 31}, 27 (2001).

\bibitem{jeckel}
G. Hager, G. Wellein, E. Jeckelmann, and H. Fehske,
Phys. Rev. B {\bf 71}, 075108 (2005).



\bibitem{dca_review}
T. Maier, M. Jarrell, T. Pruschke, and M. H. Hettler,
Rev. Mod. Phys. {\bf 77}, 1027 (2005).

\bibitem{metzner89}
W. Metzner and D.\ Vollhardt,
Phys. Rev. Lett. {\bf 62}, 324 (1989).

\bibitem{georges96}
A. Georges, G. Kotliar, W. Krauth, and M. J. Rozenberg,
Rev. Mod. Phys. {\bf 68}, 13 (1996).

\bibitem{fleck00}
M. Fleck, A. I. Lichtenstein, E. Pavarini, and A. M. Ole\'s,
Phys. Rev. Lett. {\bf 84}, 4962 (2000).

\bibitem{maier05}
T. A. Maier, M. Jarrell, T. C. Schulthess, P. R. C. Kent, and J. B. White,
Phys. Rev. Lett. {\bf 95}, 237001 (2005).

\bibitem{macridin06}
A. Macridin, M. Jarrell, T. Maier, P. R. C. Kent, and E. D'Azevedo,
Phys. Rev. Lett. {\bf 97}, 036401 (2006).

\bibitem{stanescu06}
T. D. Stanescu, M. Civelli, K. Haule, and G. Kotliar,
Ann. Physics {\bf 321}, 1682 (2006).

\bibitem{jarrell08a}
A. Macridin and M. Jarrell,
Phys. Rev. B {\bf 78}, 241101(R) (2008).

\bibitem{jarrell08b}
N. S. Vidhyadhiraja, A. Macridin, C. Sen, M. Jarrell, and M. Ma,
preprint arXiv:0809.1477.

\bibitem{haule07}
K. Haule and G. Kotliar,
Phys. Rev. B {\bf 76}, 092503 (2007).

\bibitem{civelli08}
M. Civelli, M. Capone, A. Georges, K. Haule, O. Parcollet, T. D. Stanescu,
and G. Kotliar,
Phys. Rev. Lett. {\bf 100}, 046402 (2008).

\bibitem{kent08}
P. R. C. Kent, T. Saha-Dasgupta, O. Jepsen, O. K. Andersen, A. Macridin, T. A. Maier,
M. Jarrell, and T. C. Schulthess,
Phys. Rev. B {\bf 78}, 035132 (2008).

\bibitem{domenge08}
J.-C. Domenge and G. Kotliar,
preprint arXiv:0808.2328.

\bibitem{pott03}
M. Potthoff, Eur. Phys. J B {\bf 32}, 429 (2003).

\bibitem{vca1}
M. Potthoff, M. Aichhorn, and C. Dahnken,
Phys. Rev. Lett. {\bf 91}, 206402 (2003).

\bibitem{vca2}
M. Aichhorn, E. Arrigoni, M. Potthoff, and W. Hanke,
Phys. Rev. B {\bf 74}, 024508 (2006).

\bibitem{vca3}
M. Aichhorn, E. Arrigoni, M. Potthoff, and W. Hanke,
Phys. Rev. B {\bf 74}, 235117 (2006).

\bibitem{brehm08}
S. Brehm, E. Arrigoni, M. Aichhorn, and W. Hanke,
preprint arXiv:0811.0552.


\bibitem{honer_rev}
C. Honerkamp,
Int. J. Mod. Phys. B {\bf 20}, 2636 (2006).

\bibitem{frg_ord}
M. Salmhofer, C. Honerkamp, W. Metzner, and O. Lauscher,
Prog. Theor. Phys. {\bf 112}, 943 (2004).

\bibitem{HM00}
C. J. Halboth and W. Metzner,
Phys. Rev. Lett. {\bf 85}, 5162 (2000).

\bibitem{GKW02}
I. Grote, E. K\"ording, and F. Wegner,
J. Low Temp. Phys. {\bf 126}, 1385 (2002).

\bibitem{honer01a}
C. Honerkamp,
Eur. Phys. J. B {\bf 21}, 81 (2001).

\bibitem{honer01b}
C. Honerkamp, M. Salmhofer, N. Furukawa and T. M. Rice,
Phys. Rev. B {\bf 63}, 035109 (2001).

\bibitem{metzner05}
W. Metzner, J. Reiss, and D. Rohe,
Phys. stat. sol. (b) {\bf 243}, 46 (2006).

\bibitem{ossadnik08}
M. Ossadnik, C. Honerkamp, T. M. Rice, and M. Sigrist,
Phys. Rev. Lett. {\bf 101}, 256405 (2008).

\bibitem{abdel}
M. Abdel-Jawad {\em et al.},
Nature Phys. {\bf 2}, 821 (2006).

\bibitem{fu05}
H. C. Fu, C. Honerkamp, and D.-H. Lee,
Europhys. Lett. {\bf 75}, 146 (2006).


\bibitem{pryadko}
L.~P.~Pryadko, S.~A.~Kivelson, V.~J.~Emery, Y.~B.~Bazaliy, and E.~A.~Demler,
Phys. Rev. B {\bf 60}, 7541 (1999).

\bibitem{chayes96}
L. Chayes, V. J. Emery, S. A. Kivelson, Z. Nussinov, and G. Tarjus,
Physica A {\bf 225}, 129 (1996).

\bibitem{so5stripe}
M.~Veillette, Y.~B.~Bazaliy, A.~J.~Berlinsky, and C.~Kallin,
Phys. Rev. Lett. {\bf 83}, 2413 (1999).

\bibitem{zaanen96}
J. Zaanen, M. L. Horbach, and W. van Saarloos,
Phys. Rev. B {\bf 53}, 8671 (1996).

\bibitem{hassel98}
C. Morais Smith, Y. A. Dimashko, N. Hasselmann, and A. O. Caldeira,
Phys. Rev. B {\bf 58}, 453 (1998).

\bibitem{hassel99}
N. Hasselmann, A. H. Castro Neto, C. Morais Smith, and Y. Dimashko,
Phys. Rev. Lett. {\bf 82}, 2135 (1999).

\bibitem{cherny02}
A. L. Chernyshev, S. R. White, and A. H. Castro Neto,
Phys. Rev. B {\bf 65}, 214527 (2002).

\bibitem{jpsj}
S. Sachdev and M. Vojta,
J. Phys. Soc. Jpn. {\bf 69}, Suppl. B, 1 (2000).

\bibitem{rs0}
N.~Read and S.~Sachdev,
Phys. Rev. Lett. {\bf 62}, 1694 (1989);
Phys. Rev. B {\bf 42}, 4568 (1990).


\bibitem{lavrov01}
A. N. Lavrov, Y. Ando, S. Komiya, and I. Tsukada,
Phys. Rev. Lett. {\bf 87}, 017007 (2001).



\bibitem{YK00}
H. Yamase and H. Kohno,
J. Phys. Soc. Jpn. {\bf 69}, 332 and 2151 (2000).

\bibitem{yamase06}
H. Yamase and W. Metzner,
Phys. Rev. B {\bf 73}, 214517 (2006).

\bibitem{miyanaga06}
A. Miyanaga and H. Yamase,
Phys. Rev. B {\bf 73}, 174513 (2006).

\bibitem{gros06}
B. Edegger, V. N. Muthukumar, and C. Gros,
Phys. Rev. B {\bf 74}, 165109 (2006).

\bibitem{nema_emery}
S. A. Kivelson, E. Fradkin, and T. H. Geballe,
Phys. Rev. B {\bf 69}, 144505 (2004).

\bibitem{NM03}
A. Neumayr and W. Metzner,
Phys. Rev. B {\bf 67}, 035112 (2003).

\bibitem{MRA03}
W. Metzner, D. Rohe, and S. Andergassen,
Phys. Rev. Lett. {\bf 91}, 066402 (2003).

\bibitem{OKF01}
V. Oganesyan, S. A. Kivelson, and E. Fradkin,
Phys. Rev. B {\bf 64}, 195109 (2001).



\bibitem{KKC03}
H.-Y. Kee, E. H. Kim, and C.-H. Chung,
Phys. Rev. B {\bf 68}, 245109 (2003).

\bibitem{KCOK04}
I. Khavkine, C.-H. Chung, V. Oganesyan, and H.-Y. Kee,
Phys. Rev. B {\bf 70}, 155110 (2004).

\bibitem{YOM05}
H. Yamase, V. Oganesyan, and W. Metzner,
Phys. Rev. B {\bf 72}, 035114 (2005).

\bibitem{schofield08}
J. Quintanilla, M. Haque, and A. J. Schofield,
Phys. Rev. B {\bf 78}, 035131 (2008).


\bibitem{orgad01}
D. Orgad, S. A. Kivelson, E. W. Carlson, V. J. Emery, X. J. Zhou, and Z. X. Shen
Phys. Rev. Lett. {\bf 86}, 4362 (2001).

\bibitem{castellani95}
C.~Castellani, C.~Di Castro, and M.~Grilli,
Phys. Rev. Lett. {\bf 75}, 4650 (1995).

\bibitem{caprara99}
S.~Caprara, M. Sulpizi, A. Bianconi, C. Di Castro, and M. Grilli,
Phys. Rev. B {\bf 59}, 14980 (1999).

\bibitem{zachar00}
O. Zachar,
Phys. Rev. B {\bf 62}, 13836 (2000).


\bibitem{hubb_af}
See e.g.:
F. Gebhardt,
{\em The Mott Metal-Insulator Transition--Models and Methods},
Springer, Heidelberg (1997).

\bibitem{flst}
T. Senthil, S. Sachdev, and M. Vojta,
Phys. Rev. Lett. {\bf 90}, 216403 (2003).

\bibitem{mvss06}
M. Vojta and S. Sachdev,
J. Phys. Chem. Solids. {\bf 67}, 11 (2006).

\bibitem{batista01}
C. D. Batista, G. Ortiz, and A. V. Balatsky,
Phys. Rev. B {\bf 64}, 172508 (2001).

\bibitem{scheidl03}
F. Kr\"uger and S. Scheidl,
Phys. Rev. B {\bf 67}, 134512 (2003).

\bibitem{mvtu_unpub}
M. Vojta and T. Ulbricht,
unpublished.

\bibitem{kaneshita01}
E. Kaneshita, M. Ichioka, and K. Machida,
J. Phys. Soc. Jpn. {\bf 70}, 866 (2001).

\bibitem{andersen05}
B. M. Andersen and P. Hedegard,
Phys. Rev. Lett. {\bf 95}, 037002 (2005).

\bibitem{carlson08}
D. X. Yao and E. W. Carlson,
Phys. Rev. B {\bf 77}, 024503 (2008).



\bibitem{brilee99}
J. Brinckmann and P. A. Lee,
Phys. Rev. Lett. {\bf 82}, 2915 (1999).

\bibitem{rpa1}
D. Manske, I. Eremin, and K. H. Bennemann,
Phys. Rev. B {\bf 63}, 054517 (2001).


\bibitem{eremin05b}
I. Eremin, D. K. Morr, A. V. Chubukov, K. H. Bennemann, and M. R. Norman,
Phys. Rev. Lett. {\bf 94}, 147001 (2005).

\bibitem{eremin05a}
I. Eremin and D. Manske,
Phys. Rev. Lett. {\bf 94}, 067006 (2005);
{\bf 96}, 059902(E) (2006);
{\bf 98}, 139702 (2007); 
H.-J. Zhao and J. Li,
{\em ibid.} {\bf 98}, 139701 (2007). 


\bibitem{schnyder06}
A. P. Schnyder, D. Manske, C. Mudry, and M. Sigrist,
Phys. Rev. B {\bf 73}, 224523 (2006).

\bibitem{krueger07}
F. Kr\"uger, S. D. Wilson, L. Shan, S. Li, Y. Huang, H.- H. Wen, S.-C. Zhang, P. Dai, and J.Zaanen,
Phys. Rev. B {\bf 76}, 094506 (2007).



\bibitem{zhang97}
S.-C. Zhang, Science {\bf 275}, 1089 (1997).



\bibitem{salkola96}
M. I. Salkola, V. J. Emery, and S. A. Kivelson,
Phys. Rev. Lett. {\bf 77}, 155 (1996).

\bibitem{seibold00}
G. Seibold, F. Becca, F. Bucci, C. Castellani, C. Di Castro, and M. Grilli,
Eur. Phys. J. B {\bf 13}, 87 (2000).

\bibitem{granath02}
M. Granath, V. Oganesyan, D. Orgad, and S. A. Kivelson,
Phys. Rev. B {\bf 65}, 184501 (2002).

\bibitem{granath04}
M. Granath, Phys. Rev. B {\bf 69}, 214433 (2004).

\bibitem{granath08}
M. Granath, Phys. Rev. B {\bf 77}, 165128 (2008).

\bibitem{orgad08}
S. Baruch and D. Orgad,
Phys. Rev. B {\bf 77}, 174502 (2008).

\bibitem{wollny08}
A. Wollny and M. Vojta,
preprint arXiv:0808.1163.

\bibitem{zacher00}
M. G. Zacher, R. Eder, E. Arrigoni, and W. Hanke,
Phys. Rev. Lett. {\bf 85}, 2585 (2000).

\bibitem{seibold01}
G. Seibold and M. Grilli,
Phys. Rev. B {\bf 63}, 224505 (2001).

\bibitem{wrobel06}
P. Wrobel, A. Maciag, and R. Eder,
J. Phys. Cond. Matter {\bf 18}, 9749 (2006).

\bibitem{eder04}
R. Eder and Y. Ohta,
Phys. Rev. B {\bf 69}, 100502(R) (2004).

\bibitem{tohyama99}
T. Tohyama, S. Nagai, Y. Shibata, and S. Maekawa,
Phys. Rev. Lett. {\bf 82}, 4910 (1999).


\bibitem{filter1}
J.~X.~Zhu, C.~S.~Ting, and C.~R.~Hu,
Phys. Rev. B {\bf 62}, 6027 (2000).

\bibitem{filter2}
I.~Martin, A.~V.~Balatsky, and J.~Zaanen,
Phys. Rev. Lett. {\bf 88}, 097003 (2002).

\bibitem{filter_lda}
L.-L. Wang, P. J. Hirschfeld, and H.-P. Cheng,
Phys. Rev. B {\bf 72}, 224516 (2005).

\bibitem{mcelroy05b}
K. McElroy, J. Lee, J. A. Slezak, D.-H. Lee, H. Eisaki, S. Uchida, and J. C. Davis,
Science {\bf 309}, 1048 (2005).

\bibitem{qpith1}
Q.-H. Wang and D.-H. Lee, Phys. Rev. B {\bf 67}, 020511(R) (2003).

\bibitem{qpith2}
D. Zhang and C. S. Ting, Phys. Rev. B {\bf 67}, 100506(R) (2003).

\bibitem{qpith3}
B. M. Andersen and P. Hedegard, Phys. Rev. B {\bf 67}, 172505 (2003).

\bibitem{qpith4}
L. Zhu, W. A. Atkinson, and P. J. Hirschfeld,
Phys. Rev. B {\bf 69}, 060503 (2004).



\bibitem{nunner06}
T. S. Nunner, W. Chen, B. M. Andersen, A. Melikyan, and P. J. Hirschfeld,
Phys. Rev. B {\bf 73}, 104511 (2006).

\bibitem{hirschfeld08}
B. M. Andersen and P. J. Hirschfeld,
preprint arXiv:0811.3751.

\bibitem{lawler08}
E.-A. Kim and M. J. Lawler,
preprint arXiv:0811.2242.

\bibitem{machida02}
M. Ichioka and K. Machida,
J. Phys. Soc. Jpn. {\bf 71}, 1836 (2002).

\bibitem{zhu02}
J.-X. Zhu, I. Martin, and A. R. Bishop,
Phys. Rev. Lett. {\bf 89}, 067003 (2002).

\bibitem{ting02}
Y. Chen and C. S. Ting,
Phys. Rev. B {\bf 65}, 180513(R) (2002).

\bibitem{chen02}
Y. Chen, H. Y. Chen, and C. S. Ting,
Phys. Rev. B {\bf 66}, 104501 (2002).


\bibitem{scz02}
H.-D. Chen, J.-P. Hu, S. Capponi, E. Arrigoni, and S.-C. Zhang,
Phys. Rev. Lett. {\bf 89}, 137004 (2002).

\bibitem{franz02}
M. Franz, D. E. Sheehy, and Z. Tesanovic,
Phys. Rev. Lett. {\bf 88}, 257005 (2002).

\bibitem{tolya02}
A. Polkovnikov, M. Vojta, and S. Sachdev,
Phys. Rev. B {\bf 65}, 220509(R) (2002).

\bibitem{tami08}
T. Pereg-Barnea and M. Franz,
Phys. Rev. B {\bf 78}, 020509 (2008).


\bibitem{demler1}
E.~Demler, S.~Sachdev, and Y.~Zhang,
Phys. Rev. Lett. {\bf 87}, 067202 (2001).

\bibitem{demler2}
Y.~Zhang, E.~Demler, and S.~Sachdev,
Phys. Rev. B {\bf 66}, 094501 (2002).

\bibitem{kiv_field}
S. A. Kivelson, D.-H. Lee, E. Fradkin, and V. Oganesyan,
Phys. Rev. B {\bf 66}, 144516 (2002).


\bibitem{tsvelik07}
A. V. Chubukov and A. M. Tsvelik,
Phys. Rev. B {\bf 76}, 100509 (2007).

\bibitem{so5vortex}
D. P. Arovas, A. J. Berlinsky, C. Kallin, and S.-C. Zhang,
Phys. Rev. Lett. {\bf 79}, 2871 (1997).



\bibitem{doiron07}
N. Doiron-Leyraud, C. Proust, D. LeBoeuf, J. Levallois, J.-B. Bonnemaison,
R. Liang, D. A. Bonn, W. N. Hardy, and L. Taillefer,
Nature {\bf 447}, 565 (2007).

\bibitem{yelland08}
E. A. Yelland, J. Singleton, C. H. Mielke, N. Harrison, F. F. Balakirev, B. Dabrowski, and J. R. Cooper,
Phys. Rev. Lett. {\bf 100}, 047003 (2008).

\bibitem{sebastian08}
S. E. Sebastian, N. Harrison, E. Palm, T. P. Murphy, C. H. Mielke,
R. Liang, D. A. Bonn, W. N. Hardy, and G. G. Lonzarich,
Nature {\bf 454}, 200 (2008).



\bibitem{millis_pockets}
A. J. Millis and M. R. Norman,
Phys. Rev. B {\bf 76}, 220503(R) (2007).

\bibitem{stephen92}
M. J. Stephen,
Phys. Rev. B {\bf 45}, 5481 (1992).

\bibitem{palee08}
K.-T. Chen and P. A. Lee,
preprint arXiv:0812.3351.


\bibitem{hackl_rmp}
T. P. Devereaux and R. Hackl,
Rev. Mod. Phys. {\bf 79}, 175 (2007).

\bibitem{blumberg02}
G. Blumberg, A. Koitzsch, A. Gozar, B. S. Dennis, C. A. Kendziora, P. Fournier, and R. L. Greene,
Phys. Rev. Lett. {\bf 88}, 107002 (2002).

\bibitem{hackl02}
F. Venturini, Q.-M. Zhang, R. Hackl, A. Lucarelli, S. Lupi, M. Ortolani,
P. Calvani, N. Kikugawa, and T. Fujita,
Phys. Rev. B {\bf 66}, 060502(R) (2002).

\bibitem{hackl05}
L. Tassini, F. Venturini, Q.-M. Zhang, R. Hackl, N. Kikugawa, and T. Fujita,
Phys. Rev. Lett. {\bf 95}, 117002 (2005).


\bibitem{caprara05}
S. Caprara, C. Di Castro, M. Grilli, and D. Suppa,
Phys. Rev. Lett. {\bf 95}, 117004 (2005).


\bibitem{ino97}
A. Ino, T. Mizokawa, A. Fujimori, K. Tamasaku, H. Eisaki, S. Uchida, T. Kimura, T. Sasagawa, and K. Kishio,
Phys. Rev. Lett. {\bf 79}, 2101 (1997).

\bibitem{harima01}
N. Harima, J. Matsuno, A. Fujimori, Y. Onose, Y. Taguchi, and Y. Tokura,
Phys. Rev. B {\bf 64}, 220507(R) (2001).

\bibitem{yagi06}
H. Yagi, T. Yoshida, A. Fujimori, Y. Kohsaka, M. Misawa, T. Sasagawa, H. Takagi, M. Azuma and M. Takano,
Phys. Rev. B {\bf 73}, 172503 (2006).

\bibitem{tohyama03}
T. Tohyama and S. Maekawa,
Phys. Rev. B {\bf 67}, 092509 (2003).


\bibitem{cvet07}
V. Cvetkovic, Z. Nussinov, S. Mukhin, and J. Zaanen,
Europhys. Lett. {\bf 81}, 27001 (2007).


\bibitem{yamase08}
H. Yamase,
Phys. Rev. B {\bf 79}, 052501 (2009).

\bibitem{yamase01}
H. Yamase and H. Kohno,
J. Phys. Soc. Jpn. {\bf 70}, 2733 (2001).

\bibitem{yamase07}
H. Yamase,
Phys. Rev. B {\bf 75}, 014514 (2007).



\bibitem{ddw_6vertex}
S. Chakravarty,
Phys. Rev. B {\bf 66}, 224505 (2002).

\bibitem{varma06}
C. M. Varma,
Phys. Rev. B {\bf 73}, 155113 (2006).

\bibitem{varma07a}
C. M. Varma and L. Zhu,
Phys. Rev. Lett. {\bf 98}, 177004 (2007).

\bibitem{varma07b}
V. Aji, and C. M. Varma,
Phys. Rev. B {\bf 75}, 224511 (2007).

\bibitem{varma07c}
L. Zhu, V. Aji, A. Shekhter, and C. M. Varma,
Phys. Rev. Lett. {\bf 100}, 057001 (2008).

\bibitem{varma08}
V. Aji, A. Shekhter, and C. M. Varma,
preprint arXiv:0807.3741.

\bibitem{sudbo08a}
K. Borkje and A. Sudbo,
Phys. Rev. B {\bf 77}, 092404 (2008).

\bibitem{sudbo08b}
M. S. Gronsleth, T. B. Nilssen, E. K. Dahl, C. M. Varma, and A. Sudbo,
Phys. Rev. B {\bf 79}, 094506 (2009).

\bibitem{mfl89}
C. M. Varma, P. B. Littlewood, S. Schmitt-Rink, E. Abrahams, and A. E. Ruckenstein,
Phys. Rev. Lett. {\bf 63}, 1996 (1989).

\bibitem{mfl_rev}
C. M. Varma, Z. Nussinov, and W. van Saarloos,
Phys. Rep. {\bf 361}, 267 (2002).


\bibitem{greiter07}
M. Greiter and R. Thomale,
Phys. Rev. Lett. {\bf 99}, 027005 (2007).

\bibitem{weber09}
C. Weber, A. L\"auchli, F. Mila, and T. Giamarchi,
Phys. Rev. Lett. {\bf 102}, 017005 (2009).

\bibitem{jia08}
X. Jia, P. Goswami, and S. Chakravarty,
preprint arXiv:0811.1056.

\bibitem{incommddw}
S. Chakravarty and H.-Y. Kee,
Proc. Natl. Acad. Sci. USA {\bf 105}, 8835 (2008).


\bibitem{np_feat}
J. Zaanen {\em et al.},
Nature Phys. {\bf 2}, 138 (2006).

\bibitem{mukuda08}
H. Mukuda, Y. Yamaguchi, S. Shimizu, Y. Kitaoka, P. M. Shirage, and A. Iyo,
J. Phys. Soc. Jpn. {\bf 77}, 124706 (2008).

\bibitem{wilson06}
S. D. Wilson, S. Li, H. Woo, P. Dai, H. A. Mook, C. D. Frost, S. Komiya, and Y. Ando,
Phys. Rev. Lett. {\bf 96}, 157001 (2006).

\bibitem{motoyama07}
E. M. Motoyama, G. Yu, I. M. Vishik, O. P. Vajk, P. K. Mang, and M. Greven,
Nature {\bf 445}, 186 (2007).

\bibitem{fujita08}
M. Fujita, M. Matsuda, S.-H. Lee, M. Nakagawa, and K. Yamada
Phys. Rev. Lett. {\bf 101}, 107003 (2008).


\bibitem{ssbook}
S.~Sachdev, {\it Quantum Phase Transitions},
Cambridge University Press, Cambridge (1999).

\bibitem{gurvitch87}
M. Gurvitch and A. T. Fiory,
Phys. Rev. Lett. {\bf 59}, 1337 (1987).

\bibitem{takagi92}
H. Takagi, B. Batlogg, H. L. Kao, J. Kwo, R. J. Cava, J. J. Krajewski, and W. F. Peck,Jr.,
Phys. Rev. Lett. {\bf 69}, 2975 (1992).

\bibitem{daou08}
R. Daou {\em et al.}, 
Nature Phys. {\bf 5}, 31 (2009).

\bibitem{hussey09}
R. A. Cooper {\em et al.}, 
Science {\bf 323}, 603 (2009).

\bibitem{lawler07}
M. J. Lawler and E. Fradkin,
Phys. Rev. B {\bf 75}, 033304 (2007).

\bibitem{hertz76}
J. A. Hertz, Phys. Rev. B {\bf 14}, 1165 (1976).

\bibitem{millis93}
A. J. Millis, Phys. Rev. B {\bf 48}, 7183 (1993).

\bibitem{hvl_rmp}
H. v. L\"ohneysen, A. Rosch, M. Vojta, and P. W\"olfle,
Rev. Mod. Phys. {\bf 79}, 1015 (2007).

\bibitem{DM06}
L. Dell'Anna and  W. Metzner,
Phys. Rev. B {\bf 73}, 045127 (2006).

\bibitem{DM07}
L. Dell'Anna and W. Metzner,
Phys. Rev. Lett. {\bf 98}, 136402 (2007).

\bibitem{garst09}
M. Zacharias, M. Garst, A. Rosch, and P. W\"olfle,
to be published.

\bibitem{vzs00c}
M. Vojta, Y. Zhang, and S. Sachdev,
Int. J. Mod. Phys. B {\bf 14}, 3719 (2000).

\bibitem{kivss08}
E.-A. Kim, M. J. Lawler, P. Oreto, S. Sachdev, E. Fradkin, and S. A. Kivelson,
Phys. Rev. B {\bf 77}, 184514 (2008).

\bibitem{huhss08}
Y. Huh and S. Sachdev,
Phys. Rev. B {\bf 78}, 064512 (2008).


\bibitem{abanov04}
A. Abanov and A. V. Chubukov,
Phys. Rev. Lett. {\bf 93}, 255702 (2004).

\bibitem{rosch00}
A. Rosch,
Phys. Rev. B {\bf 62}, 4945 (2000).

\bibitem{kontani99}
H. Kontani, K. Kanki, and K. Ueda,
Phys. Rev. B {\bf 59}, 14723 (1999).


\bibitem{castellani96}
A. Perali, C. Castellani, C. Di Castro, and M. Grilli,
Phys. Rev. B {\bf 54}, 16216 (1996).

\bibitem{castellani97}
C.~Castellani, C.~Di Castro, and M.~Grilli,
Z. Phys. B {\bf 103}, 137 (1997).

\bibitem{castellani98}
C. Castellani, C. Di Castro, and M. Grilli,
J. Phys. Chem. Solids {\bf 59}, 1694 (1998).

\bibitem{benfatto00}
L. Benfatto, S. Caprara, and C. Di Castro,
Eur. Phys. J. B, {\bf 17}, 95 (2000).


\bibitem{ssafm08}
A. Pelissetto, S. Sachdev, and E. Vicari,
Phys. Rev. Lett. {\bf 101}, 027005 (2008).

\bibitem{sun08}
K. Sun, B. M. Fregoso, M. J. Lawler, and E. Fradkin,
Phys. Rev. B {\bf 78}, 085124 (2008).

\bibitem{zaanen01}
J. Zaanen, O. Y. Osman, H. V. Kruis, Z. Nussinov, and J. Tworzydlo,
Phil. Mag. B {\bf 81}, 1485 (2001).

\bibitem{strain_qcp}
A. Bianconi, G. Bianconi, S. Caprara, C. Di Castro, H. Oyanagi, and N. L. Saini,
J. Phys.: Condens. Matter {\bf 12}, 10655 (2000).

\bibitem{zaanen04}
J. Zaanen and B. Hosseinkhani,
Phys. Rev. B {\bf 70}, 060509 (2004).

\bibitem{chakravarty05}
A. Kopp and S. Chakravarty,
Nature Phys. {\bf 1}, 53 (2005).

\bibitem{gauge_lin}
P. A. Lee and N. Nagaosa,
Phys. Rev. B {\bf 46}, 5621 (1992).

\bibitem{page99}
O. Parcollet and A. Georges,
Phys. Rev. B {\bf 59}, 5341 (1999).



\bibitem{phillips}
P. Phillips,
Ann. Phys. {\bf 321}, 1634 (2006).

\bibitem{preformed}
V. J. Emery and S. A. Kivelson, Nature {\bf 374}, 434 (1995),
Phys. Rev. Lett. {\bf 74}, 3253 (1995).

\bibitem{nernst_exp1}
Y. Wang, L. Li, and N. P. Ong,
Phys. Rev. B {\bf 73}, 024510 (2006).

\bibitem{nernst_exp2}
K. Behnia,
preprint arXiv:0810.3887.

\bibitem{kanigel08}
A. Kanigel, U. Chatterjee, M. Randeria, M. R. Norman, G. Koren, K. Kadowaki, and J. C. Campuzano,
Phys. Rev. Lett. {\bf 101}, 137002 (2008).

\bibitem{yazdani07}
K. K. Gomes, A. N. Pasupathy, A. Pushp, S. Ono, Y. Ando, and A. Yazdani,
Nature {\bf 447}, 569 (2007).

\bibitem{nernst_theory}
I. Ussishkin, S. L. Sondhi and D. A. Huse,
Phys. Rev. Lett. {\bf 89}, 287001 (2002).

\bibitem{nernst_diamag}
Y. Wang, . Li, M. J. Naughton, G. D. Gu, S. Uchida, and N. P. Ong,
Phys. Rev. Lett. {\bf 95}, 247002 (2005).


\bibitem{miyake86}
K. Miyake, S. Schmitt-Rink, and C. M. Varma,
Phys. Rev. B {\bf 34}, 6554 (1986).

\bibitem{scalapino86}
D. J. Scalapino, E. Loh, Jr., and J. E. Hirsch,
Phys. Rev. B {\bf 34}, 8190 (1986),
{\em ibid.} {\bf 35}, 6694 (1987).

\bibitem{antonio01}
A. H. Castro Neto,
Phys. Rev. B {\bf 64}, 104509 (2001).

\bibitem{inhom_sc03}
E. Arrigoni and S. A. Kivelson,
Phys. Rev. B {\bf 68}, 180503 (2003).

\bibitem{inhom_sc05}
I. Martin, D. Podolsky, and S. A. Kivelson,
Phys. Rev. B {\bf 72}, 060502 (2005).


\bibitem{inhom_sc08}
W.-F. Tsai, H. Yao, A. L\"auchli, and S. A. Kivelson,
Phys. Rev. B {\bf 77}, 214502 (2008).




\end{thebibliography}
\end{document}